\documentclass[12pt]{article}
\usepackage{graphicx,latexsym}
\newcommand{\bm}[1]{\mbox{\boldmath $#1$}}

\newcommand{\be}{\begin{equation}}
\newcommand{\ee}{\end{equation}}

\newcommand{\one}{\hbox{l\kern-2mm 1}}

\newcommand{\ba}{\begin{array}}
\newcommand{\ea}{\end{array}}

\def\be{\begin{equation}}
\def\ee{\end{equation}}
\def\bea{\begin{eqnarray}}
\def\eea{\end{eqnarray}}

\begin{document}
\begin{center}
{\Large{ Supersymmetry and the MSSM: An Elementary  Introduction }}

  Notes of Lectures for Graduate Students 
in Particle Physics \\ Oxford, 2004  \& 2005

\vspace{.2in}
Ian J. R. Aitchison\\
  Oxford University\\
  Department of Physics, 
 The Rudolf Peierls Centre for Theoretical Physics, \\
  1 Keble Road, Oxford OX1 3NP, UK
\end{center}

\begin{abstract}
These  notes are an  expanded version of a short course of lectures given for 
graduate students in particle physics at Oxford. The level was intended  
to be appropriate  for students in both experimental and theoretical particle physics.
The purpose is  to present an elementary  and 
self-contained  introduction to SUSY that follows on, 
relatively straightforwardly, from  
graduate-level courses in relativistic quantum mechanics  
and introductory quantum field theory. The notation adopted, at least initially,   
 is one widely used in RQM courses, rather than the `spinor calculus' 
(dotted and undotted indices) notation found in most SUSY sources, though the latter 
is introduced  in optional Asides. There is also a strong preference for a 
`do-it-yourself' constructive approach, rather than for a top-down formal deductive treatment. 
The main goal is to provide a practical understanding of how the softly broken MSSM is constructed.  
Relatively less space is devoted to phenomenology, though simple `classic' results are 
covered, including gauge unification,  the bound on the mass of the lightest Higgs boson, and sparticle mixing.  
By the end of the course students (readers) should be provided with access 
to the contemporary phenomenological literature.

\end{abstract}
\newpage 

\tableofcontents
 
 \vspace{.15in}
 
 \noindent {\bf References}

\section{Introduction and Motivation}

Supersymmetry (SUSY) - a symmetry relating bosonic and fermionic degrees of 
freedom - is a remarkable and exciting idea,  but  its implementation is technically 
pretty complicated. It can be discouraging to find that after 
standard courses on, say, the Dirac equation and quantum field theory, 
one has almost to start afresh and master a new formalism, and moreover 
one that is not fully standardized. On the other hand,  
thirty years have passed since the first 
explorations of SUSY in the early 1970's, without any direct evidence 
of its relevance to physics having been discovered. The Standard Model 
(SM) of particle physics (suitably extended to include an adequate 
neutrino phenomenology) works extremely well. So the hard-nosed seeker after 
truth may well wonder: Why spend the time learning all this intricate SUSY 
stuff? Indeed, why speculate at all about how to go `beyond' the SM, unless 
or until experiment forces us to? If it's not broken, why try and fix it?

As regards the formalism, most standard sources on SUSY  use either   
 the `dotted and undotted' 2-component spinor notation 
found in the theory of representations of the Lorentz group, or 4-component 
Majorana spinors. Neither of these is commonly included in introductory 
courses on the Dirac equation (though perhaps they should be). But it is 
of course perfectly possible to present simple aspects of SUSY using a 
notation which joins smoothly on to standard 4-component Dirac equation 
courses, and a brute force, `try-it-and-see' approach to constructing 
SUSY-invariant theories.  That is what  I aim to  do in these lectures, 
at least to start with. 
Somewhat surprisingly, it seems that such an elementary introduction is not 
available, or at least not in such detail as is given here, which is 
why these notes have been typed up. I hope that they will help to make  
 the basic nuts and bolts of SUSY   
accessible to a wider clientele. However, as we go along I shall explain 
the more compact `dotted and undotted' notation in optional Asides, 
and I'll also introduce the powerful superfield formalism; this is partly 
because the simple-minded approach becomes too cumbersome after a while, 
and partly because  contemporary discussions of the phenomenology of 
the Minimal Supersymmetric Standard Model (MSSM) make some use this more 
sophisticated notation. 

What of the need to go beyond the Standard Model?   Within the 
SM itself, there is a plausible historical answer to that question. The 
V-A current-current (four-fermion) theory of weak interactions worked 
very well for many years, when used at lowest order in perturbation theory. 
Yet Heisenberg \cite{heis} had noted as early as 1939 that problems arose if one tried to 
compute higher order effects, perturbation theory apparently breaking down 
completely at the then unimaginably high energy of some 300 GeV (the scale 
of $G_{\rm F}^{-1/2}$). Later, this became linked to the non-renormalizability 
of the four-fermion theory, a purely theoretical problem in the years 
before experiments attained the precision required for sensitivity to 
electroweak radiative corrections. This perceived disease was alleviated 
but not cured in the `Intermediate Vector Boson' model, which envisaged 
the weak force between two fermions as being mediated by massive vector bosons. 
The non-renormalizability of such a theory was recognized, but not addressed, 
 by Glashow \cite{glashow} in his 
1961 paper proposing the SU(2)$\times$U(1) structure. Weinberg \cite{wein} 
and Salam \cite{salam}, in 
their   gauge-theory models, employed the hypothesis of 
spontaneous symmetry breaking to generate masses for the gauge bosons and 
the fermions, conjecturing 
that this form of  symmetry breaking would not spoil the renormalizability possessed by the 
massless (unbroken) theory. 
When  't Hooft \cite{thooft} demonstrated this in 1971,    the Glashow-Salam-Weinberg theory 
achieved a theoretical status comparable to that of QED. In due course the precision 
electroweak experiments spectacularly confirmed the calculated radiative corrections, 
even yielding  a remarkably accurate prediction of the top quark mass, 
based on its effect as a virtual particle......but note that even this part of the 
story is not yet over, since   we have still not obtained experimental access 
to the proposed symmetry-breaking (Higgs \cite{higgs}) sector! If and when we do, it will 
surely be a remarkable vindication of theoretical pre-occupations dating back 
to the early 1960's. 

It seems fair to conclude  that 
worrying about perceived imperfections of a theory, even a phenomenologically 
very successful one, can pay off. In the case of the SM, a quite serious imperfection 
(for many theorists) is the `hierarchy problem', which we shall discuss in a moment.  SUSY 
can provide a solution to this preceived problem, provided that SUSY partners to 
known particles have masses no larger than 1-10 TeV (roughly). A lot of work has 
been done on the phenomenology of SUSY, which has influenced LHC detector design. Once 
again, it will be extraordinary if, in fact, the world turns out to be this way. 

In addition to this kind of motivation for SUSY, there are various other arguments 
which have been adduced. The rest of this section consists of a brief summary of 
 the main reasons I could find why theorists are keen on SUSY.

\subsection{The `weak scale instability problem' - also known as the 
`hierarchy problem'}

The electroweak sector of the SM (see for example Aitchison and Hey \cite{AH32}) 
contains within it a parameter with the 
dimensions of energy  (i.e. a  `weak scale'), namely  the vacuum expectation value of the 
Higgs field, 
\be
v\approx 246 \ {\rm GeV}.\label{eq:vvalue}
\ee
 This parameter sets the scale, in principle, of 
all masses in the theory. For example, the   
 mass of the ${\rm W}^{\pm}$ (neglecting radiative corrections) is 
given by
\be 
M_{\rm W}= gv/2 \sim 80 {\rm GeV},\label{eq:MW}
\ee
and the mass of the Higgs boson is
 \be 
 M_{\rm H}=v \sqrt{\frac{\lambda}{2}}, \label{eq:MH}
 \ee 
where $g$ is the SU(2) gauge coupling constant,  and $\lambda$ is the 
  strength of the Higgs self-interaction in the Higgs potential 
 \be 
 V=-\mu^2 \phi^\dagger \phi + \frac{\lambda}{4}(\phi^\dagger \phi)^2,
 \label{eq:HiggsV}
 \ee
 where $\lambda > 0$ and $\mu^2 > 0$. 
  Here $\phi$ is the SU(2) doublet field 
 \be 
 \phi = \left(\ba{c}\phi^+\\\phi^0\ea \right), \label{eq:higgsdoubSM}
 \ee
 and all fields are understood to be quantum, no `hat' being used. 
 
  Recall now that 
  the {\em negative} sign of the `${\rm mass}^2$' term $-\mu^2$ is 
 essential for the spontaneous symmetry breaking mechanism to work. With the sign as in 
 (\ref{eq:HiggsV}), the minimum of $V$ interpreted as a classical potential is at the 
 non-zero value 
 \be
  |\phi|=\sqrt{2} \mu / \sqrt{\lambda} \equiv v/ \sqrt{2}, \label{eq:v}
 \ee
 where $\mu \equiv \sqrt{\mu^2}$. This classical minimum (equilibrium value) is 
 conventionally interpreted as the expectation value of the quantum 
 field in the quantum 
 vacuum (i.e. the vev), at least at tree level. If `$-\mu^2$' in (\ref{eq:HiggsV}) 
 is replaced by the positive quantity `$\mu^2$', the classical equilibrium value is at 
 the origin in field space, which would imply $v=0$ - in which case all particles 
 would be massless. Hence it is vital to preserve the sign, and indeed magnitude, 
 of the coefficient of $\phi^\dagger \phi$ in (\ref{eq:HiggsV}). 
 
 The discussion so far has been at tree level (no loops). What happens when we include 
 loops?       The SM is 
 renormalizable, which means that finite results are obtained for all higher-order 
 (loop) corrections, even if we extend the virtual momenta in the loop integrals 
 all the way to infinity. But although this certainly implies that the theory is 
 well-defined and calculable up to infinite energies, in practice no-one  
 seriously believes  that the SM is really all there is, however high 
 we go in energy. That is to say, in loop integrals of the form 
 \be 
 \int^{\Lambda} {\rm d}^4 k \ f(k, {\rm external \ momenta})
 \ee
 we do not think that the cut-off $\Lambda$ {\em should} go to infinity, physically, even 
 though the reormalizability of the theory assures us that no inconsistency will 
 arise if it does. More reasonably, we regard the SM as part of a larger theory 
 which includes as yet unknown  `new physics' at high energy,  $\Lambda$ representing  
 the scale at which  this new physics appears, and where the SM must be modified. At the 
 very least, for instance, there surely must be some kind of new physics at the 
 scale when quantum gravity becomes important, which is believed to be indicated 
 by the Planck mass 
 \be 
 M_{\rm P}=( G_{\rm N})^{-1/2}\simeq 1.2 \times 10^{19} \ {\rm GeV}.\label{eq:planck}
 \ee 
 If this is indeed the scale of the new physics beyond the SM or, in fact, 
 if there is {\em any} scale of `new physics' even several orders of magnitude 
 different from the scale set by $v$,  then we shall see that we meet  
 a problem with  the SM, once we go beyond tree level.
 
 The  4-boson self-interaction in ({\ref{eq:HiggsV})
  generates, at one-loop order,  a  contribution to the $\phi^\dagger \phi$ 
  term, corresponding to the self-energy 
 diagram of figure 1.1  which is proportional to 
 \begin{figure}	
\begin{center}
\includegraphics[width=5.5cm]{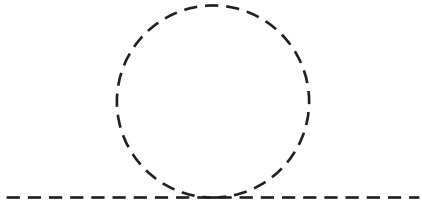}
\caption{One-loop self-energy graph in $\phi^4$ theory.}
\end{center}
\end{figure}
 \be 
 \lambda \int^\Lambda {\rm d}^4 k \ \frac{1}{k^2-M_{{\rm H}}^2}.\label{eq:higgsint}
 \ee
 This integral clearly diverges quadratically (there are four powers of $k$ 
 in the numerator, and two in the denominator), and it turns out to 
 be  {\em positive}, producing a 
 correction 
 \be \sim \lambda \Lambda^2 \phi^\dagger \phi \label{eq:higgsloop}
 \ee
  to the $-\mu^2\phi^\dagger \phi$ term  in $V$. Now 
 we know that the vev $v$  is given 
 in terms of $\mu$ by (\ref{eq:v}), and that its value  is  fixed  
 phenomenologically by (\ref{eq:vvalue}).
 Hence it seems that $\mu$ can hardly be much greater than of order 
 a few hundred GeV (or, if it is, $\lambda$ is much greater than unity - which 
 would imply that  
  we can't do perturbation theory; but since this is all we know how 
 to do, in this problem, we proceed on the assumption that $\lambda$ had better 
 be in the perturbative regime).  On the other hand,  
  if $\Lambda \sim M_{\rm P}\sim 10^{19}\  {\rm GeV}$, 
  the one-loop  quantum correction to `$-\mu^2$' is 
  then vastly greater than $\sim (100 \ {\rm GeV})^2$, and positive, so that 
   to arrive at a 
  value $\sim -(100 \ {\rm Gev})^2$ {\em after} inclusion of loop corrections would 
  seem to require that we 
  start with an equally huge but negative value of the Lagrangian parameter 
  $-\mu^2$, relying on a remarkable cancellation to get us  from $\sim -(10^{19} \ 
  {\rm GeV})^2$ to $\sim -(10^2 \ {\rm GeV})^2$. 
  
  We stress again, however, that this is {\em 
  not} a problem if the SM is treated in isolation, with the cut-off $\Lambda$ going to 
  infinity. There is then no `second scale' ($\Lambda$ as well as $v$), and 
  the Lagrangian parameter $-\mu^2$ can be chosen to depend on the cut-off $\Lambda$ 
  in just such a 
  way that, when $\Lambda \to \infty$, the final (renormalized) coefficient of 
  $\phi^\dagger \phi$ has the desired value. This is of course what happens to all 
  ordinary mass terms in renormalizable theories.     
    
    This `large cancellation' (or `fine tuning') problem involving the parameter 
    $\mu$ affects not only the mass of the Higgs particle, which is given in terms of 
    $\mu$ (combining (\ref{eq:MH}) and (\ref{eq:v})) by 
    \be 
    M_{\rm H} = \sqrt{2} \mu,
    \ee
    but also the mass of the W, 
    \be 
    M_{\rm W}=g \mu /\sqrt{\lambda},
    \ee
    and ultimately all masses in the SM,  which derive from $v$ and hence $\mu$. 
    
    But wait a minute: haven't we just admitted that something like this {\em always} 
    happens in mass terms of renormalizable theories? Why are we making a fuss about it 
    now? 
    
    Actually, it is a problem which arises in a particularly acute way in theories 
    which involve scalar particles in the Lagrangian - in contrast to theories 
    with only fermions and gauge fields in the Lagrangian, but which are capable 
    of producing scalar particles as some kind of bound states. An example of the 
    latter kind of theory would be QED, for instance. Here the analogue of figure 
    1.1 would be the one-loop process  in which an electron emits and then re-absorbs 
    a photon. This produces a correction $\delta m$ to the fermion mass  $m$ 
    in the Lagrangian,  which 
    seems to vary with the cut-off as  
    \be 
    \delta m \sim \alpha \int^\Lambda  
    \frac{{\rm d}^4 k}{{\not \!k}{k^2}} \sim \alpha \Lambda.
    \ee
    In fact, however, when the calculation is done in detail one finds 
    \be 
    \delta m \sim \alpha m \ln \Lambda,
    \ee
    so that even if $\Lambda \sim 10^{19} \ {\rm GeV}$, we have $\delta m \sim m$ and no 
    unpleasant fine-tuning is necessary after all. 
    
    Why does it happen in this case that $\delta m \sim m$?  It is because  
    the Lagrangian for QED (and the SM for that matter) has a {\em symmetry} as 
    the fermion masses go to zero, namely chiral symmetry. This is the symmetry under 
    transformations (on fermion fields) of the form 
    \be 
    \psi \to {\rm e}^{{\rm i} \alpha \gamma_5} \psi
    \ee
    in the U(1) case, or 
    \be 
    \psi \to {\rm e}^{{\rm i} {\bm \alpha} \cdot {\bm \tau}/2 \gamma_5} \psi 
    \ee
    in the SU(2) case. This symmetry guarantees that all radiative corrections to 
    $m$, computed in perturbation theory, will vanish as $m \to 0$. Hence $\delta m$ 
    must be proportional to $m$, and the dependence on $\Lambda$ is therefore 
    (from dimensional analysis) only logarithmic. 
    
    What about self-energy corrections to the masses of gauge particles?  For QED it 
    is of course the (unbroken) gauge symmetry which forces $m_\gamma =0$, to all orders 
    in perturbation theory. In other words, gauge invariance guarantees that no term 
    of the form 
    \be 
    m_\gamma^2 A^\mu A_\mu
    \ee
    can be radiatively generated in an unbroken gauge theory. On the other hand, 
    the non-zero masses of the W and Z bosons in the SM arise non-perturbatively  
    via spontaneous symmetry breaking, as we have seen - that is, via the Higgs vev $v$. 
    If $v$ is zero, the W and Z are as massless as the photon. But $v$ is proportional 
    to $\mu$, and so the masses they acquire by symmetry breaking are as sensitive to 
    $\Lambda$ as $M_{\rm H}$ is. 
    
    Can we find a symmetry which would - in a sense similar to chiral symmetry or 
    gauge symmetry -  control $\delta m^2$ for a {\rm scalar} particle appearing in 
    a Lagrangian? Well, there are also fermion loop corrections to the $-\mu^2 \phi^\dagger 
    \phi$ term, in which a $\phi$ particle turns into a fermion-antifermion 
    pair, which then annihilates back into a $\phi$ particle. This contribution behaves as 
    \be 
    \left(- g_{\rm f}^2 \int^\Lambda \frac {{\rm d}^4 k}{{\not \! k} { \not \! k}}\right)
    \phi^\dagger \phi  
     \sim - g_{\rm f}^2 \phi^\dagger \phi 
    \Lambda^2.\label{eq:fermionloop}
    \ee
    The sign here is crucial, and comes from the closed fermion loop. Combining 
    (\ref{eq:higgsloop}) and (\ref{eq:fermionloop}) we see that the total one loop 
    correction would have the form 
    \be 
    (\lambda - g_{\rm f}^2) \Lambda^2 \phi^\dagger \phi.\label{eq:totalloop}
    \ee 
    The possibility now arises that {\em if} for some reason 
    \be
    \lambda = g_{\rm f}^2 \label{eq:ccss}
    \ee  
    then this quadratic sensitivity to $\Lambda$ would not occur. Such a relation between 
    a 4-boson coupling constant and the square of a boson-fermion one is
     characteristic of SUSY, as we shall eventually see in section 8.  

    After the cancellation of the $\Lambda^2$ terms, our  two Higgs self-energy graphs 
    would contribute together something like 
    \be
    \sim \lambda (M_{\rm H}^2 -m_{{\rm f}}^2) \ln \Lambda,
    \ee
    which can be of order $M_{{\rm H}}^2$ itself (hence avoiding any fine-tuning) 
    provided all the bosons and fermions in the theory have masses no greater than, say, 
    a few TeV. The particles hypothesized to take part in this cancellation mechanism 
    need to be approximately degenerate (indicative of an approximate SUSY), 
    and not too much heavier than the scale of 
    $v$ (or $M_{{\rm H}}$), or we are back to some form of fine tuning. Essentially, 
    such a `boson $\leftrightarrow$ fermion' symmetry gives the scalar masses `protection' 
    from quadratically divergent loop corrections, by virtue of being related by symmetry 
    to the fermion masses, which are protected by chiral symmetry. Of course, how much 
    fine-tuning we are prepared to tolerate is a matter of taste. 
    
    Thus SUSY  {\em stabilizes} the hierarchy 
    $M_{{\rm H}, {\rm W}} \ll M_{{\rm P}}$, in the sense that radiative corrections will 
    not drag $M_{{\rm H},{\rm W}}$ up to the high scale $\Lambda$; and the argument implies 
    that, for the desired stabilization to occur,  SUSY should be visible at a 
    scale not too much greater than 1-10 TeV. The origin of this latter scale 
    (that of SUSY-breaking - see section 15.2)  is a separate problem.
    
    {\footnotesize{The reader should not get the impression that SUSY is the only available solution 
    to the hierarchy problem. In fact, there are several others on offer. One, which has been around 
    almost as long as SUSY, is generically called `technicolour'. It proposes \cite{weinberg} 
    \cite{susskind} that the Higgs field is not  `elementary'  but is analogous to the 
    electron-pair state in the BCS theory of superconductivity, being a bound state of new 
    doublets of  massless quarks Q and anti-quarks ${\bar{\rm Q}}$ interacting via a new strongly 
    interacting gauge theory, similar to QCD. In this case, the Lagrangian for the Higgs sector 
    is only an effective theory, valid for energies significantly below the scale at which 
    the ${\rm Q}-{\bar{\rm Q}}$ structure would be revealed - say 1 - 10 TeV. The integral in 
    (\ref{eq:higgsint}) can then only properly be extended to this scale, certainly not to a hierarchically 
    different scale such as $M_{\rm P}$. Essentially, new strong interactions enter not too far 
    above the electroweak scale. A relatively recent review is provided by Lane \cite{lane}. 
    A quite different possibility is to suggest that the gravitational (or string) scale is actually 
    very much lower than (\ref{eq:planck}) - perhaps even as low as a few TeV \cite{ant}. The hierarchy
     problem then evaporates since the ultaviolet cut-off $\Lambda$ is not much higher than the weak 
     scale itself. This miracle is worked by appealing to the notion of `large' hidden extra 
     dimensions, perhaps as large as sub-millimetre scales.  This 
     and other related ideas are discussed by Lykken \cite{lykken}, for example. Nevertheless, it 
     is fair to say that SUSY, in the form of the MSSM, is at present the most highly developed 
     framework for guiding and informing explorations of  physics `beyond the SM'. }}      
    
    \subsection{Additional positive  indications}

     (a) The precision fits to electroweak data show that $M_{{\rm H}}$ is less than about 
    200 GeV,  at 
    the $99\%$ confidence level. The `Minimal Supersymmetric Standard Model' (MSSM) (see 
    section 12), which has two Higgs doublets,  predicts (see section 16) that the lightest Higgs particle 
    should be no heavier than about 140 GeV. In the SM, by contrast, we have {\em no} 
    constraint on $M_{{\rm H}}$.\footnote{Not in quite the same sense (i.e. of a mathematical 
    bound), at any rate. One can certainly say, from (\ref{eq:MH}), that if $\lambda$ is not much 
    greater than unity, so that perturbation theory has a hope of being applicable, then $M_{{\rm H}}$ 
    can't be much greater than a few hundred GeV. For more sophisticated versions of this 
    sort of argument, see \cite{AH32} section 22.10.2.}\\ 
    (b) At one-loop order, the inverse gauge couplings $\alpha_1^{-1}(Q^2), \alpha_2^{-1}(Q^2), 
    \alpha_3^{-1}(Q^2)$ of the SM run linearly with $\ln Q^2$. Although $\alpha_1^{-1}$ decreases 
    with $Q^2$,  
    and $\alpha_2^{-1}$ and $\alpha_3^{-1}$ increase, all three tending to meet at 
    high $Q^2 \sim (10^{16} \ {\rm GeV})^2$, they do not in fact meet convincingly in the SM. 
    On the other hand, in 
    the MSSM they do, thus encouraging ideas of  unification: see section 13.\\ 
    (c) In any renormalizable theory, the mass parameters in the Lagrangian are 
    also scale-dependent (they `run'), just as the coupling parameters do. In the MSSM, 
    the evolution of a Higgs (mass)$^2$ parameter  from a typical positive value of 
    order $v^2$ at a scale of the order of $10^{16} \ {\rm GeV}$, takes it to a negative 
    value of the 
    correct order of magnitude at scales of order 100 GeV,  thus providing 
    a possible explanation for the origin of electroweak symmetry breaking, specifically 
     at those much lower scales. Actually, however, 
    this happens because the Yukawa coupling of the top 
    quark is large (being proportional to its mass), and this has a dominant effect on 
    the evolution. You might ask whether, in that case, the same result would be obtained 
    without SUSY.The answer is that it would, but the initial conditions for the evolution 
    are more naturally motivated within a SUSY theory, as discussed in section 15.

    \subsection{Theoretical considerations}

    It can certainly be plausibly argued that a dominant theme in twentieth 
    century physics was that of symmetry, the pursuit of which was heuristically 
    very successful. It is natural to ask if our current quantum field theories exploit all 
    the kinds of symmetry which could exist, consistent with Lorentz invariance. 
    Consider the symmetry `charges' that we are familiar with in the SM, for example 
    an electromagnetic charge of the form
    \be 
    Q=e \int {\rm d}^3 {\bm x} \ \psi^\dagger \psi \label{eq:U1}
    \ee
    or an SU(2) charge (isospin operator) of the form
    \be 
    {\bm T}= g \int {\rm d}^3 {\bm x} \ \psi^\dagger  ({\bm \tau}/2) \psi \label{eq:SU2}
    \ee
    where in (\ref{eq:SU2})  $\psi$ is an SU(2) doublet, and in both (\ref{eq:U1}) and 
    (\ref{eq:SU2}) $\psi$ is a fermionic field. All such symmetry operators are 
    themselves Lorentz scalars (they carry no uncontracted Lorentz indices of any kind, for 
    example vector or spinor). This implies that when they act on a state of definite spin 
    $J$, they cannot alter that spin:
    \be 
    Q | J \rangle = | \ {\rm same} \ J, \ {\rm possibly \ different \ member\ of \  
    symmetry \ multiplet}\  \rangle .\label{eq:QJ}
    \ee
    Need this be the case? 
    
    We certainly know of one vector `charge' - namely, the 4-momentum operators $P_\mu$ 
    which generate space-time displacements, and whose eigenvalues are  conserved 
    4-momenta. There are also the angular momentum operators, which belong inside 
    an antisymmetric tensor $M_{\mu \nu}$.    
    Could we, perhaps, have a conserved symmetric tensor 
    charge $Q_{\mu \nu}$? We shall provide a highly 
    simplified version (taken from Ellis \cite{ellis}) of an argument due to Coleman and Mandula 
    \cite{coleman} which 
    shows that we cannot. Consider letting such a 
    charge act on a single particle state with 4-momentum $p$:
    \be 
    Q_{\mu \nu} | p \rangle = (\alpha  p_\mu p_\nu + \beta  g_{\mu \nu} )| p \rangle ,
    \ee
    where the RHS has been written down by `covariance' arguments (i.e. the most 
    general expression with the indicated tensor transformation character, built from 
    the tensors at our disposal). Now consider a two-particle state $| p^{(1)}, p^{(2)} \rangle$, 
    and assume the $Q_{\mu \nu}$'s are additive, conserved, and act on only 
    one particle at a time, like other known charges.  Then 
    \be 
    Q_{\mu \nu} | p^{(1)}, p^{(2)} \rangle = (\alpha (p_\mu^{(1)}p_\nu^{(1)} + 
    p_\mu^{(2)}p_\nu^{(2)}) + 2 \beta  g_{\mu \nu})| p^{(1)}, p^{(2)} \rangle.
    \ee
    In an elastic scattering process  of the form $1+2 \to 3+4$ we will 
    then need (from conservation of the 
    eigenvalue) 
    \be 
    p_\mu^{(1)} p_\nu^{(1)} + p_\mu^{(2)} p_\nu^{(2)} = p_\mu^{(3)}p_\nu^{(3)} + 
    p_\mu^{(4)}p_\nu^{(4)}. \label{eq:Qcons}
    \ee
    But we also have 4-momentum conservation:
    \be 
    p^{(1)}_\mu + p^{(2)}_\mu = p^{(3)}_\mu + p^{(4)}_\mu \label{eq:4mom} 
    \ee
    The only common solution to (\ref{eq:Qcons}) and (\ref{eq:4mom}) is 
    \be 
    p^{(1)}_\mu = p^{(3)}_\mu , p^{(2)}_\mu =   p^{(4)}_\mu, \ {\rm or} \ p^{(1)}_\mu =p^{(4)}_\mu, 
    p^{(2)}_\mu =p^{(3)}_\mu, 
    \ee 
     which means that only  forward or backward scattering can occur - which is 
     obviously unacceptable. 
     
     The general message here is that there seems to be no  room for further 
     conserved operators with non-trivial Lorentz transformation  character 
     (i.e. not Lorentz scalars). The existing such operators $P_\mu$ and 
     $M_{\mu \nu}$ do allow proper scattering processes to occur, but imposing 
     any more conservation laws over-restricts the possible configurations. Such 
     was the conclusion of the Coleman-Mandula theorem \cite{coleman}. But in fact 
     their argument turns out not to exclude `charges' which transform under 
     Lorentz transformations as {\em spinors}: that is to say, things transforming 
     like a fermionic field $\psi$. We may denote such a charge by $Q_a$, the subscript 
     $a$ indicating the spinor component (we will see that we'll be dealing with 
     2-component spinors, rather than 4-component ones, for the most part). For 
     such a charge, equation (\ref{eq:QJ}) will clearly not hold; rather, 
     \be 
     Q_a | J \rangle = | J \pm 1/2 \rangle.
     \ee     
    Such an operator will not contribute to a matrix element for a two-particle 
    $\to$ two-particle elastic scattering process (in which the particle spins 
    remain the same), and consequently the above kind of `no-go' argument can't get 
    started. 
    
    The question then arises: is it  possible to include such spinorial operators 
    in a consistent algebraic scheme, along with the known conserved operators 
     $P_\mu$ and $M_{\mu \nu}$?
    The affirmative answer was first given by Gol'fand and Likhtman \cite{gol}, and the 
    most general such `supersymmetry algebra' was obtained by Haag, Lopuszanski 
    and Sohnius \cite{haag}. By `algebra' here we mean  (as usual) the set 
    of commutation relations among the `charges' - which, we recall,  are also the generators 
    of the appropriate symmetry transformations. The SU(2) algebra of the angular 
    momentum operators, which 
    are generators of rotations,  is a familiar example. The essential new feature here, 
    however, is that the charges which have a spinor character will have {\em 
    anticommutation} relations among themselves, rather than commutation relations. So such algebras 
    involve some commutation relations and some anticommutation relations. 
    
    What will such algebras look like? Since our generic spinorial charge 
    $Q_a$ is a symmetry operator, it must commute with the Hamiltonian of the 
   system, whatever it is:
   \be 
   [Q_a, H]=0,
   \ee
   and so must the anticommutator of two different components:
   \be 
   [\{Q_a, Q_b\},H]=0.\label{eq:QQH}
   \ee
   As noted above, the spinorial $Q$'s have two components, so as $a$ and $b$ vary the symmetric 
   object $\{Q_a, Q_b\}=Q_aQ_b+Q_bQ_a$ has three independent components, and we suspect that it 
   must transform as a spin-1 object (just like the symmetric combinations of two spin-1/2 
   wavefunctions). However, as usual in a relativistic theory, this spin-1 object should be 
   described by a 4-vector, not a 3-vector. Further, this 4-vector is conserved, from (\ref{eq:QQH}). 
    There is only one such conserved 4-vector operator (from the Coleman-Mandula theorem), 
    namely $P_\mu$. So the $Q_a$'s must satisfy an algebra of the form, roughly, 
    \be 
    \{Q_a, Q_b\} \sim P_\mu.
    \label{eq:QQP}
    \ee 
    Clearly (\ref{eq:QQP}) is sloppy: the indices on each side don't balance. With more than a little 
    hindsight, we might think of absorbing the `$\mbox{}_\mu$' by multiplying by $\gamma^\mu$, 
    the $\gamma$-matrix itself conveniently having two matrix indices which might correspond to 
    $a, b$. This is in fact more or less right, as we shall see in section 5, but the precise details 
    are finicky. 
    
    Accepting that (\ref{eq:QQP}) captures the essence of the matter, we can now begin to see 
    what a radical idea supersymmetry really is. Equation (\ref{eq:QQP}) says, roughly speaking, 
    that 
    if you do two SUSY transformations generated by the $Q$'s, one after the other, you get 
    the energy-momentum operator. Or, to put it even more strikingly (but quite equivalently), 
    you get the space-time translation operator, i.e. a derivative. Turning it around, the 
    SUSY spinorial $Q$'s are like square roots of 4-momentum, or square roots of 
    derivatives! It is rather like going one better than the Dirac equation, which can be 
    viewed as providing the square root of the Klein-Gordon equation: how would we take the 
    square root of the Dirac equation? 
    
    It is worth pausing to take this in properly. Four-dimensional derivatives are 
    firmly locked to our notions of a four-dimensional space-time. In now entertaining the 
    possibility that we can take square roots of them, we are effectively extending our 
    concept of space-time itself - just as, when the square root of -1 is introduced, we 
    enlarge the real axis to the complex (Argand) plane. That is to say, if we take 
    seriously an algebra involving both $P_\mu$ and the $Q$'s, we shall have to say that 
    the space-time co-ordinates are being extended to include further degrees of freedom, 
    which are acted on by the $Q$'s, and that these degrees of freedom are 
    connected  to the standard ones by means of transformations generated 
    by the $Q$'s.  These further degrees of freedom are, in fact, fermionic. So 
    we may say that SUSY invites us to contemplate `fermionic dimensions', and enlarge 
    space-time to `superspace'.    
        
     For some reason this doesn't seem to be the thing usually most emphasized about SUSY. Rather, 
     people talk much more about the fact that SUSY implies (if an exact symmetry) 
     degenerate multiplets of bosons and fermions. Of course, that aspect is 
     certainly true, and phenomenologically important, but  the fermionic enlargement 
     of space-time is arguably a more striking concept. 
     
     One final remark on motivations: if you believe in String Theory (and it still seems to be 
     the only game in town that may provide a consistent quantum theory of gravity), then 
     the phenomenologically most attractive versions incorporate supersymmetry, some 
     trace of which might  remain 
     in the theories which effectively describe physics at presently accessible energies.

\section{Spinors} 

Let's begin by recalling in outline how  symmetries, such as SU(2), are described 
in quantum field theory (see for example chapter 12 of \cite{AH32}). 
The Lagrangian involves a set of fields 
$\psi_r$ - they could be bosons or fermions - and it is taken to be invariant under an infinitesimal 
transformation on the fields of the form
\be 
\delta_\epsilon \psi_r = - {\rm i} \epsilon \lambda_{rs} \psi_s, \label{eq:epspsi}
\ee
where a summation is understood on the repeated index $s$, the  
$\lambda_{rs}$ are certain constant coefficients (for instance, the elements of the 
Pauli matrices),  and $\epsilon$ is an infinitesimal 
parameter. Supersymmetry transformations will look something like this, but they will transform 
bosonic fields into fermionic ones, for example 
\be 
\delta_\xi \phi \sim \xi \psi \label{eq:phipsi}
\ee
where $\phi$ is a bosonic (say spin-0) field, $\psi$ is a fermionic (say spin-1/2) one, and 
$\xi$ is an infinitesimal parameter (the alert reader will figure out that 
$\xi$ too has to be a spinor). In due course we shall spell 
out the details of the `$\sim$' here, but one thing should already be clear  at this stage: 
the number of (field) degrees of freedom, as between the bosonic $\phi$ fields and the 
fermionic $\psi$ fields, had better be the same in an equation of the form (\ref{eq:phipsi}), 
just as the number of fields $r=1, 2, \ldots N$ on the LHS of (\ref{eq:epspsi}) is the same as the 
number $s=1, 2 \ldots N$ on the RHS. We can't have some fields being `left out'. Now the 
simplest kind of bosonic field is of course a neutral scalar field, which has only one 
component, which is real: $\phi=\phi^\dagger$ (see \cite{AH31} chapter 5). On the other hand, 
there is {\em no} fermionic 
field with just one component: being a spinor, it has at least two. So that means that 
we must consider, at the very least, a two-degree-of-freedom bosonic field, to go 
with the spinor field, and that takes us to a complex (charged) scalar field (see chapter 7 
of \cite{AH31}).\footnote{We have been a bit slipshod here, eliding `components' with 
`degrees of freedom'. In fact, each component of a two-component spinor is complex, so there 
are 4 degrees of freedom in a two-component spinor. If the spinor is assumed to be `on-shell' - 
i.e. obeying the appropriate equation of motion -  then the number of degrees of freedom reduces 
to 2, the same as a complex scalar field. But generally  in field theory we need to go `off-shell', 
so that to match the 4 degrees of freedom in a two-component spinor we shall actually need more 
bosonic degrees of freedom than just the 2 in a complex scalar field. We shall ignore this 
complication in our first foray into SUSY, in Section 3, but return to it in Section 7. } 

But what kind of a two-component fermionic field do we `match' the complex scalar field with? 
When we learn the Dirac equation, pretty well the first result we arrive at is that fermion 
wavefunctions, or fields, have {\em four} components, not two. The simplest SUSY theory, however, 
involves a complex scalar field and a two-component fermionic field. The Dirac field actually 
uses {\em two} two-component fields, which is not the simplest case. Our first, and 
absolutely inescapable, job is therefore to `deconstruct' the Dirac field, and understand the 
nature of the two different two-component fields which together constitute it. 

This difference has to do with the different ways the two `halves' of the 4-component Dirac 
field transform under Lorentz transformations. Understanding how this works, in detail, 
is vital to being able to write down SUSY transformations which are consistent with 
Lorentz invariance. For example, the LHS of (\ref{eq:phipsi}) refers to a scalar 
(spin-0) field $\phi$; admittedly it's complex, but that just means that it has a real 
part and an imaginary part, both of which have spin-0. So it is an invariant under 
Lorentz transformations. On the RHS, however, we have the 2-component 
spinor (spin-1/2) field $\psi$, 
which is certainly not invariant under Lorentz transformations. But the parameter $\xi$ 
is also a 2-component spinor, in fact, and so we shall have to understand how to 
put $\xi$ and $\psi$ together properly so as to form a Lorentz invariant, in order to 
be consistent with the Lorentz transformation character of the LHS. While we may 
be familiar with how to do this sort of thing for 4-component Dirac spinors, we need to 
learn the corresponding tricks for 2-component ones.    
The rest of this lecture is therefore devoted to this groundwork.  

\subsection{Spinors and Lorentz Transformations}

In many branches of theoretical physics, {\em specific notation} has been invented. There are 
many reasons for this, including the following (all of which of course assume that the 
notation has been perfectly mastered): it makes the formulae more compact and less of a 
drudgery to write out (take a look at Maxwell's original paper on  Electromagnetism, 
written in 1864 before the invention of vector calculus); it can guarantee, essentially as 
an automatic consequence of writing a well-formed equation, that it incorporates some 
desired properties (for example,  4-vectors in Special Relativity, the use of which 
automatically incorporates the requirement of Lorentz covariance); and a well-conceived 
notation  lends itself to  advantageous steps in manipulations (for example, taking dot or cross 
products in equations involving vectors). Supersymmetry is no exception, there being plenty 
of specific notation available for things like spinors. 
The only problem is, that it  has not yet been standardized. 
This is very off-putting to beginners in the subject, because almost all the  introductory 
articles or books they pick up will use notation which is, to a greater or lesser extent, 
special to that source, making comparisons very frustrating. By contrast, we shall in these 
lectures not make much use of special SUSY notation. Rather, we shall aim to use a notation 
with which we assume the reader is familiar - namely, that used in standard relativistic 
quantum mechanics courses which deal with the Dirac equation. The advantage of this strategy 
is that the student doesn't, as a first task, have to learn a quite tricky new notation, and can 
gain access to the subject directly on the basis of standard courses. 
There will eventually be a price to 
pay, in the cumbersome nature of some expressions and manipulations, which could be streamlined 
using professional SUSY notation. And after all, students have, in the end, got to be able to 
read SUSY formulae when written in the  (quasi-)standard notation. So as we progress we'll 
introduce more specific notation.

We begin with the Dirac equation in momentum space, which we write as 
\be 
E \Psi=({\bm \alpha} \cdot {\bm p} + \beta m) \Psi \label{eq:dirac}
\ee
where of course we are taking $c=\hbar=1$. We shall choose the particular reprsentation 
\be 
{\bm \alpha} = \left( \begin{array}{cc} {\bm \sigma} & 0 \\ 0 & -{\bm \sigma} \end{array} \right)
\ \ \ \ \beta=\left( \begin{array}{cc} 0&1\\1&0 \end{array} \right),\label{eq:albet}
\ee 
which implies that 
\be
{\bm \gamma}=\left( \begin{array}{cc} 0 & -{\bm \sigma}\\{\bm \sigma} &0 \end{array} \right), 
\ \ \ {\rm and} \ \ \ \ \gamma_5 = \left( \begin{array}{cc} 1&0\\0&-1 \end{array} \right).
\label{eq:gammas}
\ee
 This is one of the standard representations of the Dirac matrices (see for example \cite{AH31} 
 page 91, and \cite{AH32} pages 31-2, and particularly \cite{AH32} 
 Appendix M,  section M.6). It is the one 
 which is commonly used in the `small mass' or `high energy' limit, since the (large) momentum 
 term is then (block) diagonal. As usual, ${\bm \sigma}\equiv(\sigma_x, \sigma_y, \sigma_z)$ are 
 the $2 \times 2$ Pauli matrices. 
 
 We write 
 \be 
 \Psi = \left( \begin{array}{c} \psi\\\chi \end{array}\right).\label{eq:diracpsi}
 \ee
The Dirac equation is then 
\bea 
(E-{\bm \sigma} \cdot {\bm p})\psi&=&m \chi \label{eq:Dpsi} \\
(E+{\bm \sigma} \cdot {\bm p})\chi &=& m \psi. \label{eq:Dchi}
\eea
Notice that as $ m \to 0$, (\ref{eq:Dpsi}) becomes ${\bm \sigma} \cdot {\bm p}
\psi_0=E\psi_0$, and $E \to |{\bm p}|$, so the zero mass limit of 
(\ref{eq:Dpsi}) is 
\be 
({\bm \sigma} \cdot {\bm p}/|{\bm p}|) \psi_0 = \psi_0,
\ee
which means that $\psi_0$ is an eigenstate of the helicity operator 
${\bm \sigma} \cdot {\bm p}/|{\bm p}|$ with eigenvalue +1 (`positive 
helicity'). Similarly, the zero-mass limit of (\ref{eq:Dchi}) shows that 
$\chi_0$ has negative helicity.

For $m \neq 0$, $\psi$ and $\chi$ of (\ref{eq:Dpsi}) and (\ref{eq:Dchi}) are 
plainly not helicity eigenstates: indeed the mass term (in this representation) 
`mixes' them. But, as we shall see shortly,  it is these two-component objects, 
$\psi$ and $\chi$, that have well-defined Lorentz transformation properties, 
and they are the two-component spinors we shall be dealing with. 

Although 
not helicity eigenstates, $\psi$ and $\chi$  are eigenstates of $\gamma_5$, in the sense 
that 
\be 
\gamma_5 \left( \ba{c}\psi\\0\ea \right) = \left( \ba{c} \psi \\ 0 \ea \right), 
\ \ \ {\rm and} \ \ \ \gamma_5 \left( \ba{c}0\\\chi\ea \right) 
= - \left( \ba{c} 0 \\ \chi \ea \right) .
\ee 
 These two $\gamma_5$-eigenstates can be constructed from the original $\Psi$ by 
 using the projection operators $P_{{\rm R}}$ and $P_{{\rm L}}$ defined by 
 \be 
 P_{{\rm R}}=\left(\frac{1+\gamma_5}{2}\right) = \left( \ba{cc}1&0\\0&0\ea\right)
 \ee
 and 
  \be 
 P_{{\rm L}}=\left(\frac{1-\gamma_5}{2}\right) = \left( \ba{cc}0&0\\0&1\ea\right).
 \ee    
Then 
\be 
P_{{\rm R}} \Psi= \left( \ba{c} \psi\\0\ea\right), \ \ \ P_{{\rm L}} \Psi= 
\left( \ba{c}0\\\chi\ea \right).
\ee
It is easy to check that $P_{{\rm R}} P_{{\rm L}} =0, \ P_{{\rm R}}^2=P_{{\rm L}}^2=1$. 
The eigenvalue of $\gamma_5$ is called `chirality';  $\psi$ has chirality 
+1, and $\chi$ has chirality -1. In an unfortunate terminolgy, but one now too late 
to change, `+' chirality is denoted by `R' (i.e right-handed) and `-' chirality by 
`L' (i.e. left-handed), despite the fact that (as noted above) $\psi$ and $\chi$ 
are {\em not} helicity eigenstates when $m \neq 0$. Anyway, a `$\psi$' type 
2-component spinor is often written as $\psi_{{\rm R}}$, and a `$\chi$' type one as 
$\chi_{{\rm L}}$. For the moment, we shall not use these R and L subscripts, but 
shall understand that anything called $\psi$ is an R state, and a $\chi$ is an L state. 

Now, we said above that $\psi$ and $\chi$ had well-defined Lorentz transformation 
character. Let's recall how this goes (see \cite{AH32} Appendix M, section M.6). There are 
basically two kinds of transformation: rotations and `boosts' (i.e. pure velocity 
transformations). It is sufficient to consider {\em infinitesimal} transformations, which 
we can specify by their action on a 4-vector, for example the energy-momentum 4-vector 
$(E, {\bm p})$. Under an infinitesimal 3-dimensional rotation, 
\be 
E \to E' = E, \ \ \ {\bm p} \to {\bm p}'= {\bm p} - {\bm \epsilon} \times {\bm p} 
\label{eq:rot}
\ee
where ${\bm \epsilon}=(\epsilon_1, \epsilon_2, \epsilon_3)$ are three infinitesimal 
parameters specifying the infinitesimal rotation; and under a velocity transformation 
\be 
E \to E'= E- {\bm \eta} \cdot {\bm p}, \ \ \ {\bm p} \to {\bm p}'={\bm p} - {\bm \eta} E 
\label{eq:boost}
\ee
where ${\bm \eta}=(\eta_1, \eta_2, \eta_3)$ are three infinitesimal velocities.  
Under the Lorentz transformations thus defined, $\psi$ and $\chi$ transform as follows 
(see equations (M.94) and (M.98) of \cite{AH32}, where however  the top two 
components are called `$\phi$' rather than `$\psi$'):
\be \psi \to \psi'=(1+{\rm i}{\bm \epsilon} \cdot {\bm \sigma} /2 - {\bm \eta} 
\cdot {\bm \sigma} /2) \psi \label{eq:psip}
\ee
and 
\be 
\chi \to \chi' = (1+{\rm i} {\bm \epsilon} \cdot {\bm \sigma}/2 + {\bm \eta} \cdot 
{\bm \sigma} /2) \chi. \label{eq:chip}
\ee
Equations (\ref{eq:psip}) and (\ref{eq:chip}) are extremely important equations for us. They tell 
us how to construct the spinors $\psi'$ and $\chi'$ for the rotated and boosted 
frame, in terms of the original spinors $\psi$ and $\chi$. That is to say, the $\psi'$ 
and $\chi'$ specified by (\ref{eq:psip}) and (\ref{eq:chip}) satisfy the `primed' 
analogues of (\ref{eq:Dpsi}) and (\ref{eq:Dchi}), namely
\bea 
(E'-{\bm \sigma} \cdot {\bm p}')\psi'&=&m \chi' \label{eq:Dpsip} \\
(E'+{\bm \sigma} \cdot {\bm p}')\chi' &=& m \psi'. \label{eq:Dchip}
\eea

Let's pause to check this statement in a special case, that of a pure boost. 
Define $V_{\bm \eta}=(1-{\bm \eta} \cdot {\bm \sigma}/2)$. Then since ${\bm \eta}$ 
is infinitesimal, $V_{\bm \eta}^{-1}=(1+{\bm \eta} \cdot {\bm \sigma}/2)$. Now take 
(\ref{eq:Dpsi}), multiply from the left by $V_{\bm \eta}^{-1}$, and insert the unit 
matrix $V_{\bm \eta}^{-1}V_{\bm \eta}$ as indicated:
\be 
[V_{\bm \eta}^{-1}(E-{\bm \sigma} \cdot {\bm p}) V_{\bm \eta}^{-1}]V_{\bm \eta} \psi= 
m V_{\bm \eta}^{-1} \chi.
\ee 
If (\ref{eq:psip}) is right, we have $\psi' = V_{\bm \eta} \psi$, and if
 (\ref{eq:chip}) is right we 
have $\chi'=V_{\bm \eta}^{-1} \chi$, in this pure boost case. So to establish the 
complete consistency between    (\ref{eq:psip}), (\ref{eq:chip})  
and (\ref{eq:Dpsip}), we need to show that 
\be 
V_{\bm \eta}^{-1}(E-{\bm \sigma} \cdot {\bm p})V_{\bm \eta}^{-1}=(E'-{\bm \sigma} 
\cdot {\bm p}'),
\ee  
that is 
\be 
(1+{\bm \eta} \cdot {\bm \sigma}/2)(E-{\bm \sigma} \cdot {\bm p})(1+{\bm \eta} \cdot 
{\bm \sigma}/2) = (E-{\bm \eta} \cdot {\bm p}) - {\bm \sigma} \cdot ({\bm p} -E {\bm \eta})
\label{eq:VV}
\ee
to first order in ${\bm \eta}$, 
since the RHS of (\ref{eq:VV}) is just $E'-{\bm \sigma} \cdot {\bm p}'$ from (\ref{eq:boost}).  

{\bf Exercise} Verify (\ref{eq:VV}). 

Returning now to equations (\ref{eq:psip}) and (\ref{eq:chip}), we note that $\psi$ and $\chi$ 
actually behave the same under rotations (they have spin-1/2!), but differently under boosts. 
The interesting fact is that there are {\em two} kinds of two-component spinors, distinguished 
by their different transformation character under boosts. Both are used in the Dirac 4-component 
spinor. In SUSY, however, one works with the 2-component objects $\psi$ and $\chi$ which 
(as we saw above) may also be labelled by `R' and `L' respectively. 

Before proceeding, we note another important feature of (\ref{eq:psip}) and (\ref{eq:chip}). 
Let $V$ be the transformation  matrix appearing in (\ref{eq:psip}):
\be 
V=(1+{\rm i} {\bm \epsilon} \cdot {\bm \sigma}/2 - {\bm \eta} \cdot {\bm \sigma}/2).\label{eq:V}
\ee
Then 
\be 
V^{-1}=(1-{\rm i} {\bm \epsilon} \cdot {\bm \sigma}/2 + {\bm \eta} \cdot {\bm \sigma}/2)
\ee
since we merely have to reverse the sense of the infinitesimal parameters, while  
\be 
V^\dagger=(1-{\rm i} {\bm \epsilon} \cdot {\bm \sigma}/2 - {\bm \eta} \cdot {\bm \sigma}/2)
\ee
using the Hermiticity of the ${\bm \sigma}$'s. So 
\be 
{V^\dagger}^{-1}={V^{-1}}^\dagger=(1+{\rm i} {\bm \epsilon} \cdot {\bm \sigma}/2
 + {\bm \eta} \cdot {\bm \sigma}/2), \label{eq:Vchi}
 \ee
 which is the matrix appearing in (\ref{eq:chip}). Hence we may write, compactly, 
 \be 
 \psi' = V \psi, \ \ \ \chi' = {V^\dagger}^{-1} \chi = {V^{-1}}^\dagger \chi .\label{eq:psipchip}
 \ee
 
 \subsection{Constructing invariants and 4-vectors out of 2-component spinors}

Let's start by recalling some things which should be familiar from a Dirac equation course.
 From the 4-component 
Dirac spinor we can form a Lorentz invariant 
\be 
\bar{\Psi} \Psi = \Psi^\dagger \beta \Psi,\label{eq:invt}
\ee
and a 4-vector 
\be 
\bar{\Psi} \gamma^\mu \Psi = \Psi^\dagger \beta (\beta, \beta {\bm \alpha}) \Psi = 
\Psi^\dagger (1, {\bm \alpha}) \Psi.\label{eq:4vec}
\ee
In terms of our 2-component objects $\psi$ and $\chi$  (\ref{eq:invt}) becomes 
\be 
{\rm Lorentz \ invariant} \ \ \ \ \  ( \psi^\dagger \chi^\dagger) \left( \ba{cc} 0&1\\1&0 \ea \right)
\left( \ba{c} \psi\\ \chi \ea \right)  = 
\psi^\dagger \chi + \chi^\dagger \psi.\label{eq:invpsichi}
\ee
Using (\ref{eq:psipchip}) it is easy to verify that the RHS of (\ref{eq:invpsichi}) is invariant. 
Indeed, perhaps more interestingly, each part of it is:
\be 
\psi^\dagger \chi \to {\psi^\dagger}'\chi'=\psi V^\dagger {V^\dagger}^{-1} \chi = \psi^\dagger \chi,
\label{eq:psichiinvt}
\ee
and similarly for $\chi^\dagger \psi$. Again, (\ref{eq:4vec}) becomes 
\bea 
{\rm 4-vector} \ \ \ \ (\psi^\dagger \chi^\dagger) \left[\left( \ba{cc} 1&0\\0&1 \ea \right)
, \left(  \ba{cc}{\bm \sigma} & 0\\0&-{\bm \sigma} 
\ea \right) \right] \left( \ba{c}\psi\\ \chi \ea \right) &=& (\psi^\dagger \psi +\chi^\dagger \chi, 
\psi^\dagger {\bm \sigma} \psi - \chi^\dagger {\bm \sigma} \chi ) \nonumber \\
&\equiv&\psi^\dagger \sigma^\mu \psi + \chi^\dagger {\bar{\sigma}}^\mu \chi, 
\eea
where we have introduced the important quantities 
\be 
\sigma^\mu \equiv (1, {\bm \sigma}), \ \ \ \ {\bar{\sigma}}^\mu = (1, -{\bm \sigma}).\label{eq:sigsigbar}
\ee
As with the Lorentz invariant, it is actually the case that each of $\psi^\dagger \sigma^\mu \psi$ 
and $\chi^\dagger {\bar{\sigma}}^\mu \chi$ transforms, separately, as a 4-vector. 

{\bf Exercise} Verify that last statement.

In this `$\sigma^\mu,\ {\bar{\sigma}}^\mu$' notation, the Dirac equation (\ref{eq:Dpsi}) and 
(\ref{eq:Dchi}) becomes 
\bea
\sigma^\mu p_\mu \psi&=& m \chi \label{eq:Dpsisig}\\
{\bar{\sigma}}^\mu p_\mu \chi &=& m \psi. \label{eq:Dchisig} 
\eea
So we can read off the useful news that `$\sigma^\mu p_\mu$' converts a $\psi$-type object to 
a $\chi$-type one, and ${\bar{\sigma}}^\mu p_\mu$ converts a $\chi$ to a $\psi$ - or, in 
slightly more proper language, the Lorentz transformation character of $\sigma^\mu p_\mu \psi$ is 
the same as that of $\chi$, and the LT character of ${\bar{\sigma}}^\mu p_\mu \chi$ is the same as 
that of $\psi$. 

Lastly in this re-play of Dirac stuff, the Dirac Lagrangian can be written in terms of 
$\psi$ and $\chi$:
\be 
\bar{\Psi} ( {\rm i} \gamma^\mu \partial_\mu -m) \Psi =
\psi^\dagger {\rm i}\sigma^\mu \partial_\mu \psi + 
\chi^\dagger {\rm i} {\bar{\sigma}}^\mu \partial_\mu\chi -m(\psi^\dagger \chi + \chi^\dagger \psi).
\label{eq:Ldirac}
\ee
Note how ${\bar{\sigma}}^\mu$ belongs with $\chi$, and $\sigma^\mu$ with $\psi$.   

An interesting point may have occurred to the reader here: it is possible to form 4-vectors 
using only $\psi$'s or only $\chi$'s (see the most recent Exercise), but the invariants 
introduced so far ($\psi^\dagger \chi$ and $\chi^\dagger \psi$)  make use of both. So we 
might ask: {\em can we make an invariant out of just {\mbox {$\chi$}}-
type spinors, for instance?} This is an important 
 technicality as far as SUSY is concerned, and it is  at this point that we part company with 
 what is usually contained   in standard Dirac courses. 

Another way of putting our question is this: is it possible to construct a spinor from the 
components  of, say, 
$\chi$, which has the transformation character of a $\psi$? (and of course vice versa). If it is, 
then  we can  use it, with $\chi$-type spinors, in place of $\psi$-type spinors 
 when making  invariants. The 
answer is that it is possible. Consider how the complex conjugate of $\chi$, denoted by 
$\chi^*$, transforms under Lorentz transformations.  
We have 
\be 
 \chi' = (1+{\rm i} {\bm \epsilon} \cdot {\bm \sigma}/2 + {\bm \eta} \cdot 
{\bm \sigma} /2) \chi.
\ee
Taking the complex conjugate gives 
\be 
{\chi^*}' = (1 - {\rm i} {\bm \epsilon} \cdot {\bm \sigma}^*/2 + 
{\bm \eta} \cdot {\bm \sigma}^*/2) \chi^*.\label{eq:chistar}
\ee
Now observe that $\sigma_1^*=\sigma_1, \ \sigma_2^*=-\sigma_2, \ 
\sigma_3^*=\sigma_3$, 
and that $\sigma_2 \sigma_3 = -\sigma_3 \sigma_2$ and $ \sigma_1 
\sigma_2=- \sigma_2 \sigma_1$. It follows that 
\bea
\sigma_2 {\chi^*}'&=& \sigma_2 ( 1 -{\rm i} {\bm \epsilon} \cdot 
(\sigma_1, -\sigma_2, \sigma_3)/2 + 
{\bm \eta} \cdot (\sigma_1, -\sigma_2, \sigma_3)/2) \chi^* \label{eq:chistar1}\\
&=& (1+{\rm i} {\bm \epsilon}\cdot {\bm \sigma}/2 -{\bm \eta} \cdot 
{\bm \sigma}/2) \sigma_2 \chi^* \label{eq:chistar2}\\
&=& V \sigma_2 \chi^* \label{eq:chistar3},
\eea
referring to (\ref{eq:V}) for the definition of $V$, which is precisely the 
matrix by which $\psi$ 
transforms. 

We have therefore established the important result that 
\be 
\sigma_2 \chi^* \ {\rm transforms \ like \ a \ } \psi.
\ee
So let's at once introduce `the $\psi$-like thing constructed from $\chi$' via 
the definition 
\be 
\psi_\chi \equiv {\rm i} \sigma_2 \chi^*, \label{eq:psichi1}
\ee
where the i has been put in for convenience (remember $\sigma_2$ involves i's). Then we 
are guaranteed that 
\be 
\psi_{\chi^{(1)}}^\dagger \chi^{(2)}={({\rm i} \sigma_2 \chi^{(1)*})}^{*{\rm T}}\chi^{(2)}=
{({\rm i} \sigma_2 \chi^{(1)})}^{{\rm T}}\chi^{(2)}=
\chi^{(1){\rm T}}(-{\rm i}\sigma_2)\chi^{(2)} \label{eq:chiscpr}
\ee 
where $\mbox{}^{{\rm T}}$  denotes transpose, 
is Lorentz invariant, for any two 
$\chi$-like things $\chi^{(1)}, \chi^{(2)}$,  
just as $\psi^\dagger \chi$ was. 
(Equally, so is $\chi^{(2)\dagger} \psi_{\chi^{(1)}}$.) Equation (\ref{eq:chiscpr}) is important, because  
it tells us {\em how to form the Lorentz invariant scalar product of two} $\chi$'s. This 
is the kind of product that we will need in SUSY transformations of the form 
(\ref{eq:phipsi}).

In particular, $\psi_\chi^\dagger \chi$ is Lorentz 
invariant, where the $\chi$'s are the same. This quantity is  
\be 
{({\rm i}\sigma_2 \chi^*)}^{*{\rm T}} \chi = ({\rm i} \sigma_2 \chi)^{\rm T} \chi 
= \chi^{\rm T} (-{\rm i} \sigma_2) \chi.\label{eq:chiinvt}
\ee 
Let's write it out in detail. We have 
\be 
{\rm i} \sigma_2 = \left( \ba{cc} 0&1\\-1&0 \ea \right) , \ \ \ {\rm and} \ \ \ 
\chi=\left( \ba{c}\chi_1\\
\chi_2 \ea \right),\label{eq:chi}
\ee
so that 
\be 
{\rm i} \sigma_2 \chi = \left( \ba{c} \chi_2\\-\chi_1 \ea \right), \ \ {\rm and} \ \ 
({\rm i} \sigma_2 \chi)^{\rm T} \chi = \chi_2 \chi_1 - \chi_1 \chi_2.\label{eq:chiup}
\ee
But now this seems like something of an anti-climax! It vanishes, doesn't it? Well, yes 
if $\chi_1$ and $\chi_2$ are ordinary functions, but not if they are {\em anticommuting} 
quantities, as appear in (quantized) fermionic fields. So certainly this is a satisfactory 
invariant in terms of two-component quantized fields, or in terms of Grassmann numbers 
(see Appendix O of \cite{AH32}).
\vspace{.2in}

{\footnotesize{{\bf Notational Aside (1)}. It looks as if it's going to get pretty tedious keeping track 
of which two-component spinor is a $\chi$-type one and which is \ $\psi$-type one, by 
writing things like $\chi^{(1)}, \chi^{(2)}, \dots , \psi^{(1)}, \psi^{(2)}, \ldots$, all the time, 
and (even worse) things like ${\psi^\dagger_{\chi^{(1)}}} \chi^{(2)}$. A first step in the 
direction of a more powerful notation is   
to agree that the components of $\chi$-type spinors have {\em lower indices}, as in (\ref{eq:chi}). 
That is, anything written with lower indices is a $\chi$-type spinor. So then we are free 
to name them how we please: $\chi_a, \xi_a, \ldots$, even $\psi_a$.

We can also streamline the cumbersome notation `${\psi_{\chi^{(1)}}}^\dagger \chi^{(2)}$'. 
The point here is that this  
notation was - at this stage  - introduced in order to construct invariants out of 
just $\chi$-type things. But (\ref{eq:chiscpr}) tells us how to do this, in terms of 
the two $\chi$'s involved: you take one of them, say $\chi^{(1)}$, and form 
${\rm i} \sigma_2 \chi^{(1)}$. Then you take the matrix dot product (in the sense of 
`$u^{\rm T} v$') of this quantity and the second $\chi$-type spinor. So, starting from 
a $\chi$ with lower indices, $\chi_a$, let's define a $\chi$ with {\em upper indices} via 
(see equation (\ref{eq:chiup}))   
\be 
\left( \ba{c}\chi^1\\ \chi^2 \ea \right) \equiv 
{\rm i} \sigma_2 \chi = \left(\ba{c}\chi_2\\-\chi_1 \ea \right) , 
\ee
that is, 
\be
\chi^1 \equiv \chi_2, \ \ \ \chi^2 \equiv -\chi_1.\label{eq:chiupper}
\ee 
Suppose now that $\xi$ is a second $\chi$-type spinor, and 
\be 
\xi = \left( \ba{c}\xi_1\\\xi_2 \ea \right).
\ee
Then we know that $({\rm i} \sigma_2 \chi)^{\rm T} \xi$ is a Lorentz invariant, 
and this is just 
\be 
(\chi^1 \chi^2) \left( \ba{c} \xi_1\\ \xi_2 \ea \right)= \chi^1 \xi_1 + \chi^2 \xi_2 = \chi^a 
\xi_a,\label{eq:chidot}
\ee
where $a$ runs over the values 1 and 2. Equation (\ref{eq:chidot}) is a compact notation 
for this scalar product: it is a  `spinor dot product', analogous to the 
`upstairs-downstairs' dot-products of Special Relativity, like $A^\mu B_\mu$. We can shorten the 
notation even further, indeed, to $\chi \cdot \xi$, or even to $\chi \xi$ if we know what we 
are doing. Note that if the components of $\chi$ and $\xi$ commute, then it doesn't 
matter whether we write this invariant as $\chi \cdot \xi= \chi^1 \xi_1 + \chi^2 \xi_2$ or as 
$\xi_1 \chi^1 + \xi_2 \chi^2$. But if they are anticommuting these will differ by a 
sign, and we need a convention as to which we take to be the `positive' dot 
product. It is as in (\ref{eq:chidot}), which is remembered as `summed-over 
$\chi$-type indices appear diagonally downwards, top left to bottom right'.

The 4-D Lorewntz-invariant dot product $A^\mu B_\nu$ of   Special Relativity 
 can also be written as $g^{\mu \nu}A_\nu B_\mu$, 
where $g^{\mu \nu}$ is the {\em metric tensor} of SR with components (in one common 
convention!) $g^{00}=+1, g^{11}=g^{22}=g^{33}=-1$, all others vanishing (see Appendix D of 
\cite{AH32}). In a similar way we can introduce a metric tensor $\epsilon^{ab}$ for forming 
the Lorentz-invariant spinor dot product of two two-component L-type spinors, by writing 
\be 
\chi^a = \epsilon^{ab} \chi_b \label{eq:chiupeps}
\ee
(always summing on repeated indices, of course), so that 
\be 
\chi^a \xi_a=\epsilon^{ab} \chi_b \xi_a.\label{eq:epsdotL}
\ee
For (\ref{eq:chiupeps}) to be consistent with (\ref{eq:chiupper}), we require 
\be 
\epsilon^{12}=+1, \epsilon^{21}=-1, \epsilon^{11}=\epsilon^{22}=0.
\ee
Clearly $\epsilon^{ab}$, regarded as a $2 \times 2$ matrix, is nothing but the 
matrix $ {\rm i} \sigma_2$ of (\ref{eq:chi}). We shall, however, continue to use 
the explicit `${\rm i} \sigma_2$' notation for the most part, rather than the 
`$\epsilon^{ab}$' notation. 

We can also introduce $\epsilon_{ab}$ via 
\be 
\chi_a = \epsilon_{ab} \chi^b,
\ee
which is consistent with (\ref{eq:chiupper}) if 
\be 
\epsilon_{12}=-1, \epsilon_{21}=+1, \epsilon_{11}=\epsilon_{22}=0.
\ee
Finally, you can verify that 
\be 
\epsilon_{ab} \epsilon^{bc}=\delta^c_a,
\ee
as one would expect. It is important to note that these `$\epsilon$' metrics are 
{\em antisymmetric} under the interchange of the two indices $a$ and $b$, 
whereas the SR metric $g^{\mu \nu}$ is symmetric under $\mu \leftrightarrow \nu$.

{\bf Exercise} (a) What is $\xi \cdot \chi$ in terms of $\chi \cdot \xi$ (assuming 
the components anticommute)? (b) What is $\chi_a \xi^a$ in terms of $\chi^a \xi_a$? Do 
these both by brute force via components, and by using the $\epsilon$ dot product. 

Given that $\chi$ transforms by $V^{-1 \dagger}$ of (\ref{eq:Vchi}), it is interesting to 
ask: how does the `raised-index' version, ${\rm i} \sigma_2 \chi$, transform?
 
{\bf Exercise} Show that ${\rm i} \sigma_2 \chi$   transforms by $V^*$. 

We can use the result of this Exercise to verify once more the invariance of 
$ ({\rm i} \sigma_2 \chi)^{\rm T} \xi$: $({\rm i} \sigma_2 \chi)^{\rm T} \xi \to 
({\rm i} \sigma_2 \chi)^{\prime {\rm T}} \xi' = ({\rm i} \sigma_2 \chi)^{\rm T}(V^*)^{\rm T} 
V^{-1 \dagger} 
\xi.$ But $(V^*)^{\rm T} = V^\dagger$, and so the invariance is established.

We can therefore summarize the state of play so far by saying that a downstairs 
$\chi$-type spinor transforms by $V^{-1 \dagger}$, while an upstairs $\chi$-type 
spinor transforms by $V^*$.       
}}

\vspace{.2in}

It is natural to ask: what about $\psi^*$? Performing manipulations analogous to those in 
(\ref{eq:chistar}), (\ref{eq:chistar1})-(\ref{eq:chistar3}), you can verify that 
\be 
\sigma_2 \psi^* \ \ \ {\rm transforms \ like } \ \ \ \chi.
\ee
This licenses us to introduce a $\chi$-type  object constructed from a $\psi$,  which we define by 
\be 
\chi_\psi \equiv -{\rm i} \sigma_2 \psi^*.\label{eq:chipsi}
\ee
Then for any two $\psi$'s $\psi^{(1)}, \psi^{(2)}$ say, we know that 
\be 
(-{\rm i} \sigma_2 \psi^{(1)*})^{*{\rm T}} \psi^{(2)} = (-{\rm i} \sigma_2 
\psi^{(1)})^{\rm T}\psi^{(2)} = \psi^{(1){\rm T}}{\rm i} \sigma_2 \psi^{(2)}
\label{eq:psiscpr} 
\ee
is an invariant. In particular, for the same $\psi$, the quantity 
\be 
(-{\rm i} \sigma_2 \psi)^{\rm T} \psi
\ee 
is an invariant. 
 
\vspace{.2in} 
{\footnotesize{{\bf Notational Aside (2)}. Clearly we want an `index' notation for 
$\psi$-type spinors. The general convention is that they are given `dotted indices' i.e. 
we write things like $\psi^{\dot{a}}$. By convention, also, we decide that our $\psi$-type 
thing has an {\em upstairs} index, just as it was a convention that our $\chi$-type thing had 
a downstairs index. Equation (\ref{eq:psiscpr}) tells us how to form scalar products out 
of two $\psi$-like things, $\psi^{(1)}$ and $\psi^{(2)}$, and invites us to define 
 downstairs-indexed quantities 
\be 
\left( \ba{c} \psi_{\dot{1}}\\\psi_{\dot{2}} \ea \right) \equiv - {\rm i} \sigma_2 \psi = 
\left( \ba{cc} 0&-1\\1&0 \ea \right) \left( \ba{c} \psi^{\dot{1}}\\\psi^{\dot{2}}\ea \right)
\label{eq:updowndots}
\ee
so that 
\be
\psi_{\dot{1}} \equiv -\psi^{\dot{2}}, \ \ \ \psi_{\dot{2}} \equiv \psi^{\dot{1}}.\label{eq:downdot}
\ee
Then if $\zeta$ (`zeta') is a second $\psi$-type spinor, and 
\be 
\zeta = \left( \ba{c} \zeta^{\dot{1}}\\\zeta^{\dot{2}}\ea \right),
\ee
we know that $(-{\rm i} \sigma_2 \psi)^{{\rm T}} \zeta$ is a Lorentz invariant, which is
\be
(\psi_{\dot{1}} \psi_{\dot{2}} ) \left( \ba{c} \zeta^{\dot{1}} \\ \zeta^{\dot{2}} \ea \right) 
= \psi_{\dot{1}}\zeta^{\dot{1}} + \psi_{\dot{2}} \zeta^{\dot{2}}=\psi_{\dot{a}}\zeta^{\dot{a}},
\label{eq:psizet}
\ee
where $\dot{a}$ runs over the values 1,2. Notice that with all these conventions, the `positive' scalar  
product has been defined so that the summed-over dotted 
indices appear diagonally upwards, bottom 
left to top right. 

We can introduce a metric tensor notation for  the Lorentz-invariant scalar product of two two-component R-type 
(dotted) spinors, too. We write 
\be 
\psi_{\dot{a}}=\epsilon_{\dot{a} \dot{b}} \psi^{\dot{b}}
\ee
where, to be consistent with (\ref{eq:downdot}), we need 
\be 
\epsilon_{\dot{1} \dot{2}}=-1, \epsilon_{\dot{2} \dot{1}} = +1, \epsilon_{\dot{1} \dot{1}} = 
\epsilon_{\dot{2} \dot{2}} = 0.
\ee
Then 
\be 
\psi_{\dot{a}} \zeta^{\dot{a}} = \epsilon_{\dot{a} \dot{b}} \psi^{\dot{b}} \zeta^{\dot{a}}.
\ee
We can also define 
\be 
\epsilon^{\dot{1} \dot{2}}=+1, \epsilon^{\dot{2} \dot{1}} = -1, \epsilon^{\dot{1} \dot{1}} 
= \epsilon^{\dot{2} \dot{2}} =0,
\ee
with 
\be 
\epsilon_{\dot{a} \dot{b}} \epsilon^{\dot{b} \dot{c}}=\delta_{\dot{b}}^{\dot{c}}.
\ee
Again, the $\epsilon$s with dotted indices are  antisymmetric under interchange of their indices.   

We could of course think of shortening  (\ref{eq:psizet}) further to $\psi\cdot \zeta$ or $\psi \zeta$, 
but without the dotted indices to tell us, we wouldn't in general know whether such expressions 
referred to what we have been calling $\psi$- or $\chi$-type spinors. So it is common to find people 
using a `\ $\bar{}$\ ' notation for $\psi$-type spinors. Then (\ref{eq:psizet}) would be just $\bar{\psi} 
\bar{\zeta}$.

As in the previous Aside, we can ask how (in terms of $V$) the downstairs dotted spinor 
$-{\rm i} \sigma_2 \psi$ transforms. 

{\bf Exercise} Show that $-{\rm i} \sigma_2 \psi$ transforms by $V^{-1 {\rm T}}$, and hence 
verify once again that $(-{\rm i} \sigma_2 \psi)^{\rm T} \zeta$ is invariant. 

So altogether we have arrived at four types of two-component spinor: upstairs  and 
downstairs $\chi$-type,  which transform by $V^*$ and $V^{-1 \dagger}$ respectively; 
and upstairs and downstairs $\psi$-type which transform by $V$ and $V^{-1 {\rm T}}$ 
respectively. The essential point is that invariants are formed by taking the matrix 
dot product between one quantity transforming by $M$ say, and another transforming by 
$M^{-1 {\rm T}}$.  

In the notation of this and the previous Aside, then, the familiar Dirac 4-component spinor 
(\ref{eq:diracpsi}) would be written as 
\be 
\Psi= \left( \ba{c} \psi^{\dot{a}}\\\chi_{a} \ea \right).
\ee
The conventions of different authors typically do not agree here. As far as I can tell, 
the notation I am using is the same as that of Shifman \cite{shif}, see his equation (68) on page 335. 
Other authors, for example Bailin and Love \cite{bailin}, use a choice for the Dirac matrices 
which is different from (\ref{eq:albet}) 
and (\ref{eq:gammas}), and which has the effect of interchanging the position, in 
$\Psi$, of the L (undotted)  and R (dotted)  parts - which, furthermore, they 
call `$\psi$' and `$\chi$' respectively, the opposite way round from us -  so that for them  
\be 
\Psi_{{\rm D}}=\left( \ba{c}
\psi_a \\ \chi^{\dot{a}} \ea \right).
\ee
Bailin and Love also employ the `\ $\bar{\mbox{}}$ \ ' notation, so that  
\be 
\Psi_{{\rm BL}}= \left( \ba{c} \psi_a \\ {\bar{\chi}}^{\dot{a}} \ea \right). 
\ee
{\em Note particularly, however, that this `bar' has nothing to do with the `bar' 
used in 4-component Dirac theory, as in (\ref{eq:invt}).} Also, BL's $\epsilon$ 
symbols, and hence their spinor scalar products,  have the opposite sign from   ours.

}}  
\vspace{.2in}

\subsection{Majorana fermions} 

We stated in (\ref{eq:chipsi}) that $\chi_{\psi}\equiv {\rm i} \sigma_2 \psi^*$ transforms 
like a $\chi$-type object. It follows 
that it should be perfectly consistent with Dirac theory to assemble $\psi$ and $\chi_{\psi}$ into a 4-component 
object:
\be 
\Psi^{\psi}_{{\rm M}}=\left( \ba{c} \psi\\-{\rm i} \sigma_2 \psi^*\ea \right).\label{eq:maj}
\ee
This must behave under Lorentz transformations just like an `ordinary' Dirac 4-component object $\Psi$ built 
from a $\psi$ and a $\chi$. But $\Psi^{\psi}_{{\rm M}}$ of  (\ref{eq:maj}) has {\em fewer degrees of 
freedom} than an ordinary Dirac 4-component spinor $\Psi$, since it is fully determined by the 2-component 
object $\psi$. In a Dirac spinor $\Psi$ involving a $\psi$ and a $\chi$, 
as in (\ref{eq:diracpsi}), there are two 2-component spinors, 
each of which is specified by 4 real quantities (each has two complex components), making 8 in all.  In 
$\Psi^{\psi}_{{\rm M}}$, by contrast, there are only 4 real quantities, contained in the single 
spinor $\psi$. 

What this means physically becomes clearer when we consider the operation of {\em charge conjugation}. 
On a Dirac 4-component spinor, this is defined by 
\be 
\Psi_{{\bf C}} = C_0 \Psi^* \label{eq:Cdef}
\ee
where\footnote{This choice of $C_0$ has the opposite sign from the one in equation (20.63) of \cite{AH32}
page 290; the present choice is more in conformity with SUSY conventions. We are sticking to the 
convention that the indices of the $\gamma$- matrices as defined in (\ref{eq:gammas}) appear upstairs; 
no significance should be attached to the position of the indices of the $\sigma$-matrices - it is 
common to write them downstairs.} 
\be
C_0=-{\rm i} \gamma^2 = \left( \ba{cc} 0 & {\rm i}\sigma_2\\-{\rm i} \sigma_2 &0 \ea \right).
\ee 
Then 
\be 
\Psi^{\psi}_{{\rm M},{\bf C}} = \left( \ba{cc}0&{\rm i}\sigma_2\\-{\rm i} \sigma_2 & 0 \ea \right) \left( \ba{c} \psi^* 
\\ -{\rm i} \sigma_2 \psi \ea \right) = \left( \ba{c}\psi\\-{\rm i} \sigma_2 \psi^* \ea \right) = 
\Psi^{\psi}_{{\rm M}}.\label{eq:PsiC=PsiM}
\ee
So $\Psi^{\psi}_{{\rm M}}$ describes a spin-1/2 particle which is even under charge-conjugation - that is, 
it is its own antiparticle. Such a particle is called a Majorana fermion. 

This charge-self-conjugate property is clearly the physical reason for the difference in the 
number of degrees of freedom in $\Psi^{\psi}_{{\rm M}}$ as compared with $\Psi$ of (\ref{eq:dirac}). There are 
4 physically distinguishable modes in a Dirac field, for example ${\rm e}^-_{\rm L}, {\rm e}^-_{\rm R}, 
{\rm e}^+_{\rm L}, {\rm e}^+_{\rm R}$. But   in a Majorana field one there are only two, the antiparticle 
being the same as the particle;  for example $\nu_{\rm L}, \nu_{\rm R}$ - supposing, as is possible 
(see \cite{AH32} section 20.6), that neutrinos are Majorana particles. 

We could also construct 
\be 
\Psi^\chi_{{\rm M}} = \left( \ba{c} {\rm i} \sigma_2 \chi^* \\
\chi \ea \right) \label{eq:chimaj}
\ee
which also satisfies 
\be 
\Psi^{\chi}_{{\rm M}, {\bf C}} = \Psi^{\chi}_{{\rm M}}.
\ee

Clearly a formalism using $\chi$'s only is equivalent to one using 
$\Psi^{\chi}_{{\rm M}}$'s only, and one using $\psi$'s is equivalent to 
one using $\Psi^{\psi}_{{\rm M}}$'s. A mass term of the form `$\bar{\Psi} \Psi$' 
would now be, for instance, 
\be
{\bar{\Psi}}^{\chi}_{{\rm M}} \Psi^{\chi}_{{\rm M}}= (({\rm i} \sigma_2 \chi^*)^\dagger \chi^\dagger) 
\left( \ba{cc} 0&1\\1&0 \ea \right) \left( \ba{c} {\rm i} \sigma_2 \chi^* \\ \chi \ea \right) = 
\chi^{{\rm T}}(-{\rm i} \sigma_2) \chi + \chi^\dagger ({\rm i} \sigma_2) \chi^*.\label{eq:majmass1}
\ee
The first term on the RHS of the last equality in (\ref{eq:majmass1}) we have seen 
before in (\ref{eq:chiinvt}); the second is also a possible Lorentz invariant 
formed from $\chi$'s.\footnote{Here's a useful check on why. We know from (\ref{eq:psichi1}) that 
${\rm i} \sigma_2 \chi^*$ transforms under Lorentz transformations like a $\psi$-type thing, 
which is to say it transforms by the matrix $V$ of (\ref{eq:V}). And we also know from 
(\ref{eq:psipchip}) that $\chi$ transforms by the matrix $V^{-1 \dagger}$. Hence 
$\chi^\dagger ({\rm i} \sigma_2) \chi^* \to \chi^\dagger V^{-1} V ({\rm i} \sigma_2) \chi^* = 
\chi^\dagger ({\rm i} \sigma_2) \chi^*$.}

Similarly, a  mass term made from $\Psi^{\psi}_{{\rm M}}$ would be 
\be 
 {\bar{\Psi}}^{\psi}_{{\rm M}} \Psi^{\psi}_{{\rm M}}= (\psi^\dagger (-{\rm i} \sigma_2 \psi^*)^\dagger) 
\left( \ba{cc} 0&1\\1&0 \ea \right) \left( \ba{c} \psi \\-{\rm i} \sigma_2 \psi^* \ea \right) = 
\psi^{\dagger}(-{\rm i} \sigma_2) \psi^* + \psi^{{\rm T}} ({\rm i} \sigma_2) \psi.\label{eq:majmass2}
\ee
Again, we have seen the second term on the RHS of the last equality in (\ref{eq:majmass2}) before, 
and the first is also a Lorentz invariant formed from $\psi$ (from (\ref{eq:chipsi}), it transforms 
as a `$\psi^\dagger \chi$' object, which we know from (\ref{eq:psichiinvt}) is invariant). 
Note that all the terms in 
(\ref{eq:majmass1}) and (\ref{eq:majmass2}) would vanish if the field components did not 
anticommute.

We can similarly consider the Lorentz-invariant product of two different Majorana spinors 
$\Psi_{1{\rm M}}$ and $\Psi_{2{\rm M}}$, namely 
\be 
{\bar{\Psi}}_{1{\rm M}} \Psi_{2{\rm M}}=\Psi^\dagger_{1{\rm M}}\beta \Psi_{2{\rm M}}. 
\ee
But equations (\ref{eq:Cdef}) and (\ref{eq:PsiC=PsiM})  tell us that 
\be 
\Psi_{1{\rm M}}=-{\rm i} \gamma^2 \Psi^*_{1{\rm M}},
\ee
and hence 
\be 
\Psi^\dagger_{1{\rm M}}=\Psi^{\rm T}_{1{\rm M}}(-{\rm i} \gamma^2)
\ee
using $\gamma^{2 \dagger}=-\gamma^2$. It follows that 
\be 
\Psi^\dagger_{1{\rm M}} \beta \Psi_{2{\rm M}}=\Psi^{\rm T}_{1{\rm M}}(-{\rm i} \gamma^2 \beta) \Psi_{2{\rm M}}.
\label{eq:majmet}
\ee
The matrix 
\be
-{\rm i} \gamma^2 \beta = \left( \ba{cc} {\rm i} \sigma_2 & 0 \\ 0& -{\rm i} \sigma_2 \ea \right) 
\ee
therefore acts as a metric in forming the dot product of the two $\Psi_{\rm M}$'s. It is easy to 
check that (\ref{eq:majmet}) is the same as (\ref{eq:majmass1}) when $\Psi_{1{\rm M}}=\Psi_{2{\rm M}}=\Psi^\chi_{\rm M}$, 
and the same as (\ref{eq:majmass2}) when  $\Psi_{1{\rm M}}=\Psi_{2{\rm M}}=\Psi^\psi_{\rm M}$.

\subsection{Dirac fermions using $\chi$- (or L-) type spinors only}

We noted at the beginning of Section 2 that the simplest SUSY theory 
(which is just around the corner now) involves a complex scalar field and 
a two-component spinor field. This is in fact the archetype of SUSY models 
leading to the MSSM (Minimal Supersymmetric [version of the] Standard Model). 
By convention, one uses $\chi$-type spinors, i.e. (see section 2.1) L-type 
spinors, no doubt because the V-A structure of the electroweak sector of 
the SM distinguishes the L parts of the fields, and one might as well 
 give them a privileged status. But of course there are the R parts as 
well. In a SUSY context, it is very convenient to be able to use only one 
kind of spinors, which in the MSSM is (for the reason just outlined) going to 
be L-type ones - but in that case how are we going to deal with the R parts 
of the SM fields? 

Consider for example the electron field which we write as 
\be 
\Psi^{({\rm e}^-)} = \left( \ba{c} \psi^{({\rm e}^-)}_{\rm R} \\ 
\chi^{({\rm e}^-)}_{\rm L} \ea \right).\label{eq:psiel}
\ee
Instead of using the  {\em right-handed electron} field in the top 2 components, 
we can just as well use the {\em charge conjugate of the left-handed positron} field. That is, 
instead of (\ref{eq:psiel}) we choose to write 
\be 
\Psi^{({\rm e})} = \left( \ba{c} {\rm i} \sigma_2 \chi^{({\rm e}^+)*}_{\rm L} \\
\chi^{({\rm e}^-)}_{\rm L}\ea \right).\label{eq:psipos}
\ee
A commonly used notation is to write 
\be 
\chi^{({\rm e}^+){\rm c}}_{\rm L} \equiv {\rm i} \sigma_2 \chi^{({\rm e}^+)*}_{\rm L},
\ee
or, more compactly, $e^{+{\rm c}}_{\rm L}$, accompanying the L-type electron field $e^-_{\rm L}$.

Our previous work guarantees, of course,  that the Lorentz transformation character 
  of (\ref{eq:psipos}) is OK. In terms of the choice (\ref{eq:psipos}), a mass term 
  for a (non-Majorana!) Dirac fermion is (omitting now the `L' subscripts from 
  the $\chi$'s)  
  \bea 
  {\bar{\Psi}}^{({\rm e})} \Psi^{({\rm e})} = {\Psi^{({\rm e})}}^\dagger 
 \left(  \ba{cc} 0&1\\1&0 \ea \right) \Psi^{({\rm e})}
  &=& (({\rm i} \sigma_2 \chi^{({\rm e}^+) })^{\rm T}  \chi^{({\rm e}^-) \dagger})
  \left( \ba{c} \chi^{({\rm e}^-)} \\ {\rm i} \sigma_2 \chi^{({\rm e}^+)*} \ea \right) 
  \nonumber \\
  &=& \chi^{({\rm e}^+) } \cdot \chi^{({\rm e}^-)}+
  \chi^{({\rm e}^-)\dagger} {\rm i} \sigma_2 \chi^{({\rm e}^+)*}. \label{eq:diracmass}
  \eea
In the first term on the RHS of (\ref{eq:diracmass}) we have used the quick `dot' 
notation for two $\chi$-type spinors introduced in Aside (1); see Notational 
Aside (3) for a similar treatment of the second term.  
So the `Dirac' mass has here been re-written wholly in terms of two L-type 
spinors, one associated with the ${\rm e}^-$ mode, the other with the ${\rm e}^+$ mode.

\vspace{.2in}
{\footnotesize{{\bf Notational Aside (3)} Readers who have patiently ploughed through 
Asides (1) and (2) may be beginning to think we have now got altogether too many 
different kinds of spinor in play. We previously agreed that we'd identified four 
kinds of spinor: $\chi_a$ and $\chi^a$ transforming by $V^{-1 \dagger}$ and $V^*$ 
respectively, and $\psi^{\dot{a}}$ and $\psi_{\dot{a}}$ transforming by $V$ and 
$V^{-1 {\rm T}}$. Surely $\chi^*_a$ can't be yet another kind? 
Indeed, since $\chi_a$ transforms by $V^{-1 \dagger}$, it follows that 
$\chi^*_a$ transforms by the complex conjugate of this, which is $V^{-1 {\rm T}}$. But this 
is exactly how a `$\psi_{\dot{a}}$' (or a `${\bar{\psi}}_{\dot{a}}$', using the bar notation for 
the dotted spinor) transforms. So it is consistent to {\em define} 
\be 
{\bar{\chi}}_{\dot{a}} \equiv \chi^*_a
\ee
and then raise the lower dotted index with the matrix ${\rm i} \sigma_2$, using the 
inverse of (\ref{eq:updowndots}) (remember, once we have the dotted 
indices, or the bar, to tell us what kind of spinor it is, we no longer 
care what letter we use!). Then the second term of (\ref{eq:diracmass}) becomes 
\be 
\chi^{({\rm e}^-) * {\rm T}} {\rm i} \sigma_2 \chi^{({\rm e}^+) *} = 
{\bar{\chi}}^{({\rm e}^-)}_{\dot{a}} {\bar{\chi}}^{({\rm e}^+)\dot{a}} =
{\bar{\chi}}^{({\rm e}^-)} \cdot {\bar{\chi}}^{({\rm e}^+)}.
\ee
{\bf Exercise} (a) What is $\bar{\chi}\cdot \bar{\xi}$ in terms of $\bar{\xi} \cdot \bar{\chi}$ 
(assuming the components anticommute)? (b) What is ${\bar{\chi}}_{{\dot{a}}}{\bar{\xi}}^{{\dot{a}}}$ 
in terms of ${\bar{\chi}}^{{\dot{a}}}{\bar{\xi}}_{{\dot{a}}}$? Do these by components and by using $\epsilon$ 
symbols.  
}}
\vspace{.2in}  

Now, at last, we are ready to take our first steps in SUSY. 

\section{A Simple Supersymmetric Lagrangian} 

In this section we'll look at  one of the simplest supersymmetric theories, 
one involving just two free fields: a complex spin-0 field $\phi$ and an L-type 
spinor field $\chi$ , both massless. 
 The Lagrangian (density) for this system is 
 \be 
 {\cal{L}} = \partial_\mu \phi^\dagger \partial^\mu \phi + 
 \chi^\dagger {\rm i} {\bar{\sigma}}^\mu \partial_\mu \chi.\label{eq:Lphichi}
 \ee    
The $\phi$ part is familiar from introductory qft courses: the $\chi$ bit is just the 
appropriate part of the Dirac Lagrangian (\ref{eq:Ldirac}). The equation of motion for 
$\phi$ is of course $\Box \phi=0$, while that for $\chi$ is 
${\rm i}{\bar{\sigma}}^\mu \partial_\mu \chi =0$ (compare (\ref{eq:Dchisig})). We are 
going to try and find, by `brute force', transformations in which the  change in 
$\phi$ is proportional to $\chi$ (as in (\ref{eq:phipsi})), and the  change in $\chi$ is 
proportional to $\phi$, such that ${\cal{L}}$ is invariant.\footnote{Actually we shan't 
succeed: instead, we have to settle for the Action to be invariant, which means that 
${\cal{L}}$ changes by a total derivative; it turns out that this has to do with the 
`mis-match' in the number of degrees of freedom (off-shell) in $\phi$ and $\chi$.}

As a preliminary, it is useful to get the {\em dimensions} of everything straight. The 
Action is the integral of the density ${\cal{L}}$ over all 4-dimensional space, and 
is dimensionless in units $\hbar = c= 1$. In this system, there is only one independent 
dimension left, which we take to be that of mass (or energy), M (see Appendix B of 
\cite{AH31}). Length has the same dimension as time (because $c=1$), and both 
have the dimension of ${\rm M}^{-1}$ (because $\hbar=1$). It follows that, for 
the Action to be dimensionless, 
${\cal{L}}$ has dimension ${\rm M}^4$. Since the gradients have dimension M, we can then 
read off the dimensions of $\phi$ and $\chi$ (denoted by $[\phi]$ and $[\chi]$):
\be 
[\phi]={\rm M} \ \ \ \  [\chi] = {\rm M}^{3/2}.
\ee

Now, what are the SUSY transformations linking $\phi$ and $\chi$? Several 
considerations can guide us to, if not the answer, then at least a good guess. Consider 
the  change in $\phi, \ \delta_\xi \phi$,  first where $\xi$ is a constant ($x$-independent) 
parameter. This has the form (already stated in (\ref{eq:phipsi})) 
\be 
`\mbox{ change in $\phi$ = parameter $ \xi \times$ other field $\chi$'}.
\ee 
On the LHS, we have a spin-0 field, which is invariant under Lorentz transformations. 
So we must construct a Lorentz invariant out of $\chi$ and the parameter $\xi$. One 
simple way to do this is to declare that $\xi$ is also a $\chi$- (or L-) type spinor, and 
use the invariant product (\ref{eq:chiscpr}). This gives 
\be 
\delta_\xi \phi = \xi^{\rm T} (-{\rm i} \sigma_2) \chi, \label{eq:deltaphi}
\ee
or in the notation of Aside (1)
\be 
\delta_\xi \phi =\xi^a \chi_a = \xi \cdot \chi.\label{eq:delphiprod}
\ee   

It is worth pausing to note some things about the parameter $\xi$. First, we repeat 
that it is a spinor. It doesn't depend on $x$, but  it is not an invariant under 
Lorentz transformations: it transforms as a $\chi$-type spinor, i.e. by $V^{-1\dagger}$. It has 
two components, of course, each of which is complex - hence 4 real numbers in all. These 
specify the transformation (\ref{eq:deltaphi}). Secondly, although $\xi$  doesn't 
depend on $x$, and isn't a field (operator) in that sense, we shall assume that its 
components {\em anticommute} with the components of spinor fields - that is, we assume 
they are Grassmann numbers (see \cite{AH32} Appendix O). Lastly, 
since $[\phi]={\rm M}$ and 
$[\chi]={\rm M}^{3/2}$, to make the dimensions balance on both sides of (\ref{eq:deltaphi}) 
we need to assign the dimension 
\be 
[\xi]={\rm M}^{-1/2} \label{eq:dimxi}
\ee
to $\xi$.  

Now let's think what the corresponding $\delta_\xi \chi$ might be. This has to be 
something like 
\be
\delta_\xi \chi \sim \mbox{product of $\xi$ and $\phi$}.\label{eq:deltachi1}
\ee
Now, on the LHS of (\ref{eq:deltachi1}) we have a quantity with dimensions ${\rm M}^{3/2}$, 
while on the RHS the algebraic product of $\xi$ and $\phi$ has dimensions ${\rm M}^{-1/2 +1} 
={\rm M}^{1/2}$. Hence we need to introduce something with dimensions ${\rm M}^1$ on the 
RHS. In this massless theory, there is only one possibility - the gradient operator 
$\partial_\mu$, or more conveniently the momentum operator ${\rm i} \partial_\mu$. But now 
we have a `loose' index $\mu$ on the RHS! The LHS is a spinor, and there is a spinor ($\xi$) 
also on the RHS, so we should probably get rid of the $\mu$ index altogether, by contracting 
it. We try 
\be 
\delta_\xi \chi =  ({\rm i} \sigma^\mu   \partial_\mu \phi)\  \xi \label{eq:deltachi2}
\ee
where $\sigma^\mu$ is given in (\ref{eq:sigsigbar}). Note that the $2 \times 2$ matrices 
in $\sigma^\mu$ act on the 2-component column $\xi$ to give, correctly, a 2-component column 
to match the LHS. But although both sides of (\ref{eq:deltachi2}) are 2-component column vectors, 
the RHS does not transform as a $\chi$-type spinor. If we look back at (\ref{eq:Dpsisig}) and 
(\ref{eq:Dchisig}),  we see that the combination $\sigma^\mu \partial_\mu$ acting on a $\psi$ 
transforms as a $\chi$ (and ${\bar{\sigma}}^\mu \partial_\mu$ on a $\chi$ transforms as  a $\psi$). 
 So we must let the $\sigma^\mu \partial_\mu$ in (\ref{eq:deltachi2}) act on a $\psi$-like thing, 
 not on $\xi$, in order to get something transforming as a $\chi$. But we know how to manufacture a 
 $\psi$-like thing out of $\xi$! We just take (see (\ref{eq:psichi1})) ${\rm i} \sigma_2 \chi^*$. We 
 therefore arrive at the guess
 \be 
 \delta_\xi \chi_a = A [{\rm i} \sigma^\mu({\rm i} \sigma_2 \xi^*)]_a \partial_\mu \phi \label{eq:deltachi}
 \ee
 where $A$ is some constant to be determined from the condition that ${\cal{L}}$ is invariant 
 under (\ref{eq:deltaphi}) and (\ref{eq:deltachi}), and 
 we have indicated the $\chi$-type spinor index on both sides. Note that `$\partial_\mu \phi$' 
 has no matrix structure and has been moved to the end. 
 
 Equations (\ref{eq:deltaphi}) and (\ref{eq:deltachi}) give the proposed SUSY 
 transformations for $\phi$ and $\chi$, but both are complex fields and we 
 need to be clear what the corresponding transformations are for their 
 Hermitian conjugates $\phi^\dagger$ and $\chi^\dagger$. There are 
 some notational concerns here which we shall not put in small print. First, 
 remember that $\phi$ and $\chi$ are quantum fields, even though we are not 
 explicitly putting hats on them; on the other hand, $\xi$ is not a field 
 (it's $x$-independent). In the discussion of Lorentz transformations of spinors 
 in Section 2, we used the symbol $\mbox{}^*$ to denote complex conjugation, 
 it being tacitly understood that we were dealing with wave functions rather 
 than field operators. But consider the (quantum) field $\phi$ with a mode 
 expansion 
 \be 
 \phi = \int \frac{{\rm d}^3{\bm k}}{(2 \pi)^3 \sqrt{2\omega}}\ [ a(k){\rm e}^{-{\rm i} k \cdot x} 
 + b^\dagger(k) {\rm e}^{{\rm i} k \cdot x} ].\label{eq:modephi}
 \ee   
 Here the operator $a(k)$ destroys (say) a particle with 4-momentum $k$, and 
 $b^\dagger(k)$ creates an anti-particle of 4-momentum $k$, 
 while ${\rm exp}[\pm{\rm i} k \cdot x]$ are of course ordinary wavefunctions.  For (\ref{eq:modephi}) the 
 simple complex conjugation $\mbox{}^*$ is not appropriate, since `$a^*(k)$' is not 
 defined; instead, we want `$a^\dagger(k)$'. So instead of `$\phi^*$' we  deal with 
 $\phi^\dagger$, which is defined in terms of (\ref{eq:modephi}) by (a) taking the  
 complex conjugate of the wavefunction parts and (b) taking the dagger of the mode 
 operators. This gives 
 \be 
 \phi^\dagger = \int \frac{{\rm d}^3{\bm k}}{(2 \pi)^3 \sqrt{2\omega}}\ 
 [ a^\dagger (k){\rm e}^{{\rm i} k \cdot x} 
 + b(k) {\rm e}^{-{\rm i} k \cdot x} ],\label{eq:modephiconj}
 \ee 
  the conventional definition of the Hermitian conjugate of (\ref{eq:modephi}).

For spinor fields like $\chi$, on the other hand,  the situation is slightly 
more complicated, since now in the analogue of (\ref{eq:modephi}) the scalar 
(spin-0) wavefunctions ${\rm exp}[\pm {\rm i} k \cdot x]$ will be replaced by 
(free-particle) 2-component spinors. Thus, symbolically, the first (upper) 
component of the quantum field $\chi$ will have the form 
\be 
\chi_1 \sim \mbox{mode operator $\times$ first component of free-particle spinor of $\chi$-type}
\ee  
where we are of course using the `downstairs, undotted' notation for the components of $\chi$. 
In the same way as (\ref{eq:modephiconj}) we then define 
\be 
\chi_1^\dagger \sim \mbox{(mode operator)$^\dagger$ $\times$ (
first component of free-particle spinor)$^*$}.
\ee

With this in hand, let's consider the Hermitian conjugate of (\ref{eq:deltaphi}), that is 
$\delta_\xi \phi^\dagger$. Written out in terms of components (\ref{eq:deltaphi}) is 
\be 
\delta_\xi \phi=(\xi_1 \xi_2)\left( \ba{cc} 0&-1\\1&0 \ea \right) \left( \ba{c} \chi_1\\ \chi_2 \ea \right) = 
-\xi_1\chi_2+\xi_2\chi_1.
\ee
We want to take the `dagger' of this - but we are now faced with a decision about 
how to take the dagger of products of (anticommuting) spinor components, like 
$\xi_1\chi_2$. In the case of two matrices $A$ and $B$, we know that $(AB)^\dagger=
B^\dagger A^\dagger$. By analogy, we shall {\em define} the dagger to reverse the 
order of the spinors:  
\be 
\delta_\xi \phi^\dagger = -\chi_2^\dagger \xi_1^* + \chi_1^\dagger \xi_2^*; \label{eq:delphidag1}
\ee
 $\xi$ isn't a quantum field and the `$\mbox{}^*$' notation is OK for it. But (\ref{eq:delphidag1}) 
can be written in more compact form:
\bea
\delta_\xi \phi^\dagger &=& \chi_1^\dagger \xi_2^* - \chi_2^\dagger \xi_1^* \nonumber \\
&=& (\chi_1^\dagger \chi_2^\dagger) \left( \ba{cc} 0&1\\-1&0 \ea \right) \left( \ba{c} \xi_1^* \\
\xi_2^* \ea \right) \nonumber \\
&=& \chi^\dagger ({\rm i} \sigma_2) \xi^*, \label{eq:deltaphidag}
\eea
where in the last line the $\mbox{}^\dagger$ symbol, as applied to the 
two-component spinor field $\chi$, is understood in a matrix sense as well: that is 
\be
\chi^\dagger=\left( \ba{c} \chi_1 \\
\chi_2 \ea \right)^\dagger = (\chi_1^\dagger \chi_2^\dagger). \label{eq:chidagger}
\ee
Equation (\ref{eq:deltaphidag}) is a satisfactory outcome of these rather fiddly 
considerations because (a) we have seen exactly this spinor structure before, 
in (\ref{eq:diracmass}), and we are assured its Lorentz transformation character is OK, and  
(b) it is nicely consistent with `naively' taking the dagger of (\ref{eq:deltaphi}), 
treating it like a matrix product. In particular, the RHS of the last line of 
(\ref{eq:deltaphidag}) can be written in the notation of Aside (3) as $\bar{\chi} \cdot 
\bar{\xi}$ or equally, using the Exercise in Aside (3), as  $\bar{\xi} \cdot \bar{\chi}$. 
Referring to (\ref{eq:delphiprod}) we therefore note the useful result 
\be 
(\xi \cdot \chi)^\dagger = (\chi \cdot \xi)^\dagger = 
\bar{\xi} \cdot \bar{\chi} = \bar{\chi} \cdot \bar{\xi}.\label{eq:hermprod}
\ee

In the same way, therefore, we can take the dagger of (\ref{eq:deltachi}) to obtain 
\be
\delta_\xi \chi^\dagger= A \partial_\mu \phi^\dagger \xi^{\rm T} {\rm i} \sigma_2 {\rm i} \sigma^\mu,
\label{eq:deltachidag}
\ee
where for later convenience we have here moved the $\partial_\mu \phi^\dagger$ to the front, 
and we have taken $A$ to be real (which will be sufficient, as we'll see).  
We are now ready to see if we can choose $A$ so as to make ${\cal{L}}$ invariant under (\ref{eq:deltaphi}), 
(\ref{eq:deltachi}), (\ref{eq:deltaphidag}) and (\ref{eq:deltachidag}).

We have 
\bea
\delta_\xi {\cal{L}} &=& \partial_\mu (\delta_\xi \phi^\dagger) \partial^\mu \phi + \partial_\mu \phi^\dagger 
\partial^\mu (\delta_\xi \phi) 
 +(\delta_\xi \chi^\dagger) {\rm i} {\bar{\sigma}}^\mu \partial_\mu \chi + 
\chi^\dagger {\rm i} {\bar{\sigma}}^\mu \partial(\delta_\xi \chi) \nonumber \\
&=&\partial_\mu (\chi^\dagger {\rm i} \sigma_2 \xi^*) \partial^\mu \phi + \partial_\mu \phi^\dagger 
\partial^\mu(\xi^{\rm T} (-{\rm i} \sigma_2) \chi) \nonumber 
\eea
\be
+A (\partial_\mu \phi^\dagger \xi^{\rm T} {\rm i} \sigma_2 {\rm i} \sigma^\mu) {\rm i} 
{\bar{\sigma}}^\nu \partial_\nu \chi +A \chi^\dagger {\rm i} {\bar{\sigma}}^\nu \partial_\nu({\rm i} 
\sigma^\mu {\rm i} \sigma_2 \xi^*) \partial_\mu \phi. \label{eq:deltaL1}
\ee  
Inspection of (\ref{eq:deltaL1}) shows that there are two types of term, one involving the parameters  
$\xi^*$ and the other the parameters $\xi^{\rm T}$. Consider the term involving $A\xi^*$. 
In it there appears the combination (pulling $\partial_\mu$ through the constant $\xi^*$) 
\be 
{\bar{\sigma}}^\nu \partial_\nu \sigma^\mu \partial_\mu=(\partial_0+{\bm \sigma} \cdot {\bm \nabla})
(\partial_0-{\bm \sigma} \cdot {\bm \nabla}) = \partial_0^2-{\bm \nabla}^2=\partial_\mu \partial^\mu.
\label{eq:sigsigbar1}
\ee
We can therefore combine this and the other term in $\xi^*$ from (\ref{eq:deltaL1}) to give 
\be
\delta_\xi {\cal{L}}|_{\xi^*}= \partial_\mu \chi^\dagger {\rm i} \sigma_2 \xi^* \partial^\mu \phi - 
{\rm i} A \chi^\dagger \partial_\mu \partial^\mu \sigma_2 \xi^* \phi.\label{eq:deltaL2}
\ee
This represents a change in ${\cal{L}}$ under our transformations, so it 
seems we have not succeeded in finding an invariance (or symmetry), since we 
cannot hope to cancel this change against the term involving $\xi^{\rm T}$, which 
involves quite independent parameters. However, we 
must remember that the Action is the space-time integral of ${\cal{L}}$, and this will 
be invariant if we can arrange for the change in ${\cal{L}}$ to be a 
{\em total derivative} (assuming as usual that the expression obtained by 
integrating it vanishes at the boundaries of space-time). Since $\xi$ does not depend 
on $x$, we can indeed write (\ref{eq:deltaL2}) as a total derivative 
\be 
\delta_\xi {\cal{L}}|_{\xi^*}= \partial_\mu(\chi^\dagger {\rm i} 
\sigma_2 \xi^* \partial^\mu \phi) \label{eq:delLxi*}
\ee 
provided that 
\be 
A=-1.\label{eq:A}
\ee
Similarly, if $A=-1$ the terms in $\xi^{\rm T}$ combine to give 
\be 
\delta_\xi{\cal{L}}|_{\xi^{\rm T}}= \partial_\mu \phi^\dagger \partial^\mu (\xi^{\rm T} (-{\rm i}\sigma_2) \chi) + 
\partial_\mu \phi^\dagger 
\xi^{\rm T} {\rm i} \sigma_2 \sigma^\mu {\bar{\sigma}}^\nu \partial_\nu \chi.\label{eq:B}
\ee
The second  term in (\ref{eq:B}) we can write as 
\be 
\partial_\mu ( \phi^\dagger \xi^{\rm T} {\rm i} \sigma_2 \sigma^\mu {\bar{\sigma}}^\nu \partial_\nu \chi) 
+ \phi^\dagger \xi^{\rm T} (-{\rm i} \sigma_2) \sigma^\mu {\bar{\sigma}}^\nu \partial_\mu \partial_\nu \chi \label{eq:C}
\ee
\be
=\partial_\mu ( \phi^\dagger \xi^{\rm T} {\rm i} \sigma_2 \sigma^\mu {\bar{\sigma}}^\nu \partial_\nu \chi) 
+\phi^\dagger \xi^{\rm T} (- {\rm i} \sigma_2) \partial_\mu \partial^\mu \chi. \label{eq:D}
\ee
The second term of (\ref{eq:D}) and the first term of (\ref{eq:B}) now combine to give the total 
derivative 
\be
\partial_\mu (\phi^\dagger \xi^{\rm T} (-{\rm i} \sigma_2) \partial^\mu \chi),
\ee 
so that finally 
\be 
\delta_\xi {\cal{L}}|_{\xi^{\rm T}}=\partial_\mu (\phi^\dagger \xi^{\rm T} (-{\rm i} \sigma_2) \partial^\mu \chi)
+\partial_\mu ( \phi^\dagger \xi^{\rm T} {\rm i} \sigma_2 \sigma^\mu {\bar{\sigma}}^\nu \partial_\nu \chi),
\ee
which is also a total derivative. In summary, we have shown that under (\ref{eq:deltaphi}), (\ref{eq:deltachi}), 
(\ref{eq:deltaphidag}) and (\ref{eq:deltachidag}), with $A=-1$, ${\cal{L}}$ changes by a total 
derivative:
\be
\delta_\xi {\cal{L}}= \partial_\mu(\chi^\dagger {\rm i} \sigma_2 \xi^* \partial^\mu \phi + 
\phi^\dagger \xi^{\rm T}(-{\rm i} \sigma_2) \partial^\mu \chi + \phi^\dagger \xi^{\rm T} 
{\rm i} \sigma_2 \sigma^\mu {\bar{\sigma}}^\nu \partial_\nu \chi) \label{eq:delL1}
\ee
and the Action is therefore invariant: we have a SUSY theory, in this sense. As we 
shall see in Section 6, the pair ($\phi$, spin-0) and ($\chi$, L-type spin-1/2) constitute 
a {\em chiral supermultiplet} in SUSY.

{\bf Exercise} Show that (\ref{eq:delL1}) can also be written as 
\be 
\delta_\xi {\cal{L}} = \partial_\mu(\chi^\dagger {\rm i} \sigma_2 \xi^* \partial^\mu \phi +
\xi^{\rm T} {\rm i} \sigma_2 \sigma^\nu {\bar{\sigma}}^\mu \chi \partial_\nu \phi^\dagger +
\xi^{\rm T} (-{\rm i} \sigma_2) \chi \partial^\mu \phi^\dagger). \label{eq:delL2}
\ee 

The reader may well feel that it's been pretty heavy going, considering 
especially the simplicity, triviality almost, of the Lagrangian (\ref{eq:Lphichi}). A more 
professional notation would have been more efficient, of course, but there is a lot 
to be said for doing it the most explicit and straightforward way, first time 
through. As we proceed, we shall speed up the notation. In fact, interactions 
don't constitute an order of magnitude increase in labour, and the manipulations 
gone through in this simple example are quite representative.

\section{A First Glance at the MSSM} 

Before ploughing on with more formal work, let's consider how the SUSY idea might 
relate to particle physics. All we have so far, of course, is 1 complex scalar 
field and one L-type fermion field, and they aren't even interacting. All the same, 
let's see how we might apply it to physics. One important point to realise is 
that SUSY transformations do not act on the SU(3)$_{\rm c}$, SU(2)$_{{\rm L}}$ or 
U(1)$_{{\rm em}}$ degrees of freedom. Consider for example the left-handed 
lepton fields, e.g. the electron one $e_{\rm L}$. This is in an SU(2)$_{\rm L}$ 
doublet, the partner field being $\nu_{{\rm e L}}$:
\be 
\left( \ba{c} \nu_{\rm e  L}\\
e_{\rm L}\ea \right). \label{eq:lepdoub}
\ee
These need to be partnered, in a SUSY theory, by spin-0 bosons forming another 
SU(2)$_{\rm L}$ doublet, presumably. Indeed there is such a doublet in the SM, 
the Higgs doublet 
\be 
\left( \ba{c} \phi^+ \\
\phi^0 \ea \right) \label{eq:higgs} 
\ee 
or its charge-conjugate doublet 
\be 
\left( \ba{c} {\bar{\phi}}^0 \\ \phi^- \ea \right).\label{eq:chiggs}
\ee
But these Higgs doublets don't carry lepton number (which we shall assume to be 
conserved), and we can't have some particles in a symmetry (SUSY) multiplet 
carrying a conserved quantum number, and others not. So we seem to need {\em new} 
particles to go with our doublet (\ref{eq:lepdoub}): 
\be 
 \left( \ba{c} \nu_{\rm e L} \\ e_{\rm L}\ea \right) \ \mbox{partnered by} 
 \ \left( \ba{c} {\tilde{\nu}}_{\rm e  L}\\
{\tilde{e}}_{\rm L}\ea \right)
\ee
where `${\tilde{\nu}}$' is a scalar partner for the neutrino (`sneutrino'), 
and `${\tilde{e}}$' is a scalar partner for the electron (`selectron'). Similarly, 
we'd have smuons and staus, and their sneutrinos. These are all in chiral supermultiplets, 
and SU(2)$_{\rm L}$ doublets. 

What about quarks? They are a triplet of the SU(3)$_{\rm c}$ colour gauge group, 
and no other SM particles are colour triplets. So we will need new (scalar) partners 
for the quarks too, called squarks, which are colour triplets, and also in chiral 
supermultiplets.  

The electroweak interactions of both leptons and quarks are `chiral', which means 
that the `L' parts of the fields interact differently from the `R' parts. The L 
parts belong to SU(2)$_{\rm L}$ doublets, as above, while the R parts are SU(2)$_{\rm L}$ 
singlets. So we need to arrange for scalar partners for the L and R parts separately: for 
example  ($e_{\rm R}, {\tilde{e}}_{\rm R}$),  
($u_{\rm R}, {\tilde{u}}_{\rm R}$), ($d_{\rm R}, {\tilde{d}}_{\rm R}$), etc; and 
\be 
\left( \ba{c}u_{\rm L}\\ d_{\rm L}\ea\right), \ \left( \ba{c} {\tilde{u}}_{\rm L} \\ {\tilde{d}}_{\rm L} 
\ea \right)
\ee
and so on.\footnote{As noted in Section 2.4, the `particle R-parts' will actually be 
represented by the charge conjugates of the `antiparticle L-parts'.} 

We haven't yet learned about SUSY for spin-1 fields, but we shall see in Section 10 that 
there is a {\em vector} (or {\em gauge}) supermultiplet, which associates a massless 
vector field (which has two on-shell degrees of freedom) with an L fermion, the latter 
being called generically a `gaugino'. Once again, the gauge group quantum numbers for the 
gauginos have to be the same as for the gauge bosons - i.e. we need a colour octet of 
`gluinos' to supersymmetrize QCD, plus an SU(2)$_{\rm L}$ triplet of -inos and a 
U(1)$_{\rm em}$ -ino for the SUSY electroweak theory. After SU(2)$_{\rm L}$ 
symmetry-breaking (a la Higgs) we'll have three fermionic partners for the ${\rm W}^{\pm}, {\rm Z}^0$, 
namely ${\tilde{\rm W}}^{\pm}, {\tilde{Z}}^0$, and the photino ${\tilde{\gamma}}$.

Finally, the Higgs sector: we haven't been able to partner the Higgs doublets with 
any known fermion, so they need their own `higgsinos', i.e. fermionic analogues 
forming chiral supermultiplets. In fact, a crucial consequence of making the SM 
supersymmetric, in the MSSM, is - as we shall see in Section 8  - that {\em two 
separate} Higgs doublets are required: whereas in the SM itself the doublet (\ref{eq:chiggs}) 
can be satisfactorily represented as the charge conjugate of the doublet (\ref{eq:higgs}), 
this is not possible in the SUSY version. So we need 
\be 
H_{\rm u}: \ \ \left( \ba{c} H^+_{\rm u} \\ H^0_{\rm u} \ea \right), \ \ \left( \ba{c}
{\tilde{H}}^+_{\rm u}\\ {\tilde{H}}^0_{\rm u} \ea \right)
\ee
and 
\be 
H_{\rm d} : \ \ \left( \ba{c} H^0_{\rm d} \\ H^-_{\rm d} \ea \right), \ \ \left( \ba{c} 
{\tilde{H}}^0_{\rm d} \\ {\tilde{H}}^-_{\rm d} \ea \right).
\ee

The chiral and gauge supermultiplets introduced here constitute the `minimal' extension of the SM 
field content which is required to make it supersymmetric. The full theory, including supersymmetric 
interactions, is called the minimal supersymmetric standard model (MSSM). It has been 
around for over 20 years: early reviews are given  in  \cite{nilles} and \cite{HK}; 
a more recent and very helpful `supersymmetry primer' was provided by Martin \cite{martin}, 
to which we shall make quite frequent reference in what follows.  A recent and very comprehensive review 
may be found in \cite{CEKKLW}.

We'll return to the MSSM in Section 12. For the moment, we should simply note that (a) none 
of the `superpartners' has yet been seen experimentally, in particular they certainly cannot 
have the same mass as their SM partner states (as would normally be expected for a symmetry 
multiplet), so that (b) SUSY - as applied in the MSSM - must be broken somehow. We'll include 
a brief discussion of SUSY breaking in section 15, but a more detailed treatment is well beyond  
the scope of these lectures. 

\section{Towards a Supersymmetry Algebra}

A fundamental aspect of any symmetry (other than a U(1) symmetry) is the {\em algebra} 
associated with the {\em symmetry generators} - see for example Appendix M of 
\cite{AH32}. For example, the generators $T_i$ of SU(2) satisfy the commutation 
relations 
\be 
[T_i, T_j]={\rm i} \epsilon_{i j k} T_k \label{eq:SU(2)}
\ee
where $i, j$ and $k$ run over the values 1, 2 and 3, and where the repeated 
index $k$ is summed over; $\epsilon_{i j k}$ is the totally antisymmetric symbol 
such that $\epsilon_{123}= +1, \ \epsilon_{213}=-1$, etc. The 
commutation relations summarized in  (\ref{eq:SU(2)}) constitute the `SU(2) alegra', and it 
is of course exactly that of the angular momentum operators in quantum mechanics, 
in units $\hbar=1$. Readers will be familiar with the way in which the whole 
theory of angular momentum in quantum mechanics can be developed just from 
these commutation relations. In the same way, in order to proceed in a reasonably 
systematic way with SUSY, we must know what the SUSY algebra is. 
In Section 1.3, we introduced the idea of generators of 
SUSY transformations, $Q_a$, and their associated algebra - 
which now involves anticommutation relations - was roughly indicated in (\ref{eq:QQP}). 
The purpose of this section is to find the actual SUSY algebra, by a `brute force' 
method once again, making use of what we have learned in Section 2. 

\subsection{One way of obtaining the SU(2) algebra}
    
 In Section 2, we arrived at recipes for SUSY transformations of spin-0 
 fields $\phi$ and $\phi^\dagger$, and spin-1/2 fields $\chi$ and $\chi^\dagger$. 
 From these transformations, the algebra of the SUSY generators can be deduced. 
 To understand the method, it is helpful to see it in action in a more 
 familiar context, namely that of SU(2), as we now discuss. Readers should skip 
 this subsection if they've seen it all before. 
 
 Consider an SU(2) doublet of fields 
 \be 
 q=\left( \ba{c} u \\ d \ea \right) 
 \ee
 where $u$ and $d$ have equal mass, and identical interactions, so that the 
 Lagrangian is invariant under (infinitesimal) transformations of the form (see 
 for example equation (12.95) of \cite{AH32})
 \be 
 q \to q' = (1- {\rm i} {\bm \epsilon} \cdot {\bm \tau}/2) q \equiv q + 
 \delta_{\bm \epsilon} q \label{eq:qSU(2)}
 \ee
 where 
 \be 
 \delta_{{\bm \epsilon}} q = - {\rm i} {\bm \epsilon} \cdot {\bm \tau}/2 \ q.\label{eq:deltaq}
 \ee
 Here, as usual, the three matrices ${\bm \tau}=(\tau_1, \tau_2, \tau_3)$ are the 
 same as the Pauli ${\bm \sigma}$ matrices, and ${\bm \epsilon}=(\epsilon_1, 
 \epsilon_2, \epsilon_3)$ are three real infinitesimal parameters specifying the 
 transformation. The transformed fields $q'$ satisfy the same anticommutations 
 relations as the original fields $q$, so that $q'$ and $q$ are related by a 
 unitary transformation 
 \be 
 q'=U q U^\dagger. \label{eq:SU(2)U}
 \ee
 For infinitesimal transformations, $U$ has the general form 
 \be 
 U_{\rm infl}=(1+{\rm i} {\bm \epsilon} \cdot {\bm T})\label{eq:SU(2)inf}
 \ee
 where 
 \be 
 {\bm T}=(T_1, T_2, T_3)
 \ee
 are the {\em generators} of infinitesimal SU(2) transformations; the unitarity 
 of $U$ implies that the ${\bm T}$'s are Hermitian. For infinitesimal transformations, 
 therefore, we have (from (\ref{eq:SU(2)U}) and (\ref{eq:SU(2)inf})) 
 \bea 
 q'&=& (1+{\rm i} {\bm \epsilon} \cdot {\bm T}) q (1-{\rm i} {\bm \epsilon} \cdot {\bm T})
 \nonumber \\
 &=& q  + {\rm i} {\bm \epsilon} \cdot {\bm T} q - {\rm i} {\bm \epsilon} \cdot q {\bm T} 
 \ \ \mbox{to first order in ${\bm \epsilon}$} \nonumber \\
 &=& q + {\rm i} {\bm \epsilon} \cdot [{\bm T}, q] 
 \eea
 Hence from (\ref{eq:qSU(2)}) and (\ref{eq:deltaq}) we deduce (see equation (12.100) of 
 \cite{AH32}) 
 \be
 \delta_{\bm \epsilon} q = {\rm i} {\bm \epsilon} \cdot [{\bm T}, q]=-
 {\rm i} {\bm \epsilon} \cdot {\bm \tau}/2 \ q, \label{eq:delqboth}
 \ee
 It is important to realise that the ${\bm T}$'s are themselves quantum 
 field operators, constructed from the fields of the Lagrangian; for example 
 in this simple case they would be 
 \be 
 {\bm T} = \int q^\dagger ({\bm \tau}/2) q \ {\rm d}^3 {\bm x} \label{eq:SU(2)qs}
 \ee
 as explained for example in section 12.3 of \cite{AH32}. 
 
 Given an explicit formula for the generators, such as (\ref{eq:SU(2)qs}), we can 
 proceed to calculate the commutation relations of the ${\bm T}$'s, knowing 
 how the $q$'s anticommute. The answer is that the ${\bm T}$'s obey the 
 relations (\ref{eq:SU(2)}). However, there is another way to get these 
 commutation relations, just by considering the small changes in the fields, as  
 given by (\ref{eq:delqboth}). Consider two such transformations 
 \be 
 \delta_{\epsilon_1} q = {\rm i} \epsilon_1 [T_1, q] = -{\rm i} \epsilon_1 (\tau_1/2) q
 \label{eq:delq1}
 \ee
 and 
 \be  
 \delta_{\epsilon_2} q = {\rm i} \epsilon_2 [T_2, q] = -{\rm i} \epsilon_2 (\tau_2/2) q .
 \label{eq:delq2}
 \ee
 We shall calculate the {\em difference} $(\delta_{\epsilon_1}\delta_{\epsilon_2} - 
 \delta_{\epsilon_2} \delta_{\epsilon_1}) q$ two different ways: first via the second 
 equality in (\ref{eq:delq1}) and (\ref{eq:delq2}), and then via the first equalities. Equating 
 the two results will lead us to the algebra (\ref{eq:SU(2)}).

  First, then, we use the second equality of (\ref{eq:delq1}) and (\ref{eq:delq2}) to obtain  
  \bea 
  \delta_{\epsilon_1}\delta_{\epsilon_2} q &=& \delta_{\epsilon_1}\{-{\rm i} 
  \epsilon_2 (\tau_2/2\} q \nonumber \\
  &=& -{\rm i} \epsilon_2 (\tau_2/2) \delta_{\epsilon_1}q \nonumber \\
  &=& -{\rm i} \epsilon_2 (\tau_2/2). -{\rm i} \epsilon_1 (\tau_1/2) q \nonumber \\
  &=& -(1/4) \epsilon_1 \epsilon_2 \tau_2 \tau_1 q.
  \eea
 Note that in the last line we have changed the order of the $\epsilon$ parameters as we 
 are free to do since they are ordinary numbers, but we cannot alter the order of the 
 $\tau$'s since they are matrices which don't commute. Similarly,
 \bea
   \delta_{\epsilon_2}\delta_{\epsilon_1} q &=& \delta_{\epsilon_2}\{-{\rm i} 
  \epsilon_1 (\tau_1/2\} q  \nonumber \\
   &=& -{\rm i} \epsilon_1 (\tau_1/2) \delta_{\epsilon_2}q \nonumber \\
   &=& -(1/4) \epsilon_1 \epsilon_2 \tau_1 \tau_2 q.
   \eea
   Hence 
   \bea 
   (\delta_{\epsilon_1}\delta_{\epsilon_2} - 
 \delta_{\epsilon_2} \delta_{\epsilon_1}) q &=& \epsilon_1 \epsilon_2 [\tau_1/2, \tau_2/2] q 
 \nonumber \\
 &=& \epsilon_1 \epsilon_2 {\rm i} (\tau_3/2) q \nonumber \\
 &=& -{\rm i} \epsilon_1 \epsilon_2 [T_3, q] \label{eq:t3q} 
 \eea
 where the second line follows from the fact that the $\tau$'s, as matrices, satisfy 
 the algebra (\ref{eq:SU(2)}), and the third line results from the `$\mbox{}_3$' analogue of 
 (\ref{eq:delq1}) and (\ref{eq:delq2}).
 
 Now we calculate $  (\delta_{\epsilon_1}\delta_{\epsilon_2} - 
 \delta_{\epsilon_2} \delta_{\epsilon_1}) q$ using the first equality of (\ref{eq:delq1}) and 
 (\ref{eq:delq2}). We have 
 \bea
 \delta_{\epsilon_1} \delta_{\epsilon_2} q &=& \delta_{\epsilon_1} \{{\rm i} \epsilon_2 [T_2, q]\} \nonumber \\
   &=& {\rm i} \epsilon_2 \delta_{\epsilon_1} \{[T_2,q]\} \nonumber \\
   &=& {\rm i} \epsilon_1 {\rm i} \epsilon_2 [T_1, [T_2, q]].
   \eea
 Similarly, 
 \be
  \delta_{\epsilon_2} \delta_{\epsilon_1} q ={\rm i} \epsilon_1 {\rm i} \epsilon_2 [T_2, [T_1, q]].
  \ee
 Hence 
 \be
 (\delta_{\epsilon_1}\delta_{\epsilon_2} - 
 \delta_{\epsilon_2} \delta_{\epsilon_1}) q = - \epsilon_1 \epsilon_2 \{ [T_1, [T_2, q]]-[T_2,[T_1,q]]\}.
 \label{eq:t1t2q}
 \ee
 Now we can rearrange the RHS of this equation by using the identity (which is easily checked by 
 multiplying it all out) 
 \be 
 [A, [B,C]] + [B, [C,A]] + [C, [A, B]] =0.\label{eq:jac}
 \ee
 We first write 
 \be 
 [T_2, [T_1, q]] = -[T_2, [q, T_1]]
 \ee 
 so that the two double commutators in (\ref{eq:t1t2q}) become 
 \be  
 [T_1,[T_2, q]] -[T_2, [T_1, q]] = [T_1, [T_2, q]] + [T_2, [q, T_1]] = -[q, [T_1, T_2]]
 \label{eq:comred}
 \ee
 where the last step follows by use of (\ref{eq:jac}). Finally, then,  (\ref{eq:t1t2q}) can 
 be written as 
 \be 
 (\delta_{\epsilon_1}\delta_{\epsilon_2} - 
 \delta_{\epsilon_2} \delta_{\epsilon_1}) q = - \epsilon_1 \epsilon_2 [[T_1, T_2],q] \label{eq:deldel3}
 \ee
 which can be compared with (\ref{eq:t3q}). We deduce 
 \be
 [T_1, T_2] = {\rm i} T_3 \label{eq:SU(2)p}
 \ee
 exactly as stated in (\ref{eq:SU(2)}).
 
  This is the method we shall use to find the SUSY 
 algebra, at least as far as it concerns the transformations for scalar and spinor fields 
 found in Section 2. 
 
 \subsection{Supersymmetry generators (`charges') and their algebra}
 
 In order to apply the preceding method, we need the SUSY analogue of 
 (\ref{eq:delqboth}). Equations (\ref{eq:deltaphi}) and (\ref{eq:deltachi}) (with $A=-1$) 
 provide us with the analogue of the second equality in (\ref{eq:delqboth}), for $\delta_\xi \phi$ 
 and for $\delta_\xi \chi$; what about the first? We want to write something like 
 \be 
 \delta_\xi \phi \sim {\rm i} [\xi  Q, \phi] = \xi^{\rm T} (-{\rm i} \sigma_2) \chi. \label{eq:delphiboth1}
 \ee
 where $Q$ is a SUSY generator. 
  In the first (tentative) equality in (\ref{eq:delphiboth1}), we must remember that 
  $\xi$ is a $\chi$-type spinor quantity, and so it is clear that  
  $Q$ must be a spinor quantity also, or else one side of the equality would be 
  bosonic and the other fermionic. In fact, since $\phi$ is a Lorentz scalar, we 
  must combine $\xi$ and $Q$ into a Lorentz invariant. Let us suppose that $Q$ 
  transforms as a $\chi$-type spinor also: then we know that $\xi^{\rm T} (-{\rm i} 
  \sigma_2) Q$ is Lorentz invariant. So we shall write  
  \be 
  \delta_\xi \phi= {\rm i} [\xi^{\rm T} (-{\rm i} \sigma_2) Q,\phi]=\xi^{\rm T}
   (-{\rm i} \sigma_2) \chi \label{eq:deltaphiboth}
   \ee 
  or in the faster notation of Aside (1) 
  \be 
  \delta_\xi \phi = {\rm i} [\xi \cdot Q, \phi] = \xi \cdot \chi.
  \ee
  We are going to calculate $(\delta_\eta \delta_\xi - \delta_\xi \delta_\eta) \phi$, so 
  (since $\delta \phi \sim \chi$) we shall  need (\ref{eq:deltachi}) as well. This involves 
  $\xi^*$, so to get the complete analogue of `${\rm i} {\bm \epsilon} \cdot {\bm T}$'
   we shall need 
  to extend `${\rm i} \xi \cdot  Q$' to 
  \be 
  {\rm i}(\xi^{\rm T} (-{\rm i} \sigma_2) Q + \xi^\dagger ({\rm i} \sigma_2) Q^*) =
  {\rm i} (\xi \cdot Q + \bar{\xi} \cdot \bar{Q}).
  \ee
  
  We first calculate $(\delta_\eta \delta_\xi - \delta_\xi \delta_\eta) \phi$ using 
  (\ref{eq:deltaphi}) and (\ref{eq:deltachi}) (with $A=-1$):
  \bea
  (\delta_\eta \delta_\xi - \delta_\xi \delta_\eta) \phi &=& \delta_\eta (\xi^{\rm T} (-{\rm i} 
  \sigma_2 \chi) - (\eta \leftrightarrow \xi) \nonumber \\
  &=& \xi^{\rm T} (-{\rm i} \sigma_2) {\rm i} \sigma^\mu (-{\rm i} 
  \sigma_2) \eta^* \partial_\mu \phi - (\eta \leftrightarrow \xi) \nonumber \\
  &=& (\xi^{\rm T} c  \sigma^\mu c \eta^* - 
  \eta^{\rm T} c  \sigma^\mu c \xi^*) {\rm i} \partial_\mu \phi, \label{eq:deldelphi}
    \eea 
  where (none too soon) we have introduced the notation 
  \be 
  c \equiv -{\rm i} \sigma_2 = \left( \ba{cc} 0&-1\\1&0 \ea \right).
  \ee
  (\ref{eq:deldelphi}) is sometimes written more compactly by  using 
  \be 
  c \sigma^\mu c  = - {\bar{\sigma}}^{\mu {\rm T}} \label{eq:csigc}
  \ee
   (see (\ref{eq:sigsigbar}) for the definition of $\sigma^\mu$ and ${\bar{\sigma}}^\mu$). 
  Now  $\xi^{\rm T} {\bar{\sigma}}^{\mu {\rm T}} \eta^*$ is a single quantity (row 
  vector times matrix times column vector) so it must equal its formal transpose, 
  apart from a minus sign due to interchanging the order of anticommuting variables.\footnote{
  Check this statement by looking at $(\eta^{\rm T} (-{\rm i} \sigma_2) \xi)^{\rm T}$, for 
  instance.} Hence 
  \be 
  (\delta_\eta \delta_\xi - \delta_\xi \delta_\eta) \phi=  ( \eta^\dagger 
  {\bar{\sigma}}^\mu \xi - \xi^\dagger {\bar{\sigma}}^\mu \eta) {\rm i} \partial_\mu \phi.
  \ee

   On the other hand, we also have 
   \be 
   \delta_\xi \phi = {\rm i} [\xi \cdot Q + {\bar{\xi}} \cdot {\bar{Q}}, \phi]
   \ee
   and so 
   \be 
   (\delta_\eta \delta_\xi - \delta_\xi \delta_\eta) \phi = - 
   \{ [\eta \cdot Q + {\bar{\eta}}\cdot {\bar{Q}},[\xi \cdot Q + 
   {\bar{\xi}} \cdot {\bar{Q}}, \phi]] - [\xi \cdot Q + 
   {\bar{\xi}} \cdot {\bar{Q}}, [ \eta \cdot Q + {\bar{\eta}} \cdot {\bar{Q}}, \phi]] \}.
   \label{eq:deldel2}
   \ee
   Just as in (\ref{eq:comred}), the RHS of (\ref{eq:deldel2}) can be rearranged using 
   (\ref{eq:jac}) and we obtain
   \bea 
   [[\eta \cdot Q + {\bar{\eta}} \cdot {\bar{Q}}, \xi \cdot Q + {\bar{\xi}} \cdot 
   {\bar{Q}}], \phi] &=& (\eta^{\rm T} c \sigma^\mu c \xi^* - \xi^{\rm T} c \sigma^\mu c 
   \eta^*) {\rm i} \partial_\mu \phi \nonumber \\
   &=& -(\eta^{\rm T} c \sigma^\mu c \xi^* - \xi^{\rm T} c \sigma^\mu c 
   \eta^*) [P_\mu , \phi] \ \label{eq:dblecom}
   \eea 
   where in the last step we have introduced the 4-momentum operator $P_\mu$, 
   which is also the generator of translations, such that 
   \be 
   [P_\mu, \phi] = -{\rm i} \partial_\mu \phi \label{eq:pmuphi}
   \ee
   (we shall recall the proof of this equation in section 9 - see (\ref{eq:transphi})).
   
   It is {\em tempting} now to conclude that, just as in going from 
   (\ref{eq:t3q}) and (\ref{eq:deldel3}) to (\ref{eq:SU(2)p}), we can infer from 
   (\ref{eq:dblecom}) the result 
   \be 
   [\eta \cdot Q + {\bar{\eta}} \cdot {\bar{Q}}, \xi \cdot Q + {\bar{\xi}} \cdot 
   {\bar{Q}}] = -(\eta^{\rm T} c \sigma^\mu c \xi^* - \xi^{\rm T} c \sigma^\mu c 
   \eta^*) P_\mu.\label{eq:QQP1}
   \ee
   But we have, so far, only established the RHS of (\ref{eq:dblecom}) by considering 
   the difference $\delta_\eta \delta_\xi - \delta_\xi \delta_\eta$ acting on $\phi$ 
   (see (\ref{eq:deldelphi})). 
   Is it also true that   
   \be 
    (\delta_\eta \delta_\xi -\delta_\xi \delta_\eta) \chi  = 
    (\xi^{\rm T} c \sigma^\mu c \eta^* - \eta^{\rm T} c \sigma^\mu c 
   \xi^*) {\rm i} \partial_\mu \chi \ ? \label{eq:deldelchi}
   \ee
    Unfortunately, the answer to this is {\em no}, as we shall see in Section 7, where 
   we shall also learn how to repair the situation. For the moment, we proceed on the 
   basis of (\ref{eq:QQP1}). 
   
   In order to obtain, finally, the (anti)commutation relations of the $Q$'s from 
   (\ref{eq:QQP1}), we need to get rid of the parameters $\eta$ and $\xi$ on both 
   sides. First of all, we note that since the RHS of (\ref{eq:QQP1}) has no terms in 
   $\eta\ldots\xi$ or $\eta^* \ldots \xi^*$ we can deduce
   \be 
   [\eta \cdot Q, \xi \cdot Q] = [{\bar{\eta}} \cdot {\bar{Q}}, {\bar{\xi}} \cdot {\bar{Q}}]=0.\label{eq:comQQbar}
   \ee
   The first commutator is 
   \bea 
   0=(\eta^1 Q_1 + \eta^2 Q_2)(\xi^1 Q_1 + \xi^2 Q_2) - (\xi^1 Q_1 + \xi^2 Q_2)(\eta^1 Q_1 + \eta^2 Q_2)  && 
   \nonumber \\
   =-\eta^1\xi^1(2Q_1Q_1)-\eta^1\xi^2(Q_1Q_2+Q_2Q_1)&&\nonumber \\
   -\eta^2\xi^1(Q_2Q_1+Q_1Q_2) -\eta^2\xi^2(2Q_2Q_2),&&
   \eea 
   remembering that all quantities anticommute. Since all these combinations of parameters are 
   independent, we can deduce
   \be 
   \{Q_a, Q_b\}=0, \label{eq:QQcom}
   \ee
   and similarly 
   \be 
   \{ Q^*_a, Q^*_b\}=0.\label{eq:QQstarcom}
   \ee
   Notice how, when the anti-commuting quantities $\xi$ and $\eta$ are `stripped away' from 
   the $Q$ and $\bar{Q}$, the commutators in (\ref{eq:comQQbar}) become {\em anti}-commutators in 
   (\ref{eq:QQcom}) and (\ref{eq:QQstarcom}).  
   
   Now let's look at the $[\eta \cdot Q, {\bar{\xi}} \cdot {\bar{Q}}]$ term in (\ref{eq:QQP1}). 
   Writing everything out long-hand, we have 
   \be 
   {\bar{\xi}} \cdot {\bar{Q}} = \xi^\dagger {\rm i} \sigma_2 Q^* = \xi^*_1Q^*_2-\xi^*_2Q^*_1
   \ee
   and
   \be 
   \eta \cdot Q = -\eta_1Q_2+\eta_2Q_1.
   \ee
   So 
   \bea
   [\eta \cdot Q, {\bar{\xi}} \cdot {\bar{Q}}]= \eta_1\xi^*_1(Q_2Q_2^*+Q_2^*Q_2) - 
   \eta_1\xi^*_2(Q_2Q^*_1+Q^*_1Q_2)  &&\nonumber\\
   -\eta_2\xi^*_1(Q_1Q^*_2+Q^*_2Q_1)+\eta_2\xi^*_2(Q_1Q^*_1+Q^*_1Q_1).\label{eq:QQ1}
   \eea
   Meanwhile, the RHS of (\ref{eq:QQP1}) is 
   \bea
   -(\eta_1 \eta_2) \left( \ba{cc}0&-1\\ 1&0 \ea \right) \sigma^\mu \left( \ba{cc}0&-1 \\ 1&0 \ea \right) 
   \left( \ba{c} \xi^*_1 \\ \xi^*_2 \ea \right) P_\mu  && \nonumber \\
   = -(\eta_2 -\eta_1) \sigma^\mu \left( \ba{c} -\xi^*_2 \\ \xi^*_1 \ea \right) P_\mu && \nonumber \\
   =[\eta_2\xi^*_2(\sigma^\mu)_{11} - \eta_2\xi^*_1(\sigma^\mu)_{12} - \eta_1\xi^*_2(\sigma^\mu)_{21} 
   + \eta_1\xi^*_1(\sigma^\mu)_{22}]\ P_\mu, \label{eq:QQRHS}
   \eea
   where the subscripts on the matrices $\sigma^\mu$ denote the particular 
   element of the matrix, as usual. Comparing (\ref{eq:QQ1}) and (\ref{eq:QQRHS}) we deduce 
   \be 
   \{Q_a, Q^*_b\} = (\sigma^\mu)_{ab} P_\mu.\label{eq:QQstar}
   \ee
   We have been writing $Q^*$ throughout, like $\xi^*$ and $\eta^*$, but 
   the $Q$'s are quantum field operators and so (in accord with the discussion 
   in section 3) we should more properly write (\ref{eq:QQstar}) as 
   \be 
   \{ Q_a, Q^\dagger_b \} = (\sigma^\mu)_{ab} P_\mu. \label{eq:QQstar1}
   \ee
   Once again, the commutator in (\ref{eq:QQ1}) has led to  an anticommutator in (\ref{eq:QQstar1}).

  Equation (\ref{eq:QQstar1}) is the main result of this section, and is a most 
  important equation;
  it provides the `proper' version of (\ref{eq:QQP}). Although we have 
  derived it by our customary brute force methods as applied to a particular 
  (and very simple) case, it must be emphasized that equation (\ref{eq:QQstar1}) is 
  indeed the correct SUSY algebra  (up to normalization 
  conventions\footnote{Many authors normalize the SUSY charges 
  differently, so that they get a `2' on the RHS. For completeness, we take the opportunity of this 
  footnote to mention that more general SUSY algebras also exist, in which the single generator 
  $Q_a$ is replaced by $N$ generators $Q^A_a (A=1, 2, ..... N)$. Equation  (\ref{eq:QQstar1}) is 
  then replaced by $\{Q_a^A,Q_b^{B\dagger}\}=\delta^{AB} (\sigma^\mu)_{ab} P_\mu $. The more 
  significant change occurs in the commutator (\ref{eq:QQcom}), which becomes $\{Q_a^A, Q_b^B\}= 
  \epsilon_{ab} Z^{AB}$, where $\epsilon_{12}=-1, \epsilon_{21}=+1, \epsilon_{11}=\epsilon_{22}=0$ 
  and the `central charge' $Z^{AB}$ is antisymmetric under $A \leftrightarrow B$.  
    See footnote  9 for why only the $N=1$ case 
  seems to have any immediate physical relevance.}). Equation (\ref{eq:QQstar1}) shows   
  (to repeat what was said in Section 1) that the SUSY generators are directly 
  connected to the energy-momentum operator, which is the generator of space-time 
  displacements. So it is justified to regard SUSY as some kind of extension of 
  space-time symmetry. We shall see further aspects of this in section 9. 
  
  We note finally that the commutator of two $P$'s is zero (translations 
  commute), and that the commutator of a $Q$ and a $P$ also vanishes, since 
  the $Q$'s are independent of $x$. So all the commutation relations between 
  $Q$'s, $Q^\dagger$'s, and $P$'s are now defined, and they involve only these 
  quantities; we say that `the supertranslation algebra is closed'. 
  
   \vspace{.2in}
 {\footnotesize{{\bf Appendix to Section 5: The Supersymmetry Current} 
 
 In the case of ordinary 
 symmetries, the invariance of a Lagrangian under a transformation of the fields (characterised 
 by certain parameters) implies the existence of a  4-vector $j^\mu$ (the `symmetry current'), 
 which is conserved: $\partial_\mu j^\mu=0$. The generator of the symmetry is the `charge' associated 
 with this current, namely the spatial integral of $j^0$. An expression for $j^\mu$ is easily found 
 (see for example \cite{AH32} section 12.3.1). Suppose the Lagrangian ${\cal{L}}$ is invariant under the 
 transformation 
 \be 
 \phi_r \to \phi_r + \delta \phi_r
 \ee
 where `$\phi_r$' stands generically for any field in ${\cal{L}}$, 
 having several components labelled by $r$.  Then 
 \be 
 0=\delta {\cal{L}} = \frac{\partial {\cal{L}}}{\partial \phi_r} \delta \phi_r + \frac{\partial {\cal{L}}} 
 {\partial(\partial^\mu \phi_r)} \partial^\mu(\delta \phi_r) + \mbox{hermitian conjugate}.\label{eq:delsymL}
 \ee 
 But the equation of motion for $\phi_r$ is 
 \be 
 \frac{\partial {\cal{L}}}{\partial \phi_r} = \partial_\mu \left( \frac{\partial {\cal{L}}}
 {\partial(\partial_\mu \phi_r)} \right).\label{eq:EL}
 \ee
 Using (\ref{eq:EL}) in (\ref{eq:delsymL}) yields 
 \be 
 \partial_\mu j^\mu = 0 
 \ee
 where 
 \be 
 j^\mu= \frac{\partial {\cal{L}}}{\partial(\partial_\mu \phi_r)} \delta \phi_r + 
 \mbox{hermitian conjugate}. \label{eq:jdef}
 \ee
 For example, consider the Lagrangian 
 \be 
 {\cal{L}}= {\bar{q}}({\rm i} {\not \! \partial}-m ) q \label{eq:SU(2)L}
 \ee
 where 
 \be 
 q=\left( \ba{c} u \\ d \ea \right).
 \ee 
 This is invariant under the SU(2) transformation (\ref{eq:deltaq}), which is 
 characterised by three independent infinitesimal parameters, so there are three 
 independent symmetries, three currents, and three generators (or charges). Consider 
 for instance a transformation involving $\epsilon_1$ alone. Then 
 \be 
 \delta q = -{\rm i} \epsilon_1 (\tau_1/2) q,\label{eq:delq3}
 \ee
 while from (\ref{eq:SU(2)L}) we have 
 \be 
 \frac{\partial {\cal{L}}}{\partial(\partial_\mu q)}=\bar{q} {\rm i} \gamma^\mu.
 \ee
 Hence from (\ref{eq:jdef}) and (\ref{eq:delq3}) we obtain the corresponding current as 
 \be 
 \epsilon_1 \bar{q}\gamma^\mu (\tau_1/2) q.
 \ee
 Clearly the constant factor $\epsilon_1$ is irrelevant and can be dropped. Repeating the 
 same steps for transformations associated with $\epsilon_2$ and $\epsilon_3$ we deduce 
 the existence of the {\em isospin currents} 
 \be 
 {\bm j}^\mu = \bar{q} \gamma^\mu ({\bm \tau}/2) q
 \ee
 and charges (generators) 
 \be 
 {\bm T} = \int q^\dagger ({\bm \tau}/2) q \ {\rm d}^3 {\bm x} 
 \ee
 just as stated in (\ref{eq:SU(2)qs}).

 We can apply the same procedure to find the {\em supersymmetry curent} associated 
 with the supersymmetry exhibited by the simple model considered in section 3. However, 
 there is an important difference between this example and the SU(2) model just considered: 
 in the latter, the Lagrangian is indeed invariant under the transformation (\ref{eq:qSU(2)}), 
 but in the SUSY case we were only able to ensure that the Action was invariant, the 
 Lagrangian changing by a total derivative, as given in (\ref{eq:delL1}) or (\ref{eq:delL2}). In 
 this case, the `0' on the LHS of (\ref{eq:delsymL}) must be replaced by $\partial_\mu K^\mu$ say, 
 where $K^\mu$ is the expression in brackets in (\ref{eq:delL1}) or (\ref{eq:delL2}). 
 
 Furthermore, since the SUSY charges are spinors $Q_a$, we anticipate that the associated 
 currents carry a spinor index too, so we write them as $J^\mu_a$, where $a$ is a spinor index. 
 These will be associated with transformations characterised by the usual spinor parameters 
 $\xi$. Similarly, there will be the Hermitian conjugate currents associated with the parameters 
 $\xi^*$. 
 
 Altogether, then, we can write (forming Lorentz invariants in the now familiar way)  
 \bea 
 \xi^{\rm T} (-{\rm i} \sigma_2) J^\mu  + \xi^\dagger {\rm i} \sigma_2 J^{\mu \dagger} &=& 
 \frac{\partial {\cal{L}}}{\partial(\partial_\mu \phi)}\delta \phi + \delta \phi^\dagger 
 \frac{\partial{\cal{L}}}{\partial(\partial_\mu \phi^\dagger)} +
 \frac{\partial {\cal{L}}}{\partial(\partial_\mu \chi)} \delta \chi -K^\mu.\nonumber \\
 &=& \partial^\mu \phi^\dagger \xi^{\rm T} (-{\rm i} \sigma_2) \chi + \chi^\dagger {\rm i} \sigma_2 
 \xi^* \partial^\mu \phi + \chi^\dagger {\rm i} {\bar{\sigma}}^\mu (-{\rm i} \sigma^\nu) {\rm i} 
 \sigma_2 \xi^* \partial_\nu \phi \nonumber \\
  &&- (\chi^\dagger {\rm i} \sigma_2 \xi^* \partial^\mu \phi + \xi^{\rm T} {\rm i} \sigma_2 
 \sigma^\nu {\bar{\sigma}}^\mu \chi \partial_\nu \phi^\dagger + \xi^{\rm T} (-{\rm i} \sigma_2) \chi 
 \partial^\mu \phi^\dagger) \nonumber \\
 &=& \chi^\dagger {\bar{\sigma}}^\mu \sigma^\nu {\rm i} \sigma_2 \xi^* \partial_\nu \phi + 
 \xi^{\rm T} (-{\rm i} \sigma_2) \sigma^\nu {\bar{\sigma}}^\mu \chi \partial_\nu \phi^\dagger,  
 \eea 
 whence we read off the SUSY current as  
 \be 
 J^\mu =  \sigma^\nu {\bar{\sigma}}^\mu \chi \partial_\nu \phi^\dagger 
 \label{eq:susyj}.
 \ee
 As expected, this current has two spinorial components, and it contains an unpaired 
 fermionic operator $\chi$. 
  
 }}

  \section{Supermultiplets}  
  
  We proceed to extract some physical consequences of (\ref{eq:QQstar1}). In a theory 
  which is supersymmetric, the operators $Q$ - being generators of the symmetry - will 
  commute with the Hamiltonian $H$:
  \be 
  [Q_a, H] = [Q^\dagger_a, H] =0.
  \ee
  So acting on one state of mass $M$ the $Q$'s will create another state also of mass 
  $M$, but since they are spinor operators this other state will not have the same 
  spin $j$ as the first. In fact, we know that under rotations (compare equation 
  (\ref{eq:delqboth}) for the case of isospin rotations, 
  and equations (\ref{eq:deltaphi3}) and (\ref{eq:transdiff}) below for spatial translations)   
  \be 
  \delta Q = -({\rm i} {\bm \epsilon} \cdot {\bm \sigma}/2) Q = {\rm i} {\bm \epsilon} \cdot [{\bm J},Q],
  \ee
  where the $J$'s are the generators of rotations (i.e. angular momentum operators). 
  For example, for a rotation about the 3-axis, 
  \be
  -\frac{1}{2} \sigma_3 Q = [J_3, Q],
  \ee
  which implies that 
  \be 
  [J_3, Q_1] = -\frac{1}{2}Q_1, \ \ \ [J_3, Q_2]=  \frac{1}{2} Q_2. 
  \ee
  It follows that if $|jm\rangle$ is a spin-$j$ state with $J_3=m$, then 
  \be 
  (J_3 Q_1 - Q_1 J_3) | jm \rangle = -\frac{1}{2} Q_1 | jm \rangle,
  \ee
  whence 
  \be 
  J_3 (Q_1 |jm \rangle)= (m-\frac{1}{2}) Q_1 | jm \rangle,
  \ee
  showing that $Q_1 |jm\rangle$ has $J_3=m-\frac{1}{2}$ - 
  that is, $Q_1$ lowers the $M$-value by $\frac{1}{2}$ (like an annihilation 
  operator for a `u'-state). Similarly, 
  $Q_2$ raises it by $\frac{1}{2}$ (like an annihilation operator for a `d'-state).  Likewise, since 
  \be 
  [J_3, Q^\dagger_1] = \frac{1}{2} Q^\dagger_1,
  \ee
  we find that $Q^\dagger_1$ raises the $m$-value by $\frac{1}{2}$; 
  and by the same token $Q^\dagger_2$ lowers it by $\frac{1}{2}$.
  
   We now want to find the nature of the  states which are `connected' to each other by the 
   application of the operators $Q_a$ and $Q_a^\dagger$ - that is, the analogue of the 
   $(2j+1)$-fold multiplet structure familiar in angular momentum theory. Our states 
   will be labelled as $|p, \lambda \rangle$, where we take the 4-momentum to be 
   $p=(E, 0, 0, E)$ (since the fields are massless), and where $\lambda$ is a helicity label. 
   Let's choose $|p, \lambda \rangle$ such that 
   \be 
   Q_a^\dagger |p, \lambda \rangle =0 \ \ \ {\mbox{for $a$=1,2}}.
   \ee
   Note that this is always possible: for if we started with a state $|p, \lambda \rangle^\prime$ 
   which did not satisfy this condition, then we could choose instead the state $Q_a^\dagger |p, \lambda 
   \rangle^\prime$ which does, because $Q_a^\dagger Q_a^\dagger =0$ (which follows either from 
   (\ref{eq:QQstarcom}) with $a=b$ or simply from the fact that  the $Q$'s are fermionic operators). There are then   
    only two states `connected' to $|p, \lambda \rangle$, namely $Q_1 | p, \lambda \rangle$ 
    and   $Q_2 | p, \lambda \rangle$. The first of these is not an acceptable state since its 
    norm is zero. This follows by considering the SUSY algebra (\ref{eq:QQstar1}) with $a=b=1$:     
  \be 
  Q^\dagger_1Q_1 + Q_1 Q^\dagger_1 = (\sigma^\mu)_{11} P_\mu.
  \ee
   The only components of $\sigma^\mu$ to have a non-vanishing `11' entry are 
  $(\sigma^0)_{11}=1$ and $(\sigma^3)_{11}=1$, so we have 
  \be 
  Q^\dagger_1 Q_1 +Q_1 Q^\dagger_1 = P_0 + P_3 = P^0-P^3.
  \ee
  Hence, taking the expectation value in the state $|p, \lambda \rangle$, we find 
  \be 
  \langle p, \lambda | Q^\dagger_1 Q_1 + Q_1 Q^\dagger_1 | p, \lambda \rangle = 0 
  \ee 
  since the eigenvalue of $P^0-P^3$ vanishes in this state. But also (by choice) $Q^\dagger_1 |p, \lambda \rangle =0$, 
  from which we deduce 
  \be 
  \langle p, \lambda | Q^\dagger_1 Q_1 | p, \lambda \rangle =0,
  \ee
  which shows that the norm of $Q_1 | p, \lambda \rangle$  is zero, as claimed. 
  
  This leaves just one connected state, namely 
  \be 
  Q_2 |p, \lambda \rangle.
  \ee
  We know that $Q_2$ raises $\lambda$ by 1/2, and hence 
  \be 
  Q_2 | p, \lambda \rangle \propto | p, \lambda + \frac{1}{2} \rangle.\label{eq:Q2lam}
  \ee
  We expect that the application of $Q^\dagger_2$ to this second state will take us back to the one we 
  started from, and this is correct:
  \be 
  Q^\dagger_2 | p, \lambda+\frac{1}{2} \rangle \propto Q^\dagger_2 Q_2 | p, \lambda \rangle \propto 
  (2E-Q_2 Q^\dagger_2) | p, \lambda \rangle \propto | p, \lambda \rangle,\label{eq:Q2dag}
  \ee
  where we have used (\ref{eq:QQstar1}) with $a=b=2$. 
  So there are just two distinct states, degenerate in mass, and linked by the operators 
  $Q_2$ and $Q_2^\dagger$. Equations (\ref{eq:Q2lam}) and (\ref{eq:Q2dag}) are suitable  for the case 
  $\lambda=-1/2$ (L-type); clearly a separate, but 
   analogous,  choice may be made for the case $\lambda=+1/2$ (R-type).\footnote{In $N=2$ SUSY 
   (see footnote 8) the corresponding supermultiplet contains 4 states: $\lambda=+1/2, \lambda=-1/2$ 
   and two states with $\lambda = 0$. The problem phenomenologically with this is that the R 
   ($\lambda=+1/2$) and L ($\lambda=-1/2$) states must transform in the same way under any gauge 
   symmetry. (Similar remarks hold for all $N \geq 1$ supermultiplets.) But we know that the  
     SU(2)$_{\rm L}$ gauge symmetry of the SM treats the L and R components of quark and 
     lepton fields differently. So if we want to make a SUSY extension of the SM, it had better be 
     the simple $N=1$ SUSY, where we are free to treat the supermultiplet ($\lambda=-1/2, \lambda=0$) 
     differently from the supermultiplet ($\lambda=0, \lambda=+1/2$). See \cite{bailin} section 1.6.}

   In terms of particle states, a {\em supermultiplet} contains just two types of 
   particles, differing by a 1/2-unit of helicity. The free SUSY theory of Section 3 is 
   an example of a {\em left chiral supermultiplet} containing a complex scalar field and 
   a single 2-component fermion of L-type. Later, we shall learn how to include interactions 
   (Section 8). We shall also need to consider the {\em vector, or gauge supermultiplet}, 
   which contains a  gauge field and a two-component spinor (see Section 11). 
   Finally there is the {\em gravity supermultiplet} (which we shall not present), 
   containing a spin-2 graviton field and a spin-3/2 gravitino field. 
   
   We must now take up an issue raised after (\ref{eq:dblecom}).
   
   \section{A Snag, and the Need for a Significant Complication} 
   
   In Section 5.2 we arrived at the SUSY algebra by calculating the difference 
   $\delta_\eta \delta_\xi - \delta_\xi \delta_\eta$ two different ways. We 
    explicitly evaluated this difference as applied to $\phi$, but in 
   deducing the  operator relation (\ref{eq:QQP1}),  it is crucial that a 
   consistent result be obtained when  $\delta_\eta \delta_\xi - \delta_\xi \delta_\eta$ 
   is applied to $\chi$. In fact, as noted after (\ref{eq:deldelchi}), it is not, as we 
   now show. This will necessitate a significant modification of the SUSY 
   transformations given so far, in order to bring about this desired consistency.     
   
   Consider first $\delta_\eta \delta_\xi \chi_a$, where we are indicating the 
   spinor component explicitly:
   \bea 
   \delta_\eta \delta_\xi \chi_a&=& \delta_\eta ( -{\rm i} \sigma^\mu ({\rm i} \sigma_2 \xi^*))_a 
   \partial_\mu \phi \nonumber \\
   &=&({\rm i} \sigma^\mu (-{\rm i} \sigma_2 \xi^*))_a \partial_\mu \delta_\eta \phi \nonumber \\
   &=&({\rm i} \sigma^\mu (-{\rm i} \sigma_2 \xi^*))_a (\eta^{\rm T} (-{\rm i} \sigma_2) \partial_\mu \chi).
   \label{eq:ddchi1}
   \eea
   There is an important identity involving products of three spinors, which we can use 
   to simplify (\ref{eq:ddchi1}). The identity reads, for any three spinors $\lambda$, $\zeta$ and $\rho$, 
   \be 
   \lambda_a(\zeta^{\rm T}(-{\rm i} \sigma_2) \rho) +\zeta_a(\rho^{\rm T}(-{\rm i} \sigma_2) \lambda) +
   \rho_a(\lambda^{\rm T} (-{\rm i} \sigma_2) \zeta) =0,\label{eq:3spinid}
   \ee
   or in the faster notation 
   \be 
   \lambda_a(\zeta \cdot \rho) + \zeta_a(\rho \cdot \lambda) + \rho_a( \lambda \cdot \zeta) = 0.
   \ee
  
  {\bf Exercise} Check the identity (\ref{eq:3spinid}). 
  
  We take, in (\ref{eq:3spinid}),  
  \be 
 \lambda_a = (\sigma^\mu(-{\rm i} \sigma_2) \xi^*)_a, \ \ \ \zeta_a = \eta_a, 
 \ \ \ \rho_a = \partial_\mu \chi_a.
 \ee
 The RHS of (\ref{eq:ddchi1}) is then equal to 
 \be
 -{\rm i} \{ \eta_a \partial_\mu \chi^{\rm T}(-{\rm i} \sigma_2) \sigma^\mu (-{\rm i} \sigma_2) 
 \xi^* + \partial_\mu \chi_a ( \sigma^\mu (-{\rm i} \sigma_2 \xi^*))^{\rm T} (-{\rm i} \sigma_2) \eta.\}
 \label{eq:ddchi2}
 \ee 
  But we know from (\ref{eq:csigc}) that the first term in (\ref{eq:ddchi2}) can be written as 
  \be 
  {\rm i} \eta_a (\partial_\mu \chi^{\rm T} {\bar{\sigma}}^{\mu {\rm T}} \xi^*)=
  -{\rm i}\eta_a (\xi^\dagger {\bar{\sigma}}^{\mu} \partial_\mu \chi), \label{eq:ddchi21}
  \ee
 where to reach the second equality in (\ref{eq:ddchi21}) we have taken the formal 
 transpose of the quantity in brackets, remembering the sign change from re-ordering the 
 spinors. As regards the second term in (\ref{eq:ddchi2}), we again take the transpose of 
 the quantity multiplying $\partial_\mu \chi_a$, so that it becomes 
 \be 
 -{\rm i} \partial_\mu \chi_a (-\eta^{\rm T} {\rm i} \sigma_2 \sigma^\mu (-{\rm i} \sigma_2) \xi^* =
 -{\rm i} \eta^{\rm T} c \sigma^\mu c \xi^* \partial_\mu \chi_a.
 \label{eq:ddchi22}
 \ee 
After these manipulations, then, we have arrived at 
\be 
\delta_\eta \delta_\xi \chi_a =-{\rm i} \eta_a(\xi^\dagger {\bar{\sigma}}^\mu \partial_\mu \chi) -
{\rm i} \eta^{\rm T} c \sigma^\mu c \xi^* \partial_\mu \chi_a,\label{eq:ddchi3}
\ee
and so
\bea 
(\delta_\eta \delta_\xi -\delta_\xi \delta_\eta) \chi_a = (\xi^{\rm T} c \sigma^\mu c \eta^* - 
\eta^{\rm T} c \sigma^\mu c \xi^*) {\rm i} \partial_\mu \chi_a && \nonumber \\
+{\rm i} \xi_a(\eta^\dagger {\bar{\sigma}}^\mu \partial_\mu \chi) -{\rm i} \eta_a 
(\xi^\dagger {\bar{\sigma}}^\mu \partial_\mu \chi).&& \label{eq:ddchifull}
\eea

We now see the difficulty: the first term on the RHS of (\ref{eq:ddchifull}) is indeed 
exactly the same as (\ref{eq:deldelphi}) with $\phi$ replaced by $\chi$, as 
hoped for in (\ref{eq:deldelchi}),  {\em but there are in 
addition two unwanted terms}. 

The two unwanted terms vanish when the equation of motion ${\bar{\sigma}}^\mu \partial_\mu \chi=0$ 
is satisfied (for a massless field) - i.e. `on-shell'. But this is not good enough - we 
want a symmetry that applies for the internal (off-shell) lines in Feynman graphs, as well 
as for the on-shell external lines. Actually, we should not be too surprised that our 
naive SUSY of Section 5.2 has failed off-shell, for a reason that has already been touched upon: 
the numbers of degrees of freedom in $\phi$ and $\chi$ don't match up properly, the former 
having two (one complex field)  and the latter four (two complex components). This suggests 
that we need to introduce another two degrees of freedom to supplement the two in $\phi$ - 
say a second scalar field $F$. We do this in the `cheapest' possible way (provided it 
works), which is simply to add a term $F^\dagger F$ to the Lagrangian (\ref{eq:Lphichi}), so 
that $F$ has no kinetic term, and therefore no propagator:
\be 
 {\cal{L}}_F=  
\partial_\mu \phi^\dagger \partial^\mu \phi + \chi^\dagger {\rm i} {\bar{\sigma}}^\mu \partial_\mu \chi 
+ F^\dagger F. \label{eq:LphichiF}
\ee
The strategy now is to invent a SUSY transformation for the {\em auxiliary field} $F$, 
and the existing fields $\phi$ and $\chi$,  such 
that (a) ${\cal{L}}_F$ is invariant, at least up to a total derivative, and 
(b) the unwanted terms in $(\delta_\eta \delta_\xi -\delta_\xi \delta_\eta)\chi$ are removed. 

We  note that $F$ has dimension ${\rm M}^2$, suggesting that $\delta_\xi F$  should 
probably be of the form 
\be 
\delta_\xi F \sim \xi \partial_\mu \chi,\label{eq:delF1}
\ee
which is consistent dimensionally. But as usual we need to ensure Lorentz covariance, 
and in this case that means that the RHS of (\ref{eq:delF1}) must be a Lorentz invariant. We know that 
${\bar{\sigma}}^\mu \partial_\mu \chi$ transforms as a `$\psi$'-type spinor (see 
(\ref{eq:Dchisig})), and we know that 
an object of the form `$\xi^\dagger \psi$ is Lorentz invariant (see (\ref{eq:invpsichi})). So we try 
\be 
\delta_\xi F = -{\rm i} \xi^\dagger {\bar{\sigma}}^\mu \partial_\mu \chi \label{eq:delxiF}
\ee
and correspondingly
\be 
\delta_\xi F^\dagger = {\rm i} \partial_\mu \chi^\dagger {\bar{\sigma}}^\mu \xi.
\ee
The fact that these changes vanish if the equation of motion for $\chi$ is imposed 
(the on-shell condition) suggests that they might be capable of cancelling the 
unwanted terms in (\ref{eq:ddchifull}). Note also that, since $\xi$ is independent 
of $x$, the changes in $F$ and $F^\dagger$ are total derivatives: this will be important 
later (see the end of section 9.3).

We must first ensure that the enlarged Lagrangian (\ref{eq:LphichiF}) - or 
at least the corresponding Action - remains SUSY-invariant. 
 Under these changes, the $F^\dagger F$ term in (\ref{eq:LphichiF}) changes by 
 \be 
 (\delta_\xi F^\dagger)F + F^\dagger (\delta_\xi F) = ({\rm i} \partial_\mu \chi^\dagger 
 {\bar{\sigma}}^\mu \xi)F -F^\dagger({\rm i} \xi^\dagger {\bar{\sigma}}^\mu \partial_\mu \chi).
 \label{eq:delFFdag}
 \ee
 These terms have a structure very similar to the $\chi$ term in (\ref{eq:LphichiF}), which 
 changes by 
 \be
 \delta_\xi(\chi^\dagger {\rm i} {\bar{\sigma}}^\mu \partial_\mu \chi) = 
 (\delta_\xi \chi^\dagger) {\rm i} {\bar{\sigma}}^\mu \partial_\mu \chi + \chi^\dagger {\rm i} 
 {\bar{\sigma}}^\mu \partial_\mu (\delta_\xi \chi). \label{eq:delchiterm}
 \ee
 We see that if we choose 
 \be 
 \delta_\xi \chi^\dagger_a = \mbox{previous change} + F^\dagger \xi^\dagger_a.\label{eq:delchiFdag}
 \ee
 then the $F^\dagger$ part of the first term in  (\ref{eq:delchiterm}) cancels the 
 second term in (\ref{eq:delFFdag}). As regards the second term in (\ref{eq:delchiterm}), 
 we write it as 
 \bea
 \chi^\dagger {\rm i} {\bar{\sigma}}^\mu \partial_\mu \delta_\xi \chi &=&  
 \partial_\mu (\chi^\dagger {\rm i} {\bar{\sigma}}^\mu \delta_\xi \chi) - \partial_\mu \chi^\dagger 
 {\rm i} {\bar{\sigma}}^\mu \delta_\xi \chi \nonumber \\
 &=& \partial_\mu (\chi^\dagger {\rm i} {\bar{\sigma}}^\mu \xi F) - \partial_\mu \chi^\dagger 
 {\rm i} {\bar{\sigma}}^\mu \xi F \label{eq:secterm}
 \eea
 where we have used the dagger of (\ref{eq:delchiFdag}), namely 
 \be 
 \delta_\xi \chi_a = \mbox{previous change} + \xi_a F.\label{eq:delchiF}
 \ee 
 The first term of (\ref{eq:secterm}) is a total derivative, leaving the Action 
 invariant, while the second cancels the first term in (\ref{eq:delFFdag}). This has 
 been achieved by allowing $\chi$ to `mix',  under SUSY transformations,  with the  
 auxiliary field $F$, while the transformation of $\phi$ is unaltered.

 Let us now re-calculate $(\delta_\eta \delta_\xi - \delta_\xi \delta_\eta)\chi$, 
 including the new terms involving the auxiliary field $F$. Since the transformation 
 of $\phi$ is unaltered, $\delta_\eta \delta_\xi \chi$ will be the same as 
 before, in (\ref{eq:ddchi3}), together with an extra term 
 \be 
 \delta_\eta(\xi_a F) = -{\rm i} \xi_a ( \eta^\dagger {\bar{\sigma}}^\mu \partial_\mu \chi).
 \ee 
 So $(\delta_\eta \delta_\xi - \delta_\xi \delta_\eta)\chi$ will be as before, 
 in (\ref{eq:ddchifull}), together with the extra terms 
 \be 
 {\rm i} \eta_a (\xi^\dagger {\bar{\sigma}}^\mu \partial_\mu \chi) -
 {\rm i} \xi_a (\eta^\dagger {\bar{\sigma}}^\mu \partial_\mu \chi). \label{eq:extra}
 \ee
 These extra terms precisely cancel the unwanted terms in (\ref{eq:ddchifull}), as required.   
 Similar results hold for the action of $(\delta_\eta \delta_\xi - \delta_\xi \delta_\eta)$
 on $\phi$ and on $F$, and so with this enlarged structure including $F$ we can indeed 
 claim that (\ref{eq:QQP1}) holds as an operator relation, being true when acting on any field 
 of the theory.

 \section{Interactions: the Wess-Zumino model} 
 
 The Lagrangian (\ref{eq:LphichiF}) describes a free (left) chiral supermultiplet, with a massless spin-0 
 field $\phi$, a massless L-type  spinor field $\chi$, and a non-propagating field $F$. 
 As we saw in Section 4, we have to put  
  the quarks, leptons and Higgs bosons of the SM,  labelled by gauge and flavour degrees of freedom, into    
  chiral supermultiplets, partnered by the appropriate s-particle.
  So we shall generalize (\ref{eq:LphichiF}) to 
 \be 
 {\cal{L}}_{{\rm free \; WZ}}= \partial_\mu \phi^\dagger_i \partial^\mu \phi_i +
 \chi^\dagger_i {\rm i} {\bar{\sigma}}^\mu \partial_\mu \chi_i + F^\dagger_iF_i
 \label{eq:LphichiFi}
 \ee
 where the summed-over index $i$ runs over internal degrees of freedom (e.g. flavour, and 
 eventually gauge - see section 10),   
 and is not to be confused (in the case of $\chi_i$) with the spinor component index.  The 
 corresponding Action is invariant under the SUSY transformations 
 \be 
 \delta_\xi \phi_i = \xi \cdot \chi_i, \ \ \delta_\xi \chi_i = -{\rm i} \sigma^\mu {\rm i} \sigma_2 \xi^* 
 \partial_\mu \phi_i + \xi F_i, \ \ \ \delta_\xi F_i=-{\rm i} \xi^\dagger {\bar{\sigma}}^\mu \partial_\mu \chi_i,
 \label{eq:SUSYi}
 \ee
 together with their hermitian conjugates.The obvious next step 
 is to introduce interactions in such a way as to preserve  SUSY -  
 that is, invariance of the Lagrangian (or the Action) under the transformations 
 (\ref{eq:SUSYi}).  This was first done 
 (for this type of theory, in four dimensions) by Wess and Zumino \cite{WZ1}, so the 
 resulting model is called the Wess-Zumino model. We shall largely follow the 
 account given by \cite{martin}, section 3.2. 
 
 We shall impose the important condition that the interactions  should be 
 renormalizable. This means that the mass dimension of all interaction terms must not 
 be greater than 4 - or, equivalently, that the coupling constants in the interaction 
 terms should be dimensionless, or have positive dimension  (see \cite{AH31} section 11.8). 
 The most general possible set of renormalizable interactions among the fields $\phi, \chi$ and $F$ is, 
 in fact, rather simple:
 \be 
 {\cal{L}}_{{\rm int}}= W_i(\phi, \phi^\dagger) F_i - \frac{1}{2}W_{ij}(\phi, \phi^\dagger) 
 \chi_i \cdot \chi_j + \mbox{hermitian conjugate} \label{eq:Lsusyint}
 \ee
 where there is a sum on $i$ and on $j$. Here $W_i$ and $W_{ij}$ are - for the moment - 
 arbitrary functions of the bosonic fields; we shall  see that they are actually 
 related, and have a simple form. There is no term in the $\phi_i$'s alone, because 
 under the transformation (\ref{eq:SUSYi}) this would become some function of the $\phi_i$'s 
 multiplied by $\delta_\xi \phi_i= \xi \cdot \chi_i$ or 
 $\delta_\xi \phi_i^\dagger = \bar{\xi} \cdot \bar{\chi}$; but these  terms do not  include 
 any derivatives $\partial_\mu$,  or $F_i$ or $F^\dagger_i$ fields, and it is clear by 
 inspection of (\ref{eq:SUSYi}) that they couldn't be cancelled by the transformation of 
 any other term.    
 
 As regards $W_i$ and $W_{ij}$, we first note that since $F_i$ has dimension 2, $W_i$ 
 cannot depend on $\chi_i$, which has dimension 3/2, nor on any power of $F_i$ other 
 than the first, which is already included in (\ref{eq:LphichiFi}). Indeed, $W_i$ 
 can involve no higher powers of  $\phi_i$ and $\phi_i^\dagger$ than the second. Similarly, since 
 $\chi_i \cdot \chi_j$ has dimension 3, $W_{ij}$ can only depend on  $\phi_i$ and $\phi_i^\dagger$, and contain 
 no powers higher than the first. Also, since $\chi_i \cdot \chi_j = \chi_j \cdot \chi_i$ 
 (see the first Exercise in Aside (1)), $W_{ij}$ must be symmetric in the indices $i$ and $j$.
 
 Since we know that the Action for the `free' part  (\ref{eq:LphichiFi}) is invariant under 
 (\ref{eq:SUSYi}), we consider now only the change  in ${\cal{L}}_{{\rm int}}$ under 
 (\ref{eq:SUSYi}), namely  $\delta_\xi {\cal{L}}_{{\rm int}} $. First, 
 consider the part involving four spinors, which is 
 \be 
 -\frac{1}{2} \frac{\partial W_{ij}}{\partial \phi_k} (\xi \cdot \chi_k)(\chi_i \cdot \chi_j) - 
 \frac{1}{2} \frac{\partial W_{ij}}{\partial \phi_k^\dagger}(\bar{\xi} \cdot {\bar{\chi}}_k)
 (\chi_i \cdot \chi_j) + \mbox{hermitian conjugate}. \label{eq:del4spin}
 \ee
Neither of these terms can be cancelled by the variation of any other term. However, the first term 
will vanish provided that 
\be 
 \frac{\partial W_{ij}}{\partial \phi_k} \ \mbox{is symmetric in $i, j$ and $k$}. \label{eq:Wsym}
 \ee
 The reason is that the identity (\ref{eq:3spinid}) (with $\lambda \to \chi_k, \zeta \to \chi_i, \rho 
 \to \chi_j$) implies 
 \be 
 (\xi \cdot \chi_k)(\chi_i \cdot \chi_j) + (\xi \cdot \chi_i)( \chi_j \cdot \chi_k) + 
 (\xi \cdot \chi_j) (\chi_k \cdot \chi_i) = 0, 
 \ee
 from which it follows that if (\ref{eq:Wsym}) is true, then the first term in (\ref{eq:del4spin}) 
 will vanish identically. However, there is no corresponding identity for the 4-spinor 
 product in the second term of (\ref{eq:del4spin}). The only way to get rid of this second 
 term, and thus preserve SUSY for such interactions, is to say that $W_{ij}$ cannot depend 
 on $\phi^\dagger_k$, only on $\phi_k$.\footnote{ This is a point of great importance for the MSSM: 
  the SM  uses both the Higgs field $\phi$ and its charge conjugate, which is related to 
  $\phi^\dagger$, but in the MSSM we shall need to have two separate $\phi$'s.} Thus we now know 
  that $W_{ij}$ must have the form 
  \be 
  W_{ij}=M_{ij} + y_{ijk} \phi_k \label{eq:linW}
  \ee
  where the matrix $M_{ij}$ (which has the dimensions - and significance - of a mass) is 
  symmetric in $i$ and $j$, and where the `Yukawa couplings' $y_{ijk}$ are symmetric in 
  $i$, $j$ and $k$. It is convenient to write (\ref{eq:linW}) as 
  \be 
  W_{ij} = \frac{\partial^2 W}{\partial \phi_i \partial \phi_j} \label{eq:Wijderiv}
  \ee
  which is automatically symmetric in $i$ and $j$, and  
  where\footnote{A linear term of the form $A_l \phi_l$ could be 
  added to (\ref{eq:W}), consistently with (\ref{eq:Wijderiv}) and (\ref{eq:linW}). This is 
  relevant to one model of SUSY breaking - see Section 14.}    
   (bearing in mind the symmetry properties of $W_{ij}$) 
  \be 
  W= \frac{1}{2}M_{ij} \phi_i \phi_j + \frac{1}{6} y_{ijk} \phi_i \phi_j \phi_k.\label{eq:W}
  \ee
  
  {\bf Exercise} Justify (\ref{eq:W}).

   Next, consider those parts of $\delta_\xi {\cal{L}}_{{\rm int}}$ which contain one derivative $\partial_\mu$. These 
   are (recall $c=-{\rm i} \sigma_2$)
   \be
   W_i(-{\rm i} \xi^\dagger {\bar{\sigma}}^\mu \partial_\mu \chi_i) -\frac{1}{2}W_{ij}\{\chi^{\rm T}_i 
   c {\rm i} \sigma^\mu c \xi^*\} \partial_\mu \phi_j 
   +\frac{1}{2}W_{ij} \xi^\dagger c {\rm i} \sigma^{{\rm T}\mu} \partial_\mu\phi_i \label{eq:delLgrad}
   c\chi_j.
   \ee 
   Consider the expression in curly brackets, $\{\chi_i^{\rm T} \ldots \xi^*\}$. Since this is a 
   single quantity (after evaluating the matrix products), it is equal to its transpose, which is 
   \be
   -\xi^\dagger c {\rm i} \sigma^{\mu {\rm T}}c \chi_i = \xi^\dagger {\rm i} {\bar{\sigma}}^\mu \chi_i
   \ee
   where the first minus sign comes from interchanging two fermionic quantities, and the 
   second equality uses the result $c \sigma^{\mu{\rm T}}c=-{\bar{\sigma}}^\mu$ (c.f. (\ref{eq:csigc})). 
   So the second term in  (\ref{eq:delLgrad}) is 
   \be -\frac{1}{2}W_{ij} {\rm i} \xi^\dagger {\bar{\sigma}}^\mu \chi_i \partial_\mu \phi_j,
   \ee
     and the third term is 
     \be
     \frac{1}{2}W_{ij}\xi^\dagger c {\rm i} \sigma^{\mu {\rm T}} c \chi_j \partial_\mu \phi_i 
     = - \frac{1}{2} W_{ij}{\rm i} \xi^\dagger {\bar{\sigma}}^\mu \chi_j \partial_\mu \phi_i.
     \ee
     So these two terms add to give 
     \be -W_{ij} {\rm i} \xi^\dagger {\bar{\sigma}}^\mu \chi_i \partial_\mu \phi_j = 
     -{\rm i} \xi^\dagger {\bar{\sigma}}^\mu \chi_i \partial_\mu \left( \frac{\partial W}{\partial \phi_i} 
     \right),
     \ee
     where in the second equality we have used 
     \be 
     \partial_\mu \left( \frac{\partial W}{\partial \phi_i} \right) = \frac{\partial^2 W} 
     {\partial \phi_i \partial \phi_j} \partial_\mu \phi_j = W_{ij}\partial_\mu \phi_j. 
     \ee
     Altogether, then, (\ref{eq:delLgrad}) has become 
     \be 
     -{\rm i} W_i \xi^\dagger {\bar{\sigma}}^\mu \partial_\mu \chi_i - {\rm i} \xi^\dagger 
     {\bar{\sigma}}^\mu \chi_i \partial_\mu \left( \frac{\partial W}{\partial \phi_i} \right).
     \label{eq:delLgrad2}
     \ee  
     This variation cannot be cancelled by anything else, and our only chance of saving SUSY 
     is to have it equal a total derivative (giving an invariant Action, as usual). The condition 
     for (\ref{eq:delLgrad2}) to be a total derivative is that $W_i$ should have the form 
     \be 
     W_i = \frac{\partial W}{\partial \phi_i},\label{eq:Wcond}
     \ee
     in which case (\ref{eq:delLgrad2}) becomes 
     \be \partial_\mu \{ \frac{\partial W}{\partial \phi_i}( -{\rm i} \xi^\dagger {\bar{\sigma}}^\mu \chi_j)\}.
     \ee
  Referring to (\ref{eq:W}), we see that the condition (\ref{eq:Wcond}) implies 
  \be 
  W_i = M_{ij} \phi_j + \frac{1}{2} y_{ijk} \phi_j \phi_k \label{eq:Wi}
  \ee
  together with a possible constant term $A_i$ (see footnote 10). 
  
  {\bf Exercise} Verify that the remaining terms in $\delta_\xi {\cal{L}}$ do cancel. 
  
  In summary, we have found conditions  on $W_i$ and 
  $W_{ij}$ (namely equations(\ref{eq:Wcond}) and (\ref{eq:Wijderiv}) with $W$ given by (\ref{eq:W})) 
  such that the interactions (\ref{eq:Lsusyint}) give an Action 
  which is invariant under the SUSY transformations (\ref{eq:SUSYi}). Consider now the part 
  of the complete Lagrangian (including (\ref{eq:LphichiFi}))  containing $F_i$ and $F^\dagger_i$, 
  which is just $F_iF^\dagger_i + W_i F_i + W^\dagger_i F^\dagger_i$. Since this contains no gradients, 
  the Euler-Lagrange equations for $F_i$ and $F^\dagger_i$ are simply 
  \be 
  \frac{\partial {\cal{L}}}{\partial F_i} = 0, \ {\rm or} \  F^\dagger_i +W_i =0.\label{eq:Feqnmot}
  \ee
  Hence $F_i=-W_i^\dagger$, and similarly $F_i^\dagger = -W_i$. These relations, coming 
  from the E-L equations, involve (again) no derivatives, and hence the canonical 
  commutation relations will not be affected, and it is permissible to replace $F_i$ and 
  $F^\dagger_i$ in the Lagrangian by these values determined from the E-L equations. This 
  results in the complete (Wess-Zumino \cite{WZ1}) Lagrangian now having the form 
  \be 
  {\cal{L}}_{\rm WZ}={\cal{L}}_{{\rm free \; WZ}} -|W_i|^2 -\frac{1}{2}\{W_{ij} \chi_i \cdot \chi_j 
  + {\rm h.c}\} \label{eq:WZ2}
  \ee
  where `h.c.' means hermitian conjugate. 
  
  It is worth spending a little time looking in more detail at the model of (\ref{eq:WZ2}). 
  For simplicity we shall discuss just one supermultiplet, dropping the indices $i$ and $j$. 
  First, consider   
  the terms which are quadratic in the fields $\phi$ and $\chi$, 
  which correspond to kinetic and mass terms (rather than interactions proper). This will give us 
  an opportunity to learn about mass terms for two-component spinors. The quadratic terms for a 
  single supermultiplet are  
  \be 
  {\cal{L}}_{{\rm WZ, quad}}= \partial_\mu \phi^\dagger \partial^\mu \phi + 
  \chi^\dagger {\rm i} {\bar{\sigma}}^\mu \partial_\mu \chi - MM^* \phi^\dagger \phi -
  \frac{1}{2}M \chi^{\rm T} (-{\rm i} \sigma_2) \chi - \frac{1}{2}M^*\chi^\dagger ({\rm i} 
  \sigma_2)\chi^{\dagger{\rm T}} \label{eq:WZquad}
  \ee 
  where we have reverted to the explicit forms of the spinor products. In (\ref{eq:WZquad}), $\chi^\dagger$ 
  is as given in  (\ref{eq:chidagger}), while evidently 
  \be 
  \chi^{\dagger{\rm T}}= \left( \ba{c} \chi^\dagger_1 \\ \chi^\dagger_2 \ea \right)
  \ee
  where `1' and `2', of course, label the spinor components. The E-L equation for $\phi^\dagger$ is 
  \be 
 \partial_\mu \left( \frac{\partial {\cal{L}}}{\partial(\partial_\mu \phi^\dagger)}\right)  - 
 \frac{\partial {\cal{L}}}{\partial \phi^\dagger} = 0,
 \ee
 which leads immediately to 
 \be 
 \partial_\mu \partial^\mu \phi + |M|^2 \phi = 0, 
 \ee
 which is just the standard free Klein-Gordon equation for a spinless field of mass 
 $|M|$. 
 
 In considering the analogous E-L equation for (say) $\chi^\dagger$, we need 
 to take care in evaluating (functional) derivatives of ${\cal{L}}$ with respect to 
 fields such as $\chi$ or $\chi^\dagger$ which anticommute. 
 Consider the term $-(1/2) M \chi \cdot \chi$ in (\ref{eq:WZquad}), which is 
 \be 
 - \frac{1}{2} M ( \chi_1 \chi_2) \left( \ba{cc} 0&-1\\ 1 & 0 \ea \right) \left( \ba{c} 
 \chi_1 \\ \chi_2 \ea \right) = -\frac{1}{2}M (-\chi_1 \chi_2 + \chi_2 \chi_1) 
 = -M \chi_2 \chi_1 =+M \chi_1 \chi_2.
 \ee  
 We  define 
 \be 
 \frac{\partial}{\partial \chi_1} (\chi_1 \chi_2) = \chi_2,
 \ee
 and then necessarily 
 \be 
 \frac{\partial}{\partial \chi_2} (\chi_1 \chi_2) = -\chi_1. 
 \ee  
 Hence 
 \be 
 \frac{\partial}{\partial \chi_1} \{ - \frac{1}{2} M \chi \cdot \chi \} = M \chi_2, 
 \label{eq:dchi1}
 \ee
 and 
 \be 
 \frac{\partial}{\partial \chi_2}\{ - \frac{1}{2} M \chi \cdot \chi \} = -M \chi_1.
 \label{eq:dchi2}
 \ee  
 Equations (\ref{eq:dchi1}) and (\ref{eq:dchi2}) can be combined as 
 \be 
 \frac{\partial}{\partial \chi_a}\{ -\frac{1}{2}M \chi \cdot \chi\}= M ({\rm i} \sigma_2 \chi)_a.
 \label{eq:dchimass}
 \ee 
  
  {\bf Exercise} Show similarly that 
  \be 
  \frac{\partial}{\partial \chi^\dagger_a}\{ - \frac{1}{2} M^* \chi^\dagger 
  {\rm i} \sigma_2 \chi^{\dagger {\rm T}}\} = M^* (-{\rm i} \sigma_2 \chi^{\dagger {\rm T}})_a.
  \label{eq:dchidagmass}
  \ee
  
  We are now ready to consider the E-L equation for $\chi^\dagger$, which is 
  \be 
  \partial_\mu \left( \frac{\partial {\cal{L}}}{\partial(\partial_\mu \chi^\dagger_a)} \right) 
  - \frac{\partial {\cal{L}}}{\partial \chi^\dagger_a}=0.
  \ee
  Using just the quadratic parts (\ref{eq:WZquad}) this yields 
  \be 
  {\rm i} {\bar{\sigma}}^\mu \partial_\mu \chi = M^* {\rm i} \sigma_2 \chi^{\dagger {\rm T}}.
  \label{eq:chieq}
  \ee
  
 {\footnotesize{From Notational Aside (3) in section 2.4 we know that $\chi$ transforms by 
 $V^{-1 \dagger}$, and hence $\chi^{\dagger {\rm T}}$ transforms by $V^{-1 {\rm T}}$, which is 
 the same as a `lower dotted' spinor of type $\psi_{\dot{a}}$. The lower dotted index is raised 
 by the matrix ${\rm i} \sigma_2$. Hence the RHS of (\ref{eq:chieq}) transforms like a $\psi^{\dot{a}}$ 
 spinor, and this is consistent with the LHS, by (\ref{eq:Dchisig}).}} 
 
 {\bf Exercise} Similarly, show that 
 \be 
 {\rm i} \sigma^\mu \partial_\mu ( {\rm i} \sigma_2 \chi^{\dagger {\rm T}}) = M \chi.
 \label{eq:chidageq}
 \ee
 It follows from (\ref{eq:chieq}) and (\ref{eq:chidageq}) that 
 \bea
 {\rm i} \sigma^\mu \partial_\mu ( {\rm i} {\bar{\sigma}}^\nu \partial_\nu \chi) &=& 
 {\rm i} \sigma^\mu \partial_\mu ( M^* {\rm i} \sigma_2 \chi^{\dagger {\rm T}}) \nonumber \\
 &=& |M|^2 \chi.
 \eea
 So, using (\ref{eq:sigsigbar1}) on the LHS we have simply 
 \be
 \partial_\mu \partial^\mu \chi + |M|^2 \chi =0,
 \ee  
  which shows that the $\chi$ field also has mass $|M|$. So we have verified that the quadratic 
  parts (\ref{eq:WZquad}) describe a free spin-0 and spin-1/2 field which are degenerate, both 
  having mass $|M|$. It is perhaps worth pointing out that, although we started 
  (for simplicity) with massless fields, we now see that it is perfectly possible 
  to have massive supersymmetric theories, the bosonic and fermionic superpartners 
  having (of course) the same mass.

  {\footnotesize{{\bf Aside: Majorana version} This seems as good a moment as any to say a 
  few words about a possible alternative formalism, in which one uses 4-component Majorana 
  spinor fields (see section 2.3) rather than the 2-component L- or R-spinor fields we have been using 
  up till now (and will continue to use). First recall from section 2.3 that the number 
  of degrees of freedom is the same, because the Majorana spinor $\Psi^\chi_{\rm M}$ 
  of equation (\ref{eq:chimaj}), for example, is 
  constructed explicitly from $\chi$; so we expect that it must be possible to re-write 
  everything involving $\chi$'s in terms of $\Psi^\chi_{\rm M}$'s. To check this, consider 
  first the fermion mass term in (\ref{eq:WZquad}). We may take $M$ to be real, by absorbing 
  any phase into the undefined phase of $\chi$. It then follows at once from equation (\ref{eq:majmass1}) that 
  \be 
  -\frac{1}{2}M[\chi^{\rm T}(-{\rm i}\sigma_2)\chi + \chi^\dagger ({\rm i} \sigma_2) \chi^{\dagger {\rm T}}] 
  =-\frac{1}{2}M {\bar{\Psi}}^\chi_{\rm M} \Psi^\chi_{\rm M}.
  \ee
  The kinetic term is a little more involved. We have 
  \bea
  \Psi^{\chi \dagger}_{\rm M} \beta {\rm i} \gamma^\mu \partial_\mu \Psi^\chi_{\rm m} &=&
  (\chi^{\rm T}(-{\rm i}\sigma_2) \, \, \,  \chi^\dagger)\left( \ba{cc} {\rm i} \sigma^\mu \partial_\mu & 0 \\
  0 & {\rm i}{\bar{\sigma}}^\mu \partial_\mu \ea \right) \left( \ba{c} {\rm i} \sigma_2 \chi^{\dagger {\rm T}} \\
  \chi \ea \right) \nonumber \\
  &=&[\chi^{\rm T}(-{\rm i}\sigma_2) {\rm i} \sigma^\mu \partial_\mu {\rm i} \sigma_2 \chi^{\dagger {\rm T}} ] 
  + \chi^\dagger {\rm i} {\bar{\sigma}}^\mu \partial_\mu \chi .
  \eea
  The first term can be maniplulated in the now familiar way, by taking its transpose and using (\ref{eq:csigc}); 
  one finds that it is just equal to the second term, and so 
  \be 
  \chi^\dagger {\rm i} {\bar{\sigma}}^\mu \partial_\mu \chi = \frac{1}{2} {\bar{\Psi}}^\chi_{\rm M} 
  {\rm i} \gamma^\mu \partial_\mu \Psi^\chi_{\rm M}.
  \ee
  Thus the kinetic term and the mass terms have been re-written in terms of the Majorana spinor 
  $\Psi^\chi_{\rm M}$; a factor of `1/2' appears in the Majorana version 
  of both the kinetic and the mass terms. The other terms can be treated similarly. 
  
  One might wonder about a Majorana version of the SUSY algebra. Just as we can construct a 4-component Majorana 
  spinor from a $\chi$ (or of course a $\psi$), so we can make a 4-component Majorana spinor charge $Q_{\rm M}$ 
  from our L-type spinor charge Q,  by setting  (c.f. (\ref{eq:chimaj})) 
  \be 
  Q_{\rm M}= \left( \ba{c} {\rm i} \sigma_2 Q^{\dagger {\rm T}} \\ Q \ea \right) = 
  \left( \ba{c} Q^\dagger_2 \\-Q^\dagger_1 \\ Q_1 \\ Q_2 \ea \right).\label{eq:QMdef}
  \ee 
  Let's call the components of this $Q_{{\rm M} \alpha}$, so that $Q_{{\rm M}1}=Q^\dagger_2, \ 
  Q_{{\rm M}2}=-Q^\dagger_1$, etc. It is not completely obvious (to me, at any rate) what the 
  anticommutation relations of the $Q_{{\rm M} \alpha}$'s ought to be (given those of the $Q_a$'s 
  and $Q^\dagger_a$'s), but the answer turns out to be 
  \be 
  \{Q_{{\rm M} \alpha}, Q_{{\rm M} \beta} \} = (\gamma^\mu({\rm i} \gamma^2 \gamma^0))_{\alpha \beta} P_\mu,
  \label{eq:QMcom}
  \ee
  as can be checked with the help of (\ref{eq:QQcom}), (\ref{eq:QQstarcom}), (\ref{eq:QQstar1}) 
  and (\ref{eq:QMdef}). Note that `$-{\rm i} \gamma^2 \gamma^0$' is the `metric' we met in section 2.3. The 
  anticommutator (\ref{eq:QMcom}) can be re-written rather more suggestively as 
  \be \{ Q_{{\rm M} \alpha} , {\bar{Q}}_{{\rm M} \beta} \} = (\gamma^\mu)_{\alpha \beta} P_\mu 
  \label{eq:QMQMbarcom}
  \ee
  where (compare (\ref{eq:majmet})) 
  \be 
  {\bar{Q}}_{{\rm M} \beta}=(Q^{\rm T}_{\rm M}(-{\rm i} \gamma^2 \gamma^0))_{\beta} = 
  (Q^\dagger_{\rm M} \gamma^0)_\beta.
  \ee

  }}

  Next, let's consider briefly the interaction terms in (\ref{eq:WZ2}), again just for the 
  case of one chiral superfield. These terms are 
  \be 
  -|M \phi + \frac{1}{2} y \phi^2|^2 -\frac{1}{2}\{(M+y\phi)\chi \cdot \chi + \mbox{h.c.}\}.
  \label{eq:WZint}
  \ee
  In addition to the quadratic parts $|M|^2 \phi^\dagger \phi$ and $-(1/2) M \chi \cdot \chi + 
  \mbox{h.c.}$ which 
  we have just discussed, (\ref{eq:WZint}) contains three true interactions, namely
  
  (i) a `cubic' interaction among the $\phi$ fields,
  \be 
  -\frac{1}{2}(My^* \phi \phi^{\dagger 2} + M^* y \phi^2 \phi^\dagger);\label{eq:cubic}
  \ee
  
  (ii) a `quartic' interaction among the $\phi$ fields, 
  \be 
  -\frac{1}{4} |y|^2 \phi^2 \phi^{\dagger 2};\label{eq:quartic}
  \ee 
  
  (iii) a Yukawa-type coupling between the $\phi$ and $\chi$ fields, 
  \be 
  -\frac{1}{2}\{ y \phi \chi \cdot \chi + \mbox{h.c.}\}.\label{eq:yuk}
  \ee
  It is noteworthy that the same coupling parameter $y$ enters into the cubic and quartic 
  bosonic interactions (\ref{eq:cubic}) and (\ref{eq:quartic}), as well as the Yukawa-like 
  fermion-boson interaction (\ref{eq:yuk}). In particular, the quartic coupling constant 
  appearing in (\ref{eq:quartic}) is equal to the square of the Yukawa coupling in (\ref{eq:yuk}). This 
  is exactly the relationship noted in (\ref{eq:ccss}), as being required for a cancellation between 
  the bosonic self-energy graph of figure 1, and the fermion-antifermion loop contribution to this 
  self energy (see Peskin \cite{pes} section 3.3 for details of the calculation in the massless case).   
     
  The cancellation of radiative (loop) corrections in models of this type is actually 
  a more general phenomenon: the only non-vanishing radiative corrections to the interaction 
  terms (including masses) are field rescalings (\cite{IZ}, \cite{GSR}).

  Thus far in these lectures we have adopted (pretty much) a `brute force', or 
  `do-it-yourself' approach,  retreating quite often to  explicit matrix expressions, 
  and arriving at SUSY-invariant Lagrangians by direct construction. We might well 
  wonder whether there is not a more general procedure which would somehow automatically 
  generate  SUSY-invariant interactions. Such a procedure is indeed available within the 
  superfield approach, to which we now turn. This formalism has other advantages too. First, 
  it gives us more insight into SUSY 
  transformations, and their linkage with space-time translations. Second, the appearance of the 
  auxiliary field $F$ is better motivated. And finally, and in practice rather importantly, 
  the superfield notation is widely used in discussions of the MSSM.

  \section{Superfields}
  
  \subsection{SUSY transformations on fields}
  
 By way of a warm-up exercise, let's recall some things about space-time 
 translations. A translation of coordinates takes the form 
 \be 
 x^{\prime \mu}=x^\mu + a^\mu
 \ee
 where $a^\mu$ is a constant 4-vector. In the unprimed coordinate frame, observers 
 use states $| \alpha \rangle, | \beta \rangle, \ldots$, 
 and deal with amplitudes of the form $\langle \beta | \phi(x) | \alpha \rangle$, where 
 $\phi(x)$ is \ scalar field.  In the primed frame, 
 observers evaluate $\phi$ at $x'$, and use states $| \alpha {\rangle}^\prime = 
 U | \alpha \rangle, \ldots$, where $U$ is unitary, in such a way that their matrix elements (and hence 
 transition probabilities) are equal to those calculated in the unprimed frame:
 \be 
 \langle \beta | U^{-1} \phi(x') U | \alpha \rangle = \langle \beta | \phi(x) | \alpha \rangle.
 \ee
 Since this has to be true for all pairs of states, we can deduce 
 \be 
 U^{-1} \phi(x') U = \phi(x)
 \ee
 or 
 \be 
 U \phi(x) U^{-1} = \phi(x') = \phi(x+a).\label{eq:Uphi}
 \ee
 For an infinitesimal translation, $x^{\prime \mu} = x^\mu +\epsilon^\mu$, 
 we may write 
 \be 
 U= 1 + {\rm i} \epsilon_\mu P^\mu
 \ee
 where the four operators $P^\mu$ are the {\em generators} of this transformation (c.f. (\ref{eq:SU(2)inf}));
  (\ref{eq:Uphi}) then becomes 
 \bea
 (1+{\rm i} \epsilon_\mu P^\mu) \phi(x) (1-{\rm i} \epsilon_\mu P^\mu) &=& 
 \phi(x^\mu + \epsilon^\mu) \nonumber \\
 &=& \phi(x^\mu) + \epsilon^\mu \frac{\partial \phi}{\partial x^\mu};
 \eea
 that is, 
 \be 
 \phi(x) + \delta \phi(x) = \phi(x) + \epsilon^\mu \partial_\mu \phi(x),
 \ee
 where (c.f. (\ref{eq:delqboth}))
 \be 
 \delta \phi(x) = {\rm i} \epsilon_\mu [P^\mu,\phi(x)] = \epsilon_\mu \partial^\mu \phi(x).
 \label{eq:deltaphi3}
 \ee
 We therefore obtain the fundamental relation 
 \be 
 {\rm i} [ P^\mu, \phi(x)] = \partial^\mu \phi(x).\label{eq:transphi}
 \ee
 In (\ref{eq:transphi}) the $P^\mu$ are constructed from field operators - for 
 example $P^0$ is the Hamiltonian, which is the spatial integral of the appropriate 
 Hamiltonian density - and the canonical commutation relations of the fields must 
 be consistent with (\ref{eq:transphi}). We used (\ref{eq:transphi}) in section 5.2 - 
 see (\ref{eq:pmuphi}). 
 
 But we can also look at (\ref{eq:deltaphi3}) another way: we can say 
 \be 
 \delta \phi = \epsilon^\mu \partial_\mu \phi = (1-{\rm i} \epsilon^\mu {\hat{P}}_\mu ) \phi,
 \label{eq:transdiff}
 \ee
 where ${\hat{P}}_\mu$ is a  {\em differential operator} acting on the {\em argument} of $\phi$. 
 Clearly ${\hat{P}}^\mu = {\rm i} \partial^\mu$ as usual.   
   
 We are now going to carry out analogous steps using SUSY transformations. This will entail   
   enlarging   the space of  coordinates $x^\mu$ on which the fields can depend to include  
 also {\em fermionic} degrees of freedom - 
 specifically, spinor degrees of freedom $\theta$ and $\theta^*$.  Fields which  
 depend on these spinorial degrees of freedom as well as on $x$ are called 
 {\em superfields}, and the extended space of $x^\mu, \theta$ and $\theta^*$ is called 
 {\em superspace}. Just as 
 the operators $P^\mu$ generate  (via the unitary operator $U$ of (\ref{eq:Uphi})) a shift in the 
 space-time argument 
 of $\phi$, so we expect to be able to construct analogous unitary operators 
 from $Q$ and $Q^\dagger$ which should similarly  effect  shifts in the spinorial arguments of the 
 field. Actually, we shall see that the matter is rather more interesting than that, because 
 a shift will also be induced in the space-time argument $x$; this is actually to be 
 expected, given the link between the SUSY generators and the space-time translation 
 generators $P^\mu$ embodied in the SUSY algebra (\ref{eq:QQstar1}).   
 Having constructed these operators and seen what shifts they induce, 
 we shall then look at the analogue of (\ref{eq:transdiff}), and arrive at a differential 
 operator representation of the SUSY generators, say ${\hat{Q}}$ and ${\hat{Q}}^\dagger$, 
   the differentials in this case being with 
 respect to the spinor degrees of freedom of superspace (i.e. $\theta$ and $\theta^*$). We 
 can close the circle by checking that the generators ${\hat{Q}}$ and ${\hat{Q}}^\dagger$ 
 defined this way do indeed satisfy the SUSY algebra (\ref{eq:QQstar1}) (this step 
 being analogous to checking that the angular momentum operators ${\hat{\bm L}}= 
  -{\rm i} {\bm x} \times  {\bm \nabla}$ obey 
 the SU(2) algebra).     
   
   The basic idea is simple. We may write (\ref{eq:Uphi}) as 
   \be
   {\rm e}^{{\rm i} x \cdot P} \phi (0) {\rm e}^{-{\rm i} x \cdot P} = \phi(x).  
  \ee
  In analogy to this, let's consider a `$U$' for a SUSY transformation which has 
  the form 
  \be 
  U(x, \theta, \theta^*)= {\rm e}^{{\rm i} x \cdot P} {\rm e}^{{\rm i} \theta \cdot Q} 
  {\rm e}^{{\rm i} {\bar{\theta}} \cdot {\bar{Q}}}.\label{eq:Ususy}
  \ee
  Here $Q$ and $Q^*$ (or $Q^{\dagger {\rm T}}$) are the (spinorial) SUSY generators met in section 5.2, and 
  $\theta$ and $\theta^*$ are spinor degrees of freedom associated with these 
  SUSY `translations'. Note that, as usual, 
  \be 
  \theta \cdot Q \equiv \theta^{\rm T} (-{\rm i} \sigma_2) Q,
  \ee
  and 
  \be 
  {\bar{\theta}} \cdot {\bar{Q}}  \equiv \theta^\dagger ({\rm i} 
  \sigma_2) Q^{\dagger {\rm T}}.
  \ee 
  When the  field $\phi(0)$ is transformed via `$U(x, \theta, \theta^*)\phi(0)U^{-1}(x, \theta, \theta^*)$', 
  we expect to obtain 
  a $\phi$ which is a function of $x$, but also now of the `fermionic coordinates' $\theta$ 
  and $\theta^*$, so we shall write it as $\Phi$, a  superfield:
  \be 
  U(x, \theta, \theta^*) \Phi(0) U^{-1}(x, \theta, \theta^*)= \Phi(x, \theta, \theta^*). 
  \ee 
  
  Now consider the product of two ordinary spatial translation operators:
  \be 
  {\rm e}^{{\rm i} x \cdot P} {\rm e}^{{\rm i} a \cdot P} = {\rm e}^{{\rm i} (x+a) \cdot P},
  \label{eq:prodtrans}
  \ee
   since all the components of $P$ commute.  We  say that this product of translation 
   operators `induces the transformation $x \to x+a $ in parameter (coordinate) space'. We are 
   going to generalize this by  
   multiplying  two $U$'s of the form (\ref{eq:Ususy}) together, and asking: {\em what transformations 
   are induced in the space-time coordinates,  and in the spinorial degrees of freedom}? 
   
   Such a  product is 
   \be 
   U(a,\xi, \xi^*)U(x, \theta, \theta^*)= 
   {\rm e}^{{\rm i} a \cdot P} {\rm e}^{{\rm i} \xi \cdot Q} {\rm e}^{{\rm i}{\bar{\xi}} \cdot {\bar{Q}}} 
   {\rm e}^{{\rm i} x \cdot P} {\rm e}^{{\rm i} \theta \cdot Q} {\rm e}^{{\rm i} {\bar{\theta}} \cdot {\bar{Q}}}.
   \label{eq:twoUs}
   \ee
   Unlike in (\ref{eq:prodtrans}), 
   it is {\em not} possible simply to combine all the exponents here, because the operators 
   $Q$ and $Q^\dagger$ do not commute - rather, they satisfy the algebra (\ref{eq:QQstar1}). However, 
   as noted in section 5.2, the components of $P$ do commute with those of $Q$ and $Q^\dagger$, 
   so we can freely move the operator ${\exp}[{\rm i} x \cdot P]$ through the operators to the left of it, 
   and combine it with ${\rm exp}[{\rm i} a \cdot P]$ to yield ${\rm exp}[{\rm i} (x+a) \cdot P]$, 
   as in (\ref{eq:prodtrans}). The non-trivial part is 
   \be 
   {\rm e}^{{\rm i} \xi \cdot Q} {\rm e}^{{\rm i} {\bar{\xi}} \cdot {\bar{Q}}} {\rm e}^{{\rm i} \theta 
   \cdot Q} {\rm e}^{{\rm i} {\bar{\theta}} \cdot {\bar{Q}}}.\label{eq:4exp}
   \ee
   To simplify this, we use the Baker-Campbell-Hausdorff identity:
   \be 
   {\rm e}^{A} {\rm e}^{B} = {\rm e}^{A+B+\frac{1}{2}[A,B] + \frac{1}{6}[[A,B],B] + \ldots}.
   \label{eq:BCH}
   \ee
   Let's apply (\ref{eq:BCH}) to the first two products in (\ref{eq:4exp}), taking 
   $A={\rm i} \xi \cdot Q$ and $B= {\rm i} {\bar{\xi}} \cdot {\bar{Q}}$. We get 
   \be 
   {\rm e}^{{\rm i} \xi \cdot Q} {\rm e}^{{\rm i} {\bar{\xi}} \cdot {\bar{Q}}} = 
   {\rm e}^{{\rm i} \xi \cdot Q +{\rm i} {\bar{\xi}} \cdot {\bar{Q}} - \frac{1}{2}[ \xi \cdot Q, 
   {\bar{\xi}} \cdot {\bar{Q}}] + \ldots}
   \ee
   Writing out the commutator in detail, we have 
   \bea 
   [\xi \cdot Q, {\bar{\xi}} \cdot {\bar{Q}}]&=& [\xi^1Q_1+\xi^2Q_2, \xi^*_1Q^\dagger_2-\xi^*_2Q^\dagger_1]
   \nonumber \\
   &=& [ \xi^1Q_1+\xi^2Q_2, -\xi^{2*}Q^\dagger_2 - \xi^{1*}Q^\dagger_1] \nonumber \\
   &=& [\xi^aQ_a, -\xi^{b*} Q^\dagger_b] \nonumber \\
   &=& -\xi^aQ_a\xi^{b*}Q^\dagger_b + \xi^{b*} Q^\dagger_b\xi^a Q_a \nonumber \\
   &=& \xi^a \xi^{b*} (Q_a Q^\dagger_b + Q^\dagger_b Q_a) \nonumber \\
   &=& \xi^a \xi^{b*} (\sigma^\mu)_{ab} P_\mu \label{eq:QQdagcom}
   \eea
   using (\ref{eq:QQstar1}). This means that life is not so bad after all: since $P$ commutes with 
   $Q$ and $Q^\dagger$, there are no more terms in the B-C-H identity to calculate, and we have 
   established the result 
   \be 
   {\rm e}^{{\rm i} \xi \cdot Q} {\rm e}^{{\rm i} {\bar{\xi}} \cdot {\bar{Q}}} = 
   {\rm e}^{{\rm i} A \cdot P} {\rm e}^{{\rm i} ( \xi \cdot Q + {\bar{\xi}} \cdot {\bar{Q}})},
   \label{eq:bchred1} 
   \ee
   where 
   \be 
   A^\mu = \frac{1}{2}  {\rm i} \xi^a (\sigma^\mu)_{ab} \xi^{b*}.
   \ee
   Note that we have moved the ${\rm exp}[{\rm i} A \cdot P]$ expression to the front, using the 
   fact that $P$ commutes with $Q$ and $Q^\dagger$. 
   
   We pause in the development to comment immediately on (\ref{eq:bchred1}): under this kind 
   of transformation, the spacetime coordinate acquires an additional shift, namely $A^\mu$, 
   which is built out of the spinor parameters $\xi$ and $\xi^*$. 
   
   {\bf Exercise} Explain why $ \xi^a (\sigma^\mu)_{ab} \xi^{b*}$ is a 4-vector. 
   
   Continuing on with the reduction of (\ref{eq:4exp}), we consider
   \be 
     {\rm e}^{{\rm i} \xi \cdot Q} {\rm e}^{{\rm i} {\bar{\xi}} \cdot {\bar{Q}}} {\rm e}^{{\rm i} \theta 
   \cdot Q} {\rm e}^{{\rm i} {\bar{\theta}} \cdot {\bar{Q}}}= 
   {\rm e}^{{\rm i} A \cdot P} {\rm e}^{{\rm i} ( \xi \cdot Q + {\bar{\xi}} \cdot {\bar{Q}})}
   {\rm e}^{{\rm i} \theta \cdot Q} {\rm e}^{{\rm i} {\bar{\theta}} \cdot {\bar{Q}}},
   \ee
   and apply B-C-H to the second and third terms in the product on the RHS:
   \bea 
   {\rm e}^{{\rm i} ( \xi \cdot Q + {\bar{\xi}} \cdot {\bar{Q}})}{\rm e}^{{\rm i} \theta \cdot Q}&=&
    {\rm e}^{{\rm i} ( \xi \cdot Q + {\bar{\xi}} \cdot {\bar{Q}} + \theta \cdot Q) -\frac{1}{2} [\xi \cdot Q 
    + {\bar{\xi}} \cdot {\bar{Q}}, \theta \cdot Q] + \ldots }\nonumber \\
    &=& {\rm e}^{{\rm i} ( \xi \cdot Q + {\bar{\xi}} \cdot {\bar{Q}} + \theta \cdot Q) +\frac{1}{2}
    \theta^a  (\sigma^\mu)_{ab}\xi^{b*} P_\mu}, 
    \eea
    using (\ref{eq:QQdagcom}) and (\ref{eq:comQQbar}). The expression 
    (\ref{eq:4exp}) is now 
    \be 
    {\rm e}^{- \frac{1}{2}   \xi^a (\sigma^\mu)_{ab} \xi^{b*} P_\mu + \frac{1}{2} 
     \theta^a  (\sigma^\mu)_{ab}\xi^{b*} P_\mu} {\rm e}^{{\rm i}(\xi \cdot Q + {\bar{\xi}} \cdot {\bar{Q}} 
     + \theta \cdot Q)} {\rm e}^{{\rm i} {\bar{\theta}} \cdot {\bar{Q}}}.
     \ee
     We now apply B-C-H `backwards' to the penultimate factor:
     \be 
     {\rm e}^{{\rm i}(\xi \cdot Q + {\bar{\xi}} \cdot {\bar{Q}} + \theta \cdot Q)}= 
     {\rm e}^{{\rm i}(\xi +\theta) \cdot Q} {\rm e}^{{\rm i} {\bar{\xi}} \cdot {\bar{Q}}} {\rm e}^{\frac{1}{2}
     [(\xi + \theta) \cdot Q, {\bar{\xi}} \cdot {\bar{Q}}]}.
     \ee
     Evaluating the commutator as before leads to the final result 
     \be 
       {\rm e}^{{\rm i} \xi \cdot Q} {\rm e}^{{\rm i} {\bar{\xi}} \cdot {\bar{Q}}} {\rm e}^{{\rm i} \theta 
   \cdot Q} {\rm e}^{{\rm i} {\bar{\theta}} \cdot {\bar{Q}}}= {\rm e}^{{\rm i}[-{\rm i} \theta^a 
   (\sigma^\mu)_{ab} \xi^{b*} P_\mu ]} {\rm e}^{{\rm i} (\xi+\theta) \cdot Q } {\rm e}^{{\rm i} 
   ({\bar{\xi}} + {\bar{\theta}}) \cdot {\bar{Q}}} \label{eq:4expred}
   \ee
   where in the final product we have again used (\ref{eq:comQQbar}) to 
   add the exponents. 
   
   {\bf Exercise} Check (\ref{eq:4expred}).
   
   Inspecting (\ref{eq:4expred}), we infer that the product $U(a, \xi, \xi^*) U(x, \theta, \theta^*)$  
    induces the transformations
    \bea 
   0 &\to&  \theta \to \theta + \xi \nonumber \\
    0 &\to& \theta^*  \to  \theta^* + \xi^* \nonumber \\
   0 &\to&  x^\mu \to  x^\mu +a^\mu -{\rm i} \theta^a (\sigma^\mu)_{ab} \xi^{b*}.\label{eq:UU}
    \eea 
    That is to say, 
    \be
    U(a, \xi,  \xi^*) U(x, \theta, \theta^*) \Phi(0)U^{-1}(x, \theta, \theta^*) U^{-1}(a, \xi, \xi^*) =
    U(a, \xi, \xi^*) \Phi(x, \theta, \theta^*) U^{-1}(a, \xi, \xi^*) \nonumber
    \ee
    \be
    = \Phi(x^\mu +a^\mu -{\rm i} \theta^a (\sigma^\mu)_{ab} \xi^{b*}, \theta +\xi, \theta^* + \xi^*).
    \label{eq:susydisp}
    \ee

   We now proceed with the second part of our SUSY extension of  ordinary translations, namely 
   the analogue of equation (\ref{eq:transdiff}). 
   
   \subsection{A differential operator representation of the SUSY generators}
   
   Equation (\ref{eq:transdiff}) provided us with a differential operator representation 
   of the generators of translations, by considering an infinitesimal displacement (the reader 
   might care to recall similar steps for infinitesimal rotations, which lead to the usual 
   representation of the angular momentum operators as ${\hat{\bm L}}= - {\rm i} {\bm x} \times {\bm \nabla}$). 
   Analogous steps applied to (\ref{eq:susydisp}) will lead to an explicit representation of the 
   SUSY generators as certain differential operators. We will then check that they satisfy the 
   anticommutation relations (\ref{eq:QQstar1}), just as the angular momentum operators 
   satisfy the familiar SU(2) algebra.  
   
   We regard (\ref{eq:susydisp}) as the result of applying the transformation parametrized 
   by $a, \xi, \xi^*$ to the field $\Phi(x, \theta, \theta^*)$. For an infinitesimal such 
   transformation, the change in $\Phi$ is 
   \be 
   \delta \Phi = -{\rm i} \theta^a (\sigma^\mu)_{ab} \xi^{b*} \partial_\mu \Phi +
    \xi^a \frac{\partial \Phi}{\partial \theta^a} + \xi^*_a \frac{\partial \Phi}{\partial \theta^*_a}.
    \label{eq:inflsusy}
    \ee 
    
    {\footnotesize{{\bf Notation check} In Notational Aside (1), section 2.2, we stated the 
    convention for summing over undotted labels, which was `diagonally from top left to 
    bottom right, as in $\xi^a \chi_a$'. For (\ref{eq:inflsusy}) to be consistent with this 
    convention, it should be the case that the derivative $\partial / \partial \theta^a$ 
    behaves as a `$\chi_a$'-type object. A quick way of seeing that this is likely to be OK is 
    simply to calculate 
    \be 
    \frac{\partial}{\partial \theta^a} (\theta^b \theta_b) 
    \ee
    Consider $a=1$. Now $\theta^b \theta_b = -2\theta^1 \theta^2$ and so 
    \be 
    \frac{\partial}{\partial \theta^1} (\theta^b \theta_b) = -2 \theta^2 = 2 \theta_1.
    \ee 
    Similarly, 
    \be 
    \frac{\partial}{\partial \theta^2} (\theta^b \theta_b) = 2 \theta_2,
    \ee
    or generally 
    \be 
    \frac{\partial}{\partial \theta^a} (\theta \cdot \theta) = 2 \theta_a,\label{eq:dtheta2}
    \ee
    which at least checks the claim in this simple case. Similarly, in Notational 
    Aside (2), we stated the convention for products of dotted indices as 
    $\psi_{\dot{a}}\zeta^{\dot{a}}$, and in Aside (3) we related dotted-index 
    quantities to complex conjugated quantities, via ${\bar{\chi}}_{\dot{a}} \equiv \chi^*_a$. 
    Consider the last term in (\ref{eq:inflsusy}): since $\xi^*_a \equiv {\bar{\xi}}_{\dot{a}}$, 
    it should be the case that $\partial / \partial \theta^*_a$ behaves as a `${\bar{\zeta}}^{\dot{a}}$'-type 
    (or equivalently as a `$\zeta^{a *}$') object.
    
    {\bf Exercise} Check this by considering $\partial / \partial \theta^*_a ({\bar{\theta}} \cdot 
    {\bar{\theta}})$.
    
    }}

    In analogy with (\ref{eq:transdiff}), we want to write (\ref{eq:inflsusy}) as 
    \be 
    \delta \Phi = (-{\rm i} \xi \cdot \hat{Q} -{\rm i} {\bar{\xi}} \cdot {\bar{\hat{Q}}}) \Phi
     = (-{\rm i} \xi^a 
    {\hat{Q}}_a   - {\rm i} \xi^*_a {\hat{Q}}^{\dagger a}) \Phi.\label{eq:infsusyQ}
    \ee 
    Comparing (\ref{eq:inflsusy}) with (\ref{eq:infsusyQ}), it is easy to identify ${\hat{Q}}_a$ as 
    \be 
    {\hat{Q}}_a = {\rm i} \frac{\partial}{\partial \theta^a}.\label{eq:hatQ}
    \ee
    There is a similar term in  ${\hat{Q}}^{\dagger a}$, namely   
    \be 
    {\hat{Q}}^{\dagger a} = {\rm i} \frac{\partial}{\partial \theta^*_a},\label{eq:Qdaghat1}
    \ee
  and in addition another contribution  given by 
  \be 
  -{\rm i} \xi^*_a {\hat{Q}}^{\dagger a} \Phi = -{\rm i} \theta^a (\sigma^\mu)_{ab} \xi^{b*} \partial_\mu 
  \Phi.\label{eq:Qdaghat2}
  \ee
   Our present objective is to verify that these ${\hat{Q}}$ operators satisfy the SUSY anticommutation 
   relations (\ref{eq:QQstar1}). To do this, we need to deal with the lower-index operators 
   ${\hat{Q}}^\dagger_a$ rather than  ${\hat{Q}}^{\dagger a}$. 
   
   {\bf Exercise} Check that (\ref{eq:Qdaghat1}) can be converted to
   \be 
   {\hat{Q}}^\dagger_a = -{\rm i} \frac{\partial}{\partial \theta^{a*}}.
   \ee
   
   As regards (\ref{eq:Qdaghat2}), we use $\xi^*_a {\hat{Q}}^{\dagger a} = - \xi^{a*} 
   {\hat{Q}}^\dagger_a$ (see Exercise (b) in Notational Aside (3)) 
   , and $\theta^a \xi^{b*} = -\xi^{b*} \theta^a$, followed by 
   an interchange of the indices $a$ and $b$ to give finally
   \be 
   {\hat{Q}}^\dagger_a = - {\rm i} \frac{\partial}{\partial \theta^{a *}} + \theta^b 
   (\sigma^\mu)_{ba} \partial_\mu.\label{eq:hatQdag}
   \ee 
    It is now a useful {\bf Exercise} to check that the explicit representations 
    (\ref{eq:hatQ}) and (\ref{eq:hatQdag}) do indeed result in the required relations 
    \be 
    [{\hat{Q}}_a, {\hat{Q}}^\dagger_b] = {\rm i} (\sigma^\mu)_{ab} \partial_\mu = 
    (\sigma^\mu)_{ab} {\hat{P}}_\mu,  
    \ee 
   as well as $[{\hat{Q}}_a, {\hat{Q}}_b] = [{\hat{Q}}^\dagger_a, {\hat{Q}}^\dagger_b]=0$. We 
   have therefore produced a representation of the SUSY generators in terms of fermionic 
   parameters, and derivatives with respect to them, which satisfies the SUSY algebra (\ref{eq:QQstar1}).  
   
   \subsection{Chiral superfields, and their (chiral) component fields}
   
   Suppose now that a superfield $\Phi(x, \theta, \theta^*)$  does not in fact 
   depend on $\theta^*$, only on $x$ and $\theta$: $\Phi(x, \theta)$\footnote{Such a superfield is usually 
   called a `left-chiral superfield' and denoted by $\Phi_{\rm L}$, because 
   (see (\ref{eq:chisupexp})) it contains only the L-type spinor $\chi$, and not the 
   R-type spinor $\psi$. By the same token, the transformation (\ref{eq:Ususy}) could be 
   denoted by $U_{\rm L}$, while the representations (\ref{eq:hatQ}) and (\ref{eq:hatQdag}) of 
   the generators ${\hat{Q}}_a$ and ${\hat{Q}}^\dagger_a$ could be called the L-representation of 
   these operators. For more on this see the comment in small print soon after (\ref{eq:deltaF1}). 
   We shall only be dealing with left-chiral superfields and we shall therefore omit the L-subscript.}.  
   Consider the expansion of  
   such a $\Phi$  in powers of $\theta$. Because of the fermionic nature of the 
   variables $\theta$, which implies that $(\theta_1)^2 = (\theta_2)^2=0$, there 
   will only be three terms in the expansion, namely a term independent of $\theta$, 
   a term linear in $\theta$ and a term involving $\frac{1}{2} \theta \cdot \theta = 
   -\theta_1 \theta_2$:
  \be
  \Phi(x, \theta)= \phi(x) + \theta \cdot \chi(x) + \frac{1}{2} \theta \cdot \theta F(x).
  \label{eq:chisupexp} 
  \ee
  This is the most general form of such a superfield (which depends only on $x$ and 
  $\theta$), and it depends on three {\em component} fields, $\phi, \chi$ and $F$. We have 
  of course deliberately given  these component fields the same names as those in 
  our previous chiral supermultiplet. We shall now verify that the transformation law 
  (\ref{eq:infsusyQ}) for the superfield $\Phi$, 
  with ${\hat{Q}}$ given  by (\ref{eq:hatQ}) and ${\hat{Q}}^\dagger$ by (\ref{eq:hatQdag}), 
   implies precisely the previous transformations 
  (\ref{eq:SUSYi}) for the component fields $\phi$, $\chi$ and $F$, 
  thus justifying this identification.

   We have 
   \bea 
   \delta \Phi &=& (-{\rm i} \xi^a {\hat{Q}}_a -{\rm i} \xi^*_a {\hat{Q}}^{\dagger a}) \Phi = 
   (-{\rm i} \xi^a {\hat{Q}}_a +{\rm i} \xi^{a *} {\hat{Q}}^\dagger_a)\Phi \nonumber \\
   &=& (\xi^a \frac{\partial}{\partial \theta_a} + \xi^{a*} \frac{\partial}{\partial \theta^{a*}} 
   +{\rm i} \xi^{a*} \theta^b (\sigma^\mu)_{ba} \partial_\mu) [\phi(x) + \theta^c \chi_c 
   +\frac{1}{2} \theta \cdot \theta F] \nonumber \\
   &\equiv& \delta_\xi \phi + \theta^a \delta_\xi \chi_a + \frac{1}{2} \theta \cdot \theta \delta_\xi F.
   \label{eq:delPhiinf} 
   \eea
  We evaluate the derivatives in the second line as follows. First, we have  
  \be 
  \frac{\partial}{\partial \theta^a} [\theta^c \chi_c + \frac{1}{2} \theta \cdot \theta] = 
  \chi_a +  \theta_a, 
  \ee
  using (\ref{eq:dtheta2}), so that  the $\xi^a \partial / \partial \theta^a$ term yields 
  \be 
  \xi^a \chi_a + \theta^a \xi_a F. 
  \ee
  Next, the term in $\partial/ \partial \theta^{a*}$ vanishes since 
  $\Phi$ doesn't depend on $\theta^*$. The remaining term is 
  \be 
  {\rm i} \xi^{a*} \theta^b (\sigma^\mu)_{ba} \partial_\mu \phi +
  {\rm i} \xi^{a*} \theta^b (\sigma^\mu)_{ba} \theta^c \partial_\mu \chi_c; \label{eq:X}
  \ee
  note that the fermionic nature of $\theta$ precludes any cubic term in $\theta$.  
  The first  term in (\ref{eq:X}) can alternatively be written as 
  \be 
  -{\rm i} \theta^b (\sigma^\mu)_{ba} \xi^{a*} \partial_\mu \phi.
  \ee
  Referring to (\ref{eq:delPhiinf}) we can therefore identify the part independent of 
  $\theta$ as 
  \be 
  \delta_\xi \phi = \xi^a \chi_a,\label{eq:delphi4}
  \ee
  and the part linear in $\theta$ as 
  \be 
  \theta^a \delta_\xi \chi_a = \theta^a (\xi_a F -{\rm i} (\sigma^\mu)_{ab} \xi^{b*} \partial_\mu \phi).
  \label{eq:delchi4}
  \ee 
  Since (\ref{eq:delchi4}) has to be true for all $\theta$ we can remove the $\theta^a$ 
  throughout, and then (\ref{eq:delphi4}) and (\ref{eq:delchi4}) indeed reproduce 
  (\ref{eq:SUSYi}) for the fields $\phi$ and $\chi$ (recall that $({\rm i} \sigma_2 \xi^*)_b
  =\xi^{b*}$).
  
  We are left with the second term of (\ref{eq:X}), which is bilinear in $\theta$, and which 
  ought to yield $\delta_\xi F$. We manipulate this term as follows. First, we write 
  the general product $\theta^a \theta^b$ in terms of the scalar product $\theta \cdot \theta$ 
  by using the result of this exercise: 
  
  {\bf Exercise} Show that $\theta^a \theta^b = -\frac{1}{2} \epsilon^{ab} \theta \cdot \theta$, 
  where $\epsilon^{12}=1, \epsilon^{21}=-1, \epsilon^{11}=\epsilon^{22}=0$.
  
  The second term in (\ref{eq:X}) is then 
  \be 
  -{\rm i} \xi^{a*} (\sigma^{\mu {\rm T} })_{ab} \epsilon^{bc} \partial_\mu \chi_c \frac{1}{2}\theta 
  \cdot \theta.\label{eq:333}
  \ee
  Comparing this with (\ref{eq:delPhiinf}) we deduce 
  \be 
  \delta_\xi F= -{\rm i} \xi^{a*} (\sigma^{\mu {\rm T} })_{ab} \epsilon^{bc} \partial_\mu \chi_c.
  \label{eq:deltaF1} 
  \ee
  
  {\bf Exercise} Verify that this is in fact the same as the $\delta_\xi F$ given in (\ref{eq:SUSYi})
   (remember that `$\xi^\dagger$' means $(\xi^*_1,  \xi^*_2)$, not $(\xi^{1*}, \xi^{2*}$).  
  
    So the chiral superfield $\Phi(x, \theta)$ of  (\ref{eq:chisupexp}) contains the 
    component fields $\phi$, $\chi$ and $F$ transforming correctly under SUSY 
    transformations; we say that the chiral superfield provides a linear 
    representation of the SUSY algebra. Note that {\em three} component fields ($\phi$, $\chi$ and 
    $F$) are required for this result: here is a more `deductive' justification for the introduction 
    of the field $F$.  
    
    {\footnotesize{ The thoughtful reader may be troubled by the following thought. 
    Our development has been based on the form (\ref{eq:Ususy}) for the unitary operator 
     associated with finite SUSY transformations. But we could have started, instead,  from 
     \be 
     U_{\rm red}(x, \theta, \theta^*)={\rm e}^{{\rm i} x \cdot P} {\rm e}^{{\rm i} [\theta \cdot Q + 
     {\bar{\theta}} 
     \cdot {\bar{Q}}]},  \label{eq:U11}
     \ee
      and since $Q$ and $Q^\dagger$ don't commute, (\ref{eq:U11}) is not the same 
     as (\ref{eq:Ususy}). Indeed, (\ref{eq:U11}) might be regarded as more natural -  and 
     certainly more in line with the angular momentum case, which also involves non-commuting 
     generators, and where the corresponding unitary operator is 
     ${\rm exp}[{\rm i} {\bm \alpha} \cdot {\bm J}]$. In the case of (\ref{eq:U11}), the 
     induced transformation corresponding to (\ref{eq:UU}) is 
     \bea
     0 &\to& \theta \to \theta + \xi \nonumber \\
     0 &\to& \theta^* \to \theta^* + \xi^* \nonumber \\
     0 &\to&  x^\mu \to x^\mu + a^\mu +\frac{1}{2}{\rm i} \xi^a (\sigma^\mu)_{ab} \theta^{b*} 
     -\frac{1}{2}{\rm i} \theta^a (\sigma^\mu)_{ab}\xi^{b*}.\label{eq:UU11}
     \eea 
     We can again find differential operators representing the SUSY generators by 
     expanding the change in the  field 
      up to first order in $\xi$ and $\xi^*$, as in (\ref{eq:inflsusy}), and this will 
     lead to different expressions from those given in (\ref{eq:hatQ}) and (\ref{eq:hatQdag}). 
     However, the  new operators will be found to satisy the {\em same} SUSY algebra 
     (\ref{eq:QQstar1}). We could also imagine using 
     \be 
     U_{\rm R}(x, \theta, \theta^*)={\rm e}^{{\rm i} x \cdot P}{\rm e}^{{\rm i} {\bar{\theta}} \cdot 
     {\bar{Q}}} {\rm e}^{{\rm i} \theta \cdot Q},
     \ee
     which is not the same either, and for which the induced transformation is 
      \bea
     0 &\to& \theta \to \theta + \xi \nonumber \\
     0 &\to& \theta^* \to \theta^* + \xi^* \nonumber \\
      0 &\to& x^\mu \to x^\mu + a^\mu +{\rm i} \xi^a (\sigma^\mu)_{ab} \theta^{b*}. 
      \label{eq:UU2}
      \eea 
      Here also yet a third set of (differential operator) generators will be found, but 
      again they'll satisfy the same SUSY algebra (\ref{eq:QQstar1}). The superfields 
      produced by $U$ - or more properly in the present context $U_{\rm L}$ - 
      of (\ref{eq:Ususy}), $U_{\rm red}$ and $U_{\rm R}$ are called ` left' or `type I' 
      (our $\Phi$, or more properly $\Phi_{\rm L}$), `reducible' or `real'  ($\Phi_{\rm red}$),  
      and `right' or `type II' ($\Phi_{\rm R}$)  superfields respectively. It can be shown that 
      \be 
      \Phi_{\rm red}(x, \theta, \theta^*)= \Phi_{\rm L}(x^\mu-\frac{1}{2}
      {\rm i}\theta^a (\sigma^\mu)_{ab}\theta^{b*}, 
      \theta, \theta^*)= \Phi_{\rm R}(x^\mu +\frac{1}{2}{\rm i}\theta^a (\sigma^\mu)_{ab}\theta^{b*}, 
      \theta, \theta^*).
      \ee 
      
      We can expand $\Phi_{\rm red}(x, \theta, \theta^*)$ as a power series 
      in $\theta$ and $\theta^*$, just as we did for $\Phi(x, \theta)$. But such an expansion will 
      contain a lot more terms than (\ref{eq:chisupexp}), and will  involve  
       more component fields than $\phi$, $\chi$ and $F$. This `enlarged' superfield will again 
       provide a representation of the SUSY algebra, but it will be a {\em reducible} one, in the 
       sense that we'd find that we could pick out sets of components that only transformed 
       among themselves - such as those in a chiral supermultiplet, for example. These {\em 
       irreducible} sets of fields can be selected out from the beginning by applying a 
       suitable constraint. For example, we got straight to the irreducible left chiral 
       supermultiplet by starting with the  chiral superfield $\Phi$ and requiring it not to 
       depend on $\theta^*$. In general, the constraints which may be applied must 
       commute with the SUSY transformation when expressed in terms of differential 
       operators. See \cite{bailin} section 2.2.     
      
      }}
      
      We close this rather heavily formal section with a most important observation: 
      {\em the change in the $F$ field, (\ref{eq:deltaF1}), is actually a total derivative, 
      since the parameters $\xi$ are independent of $x$; it follows that, in 
      general, the `$F$-component' of a chiral superfield, in the sense of the expansion 
      (\ref{eq:chisupexp}), will always transform by a total derivative - and will therefore 
      automatically correspond to a SUSY-invariant Action.} 
      
      We now consider products of chiral superfields, and show how to exploit the 
      italicized remark so as to obtain SUSY-invariant interactions - in particular, those of the 
      W-Z model introduced in section 8. 
        
       \subsection{Products of chiral superfields}
       
       Let $\Phi_i$ be a chiral superfield (understood to be of `left' type) 
       where, as in section 8, the suffix $i$ labels 
       the gauge and flavour degrees of freedom of the component fields. $\Phi_i$ has an 
       expansion of the form (\ref{eq:chisupexp}):
       \be 
       \Phi_i(x, \theta)  = \phi_i(x) + \theta \cdot \chi_i(x) + \frac{1}{2}\theta \cdot \theta F_i(x). 
       \ee
       Consider now the product of two such superfields:
       \be 
       \Phi_i \Phi_j =(\phi_i + \theta \cdot \chi_i + \frac{1}{2} \theta \cdot \theta F_i)(\phi_j +
       \theta \cdot \chi_j + \frac{1}{2} \theta \cdot \theta F_j).\label{eq:PhiPhi}
       \ee
       On the RHS there are the following terms: 
       \be 
       \mbox{independent of $\theta$:}\  \phi_i \phi_j;
       \ee
       \be
       \mbox{linear in $\theta$:} \ \theta \cdot (\chi_i \phi_j + \chi_j \phi_i);  
       \ee
       \be 
       \mbox{bilinear in $\theta$:} \ \frac{1}{2}\theta \cdot \theta (\phi_iF_j + \phi_j F_i) + 
       \theta \cdot \chi_i \; \theta \cdot \chi_j.\label{eq:phiphi3}
       \ee
       In the second term of (\ref{eq:phiphi3}) we use the result given in the Exercise above 
       equation (\ref{eq:333}) to write it as 
       \bea \theta \cdot \chi_i \; \theta \cdot \chi_j &=& \theta^a \chi_{ia} \theta^b \chi_{jb} = 
       -\theta^a \theta^b \chi_{ia} \chi_{jb} \nonumber \\
       &=&\frac{1}{2}\epsilon^{ab} \theta \cdot \theta \chi_{ia} \chi_{jb} = \frac{1}{2}
       \theta \cdot \theta (\chi_{i1} \chi_{j2} 
       - \chi_{i2} \chi_{j1}) \nonumber \\
       &=& -\frac{1}{2} \theta \cdot \theta \chi_i \cdot \chi_j.\label{eq:theta2}
       \eea 
       Hence the term in the product (\ref{eq:PhiPhi}) which is bilinear in $\theta$ is 
       \be 
       \frac{1}{2}\theta \cdot \theta (\phi_i F_j + \phi_j F_i - \chi_i \cdot \chi_j).
       \ee
       
       {\bf Exercise} Show that the terms in the product (\ref{eq:PhiPhi}) which are cubic 
       and quartic in $\theta$ vanish. 
       
        Altogether, then, we have shown that if the product (\ref{eq:PhiPhi}) is itself 
        expanded in component fields via 
        \be 
        \Phi_i\Phi_j=\phi_{ij} + \theta \cdot \chi_{ij} + \frac{1}{2} \theta \cdot \theta F_{ij},
        \ee
        then 
        \be 
        \phi_{ij}=\phi_i \phi_j, \ \ \chi_{ij}=\chi_i \phi_j + \phi_j \chi_i, \ \ F_{ij} 
        = \phi_i F_j + \phi_j F_i - \chi_i \cdot \chi_j.\label{eq:sfieldcpts}
        \ee

  Suppose now that we introduce a quantity $W_{\rm quad}$ defined by 
  \be 
  W_{\rm quad}=\left. \frac{1}{2} M_{ij} \Phi_i \Phi_j \right|_{F},\label{eq:WquadPhi}
  \ee
  where `$|_F$' means `the $F$-component of' (i.e. the coefficient of $\frac{1}{2} \theta \cdot 
  \theta$ in the product). Here $M_{ij}$ is taken to be symmetric in $i$ and $j$. Then 
   \bea 
   W_{\rm quad}&=&\frac{1}{2} M_{ij} ( \phi_i F_j + \phi_j F_i - \chi_i \cdot \chi_j)\nonumber \\ 
  &=& M_{ij} \phi_i F_j - \frac{1}{2}M_{ij} \chi_i \cdot \chi_j.\label{eq:Wquad}
  \eea
  Referring back to the italicized comment at the end of the previous subsection, the fact that 
  (\ref{eq:Wquad}) is the $F$-component of a chiral superfield (which is the product of two other 
  such superfields, in this case), guarantees that the terms in (\ref{eq:Wquad}) provide a 
  SUSY-invariant Action. And in fact they are precisely the terms involving $M_{ij}$ in the W-Z 
  model of section 8: see (\ref{eq:Lsusyint}) with $W_{i}$ given by the 
  first term in (\ref{eq:Wi}),  and $W_{ij}$ given by the first term in 
  (\ref{eq:linW}). Note also that our $W_{\rm quad}$ has exactly the same form, as a function 
  of $\Phi_i$ and $\Phi_j$, as the $M_{ij}$ part of $W$ in 
   (\ref{eq:W}) had, as a function of $\phi_i$ and $\phi_j$. 
  
  Thus encouraged, let's go on to consider the product of three chiral superfields:
  \be 
  \Phi_i \Phi_j \Phi_k = [\phi_i \phi_j + \theta \cdot (\chi_i \phi_j + \chi_j \phi_i) + 
  \frac{1}{2} \theta \cdot \theta (\phi_i F_j + \phi_j F_i - \chi_i \cdot \chi_j)][
  \phi_k + \theta \cdot \chi_k + \frac{1}{2}\theta \cdot \theta F_k].\label{eq:Phi3}
  \ee 
  Because our interest is confined to obtaining candidates for SUSY-invariant Actions, we 
  shall only be interested in the $F$ component. Inspection of (\ref{eq:Phi3}) yields the 
  obvious terms 
  \be 
  \phi_i \phi_j F_k + \phi_j \phi_k F_i + \phi_k \phi_i F_j - \chi_i \cdot \chi_j \phi_k.
  \ee
  In addition, the term $\theta \cdot (\chi_i \phi_j +  \chi_j \phi_i)\theta \cdot \chi_k$ 
  can be re-written as in (\ref{eq:theta2}) to give 
  \be 
  -\frac{1}{2} \theta \cdot \theta (\chi_i \phi_j + \chi_j \phi_i) \cdot \chi_k.
  \ee  
  So altogether 
  \be 
 \left.  \Phi_i \Phi_j \Phi_k \right|_F = \phi_i \phi_j F_k + \phi_j \phi_k F_i + \phi_k \phi_i F_j  
 -\chi_i \cdot \chi_j \phi_k - \chi_j \cdot \chi_k \phi_i - \chi_i \chi_k \phi_j. \label{eq:cubicF} 
 \ee
  
  Let's now consider the cubic analogue of (\ref{eq:WquadPhi}), namely 
  \be 
  W_{\rm cubic} = \left. \frac{1}{6} y_{ijk} \Phi_i \Phi_j \Phi_k \right|_F, 
  \ee
  where the coefficients $y_{ijk}$ are totally symmetric in $i$, $j$ and $k$. Then 
  from (\ref{eq:cubicF}) we immediately obtain 
  \be 
  W_{\rm cubic} = \frac{1}{2} y_{ijk} \phi_i \phi_j F_k -\frac{1}{2} y_{ijk} \chi_i \cdot \chi_j \phi_k.
  \label{eq:cubicW} 
  \ee
    Sure enough, the first term here is precisely the first term in (\ref{eq:Lsusyint}) with 
    $W_i$ given by the second ($y_{ijk}$) term in (\ref{eq:Wi}), while the second term in 
    (\ref{eq:cubicW}) is the second term in (\ref{eq:Lsusyint}) with $W_{ij}$ given by 
    the $y_{ijk}$ term in (\ref{eq:linW}). Note, again, that our $W_{\rm cubic}$ has exactly the 
    same form, as a function of the $\Phi$'s, as the $y_{ijk}$ part of the $W$ in (\ref{eq:W}), 
    as a function of the $\phi$'s.

    Thus we have shown that  all the interactions found in  
    section 8 can be expressed as $F$-components of products of superfields, a result which 
    guarantees the SUSY-invariance of the associated Action. Of course, we must also include 
    the Hermitian conjugates of the terms considered here. Because all the interactions are 
    generated from the superfield products in $W_{\rm quad}$ and $W_{\rm cubic}$, such 
    $W$'s are called {\em superpotentials}. The full superpotential for the W-Z model is 
    thus 
    \be 
    W=\frac{1}{2}M_{ij} \Phi_i \Phi_j + \frac{1}{6}y_{ijk} \Phi_i \Phi_j \Phi_k, \label{eq:WPhis}
    \ee
    it being understood that the $F$-component is to be taken in the Lagrangian.
    
    {\footnotesize{The 
    understanding is often made explicit by integrating over $\theta_1$ and $\theta_2$. 
    Integrals over such anticommuting variables are defined by the following rules:
    \be  
    \int {\rm d} \theta_1 1 =0;  \int {\rm d} \theta_1 \, \, \theta_1 = 1; \int{\rm d} \theta_1 
    \int {\rm d} \theta_2 \, \, \theta_2 \theta_1 =1  
    \ee
     (see Appendix O of \cite{AH32} for example. These rules imply that 
     \be 
     \int {\rm d} \theta_1 \int {\rm d} \theta_2 \, \, \frac{1}{2}\theta \cdot \theta = 
     \int {\rm d} \theta_1 \int {\rm d} \theta_2 \, \, \theta_2 \theta_1 =1.  
     \ee
     On the other hand, we can write 
     \be 
     {\rm d} \theta_1 \, {\rm d} \theta_2 = - {\rm d} \theta_2 \, {\rm d} \theta_1=- \frac{1}{2} 
     {\rm d} \theta \cdot {\rm d} \theta \equiv {\rm d}^2 \theta.  
     \ee
     It then follows that 
     \be 
     \int {\rm d}^2 \theta \, \, W = {\mbox{coefficient of $\frac{1}{2}\theta \cdot \theta$ in $W$ (i.e. the 
     $F$ component)}}.  
     \ee
     Such integrals are commonly used to project out the desired parts of superfield expressions.}}
     
     As already 
    noted, the functional form of (\ref{eq:WPhis}) is the same as that of (\ref{eq:W}), 
    which is why they are both called $W$. Note, however, that the $W$ of (\ref{eq:WPhis}) 
    includes, of course, {\em all} the interactions of the W-Z model, not only those 
    involving the $\phi$ fields alone. In the MSSM, superpotentials of the 
    form (\ref{eq:WPhis}) describe the non-gauge interactions of the fields - 
    that is, in fact, interactions involving the Higgs supermultiplets; in this case 
    the quadratic and cubic products of the $\Phi$'s must be constructed so as to be 
    singlets (invariant) under the gauge groups.
    
    It is  time to consider other supermultiplets, in particular ones containing 
    gauge fields, with a view to supersymmetrizing the gauge interactions of the SM. 
    
    \section{ Vector (or Gauge) Supermultiplets}

    Having developed a certain amount of  superfield formalism, it might seem 
    sensible to use it now to discuss supermultiplets containing vector (gauge) 
    fields. But although this is of course perfectly possible (see for example 
    \cite{bailin} chapter 3), it is actually fairly complicated, and we prefer 
    the `try it and see' approach that we used in section 3, which  (as  before)  
     establishes  the appropriate SUSY transformations more intuitively.   
     We begin with a simple example, a kind of vector analogue of the model 
    of section 3. 
    
    \subsection{The free Abelian gauge supermultiplet}
    
    Consider a simple massless U(1) gauge field $A^\mu(x)$, like that of the photon. The spin 
    of such a field is 1, but on-shell it contains only two (rather than three)  degrees of freedom, 
    both transverse to the direction of propagation. As we saw in section 6, we expect that SUSY 
    will partner this field with a spin-1/2 field, also with two on-shell degrees of freedom. Such a
    fermionic partner of a gauge field is called generically a `gaugino'. This one 
    is a photino, and we'll denote its field by $\lambda$, and take it to be L-type.  Being in the same 
    multiplet as the photon, it must have the same `internal' quantum numbers as the photon, 
    in particular it must be electrically neutral. So it doesn't have any coupling to the photon. 
    The photino must also have the same mass as the photon, namely zero.  
    The Lagrangian is therefore just a sum of the Maxwell term for the photon, and the  
    appropriate free massless spinor term for the photino:
    \be 
    {\cal{L}}_{\gamma \lambda}= -\frac{1}{4}F_{\mu \nu} F^{\mu \nu} +{\rm i}
     \lambda^\dagger {\bar{\sigma}}^\mu \partial_\mu \lambda,\label{eq:Lgamlam}
     \ee
     where as usual $F^{\mu \nu}= \partial^\mu A^\nu - \partial^\nu A^\mu$. 
      We now set about investigating what might be the SUSY transformations between 
     $A^\mu$ and $\lambda$, such that the Lagrangian (\ref{eq:Lgamlam}) (or the corresponding 
     Action) is invariant.
     
     We anticipate that, as with the chiral supermultiplet, we shall not be able consistently to  
     ignore the off-shell degree of freedom of the gauge field  - but we shall 
     start by doing so.  
     First, consider $\delta_\xi A^\mu$. This has to be a 4-vector, and also a real 
     rather than complex quantity, linear in $\xi$ and $\xi^*$. We try (recalling the 
     4-vector combination from section 2.2) 
     \be 
     \delta_\xi A^\mu=\xi^\dagger {\bar{\sigma}}^\mu \lambda + \lambda^\dagger {\bar{\sigma}}^\mu \xi,
     \label{eq:delAmu}
     \ee  
    where $\xi$ is also an L-type spinor, but has dimension $M^{-1/2}$ as in (\ref{eq:dimxi}). 
    The spinor field $\lambda$ has dimension $M^{3/2}$, so (\ref{eq:delAmu}) is consistent with $A^\mu$ 
    having the desired dimension $M^1$.
    
    What about $\delta_\xi \lambda$? This must presumably be proportional to $A^\mu$ - or 
    better, since $\lambda$ is gauge-invariant, to the gauge-invariant quantity $F^{\mu \nu}$, so we try 
    \be 
    \delta_\xi \lambda \sim \xi F^{\mu \nu}.\label{eq:dellamsim}
    \ee
    Since the dimension of $F^{\mu \nu}$ is $M^2$, we see that the dimensions already 
    balance on both sides of (\ref{eq:dellamsim}), so there is no need to introduce any 
    derivatives. But we do need to absorb the two Lorentz indices $\mu$ and $\nu$ on the RHS, 
    and leave ourselves with something transforming correctly as an L-type spinor. This can be 
    neatly done by recalling (section 2.2) that the quantity ${\bar{\sigma}}^\nu \xi$ transforms as 
    an R-type spinor $\psi$, while $\sigma^\mu \psi$ transforms as an L-type spinor. So we try 
    \be 
    \delta_\xi \lambda = C \sigma^\mu {\bar{\sigma}}^\nu \xi F_{\mu \nu},
    \ee
    whee $C$ is a constant to be determined. Then we also have 
    \be 
    \delta_\xi \lambda^\dagger = C^* \xi^\dagger {\bar{\sigma}}^\nu \sigma^\mu F_{\mu \nu}.
    \ee
    
    Consider the SUSY variation of the Maxwell term in (\ref{eq:Lgamlam}). Using the antisymmetry of 
    $F^{\mu \nu}$ we have 
    \bea 
    \delta_\xi\left(-\frac{1}{4}F_{\mu \nu} F^{\mu \nu}\right)&=&-\frac{1}{2}F_{\mu \nu}
    (\partial^\mu \delta_\xi 
    A^\nu - \partial^\nu \delta_\xi A^\mu) \nonumber \\
    &=&-F_{\mu \nu} \partial^\mu \delta_\xi A^\nu \nonumber \\
    &=& - F_{\mu \nu} \partial^\mu(\xi^\dagger {\bar{\sigma}}^\nu \lambda + \lambda^\dagger 
    {\bar{\sigma}}^\nu \xi).\label{eq:delmaxlam}
    \eea
    The variation of the spinor term is 
    $$
    {\rm i} (\delta_\xi \lambda^\dagger) {\bar{\sigma}}^\mu \partial_\mu \lambda 
    +{\rm i} \lambda^\dagger {\bar{\sigma}}^\mu \partial_\mu (\delta_\xi \lambda)  
    $$
    \be  
    = {\rm i} (C^*\xi^\dagger {\bar{\sigma}}^\nu \sigma^\mu F_{\mu \nu})
     {\bar{\sigma}}^\rho \partial_\rho \lambda + {\rm i} C \lambda^\dagger {\bar{\sigma}}^\rho \partial_\rho 
     (\sigma^\mu {\bar{\sigma}}^\nu \xi F_{\mu \nu})  
     .\label{eq:dellamF}
     \ee 
    The $\xi$ part of (\ref{eq:delmaxlam}) must cancel the $\xi$ part of (\ref{eq:dellamF}) 
    (or else their sum must be expressible as a total derivative), and the same is true of the 
    $\xi^\dagger$ parts. So consider the $\xi^\dagger$ part of (\ref{eq:dellamF}). It is 
    \be 
    {\rm i} C^* \xi^\dagger {\bar{\sigma}}^\nu \sigma^\mu {\bar{\sigma}}^\rho \partial_\rho \lambda 
    F_{\mu \nu} = -{\rm i} C^* \xi^\dagger {\bar{\sigma}}^\mu \sigma^\nu {\bar{\sigma}}^\rho 
    \partial_\rho \lambda F_{\mu \nu}.\label{eq:dellamxidag}
    \ee
    Now the $\sigma$'s are just Pauli matrices, together with the 
    identity matrix, and we know that  products of two Pauli matrices  
    will give either the identity matrix or a third Pauli matrix. Hence products 
    of three $\sigma$'s as in (\ref{eq:dellamxidag}) must be expressible as a linear 
    combination of $\sigma$'s. The identity we need is
    \be
    {\bar{\sigma}}^\mu \sigma^\nu {\bar{\sigma}}^\rho=g^{\mu \nu} {\bar{\sigma}}^\rho -
    g^{\mu \rho} {\bar{\sigma}}^\nu +g^{\nu \rho}{\bar{\sigma}}^\mu -{\rm i} \epsilon^{\mu \nu 
    \rho \delta} {\bar{\sigma}}_\delta.\label{eq:3sigid}
    \ee
    When (\ref{eq:3sigid}) is inserted into (\ref{eq:dellamxidag}), some simplifications 
    occur. First, the term involving $....g^{\mu \nu}....F_{\mu \nu}$ vanishes, because 
    $g^{\mu \nu}$ is symmetric in its indices while $F_{\mu \nu}$ is antisymmetric. Next, 
    we can do a partial integration to re-write $\partial_\rho \lambda F_{\mu \nu}$ as 
    $-\lambda \partial_\rho F_{\mu \nu} = - \lambda (\partial_\rho \partial_\mu A_\nu - 
    \partial_\rho \partial_\nu A_\mu)$. The first of these two terms is symmetric under 
    interchange of $\rho$ and $\mu$, and the second is symmetric under interchange of 
    $\rho$ and $\nu$. But they are both multiplied by $\epsilon^{\mu \nu \rho \delta}$, 
    which is antisymmetric under the interchange of either of these pairs of indices. Hence 
    this whole term vanishes, and (\ref{eq:dellamxidag}) becomes 
    \be 
    -{\rm i} C^* \xi^\dagger [-{\bar{\sigma}}^\nu \partial^\mu \lambda + 
    {\bar{\sigma}}^\mu \partial^\nu \lambda]F_{\mu \nu}.
    \ee
    In the second term here, interchange the indices $\mu$ and $\nu$ throughout, 
    and then use the antisymmetry of $F_{\nu \mu}$: you find that the second term equals the 
    first, so that this `$\xi^\dagger$' part of the variation of the fermionic part of 
    ${\cal{L}}_{\gamma \lambda}$ is 
    \be 
    2{\rm i}C^* \xi^\dagger {\bar{\sigma}}^\nu \partial^\mu \lambda F_{\mu \nu}.
    \ee
    This will cancel the $\xi^\dagger$ part of (\ref{eq:delmaxlam}) if $C={\rm i}/2$, and so 
    the required SUSY transformations are (\ref{eq:delAmu}) and 
    \be 
    \delta_\xi \lambda = \frac{1}{2}{\rm i} \sigma^\mu {\bar{\sigma}}^\nu \xi F_{\mu \nu},
    \label{eq:dellam3}
    \ee
    \be 
    \delta_\xi \lambda^\dagger = -\frac{1}{2}{\rm i} \xi^\dagger 
    {\bar{\sigma}}^\nu \sigma^\mu F_{\mu \nu}.\label{eq:dellamdag3}
    \ee

        However, if we try to calculate (as in section 7) $\delta_\eta \delta_\xi-\delta_\xi \delta_\eta$ 
      as applied to the fields $A^\mu$ and $\lambda$, we shall find that consistent results are not 
      obtained unless the free-field equations of motion are assumed to hold, 
      which is not satisfactory.   Off-shell, 
       $A^\mu$ has a third degree of freedom, and so we expect to have to introduce one more 
       auxiliary field, call it $D(x)$, which is a real scalar field with one degree of freedom. 
       We add to ${\cal{L}}_{\gamma \lambda}$ the extra (non-propagating) term 
       \be 
       {\cal{L}}_{D} = \frac{1}{2}D^2. \label{eq:LD}
       \ee
       We now have to consider SUSY transformations including $D$.
       
       First note that the dimension of $D$ is $M^2$, the same as for $F$. This suggests that 
       $D$ transforms in a similar way to $F$, as given by (\ref{eq:delxiF}). However, $D$ is a real 
       field, so we modify (\ref{eq:delxiF}) by adding the hermitian conjugate term, arriving at 
       \be 
       \delta_\xi D= -{\rm i} ( \xi^\dagger {\bar{\sigma}}^\mu \partial_\mu \lambda - 
       (\partial_\mu \lambda)^\dagger {\bar{\sigma}}^\mu \xi).\label{eq:delxiD}
       \ee      
       As in the case of $\delta_\xi F$, this is also a total derivative. Analogously to  
       (\ref{eq:delchiFdag}) and (\ref{eq:delchiF}), we expect to modify (\ref{eq:dellam3}) 
       and (\ref{eq:dellamdag3}) so as to 
       include additional terms 
       \be 
       \delta_\xi \lambda = \xi D, \ \ \ \delta_\xi \lambda^\dagger = \xi^\dagger D. 
       \ee
       The variation of ${\cal{L}}_D$ is then 
       \be 
       \delta_\xi \left( \frac{1}{2}D^2\right)=D \delta_\xi D = -{\rm i} D (\xi^\dagger {\bar{\sigma}}^\mu \partial_\mu 
       \lambda - (\partial_\mu \lambda)^\dagger {\bar{\sigma}}^\mu \xi),\label{eq:delLD}
       \ee
       and the variation of the fermionic part of ${\cal{L}}_{\gamma \lambda}$ gets an additional 
       contribution which is 
       \be 
       {\rm i} \xi^\dagger {\bar{\sigma}}^\mu \partial_\mu \lambda D +{\rm i} \lambda^\dagger 
       {\bar{\sigma}}^\mu \partial_\mu \xi D.\label{eq:delLferm}
       \ee
       The first term of (\ref{eq:delLferm}) cancels the first term of (\ref{eq:delLD}), and 
       the second terms also cancel after either one  has been integrated 
       by parts.  
    
    \subsection{Non-Abelian gauge  supermultiplets}
    
    The preceding example is clearly unrealistic physically, but it will help us in guessing 
    the SUSY transformations in the physically relevant non-Abelian case. For definiteness, 
    we'll mostly consider an SU(2) gauge theory, such as occurs in the electroweak sector of the SM. We 
    begin by recalling some necessary facts about non-Abelian gauge theories.
    
    For an SU(2) gauge theory, 
    the Maxwell field strength tensor $F_{\mu \nu}$ of U(1) is generalized to 
    (see for example \cite{AH32} chapter 13)
    \be 
     F^\alpha_{\mu \nu} = \partial_\mu W^\alpha_\nu - \partial_\nu W^\alpha_\mu  -g \epsilon^{\alpha 
    \beta \gamma} W_\mu^\beta W_\nu^\gamma,\label{eq:FSU(2)}
    \ee
    where $\alpha$, $\beta$ and $\gamma$ have the values 1,2 and 3, the gauge field ${\bm W}_\mu 
    =(W_\mu^1, W_\mu^2, W_\mu^3)$ is an SU(2) triplet (or `vector', thinking of it in SO(3) terms), 
    and $g$ is the gauge coupling constant. We are writing the SU(2) indices as superscripts rather 
    than subscripts, but this has no mathematical significance; rather, it is to avoid confusion, later, 
    with the spinor index of the gaugino field $\lambda^\alpha_a$.  
    Equation (\ref{eq:FSU(2)}) can alternatively be written 
    in `vector' notation as 
    \be 
    {\bm F}_{\mu \nu} = \partial_\mu {\bm W}_\nu - \partial_\nu {\bm W}_\mu - g {\bm W}_\mu \times 
    {\bm W}_\nu.\label{eq:FSU(2)bm}
    \ee
    If the gauge group was SU(3) there would be 8 gauge fields (gluons, in the 
    QCD case), and in general for SU(N) there are $N^2 - 1$. Gauge fields always belong to a particular 
    representation of the gauge group, namely the {\em regular} or {\em adjoint} one, which has as 
    many components as there are generators of the group: see pages 400-401 of \cite{AH32}.     
    
    An infinitesimal gauge transformation on the gauge fields $W_\mu^\alpha$ takes the form 
    \be
    W^{\prime \alpha}_\mu(x)=W_\mu^{\alpha}(x) - \partial_\mu \epsilon^\alpha(x) - g \epsilon^{
    \alpha \beta \gamma} \epsilon^\beta(x) W_\mu^\gamma(x),\label{eq:gtW}
    \ee 
    where we have here indicated the $x$-dependence explicitly, to emphasize the fact that this is 
    a local transformation, in which the three infinitesimal parameters $\epsilon^\alpha(x)$ 
    depend on $x$. In U(1) we would have only one such $\epsilon(x)$,  the second term in (\ref{eq:gtW}) 
    would be absent, and the field strength tensor $F_{\mu \nu}$ would be gauge-invariant. In 
    SU(2), the corresponding tensor (\ref{eq:FSU(2)bm}) transforms by 
    \be 
     F^{\alpha \prime}_{\mu \nu}(x) = F^\alpha_{\mu \nu}(x) - g \epsilon^{\alpha \beta \gamma} 
      \epsilon^\beta (x) 
    F^\gamma_{\mu \nu}(x), \label{eq:FSU(2)trans}
    \ee
    which is nothing but the statement that ${\bm F}_{\mu \nu}$ transforms as an SU(2) triplet. Note 
    that (\ref{eq:FSU(2)trans}) involves no derivative of ${\bm \epsilon}(x)$, such as appears in 
    (\ref{eq:gtW}), even though the transformations being considered are local ones. This fact 
    shows that the simple generalization of the Maxwell Lagrangian in terms of ${\bm F}_{\mu \nu}$, 
    \be 
    -\frac{1}{4} {\bm F}_{\mu \nu} \cdot {\bm F}^{\mu \nu} = -\frac{1}{4}F^\alpha_{\mu \nu} 
    F^{\mu \nu \alpha} \label{eq:maxnonab}
    \ee
    is invariant under local SU(2) transformations - i.e. is SU(2) gauge-invariant.    
    
     We now need to generalize the simple U(1) SUSY model of the previous subsection. Clearly the 
     first step is to 
      introduce an SU(2) triplet of gauginos, ${\bm \lambda}=(\lambda^1, \lambda^2, \lambda^3)$, 
      to partner the triplet 
     of gauge fields. Under an infinitesimal SU(2) gauge transformation,  
     $ \lambda^\alpha$ transforms as in 
     (\ref{eq:FSU(2)trans}): 
     \be  
      \lambda^{\alpha \prime }(x)  =  \lambda^\alpha(x) - g \epsilon^{\alpha 
      \beta \gamma}\epsilon^\beta (x) \lambda^\gamma (x).
     \label{eq:lamSU(2)trans}
     \ee  
    The gauginos are of course 
    not gauge fields and so their transformation does not include any derivative of ${\bm \epsilon}(x)$. 
    So the straightforward generalization of (\ref{eq:Lgamlam}) would be 
    \be 
    {\cal {L}}_{ W  \lambda}= -\frac{1}{4}F^\alpha_{\mu \nu}F^{\mu \nu \alpha} + 
    {\rm i} {\lambda}^{\alpha \dagger} {\bar{\sigma}}^\mu \partial_\mu \lambda^\alpha.
    \label{eq:LWlam1} 
    \ee
    But although the first term of (\ref{eq:LWlam1}) is SU(2) gauge-invariant, the second is 
    not, because the gradient will act on the $x$-dependent parameters $\epsilon^\beta (x)$ in 
    (\ref{eq:lamSU(2)trans}) to leave uncancelled $\partial_\mu \epsilon^\beta (x)$  
    terms after the gauge transformation. The way to make this term gauge-invariant is to 
    replace the ordinary gradient in it by the appropriate {\em covariant derivative} - see 
    \cite{AH32} page 47, for instance. The general recipe is 
    \be 
    \partial_\mu \to D_\mu \equiv \partial_\mu +{\rm i} g {\bf T}^{(t)} 
    \cdot {\bm W}_\mu, \label{eq:covDNAB}
    \ee
    where the three matrices $T^{(t)\alpha}, \alpha=1, 2, 3$,  
     are of dimension $2t+1 \times 2t+1$ 
    and represent the generators of SU(2) when acting on a $2t+1$-component field, which is in the 
    representation of SU(2) characterized by the `isospin' $t$ (see \cite{AH32} section M.5). In 
    the present case, the $\lambda^\alpha$'s belong in the triplet ($t=1$) representation, for 
    which the three $3\times3$ matrices $T^{(1)\alpha}$ are 
    given by (see \cite{AH32} equation (M.70)) 
    \be 
    \left( T^{(1)\alpha}\right)_{\beta \gamma \ {\rm element}}=-{\rm i} \epsilon^{\alpha \beta \gamma}.
    \ee
     Thus, in (\ref{eq:LWlam1}), we need to make the replacement 
     \bea
     \partial_\mu \lambda^\alpha \to (D_\mu \lambda)^\alpha&=&\partial_\mu \lambda^\alpha 
     +{\rm i} g ( {\bf T}^{(1)} \cdot {\bm W}_\mu )_{\alpha \beta \ {\rm element}} 
     \lambda^\beta \nonumber \\
     &=& \partial_\mu \lambda^\alpha +{\rm i} g (-{\rm i} 
    \epsilon^{\gamma \alpha \beta} W^\gamma_\mu )\lambda^\beta \nonumber \\
    &=& \partial_\mu \lambda^\alpha +g \epsilon^{\gamma \alpha \beta}W^\gamma_\mu \lambda^\beta \nonumber \\
    &=& \partial_\mu \lambda^\alpha - g \epsilon^{\alpha \beta \gamma} W^\beta_\mu \lambda^\gamma.
    \eea
    With this replacement for $\partial_\mu \lambda^\alpha$ in (\ref{eq:LWlam1}), the resulting 
    ${\cal{L}}_{W \lambda}$ is SU(2) gauge-invariant.
    
    What about making it also invariant under SUSY transformations? From the experience of 
    the U(1) case in the previous subsection, we expect that we'll need to introduce the 
    analogue of the auxiliary field $D$. In this case, we need a triplet of $D$'s, $D^\alpha$, 
    balancing the third off-shell degree of freedom for each $W^\alpha_\mu$. So our shot at 
    a SUSY- and gauge-invariant Lagrangian for an SU(2) gauge supermultiplet is
    \be 
    {\cal{L}}_{{\rm gauge}}= -\frac{1}{4}F^\alpha_{\mu \nu}F^{\mu \nu \alpha} + 
    {\rm i} \lambda^{\alpha \dagger} {\bar{\sigma}}^\mu (D_\mu \lambda)^\alpha +
     \frac{1}{2}D^\alpha D^\alpha. \label{eq:LDWlam}
     \ee
     Confusion must be avoided as between the covariant derivative  and the auxiliary field!    
    
    What are reasonable guesses for the relevant SUSY transformations? We try the obvious 
    generalizations of the U(1) case:
    \bea
    \delta_\xi W^{\mu \alpha}&=&\xi^\dagger {\bar{\sigma}}^\mu \lambda^\alpha + \lambda^{\alpha \dagger} 
    {\bar{\sigma}}^\mu \xi, \nonumber \\
    \delta_\xi\lambda^\alpha &=& \frac{1}{2}{\rm i} \sigma^\mu {\bar{\sigma}}^\nu \xi F^\alpha_{\mu \nu} 
    + \xi D^\alpha \nonumber \\
    \delta_\xi D^\alpha &=& - {\rm i} (\xi^\dagger {\bar{\sigma}}^\mu (D_\mu \lambda)^\alpha 
    -(D_\mu \lambda)^{\alpha \dagger}{\bar{\sigma}}^\mu \xi);\label{eq:susygauge}
    \eea
    note that in the last equation we have replaced the `$\partial_\mu$' of 
    (\ref{eq:delxiD}) by `$D_\mu$', 
    so as to maintain gauge-invariance. This in fact works, just as it is! Quite 
    remarkably,  the Action 
    for (\ref{eq:LDWlam}) is invariant under the transformations (\ref{eq:susygauge}), 
    and ($\delta_\eta \delta_\xi - \delta_\xi \delta_\eta$) can be consistently applied 
    to all the fields $W_\mu^\alpha, \lambda$ and $D^\alpha$ in this gauge supermultiplet. This   
    supersymmetric gauge theory therefore has two sorts of interactions: (i) the usual 
    self-interactions among the $W$ fields as generated by the term (\ref{eq:maxnonab}); 
    and (ii) interactions between the $W$'s and the $\lambda$'s generated by the covariant 
    derivative coupling in (\ref{eq:LDWlam}). 
     We stress again that the supersymmetry requires the gaugino partners 
    to belong to the same representation of the gauge group as the gauge bosons themselves - 
    i.e. to the regular, or adjoint, representation. 
    
    We are getting closer to the MSSM at last. The next stage is to build Lagrangians containing 
    both chiral and gauge supermultiplets, in such a way that they (or the Actions) are 
    invariant under both SUSY and gauge transformations.

    \section{Combining Chiral and Gauge Supermultiplets}
    
     We do this in two steps. First we introduce - via appropriate covariant derivatives - 
     the couplings of the gauge fields to the scalars and   fermions  (`matter fields')  
     in the chiral supermultiplets. This will account for the interactions between the gauge 
     fields of the vector supermultiplets and the matter fields of the chiral supermultiplets. But 
     there are also gaugino and $D$ fields in the vector supermultiplets, and we need to consider 
     whether there are any possible renormalizable interactions between the matter fields 
     and gaugino and $D$ fields, which are both gauge- and SUSY-invariant. Including such interactions 
     is the second step in the programme of combining the two kinds of supermultiplets. 
     
     The essential points in such a construction are contained in the simplest case, namely that 
     of a single U(1) (Abelian) vector supermultiplet and a single free chiral supermultiplet, 
     the combination of which we shall now consider. 
     
     \subsection{Combining one U(1) vector supermultiplet and one free chiral supermultiplet}
     
     The first step is accomplished by taking the Lagrangian of (\ref{eq:LphichiFi}), 
     for only a single supermultiplet,  replacing 
     $\partial_\mu$ by $D_\mu$ where (compare (\ref{eq:covDNAB}))
     \be 
     D_\mu = \partial_\mu + {\rm i} q A_\mu,  \label{eq:covDU(1)}
     \ee
     where $q$ is the U(1) coupling constant (or charge), 
     and adding on the Lagrangian for the U(1) vector supermultiplet (i.e. (\ref{eq:Lgamlam}) 
     together with (\ref{eq:LD})). This produces 
     the Lagrangian 
     \be 
     {\cal{L}}=(D_\mu \phi)^\dagger (D^\mu \phi) + {\rm i} \chi^\dagger {\bar{\sigma}}^\mu D_\mu \chi 
     + F^\dagger F - \frac{1}{4}F_{\mu \nu} F^{\mu \nu} + {\rm i} \lambda^\dagger {\bar{\sigma}}^\mu 
     \partial_\mu \lambda + \frac{1}{2} D^2. \label{eq:freechigau}
     \ee
     
     We now have to consider possible  interactions between the matter 
     fields $\phi$ and $\chi$, and the other fields $\lambda$ and $D$ in the vector 
     supermultiplet. Any such interaction terms must certainly be Lorentz-invariant, 
     renormalizable (i.e. have mass dimension less than or equal to 4), and gauge-invariant. Given 
     some  terms with these characteristics, we shall then have to examine whether 
     we can  include them in a SUSY-preserving way. 
     
     Since the fields $\lambda$ and $D$ are neutral, any gauge-invariant couplings between 
     them and the charged fields $\phi$ and $\chi$ must involve neutral bilinear combinations 
     of the latter fields, namely $\phi^\dagger \phi, \ \phi^\dagger \chi, \  \chi^\dagger \phi$ 
     and $\chi^\dagger \chi$.  These have mass dimension 2, 5/2, 5/2 and 3 respectively. They have 
     to be coupled to the fields $\lambda$ and $D$ which have dimension 3/2 and 2 respectively, so 
     as to make quantities with dimension no greater than 4. This rules out the bilinear 
     $\chi^\dagger \chi$, and allows just three possible Lorentz- and gauge-invariant 
     renormalizable couplings: $(\phi^\dagger \chi) \cdot \lambda$, 
     $\chi^\dagger \cdot (\chi^\dagger \phi)$, and $\phi^\dagger \phi D$. In the first of these 
     the Lorentz invariant is formed as the `$\cdot$' product of the L-type quantity 
     $\phi^\dagger \chi$ and the L-type spinor $\lambda$, while in the second it is 
     formed as a `$\lambda^\dagger \cdot \chi^\dagger$'-type product. We take the sum of the 
     first two couplings to obtain a Hermitian interaction, and arrive at the possible 
     allowed interaction terms 
     \be 
     Aq[(\phi^\dagger \chi) \cdot \lambda + \lambda^\dagger \cdot (\chi^\dagger \phi)] + Bq \phi^\dagger 
     \phi D. \label{eq:possint}
     \ee
     The coefficients $A$ and $B$ are now to be determined by requiring that the complete 
     Lagrangian of (\ref{eq:freechigau}) together with (\ref{eq:possint}) is SUSY-invariant (note that for 
     convenience we have extracted an explicit factor of $q$ from $A$ and $B$). 
     
     To implement this programme we need to specify the SUSY transformations of the fields. At first 
     sight, this seems straightforward enough: we use (\ref{eq:delAmu}), (\ref{eq:dellam3}),  
     (\ref{eq:dellamdag3}) and (\ref{eq:delxiD}) 
     for the fields in the vector supermultiplet, and we `covariantize' the transformations used for 
     the chiral supermultiplet. For the latter, then, we provisionally assume 
     \be 
     \delta_\xi \phi = \xi \cdot \chi, \ \ \ \delta_\xi \chi=-{\rm i} \sigma^\mu ({\rm i} \sigma_2) \xi^{\dagger 
     {\rm T}} D_\mu \phi + \xi F, \ \ \  \delta_\xi F= -{\rm i} \xi^\dagger {\bar{\sigma}}^\mu D_\mu \chi,
     \label{eq:delchitry}
     \ee
     together with the analogous transformations for the Hermitian conjugate fields. 
     As we shall see, however,  there is no choice we can make for  $A$ and $B$ in (\ref{eq:possint}) 
     such that the complete Lagrangian is invariant under these transformations. One may not be too 
     surprised by this: after all, the transformations for the chiral supermultiplet were 
     found for the case $q=0$, and it is quite possible, one might think, that  
     one or more of the transformations in (\ref{eq:delchitry}) have to be modified by pieces 
     proportional to $q$. Indeed, we shall find that the transformation for $F$ does need to be so modified. 
     But there is a more important reason for the `failure' to find a suitable $A$ and $B$. The transformations 
     of (\ref{eq:delAmu}), (\ref{eq:dellam3}), (\ref{eq:dellamdag3}) 
      and (\ref{eq:delxiD}), on the one hand, and those of 
     (\ref{eq:delchitry}) on the other,  certainly do ensure the SUSY-invariance of the gauge and chiral  
     parts of (\ref{eq:freechigau}) respectively, in the limit $q=0$. But there is no {\em a priori} 
     reason - at least in our `brute-force' approach - why the `$\xi$' parameter in one set of 
     transformations should be exactly the same as that in the other. Either `$\xi$' can be rescaled by 
     a constant multiple, and the relevant sub-Lagrangian will remain invariant. However, when we 
     {\em combine} the Lagrangians and include (\ref{eq:possint}), for the case $q \neq 0$, we shall see 
     that the requirement of overall SUSY-invariance fixes the relative scale of the two 
     `$\xi$'s' (up to a sign), and without a rescaling in one or the other transformation 
     we cannot get a SUSY-invariant theory. 
     For definiteness we shall keep the `$\xi$' in (\ref{eq:delchitry}) unmodified, and introduce a real 
     scale parameter $\alpha$ into the transformations for the vector supermultiplet, so that they 
     now become 
     \bea
     \delta_\xi A^\mu &=& \alpha (\xi^\dagger {\bar{\sigma}}^\mu \lambda + \lambda^\dagger {\bar{\sigma}}^\mu 
     \xi) \\ 
     \delta_\xi \lambda &=& \frac{\alpha {\rm i}}{2}(\sigma^\mu {\bar{\sigma}}^\nu \xi) F_{\mu \nu} +\alpha 
     \xi D \\
     \delta_\xi \lambda^\dagger &=& -\frac{\alpha {\rm i}}{2}(\xi^\dagger {\bar{\sigma}}^\nu \sigma^\mu)F_{\mu 
     \nu} + \alpha \xi^\dagger D \\
     \delta_\xi D &=& - \alpha {\rm i} (\xi^\dagger {\bar{\sigma}}^\mu \partial_\mu \lambda -(\partial_\mu 
     \lambda^\dagger) {\bar{\sigma}}^\mu \xi).\label{eq:delDU(1)}
     \eea
     
     Consider first the SUSY variation of the `$A$' part of (\ref{eq:possint}). This is 
     \bea 
     && Aq[(\delta_\xi \phi^\dagger) \chi \cdot \lambda + \phi^\dagger (\delta_\xi \chi) \cdot \lambda + 
     \phi^\dagger \chi \cdot (\delta_\xi \lambda) \nonumber \\
     && (\delta_\xi \lambda^\dagger) \cdot \chi^\dagger \phi + \lambda^\dagger \cdot (\delta_\xi \chi^\dagger) 
     \phi + \lambda^\dagger \cdot \chi^\dagger (\delta_\xi \phi)].\label{eq:delApart} 
     \eea
     Among these terms there are two which are linear in $q$ and $D$, arising from 
     $\phi^\dagger \chi \cdot (\delta_\xi \lambda)$ and its Hermitian conjugate, namely  
     \be
     Aq[\alpha \phi^\dagger \chi \cdot \xi D + \alpha \xi^\dagger \cdot \chi^\dagger D \phi].\label{eq:DdelApart}
     \ee
     Similarly, the variation of the `$B$' part is 
     \bea
     &&Bq[(\delta_\xi \phi^\dagger) \phi D + \phi^\dagger (\delta_\xi \phi) D + \phi^\dagger \phi (\delta_\xi 
     D)] =Bq[\chi^\dagger \cdot \xi^\dagger \phi D +\nonumber \\
     && \phi^\dagger \xi \cdot \chi D + 
     \phi^\dagger \phi (-\alpha {\rm i}) (\xi^\dagger {\bar{\sigma}}^\mu \partial_\mu \lambda 
     -(\partial_\mu \lambda^\dagger) {\bar{\sigma}}^\mu  \xi)].\label{eq:delBpart}
     \eea
     The `$D$' part of (\ref{eq:delBpart}) will cancel the term (\ref{eq:DdelApart}) if (using 
     $\chi^\dagger \cdot \xi^\dagger = \xi^\dagger \cdot \chi^\dagger$ and $\xi \cdot \chi = \chi \cdot \xi$)  
     \be
     A \alpha = -B.\label{eq:Aalpha}
     \ee

     Next, note that the first and last terms of (\ref{eq:delApart}) produce the changes 
     \be 
     Aq[\chi^\dagger \cdot \xi^\dagger \, \chi \cdot \lambda + \lambda^\dagger \cdot \chi^\dagger \, 
     \xi \cdot \chi].\label{eq:4ferm1}
     \ee
     Meanwhile, there is a corresponding change  coming from the variation of the term $-q 
     \chi^\dagger {\bar{\sigma}}^\mu \chi A_\mu$, namely 
     \be 
     -q\chi^\dagger {\bar{\sigma}}^\mu \chi (\delta_\xi A_\mu)=-q \alpha \chi^\dagger {\bar{\sigma}}^\mu \chi 
     (\xi^\dagger {\bar{\sigma}}_\mu \lambda + \lambda^\dagger {\bar{\sigma}}_\mu \xi).\label{eq:4ferm}
     \ee
     This can be simplified with the help of the exercise:\\
     {\bf Exercise} Show that 
     \be 
     (\chi^\dagger {\bar{\sigma}}^\mu \chi)(\lambda^\dagger {\bar{\sigma}}_\mu \xi)=
     2(\chi^\dagger \cdot \lambda^\dagger) (\chi \cdot \xi).
     \ee
     So (\ref{eq:4ferm}) becomes
     \be
     -2q\alpha[ \chi^\dagger \cdot \xi^\dagger 
     \, \chi \cdot \lambda + \chi^\dagger \cdot \lambda^\dagger \, \chi \cdot \xi ],
     \ee
     which will cancel (\ref{eq:4ferm1}) if (again using 
     $\chi \cdot \xi = \xi \cdot \chi$ and $\chi^\dagger \cdot \lambda^\dagger 
     = \lambda^\dagger \cdot \chi^\dagger$)
     \be
     A=2 \alpha.\label{eq:2alpha}
     \ee

     So far, there is nothing to prevent us from choosing $\alpha=1$, say, in (\ref{eq:Aalpha}) and 
     (\ref{eq:2alpha}). However, a constraint on $\alpha$ arises when we consider the variation of the   
     $A^\mu-\phi$ interaction term in (\ref{eq:freechigau}), namely 
     \be 
    - {\rm i}q\delta_\xi(A^\mu \phi^\dagger \partial_\mu \phi - (\partial_\mu \phi)^\dagger A^\mu \phi).
    \ee 
     The terms in $\delta_\xi A^\mu$ yield a change 
     \be 
     {\rm i} q \alpha [ (\partial_\mu \phi^\dagger)  (\xi^\dagger {\bar{\sigma}}^\mu \lambda + 
     \lambda^\dagger {\bar{\sigma}}^\mu \xi) \phi - (\xi^\dagger {\bar{\sigma}}^\mu \lambda + \lambda^\dagger 
     {\bar{\sigma}}^\mu \xi) \phi^\dagger \partial_\mu \phi].\label{eq:delAphi}
     \ee 
     A similar change arises  from the terms $Aq[\phi^\dagger (\delta_\xi \chi) \cdot \lambda + 
     \lambda^\dagger \cdot (\delta_\xi \chi^\dagger) \phi]$ in (\ref{eq:delApart}), namely 
     \be 
     Aq[\phi^\dagger(-{\rm i} \sigma^\mu {\rm i} \sigma_2 \xi^{\dagger {\rm T}}\partial_\mu \phi) \cdot \lambda 
     + \lambda^\dagger \cdot \partial_\mu \phi^\dagger \xi^{\rm T}(-{\rm i} \sigma_2 {\rm i} \sigma^\mu \phi)].
     \label{eq:delAfm}
     \ee
     The first spinor dot product is 
     \be
     \xi^\dagger (-{\rm i} \sigma_2) (-{\rm i} \sigma^{\mu {\rm T}})(-{\rm i} \sigma_2) \lambda = 
     {\rm i} \xi^\dagger {\bar{\sigma}}^\mu \lambda,
     \ee
     using (\ref{eq:csigc}). The second spinor product is the Hermitian conjugate of this, so that   
     (\ref{eq:delAfm}) yields a change 
     \be 
     Aq{\rm i}[\phi^\dagger (\partial_\mu \phi) \xi^\dagger {\bar{\sigma}}^\mu \lambda -
     (\partial_\mu \phi^\dagger) \phi \, \lambda^\dagger {\bar{\sigma}}^\mu \xi].\label{eq:delAfm1}
     \ee   
     Along with (\ref{eq:delAphi}) and (\ref{eq:delAfm1}}) we must also group the last two terms in 
     (\ref{eq:delBpart}), which we write out again here for convenience
     \be 
      Bq[\phi^\dagger \phi (-\alpha {\rm i}) (\xi^\dagger {\bar{\sigma}}^\mu \partial_\mu \lambda 
     -(\partial_\mu \lambda^\dagger) {\bar{\sigma}}^\mu  \xi)],\label{eq:delBphiphi}
     \ee
     and integrate by parts to yield
     \be 
     \alpha {\rm i} B q\{[(\partial_\mu \phi^\dagger) \phi + \phi^\dagger \partial_\mu \phi](\xi^\dagger 
     {\bar{\sigma}}^\mu \lambda)-[(\partial_\mu \phi^\dagger)\phi+\phi^\dagger \partial_\mu \phi]
     (\lambda^\dagger {\bar{\sigma}}^\mu \xi)\}.\label{eq:delBlast}
     \ee
     Consider now the terms involving the quantity $\xi^\dagger {\bar{\sigma}}^\mu \lambda$ in 
     (\ref{eq:delAphi}), (\ref{eq:delAfm1}) and (\ref{eq:delBlast}), which are 
     \be
     {\rm i}q \alpha[(\partial_\mu \phi^\dagger) \phi - \phi^\dagger \partial_\mu \phi] +Aq{\rm i} 
     \phi^\dagger \partial_\mu \phi 
     +\alpha {\rm i} B q[(\partial_\mu \phi^\dagger)\phi + \phi^\dagger \partial_\mu \phi].\label{eq:alcond}
     \ee
     These will all cancel if the condition (\ref{eq:2alpha}) holds, and if in addition 
     \be 
     B=-1.\label{eq:Bcond}
     \ee
     From (\ref{eq:2alpha}) and (\ref{eq:Aalpha}) it now follows that 
     \be 
     \alpha^2=\frac{1}{2}. \label{eq:alval}
     \ee

     We conclude that, as promised, the combined Lagrangian will not be SUSY-invariant unless we 
     modify the scale of the transformations of the gauge supermultiplet, relative to those of the 
     chiral supermultiplet, by a non-trivial factor, which we choose (in 
     agreement with what seems to be the usual convention - see \cite{martin} equations 
     (3.57)-(3.59)) to be 
     \be \alpha = -\frac{1}{\sqrt{2}}.
     \ee
     With this choice, the coefficient $A$ is determined to be 
     \be 
     A=-\sqrt{2},
     \ee
     and our combined Lagrangian is fixed. 
     
     We have, of course, not given a complete analysis 
     of all the terms in the SUSY variation of our Lagrangian, an exercise we leave to the 
     dedicated reader -  who will find that (with one more adjustment to the SUSY transformations)  
     all the variations do indeed vanish (after partial integrations in some cases, as usual). 
     The need for the adjustment appears when we consider the variation associated with the terms 
      $Aq[\phi^\dagger (\delta_\xi \chi) \cdot \lambda + 
      \lambda^\dagger \cdot (\delta_\xi \chi^\dagger) \phi]$ in (\ref{eq:delApart}), 
      which includes the term 
      \be 
      Aq[\phi^\dagger \xi \cdot \lambda F + \lambda^\dagger \cdot \xi^\dagger F^\dagger \phi].
      \ee
      This cannot be cancelled by any other variation, and we therefore have to modify the transformation 
      for $F$ and $F^\dagger$ so as to generate a cancelling term from the variation of $F^\dagger F$ in 
      the Lagrangian. This requires 
      \be 
      \delta_\xi F = -\sqrt{2}q \lambda^\dagger \cdot \xi^\dagger \phi + {\mbox{previous transformation}} 
      \ee
      and 
      \be
      \delta_\xi F^\dagger = - \sqrt{2}q \xi \cdot \lambda \phi^\dagger +{\mbox{previous transformation}},
      \ee
      where we have now inserted the known value of $A$.
      
      In summary then, our SUSY-invariant combined chiral and U(1) gauge supermultiplet Lagrangian is 
      \bea 
     &&{\cal{L}}=(D_\mu \phi)^\dagger (D^\mu \phi) + {\rm i} \chi^\dagger {\bar{\sigma}}^\mu D_\mu \chi 
     + F^\dagger F - \frac{1}{4}F_{\mu \nu} F^{\mu \nu} + {\rm i} \lambda^\dagger {\bar{\sigma}}^\mu 
     \partial_\mu \lambda + \frac{1}{2} D^2 \nonumber \\
     &&-\sqrt{2}q[(\phi^\dagger \chi) \cdot \lambda + \lambda^\dagger \cdot (\chi^\dagger \phi)] 
     - q \phi^\dagger \phi D.\label{eq:susyu(1)ch}
     \eea
     Note that the terms in the last line of (\ref{eq:susyu(1)ch}) are interactions whose 
     strengths are fixed by SUSY to be proportional to the gauge coupling constant $q$, even though 
     they don't have the form of ordinary gauge interactions; the terms coupling the photino $\lambda$ 
     to the matter fields may be thought of as arising from supersymmetrizing the usual coupling of 
     the gauge field to the matter fields. 
     
     The equation of motion for the field $D$ is 
     \be 
     D=q  \phi^\dagger \phi.\label{eq:Deq}
     \ee
     Since no derivatives of $D$ enter, we may (as in the W-Z case for $F_i$ and $F^\dagger_i$, 
     c.f. equations (\ref{eq:Feqnmot}) and (\ref{eq:WZ2})) eliminate the auxliary field $D$ from the Lagrangian 
     by using (\ref{eq:Deq}). The effect of this is clearly to replace the two terms involving $D$ in 
     (\ref{eq:susyu(1)ch}) by the single term 
     \be 
     - \frac{1}{2} q^2 (\phi^\dagger \phi)^2.
     \ee
     This is a `$(\phi^\dagger \phi)^2$' type of interaction, just as in the Higgs potential 
     (\ref{eq:HiggsV}), but here appearing with a coupling constant which is not an unknown 
     parameter, but is determined by the gauge coupling $q$. In the next section we shall see 
     that the same feature persists in the more realistic non-Abelian case. Since the Higgs mass 
     is (for a fixed vev of the Higgs field) determined by the $(\phi^\dagger \phi)^2$ coupling - 
     see (\ref{eq:MH}) for example - it follows that there is likely to be less arbitrariness 
     in the mass of the Higgs in the MSSM than in the SM. We shall see in section 16, 
     when we examine the Higgs sector of the MSSM, that this is indeed the case.

     \subsection{The non-Abelian case} 
     
     Once again, we proceed in two steps. We start from the W-Z Lagrangian for     
     a collection of chiral supermultiplets labelled by $i$, and including the 
     superpotential terms:  
     \be
     \partial_\mu \phi_i^\dagger \partial^\mu \phi_i + \chi_i^\dagger {\rm i} {\bar{\sigma}}^\mu 
     \partial_\mu \chi_i + F_i^\dagger F_i  
     +\left[ \frac{\partial W}{\partial \phi_i} F_i - \frac{1}{2} \frac{\partial^2 W}{\partial \phi_i 
     \phi_j} \chi_i \cdot \chi_j + {\rm h.c.}\right] \label{eq:LWZfin}
     \ee
     into which we introduce the gauge couplings via the covariant derivatives    
     \bea
     \partial_\mu \phi_i &\to& D_\mu \phi_i = \partial_\mu \phi_i +{\rm i} g A^\alpha_\mu
     (T^{ \alpha} \phi)_i \label{eq:covDphi}\\
     \partial \chi_i &\to& D_\mu \chi_i = \partial_\mu \chi_i + {\rm i} g A^\alpha_\mu 
     (T^{ \alpha} \chi)_i \label{eq:covDchi},
     \eea
    where $g$ and $A^\alpha_\mu$ are the  gauge coupling constant and gauge fields 
    (for example, $g_{\rm s}$ and gluon fields for QCD), and the $T^{ \alpha}$  
     are the hermitian matrices representing the generators of the gauge 
    group in the representation to which, for given $i$,   $\phi_i$ and $\chi_i$ belong (for example, 
    if $\phi_i$ and $\chi_i$ are SU(2) doublets, the $T^\alpha$'s would be the $\tau^\alpha/2$, 
     with $\alpha$ running from 1 to 3). Recall that SUSY 
    requires that $\phi_i$, $\chi_i$ and $F_i$ must all be in the same representation of the relevant   
    gauge group.  Of course, 
    if - as is the case in the SM - some matter fields interact with more than one gauge field, then 
    all the gauge couplings must be included in the covariant derivatives. There is no covariant 
    derivative for the auxiliary fields $F_i$, because their ordinary derivatives don't appear 
    in (\ref{eq:LWZfin}). To (\ref{eq:covDchi}) we need to add the Lagrangian for the gauge 
    supermultiplet(s), equation (\ref{eq:LDWlam}), and then (in the second step)   additional `mixed' 
    interactions as in (\ref{eq:possint}). 
    
    	 We therefore need to construct all possible Lorentz- and gauge-invariant renormalizable 
    	interactions between the matter fields and the gaugino ($\lambda^\alpha$) and 
    	auxiliary ($D^\alpha$) fields, as in the U(1) case.    
   We have the specific particle content of the SM in mind, so  we need  
     only consider the cases in which the matter fields are either singlets under 
    the gauge group (for example, the R parts of quark and lepton fields), or belong to 
    the fundamental representation of the gauge group (that is, the triplet for SU(3) 
     and the doublet for SU(2)). For matter fields in  singlet representations, there is 
     no possible  gauge-invariant coupling between them and $\lambda^\alpha$ or $D^\alpha$, 
     which are in the regular representation. For matter fields in the fundamental 
     representation, however, we can form bilinear combinations of them which 
     transform according to the regular representation, and these bilinears can 
     be `dotted' into $\lambda^\alpha$ and $D^\alpha$ to give gauge singlets 
     (i.e. gauge-invariant couplings). We must also arrange the couplings to be 
     Lorentz invariant, of course. 
     
     The bilinear combinations of the $\phi_i$ and $\chi_i$ 
     which transform as the regular representation are (see for example \cite{AH32} sections 
     12.1.3 and 12.2)
     \be   
     \phi_i^\dagger T^\alpha \phi_i, \ \phi_i^\dagger T^\alpha \chi_i, \ \chi_i^\dagger 
     T^\alpha \phi_i, \ {\rm and}\  \chi_i^\dagger T^\alpha \chi_i,
     \ee
      where for example $T^\alpha = \tau^\alpha/2$ 
      in the case of SU(2), and where the $\tau^\alpha, \ (\alpha=1,2,3) $ are usual the Pauli matrices 
     used in the isospin context. These bilinears are the obvious analogues of the ones 
     considered in the U(1) case; in particular they have the same dimension. Following the same reasoning, 
     then, the allowed additional interaction terms are  
     \be 
      Ag[(\phi_i^\dagger T^\alpha \chi_i) \cdot \lambda^\alpha + 
      \lambda^{\alpha \dagger} \cdot (\chi_i^\dagger T^\alpha \phi_i)] +
      Bg(\phi_i^\dagger T^\alpha \phi_i)D^\alpha, \label{eq:residint}
      \ee  
      where $A$ and $B$ are coefficients to be determined by the requirement of SUSY-invariance. 
      
      In fact, however, a consideration of the SUSY transformations in this case 
      shows that they are essentially the same as in the U(1) case (apart from
       straightforward changes involving the matrices $T^\alpha$). The upshot is that, just as in the 
       U(1) case, we need to change the SUSY transformations of (\ref{eq:susygauge}) by 
       replacing $\xi$ by $-\xi/ \sqrt{2}$, and by modifying the transformation of $F_i^\dagger$ to 
       \be
       \delta_\xi F_i^\dagger =-\sqrt{2}g \phi_i^\dagger T^\alpha \xi \cdot \lambda^\alpha + {\mbox{previous 
       transformation}},
       \ee
       and similarly for $\delta_\xi F_i$. The coefficients $A$ and $B$ in (\ref{eq:residint}) are then 
       $-\sqrt{2}$ and -1 respectively, as in the U(1) case, and the combined SUSY-invariant 
       Lagrangian is   
        $$ 
       {\cal{L}}_{{\rm gauge\ + \ chiral}}={\cal{L}}_{{\rm gauge}}(\mbox{equation (\ref{eq:LDWlam})})
       $$
       $$  
       +{\cal{L}}_{{\rm W-Z, \ covariantized}}(\mbox{equation (\ref{eq:LWZfin}), with $\partial_\mu 
       \to D_\mu$ as in (\ref{eq:covDphi}) and (\ref{eq:covDchi})})
       $$
       \be 
       -\sqrt{2}g[(\phi_i^\dagger T^\alpha \chi_i) \cdot \lambda^\alpha + \lambda^{\alpha \dagger} \cdot 
       (\chi_i^\dagger T^\alpha \phi_i)] -g(\phi_i^\dagger T^\alpha \phi_i)D^\alpha.\label{eq:Lgch}
       \ee

       We draw attention to an important consequence of the terms $-\sqrt{2}g[\ldots]$ in (\ref{eq:Lgch}), 
       for the case in which the chiral multiplets ($\phi_i, \chi_i$) are the two Higgs 
       supermultiplets $H_{\rm u}$ and $H_{\rm d}$, containing higgs and higgsino fields (see Table 1 below).   
       When the scalar Higgs fields $H^0_{\rm u}$ and $H^0_{\rm d}$ acquire vevs, these terms 
       will be bilinear in the higgsino and gaugino fields, implying that mixing will occur 
       among these fields as a consequence of electroweak symmetry breaking. We shall discuss 
       this in section 17.

     The equation of motion for the field $D^\alpha$ is 
     \be 
     D^\alpha =  g \sum_i (\phi_i^\dagger T^\alpha \phi_i),\label{eq:Deqnmot}
     \ee
     where the sum over $i$ (labelling a given chiral supermultiplet) has been re-instated 
     explicitly. 
     As before, we may  
     eliminate  these auxiliary fields from the Lagrangian by using (\ref{eq:Deqnmot}). The complete 
     scalar potential (as in `${\cal{L}}={\cal{T}} - {\cal{V}}$') is then 
     \be 
     {\cal{V}}(\phi_i, \phi_i^\dagger)=|W_i|^2 + 
     \frac{1}{2}\sum_{{\rm G}}\sum_\alpha \sum_{i, j} g_{\rm G}^2 
     (\phi_i^\dagger  T_{\rm G}^\alpha \phi_i) (\phi_j^\dagger T_{\rm G}^\alpha \phi_j),\label{eq:VDF}
     \ee
     where in the summation we have recalled that more than one gauge group ${\rm G}$ 
     will enter, in general, 
     given the SU(3)$\times$SU(2)$\times$U(1) structure of the SM, with different couplings $g_{\rm G}$ 
     and generators $T_{\rm G}$. The first term in (\ref{eq:VDF}) is called the `$F$-term', for 
     obvious reasons; it is determined by the fermion mass terms $M_{ij}$ and Yukawa couplings 
     (see (\ref{eq:Wi})). The second term is called the `$D$-term', and is determined by the 
     gauge interactions. There is no room for any other scalar potential, {\em independent} of 
     these parameters appearing in other parts of the Lagrangian. It is worth emphasizing that 
     ${\cal{V}}$ is a sum of squares, and is hence always greater than or equal to zero for every 
     field configuration. We shall see in section 16 how the form of the $D$-term allows an 
     important bound to be put on the mass of the lightest Higgs boson in the MSSM.

     \section{The MSSM}
     
       We have now introduced all the interactions appearing in the MSSM, apart from 
       specifying the superpotential $W$.  We already had a brief 
       look at the particle multiplets in section 4 - let's begin by reviewing them again. 
       
       All the SM fermions - i.e. the quarks and the leptons - have the property that their L 
       (`$\chi$') parts are ${\rm SU}(2)_{\rm L}$ doublets, while their R (`$\psi$') parts 
       are ${\rm SU}(2)_{\rm L}$ singlets. So these weak gauge group properties suggest that we should  
       treat the L and R parts separately, rather than together as in a Dirac 4-component spinor. 
       The basic `building block' is therefore the chiral supermultiplet, suitably `gauged'. 
       
       We have set up the chiral multiplet to involve an L-type spinor $\chi$: this is clearly 
       fine for ${e}^-_{\rm L}, \mu^-_{\rm L}, {u}_{\rm L}, { d}_{\rm L}$, etc., but what about 
       ${ e}^-_{\rm R}, \mu^-_{\rm R}$, etc. ?  These {\em R-type particle fields} can be accommodated 
       within the `L-type' convention for chiral supermultiplets 
       by regarding them as the charge conjugates of {\em L-type 
       antiparticle fields}, which we use instead. Charge conjugation was mentioned in section 
       2.3; see also section 20.5 of \cite{AH32} (but note that we are here using $C_0=-{\rm i}
       \gamma_2$). 
       If (as is often done) we denote the field by 
       the particle name, then we have $e^-_{\rm R} \equiv \psi_{{\rm e}^-}$, 
       while $e^+_{\rm L} \equiv \chi_{{\rm e}^+}$.  On the other 
       hand, if we regard ${ e}^-_{\rm R}$ as the charge conjugate of ${ e}^+_{\rm L}$, then 
       (compare equation (\ref{eq:chimaj})) 
       \be 
       e^-_{\rm R} \equiv \psi_{{\rm e}^-} = (e^+_{\rm L})^{\rm c} \equiv {\rm i} 
       \sigma_2 \chi^{\dagger {\rm T}}_{{\rm e}^+}.
       \ee        
     To remind ourselves of how this works (see also section 2.4), consider a Dirac mass term 
     for the electron:
     \bea 
     {\bar{\Psi}}_{{\rm e}^-} \Psi_{{\rm e}^-}= 
     \psi^\dagger_{{\rm e}^-} \chi_{{\rm e}^-} + 
     \chi^\dagger_{{\rm e}^-} \psi_{{\rm e}^-} &=&
     ({\rm i} \sigma_2 \chi^{\dagger {\rm T}}_{{\rm e}^+})^\dagger \chi_{{\rm e}^-} +
     \chi^\dagger_{{\rm e}^-} {\rm i} \sigma_2 \chi^{\dagger {\rm T}}_{{\rm e}^+}\nonumber \\
     &=& \chi^{\rm T}_{{\rm e}^+} (-{\rm i} \sigma_2 )\chi_{{\rm e}^-} + \chi^\dagger_{{\rm e}^-}
     {\rm i} \sigma_2  \chi^{\dagger {\rm T}}_{{\rm e}^+}\nonumber \\
     &=& \chi_{{\rm e}^+} \cdot \chi_{{\rm e}^-} + \chi^\dagger_{{\rm e}^-} \cdot \chi^\dagger_{{\rm e}^+}.
     \label{eq:dirmass}
     \eea
   So it's all expressed in terms of $\chi$'s. It is also useful to note that 
   \be 
   {\bar{\Psi}}_{{\rm e}^-} \gamma_5 \Psi_{{\rm e}^-} = - \chi_{{\rm e}^+} \cdot 
   \chi_{{\rm e}^-} + \chi^\dagger_{{\rm e}^-} \cdot \chi^\dagger_{{\rm e}^+}. \label{eq:gam5cplng}
   \ee 
    
     In Table 1 we list the chiral supermultiplets appearing in the MSSM (our $y$ is twice that of 
     \cite{martin}, following the convention of \cite{AH32} chapter 22). Note that the `bar' on 
     the fields in this Table is merely a label, signifying `antiparticle', not (for example) 
     Dirac conjugation. The subscript ${\mbox{}}_{i}$ can be added to the names to signify 
      the family index: for example, 
      $u_{1 {\rm L}}=u_{\rm L}, u_{2 {\rm L}}= c_{\rm L}, 
      u_{3 {\rm L}}= t_{\rm L}$, and similarly for leptons. 
     \begin{table} 
     \begin{tabular}{|c|c|c|c|c|}    \hline
     \multicolumn{2}{|c|}{Names} & spin 0 & spin 1/2 & SU(3)$_{\rm c}$, SU(2)$_{\rm L}$, U(1)$_{y}$ \\
     \hline \hline
     squarks, quarks & $Q$ & ($\tilde{ u}_{\rm L},  \tilde{d}_{\rm L}$) & 
     ($u_{\rm  L},  d_{\rm L}$) & {\bf 3},\ \ \  {\bf 2},\ \ \  1/3 \\
     ($\times$ 3 families) 
     & ${\bar{u}}$ & $\tilde{\bar{u}}_{\rm L}(\tilde{u}_{{\rm R}})$ & ${\bar{u}}_{\rm L} \sim 
     (u_{\rm R})^{\rm c}$ & $\bar{{\bm 3}}$,\ \ \  {\bf 1},\ \ \  -4/3 \\
     & $\bar{d}$ & $\tilde{\bar{d}}_{\rm L}(\tilde{d}_{{\rm R}})$ & ${\bar{d}}_{\rm L} \sim 
     (d_{{\rm R}})^{\rm c}$ & $\bar{{\bm 3}}$,\ \ \  {\bf 1},\ \ \  2/3 \\
     \hline
     sleptons, leptons & $L$ & (${\tilde{\nu}}_{{\rm e}\rm L}, {\tilde{e}}_{\rm L}$) & 
     ($\nu_{{\rm e}\rm L}, { e}_{\rm L}$)& {\bf 1},\ \ \  {\bf 2},\ \ \  -1 \\
     ($\times$ 3 families)
     & ${\bar{ e}}$& ${\tilde{\bar{ e}}}_{\rm L}({\tilde{e}}_{\rm R})$ & ${\bar{ e}}_{\rm L} 
     \sim ({ e}_{\rm R})^{\rm c}$ & {\bf 1},\ \ \  {\bf 1},\ \ \  2 \\
      \hline
     higgs, higgsinos & ${ H}_{\rm u}$ & $({ H}^+_{\rm u}, { H}^0_{\rm u})$ & 
     (${\tilde{H}}^+_{\rm u}, {\tilde{H}}^0_{\rm u}$) & {\bf 1},\ \ \  {\bf 2}, \ \ \  1 \\
     & ${ H}_{\rm d}$ & (${ H}^0_{\rm d}, { H}^-_{\rm d}$) & (${\tilde{ H}}^0_{\rm d}, 
     {\tilde{ H}}^-_{\rm d}$) & {\bf 1},\ \ \  {\bf 2},\ \ \  -1 \\
     \hline
     \end{tabular}
     \caption{Chiral supermultiplet fields  in the MSSM.}
     \end{table}
     In Table 2, similarly, we list the gauge supermultiplets of the MSSM. 
     \begin{table}
     \begin{tabular}{|c|c|c|c|} \hline
     Names & spin 1/2 & spin 1 & SU(3)$_{\rm c}$, SU(2)$_{\rm L}$, U(1)$_{y}$ \\
     \hline \hline 
     gluinos, gluons & ${\tilde{{g}}}$ & ${ g}$ & {\bf 8}, \ \ \ {\bf 1}, \ \ \ 0 \\
     \hline
     winos, W bosons &$\widetilde{{{ W}}}^{\pm}, {\widetilde{{ W}}^0}$ & $W^{\pm}$, $W^0$ & 
     {\bf 1}, \ \ \  {\bf 3}, \ \ \  0 \\
     \hline
      bino, B boson & ${\tilde{{ B}}}$ & $B$ & {\bf 1}, \ \ \ {\bf 1}, \ \ \ 0 \\
      \hline 
    \end{tabular}
     \caption{Gauge supermultiplet fields  in the MSSM.}
     \end{table}
     After electroweak symmetry breaking, the ${ W}^0$ and the $B$ fields mix to produce the 
     physical ${ Z}^0$ and $\gamma$ fields, while the corresponding `s'-fields mix to produce 
     a zino (${\tilde{{ Z}}}^0$)  degenerate with the ${ Z}^0$, and a massless photino 
     $\tilde{\gamma}$. 
     
     So, knowing the gauge groups, the particle content, and the gauge transformation properties, 
     all we need to do to specify any proposed model is to give the superpotential $W$. {\em The 
     MSSM is specified by the choice} 
     \be 
     W= y_{\rm u}^{ij}{\bar{ u}}_i Q_j \cdot  H_{\rm u} - 
     y_{\rm d}^{ij}{\bar{ d}}_i Q_j \cdot H_{\rm d} 
      - y_{\rm e}^{ij} {\bar{ e}}_i L_j\cdot  H_{\rm d} + 
     \mu H_{\rm u}\cdot H_{\rm d}.\label{eq:WMSSM}
     \ee
     The fields appearing in (\ref{eq:WMSSM}) are the chiral superfields indicated under the `Names' 
     column of Table 1. We can alternatively think of $W$ as being the same function of the scalar 
     fields in each chiral supermultiplet, as explained in section 9.4. In either case, 
     the $y$'s are $3 \times 3$ matrices in family (or generation) space, {\em and are 
     exactly the same Yukawa couplings as those which enter the SM} 
     (see for example section 22.7 of \cite{AH32}).\footnote{We stress 
     once again - see section 4 and footnote 9 - 
      that whereas in the SM we can use one Higgs doublet 
      and its  charge conjugate doublet (see section 22.6 of 
     \cite{AH32}), this is not allowed in SUSY, because $W$ cannot   depend 
     on the complex conjugate of any field (which would appear in  the charge conjugate). By convention, 
     the MSSM does not include Dirac-type neutrino mass terms, neutrino masses being 
     generally regarded as `beyond the SM' physics.}These couplings give masses to 
     the quarks and leptons when the Higgs fields acquire vacuum expectation values: there are 
     no `Lagrangian' masses for the fermions, since these would explicitly break the 
     SU(2)$_{\rm L}$ gauge symmetry. The `$\cdot$' notation here means the SU(2)-invariant coupling 
     of two doublets.\footnote{To take an elementary example: Consider the isospin part of the 
     deuteron's wavefunction. It has $I=0$ - i.e. it is the SU(2)-invariant coupling of the two 
     doublets $N^{(1)}=\left( \ba{c} p^{(1)} \\ n^{(1)} \ea \right), \ \ N^{(2)} = 
     \left( \ba{c} p^{(2)} \\ n^{(2)} \ea \right).$ 
     This $I=0$ wavefunction is, as usual, $\frac{1}{\sqrt{2}}(p^{(1)} n^{(2)} - n^{(1)} p^{(2)})$, 
     which (dropping the $1/ \sqrt{2}$) 
     we may write as $N^{(1) \rm T} {\rm i} \tau_2 N^{(2)} \equiv N^{(1)} \cdot N^{(2)}$. 
     Clearly this isospin-invariant 
     coupling is basically the same as the Lorentz-invariant spinor 
     coupling `$\chi^{(1)} \cdot \chi^{(2)}$', which is why we use the same `$\cdot$' notation 
     for both - we hope without causing confusion.}Also, colour indices have been suppressed, so 
     that `${\bar{ u}}_i Q_j$', for example, is really ${\bar{ u}}_{\alpha i}Q^\alpha_j$, 
     where the upstairs $\alpha = 1,2,3$  is  a colour {\bf 3} (triplet) index, and the downstairs 
     $\alpha$ is a colour $\bar{{\bm 3}}$ (antitriplet) index.
     
     [In parenthesis, we note a possibly confusing aspect of the labelling adopted for the 
     Higgs fields. In the conventional formulation of the SM, the Higgs field $\phi = \left( 
     \begin{array}{c} \phi^+ \\ \phi^0 \end{array} \right)$ generates mass for the `down' 
     quark, say, via a Yukawa interaction of the form (suppressing family labels) 
     \be 
     b\, {\bar{q}}_{{\rm L} } \phi  d_{{\rm R}}  + {\rm h.c.} \label{eq:SMdown}
     \ee
     In this case, the SU(2) dot product is simply $q^\dagger_{\rm L} \phi$, which is 
     plainly invariant under $q_{\rm L} \to U q_{\rm L}, \ \phi \to U \phi$. Now  
     $q^\dagger_{\rm L} \phi = u^\dagger \phi^+ + d^\dagger \phi^0$; so when 
     $\phi^0$ develops a vev, (\ref{eq:SMdown}) contributes 
     \be 
     b \langle \phi^0 \rangle {\bar{d}}_{\rm L} d_{\rm R} + {\rm h.c.} \label{eq:SMdmass}
     \ee
     which is a d-quark mass. Why, then, do we label our Higgs field $\left( \begin{array}{c} 
     H^+_{\rm u}\\H^0_{\rm u} \end{array} \right)$ with a subscript `u' rather than `d'? 
     The point is that, in the SUSY version (\ref{eq:WMSSM}),  the SU(2) dot product involving 
     the superfield $H_{\rm u}$ is taken with the superfield $Q$ which has the quantum 
     numbers of the quark doublet $q_{\rm L}$ rather than  the antiquark doublet $q^\dagger_{\rm L}$.  
     If we revert for the moment to the procedure of section 8, and write $W$ just in  
     terms of the corresponding scalar fields, the first term in (\ref{eq:WMSSM}) is 
     \be 
     y_{\rm u}^{ij} {\tilde{\bar{u}}}_{{\rm L}i}({\tilde{u}}_{{\rm L}j} H^0_{\rm u} - {\tilde{d}}_{{\rm L}j} 
     H^+_{\rm u}). \label{eq:Wfirst}
     \ee
     The first term here will, via (\ref{eq:Wijderiv}) and (\ref{eq:Lsusyint}), 
     generate a term in the Lagrangian (c.f. (\ref{eq:yuk}))
     \bea 
     &&-\frac{1}{2}y_{\rm u}^{ij}(\chi_{{\bar{\rm u}}_{{\rm L}i}} \cdot \chi_{{\rm u}_{{\rm L}j}} + 
     \chi_{{\rm u}_{{\rm L}i}} \cdot \chi_{{\bar{\rm u}}_{{\rm L}j}})H^0_{\rm u} 
     +{\rm h.c.} \nonumber \\
     && = -y^{ij}_{\rm u}(\chi_{{\bar{\rm u}}_{{\rm L}i}} \cdot \chi_{{\rm u}_{{\rm L}j}})H^0_{\rm u} 
     +{\rm h.c.}
      \label{eq:umass}
     \eea
     When $H^0_{\rm u}$ develops a (real) vacuum value $v_{\rm u}$ 
     (see section 16), this  will become a Dirac-type mass   
     term for the ${\rm u}$-quark (compare (\ref{eq:dirmass})):
     \be
     -(m_{{\rm u}ij} \chi_{{\bar{\rm u}}_{{\rm L}i}} \cdot \chi_{{\rm u}_{{\rm L}j}} + {\rm h.c.})
     \label{eq:mudirmass}
     \ee
     where 
     \be 
     m_{{\rm u}ij}= v_{\rm u} y_{{\rm u}}^{ij}.\label{eq:muvy}
     \ee
     Transforming to the basis which diagonalizes the mass matrices then leads to flavour 
     mixing exactly as in the SM (see section 22.7 of \cite{AH32}, for example).]

      {\em In summary, then, at the cost of only one new parameter, $\mu$, we have got an 
      exactly supersymmetric extension of the SM.}

    The fermion masses are evidently proportional to the 
     relevant $y$ parameter, so since 
     the top, bottom and tau are the heaviest fermions in the SM, it is sometimes useful to consider 
     an approximation in which the only non-zero $y$'s are 
     \be 
     y^{33}_{\rm u} = y_{\rm t}; \ \ \ y^{33}_{\rm d}=y_{\rm b}; \ \ \ y^{33}_{\rm e}= y_\tau.
     \ee
     In terms of the SU(2)$_{\rm L}$ weak isospin fields, this gives (for the scalar fields, and 
     omitting the $\mu$ term)  
     \be 
     W \approx y_{\rm t}[{\tilde{\bar{t}}}_{{\rm L}}({\tilde{t}}_{\rm L} H^0_{\rm u} -
     {\tilde{b}}_{\rm L} H^+_{\rm u})] -
     y_{\rm b}[{\tilde{\bar{b}}}_{\rm L}({\tilde{t}}_{\rm L} H^-_{\rm d} - {\tilde{b}}_{\rm L} H^0_{\rm d})]
      -y_{\rm t}[{\tilde{\bar{\tau}}}_{\rm L}({\tilde{\nu}}_{\tau{\rm L}} H^-_{\rm d} 
     -{\tilde{\tau}}_{\rm L} H^0_{\rm d})].\label{eq:Wapprox}
     \ee
      The minus signs in $W$ have been chosen so that the terms  
      $y_{\rm t}{\tilde{\bar{t}}}_{\rm L}{\tilde{t}}_{\rm L}$, 
     $y_{\rm b}{\tilde{\bar{b}}}_{\rm L}{\tilde{b}}_{\rm L}$ and $y_{\tau}{\tilde{\bar{\tau}}}_{\rm L}
     {\tilde{\tau}}_{\rm L}$ have the 
     correct  sign to generate mass terms for the top, bottom and tau when 
     $\langle H^0_{\rm u}\rangle \neq 0$ and $\langle H^0_{\rm d} \rangle \neq 0$.
     
     It is worth recalling that in such a SUSY theory,  in addition to the Yukawa couplings 
     of the SM, which couple the Higgs fields to quarks and to leptons, there must also be 
     similar couplings between Higgsinos, squarks and quarks, and between Higgsinos, sleptons 
     and leptons (i.e. we change two ordinary particles into their superpartners). There are 
     also scalar quartic interactions with strength proportional to $y_{\rm t}^2$, as noted in 
     section 8,  arising from the term `$|W_i|^2$' in the scalar potential (\ref{eq:VDF}). In 
     addition, there are scalar quartic interactions proportional to the squares of the 
     gauge couplings $g$ and $g'$ coming from the `$D$-term' in (\ref{eq:VDF}). These include 
     (Higgs)$^4$ coupings such as are postulated in the SM, but now appearing with 
     coefficients which are determined in terms of the parameters $g$ and $g'$ already 
     present in the model. The important phenomenological consequences of this will be 
     discussed in section 16.     
     
     Although there are no conventional mass terms in (\ref{eq:WMSSM}), 
     there is one term which is quadratic in the fields, the so-called `$\mu$ term', which is the 
     SU(2)-invariant coupling of the two different Higgs superfield doublets:
     \be 
     W(\mu \ {\rm term}) = \mu H_{\rm u} \cdot H_{\rm d} = \mu (H_{{\rm u} 1}H_{{\rm d}2} 
     - H_{{\rm u} 2} H_{{\rm d} 1}) \label{eq:muterm}
     \ee
     where the subscripts 1 and 2 denote the isospinor component. 
      This is the only 
     such bilinear coupling of the Higgs fields allowed in $W$, because the other possibilities 
     $H^\dagger_{\rm u} \cdot H_{\rm u}$ and $H^\dagger_{\rm d}\cdot  H_{\rm d}$ involve  Hermitian 
     conjugate fields, which would violate SUSY (see footnotes 9 and 11). As always, we need the $F$-component 
     of (\ref{eq:muterm}), which is (see (\ref{eq:sfieldcpts})) 
     \be 
     \mu[(H^+_{\rm u} F^-_{\rm d}  -H^0_{\rm u} F^0_{\rm d} + H^0_{\rm d} 
     F^0_{\rm u}- H^-_{\rm d} F^+_{\rm u}) - ( {\tilde{H}}^+_{\rm u} \cdot {\tilde{H}}^-_{\rm d} -{\tilde{H}}^0_{\rm u} 
     \cdot {\tilde{H}}^0_{\rm d})],\label{eq:mutermF}
     \ee
      and we must include also  the Hermitian conjugate of (\ref{eq:mutermF}). The second term 
      in (\ref{eq:mutermF}) will contribute to (off-diagonal) 
      Higgsino mass terms. The first term has the general 
      form `$W_iF_i$' of section 8,   
      and hence (see (\ref{eq:WZ2})) it leads to the following term in the scalar potential, 
      involving the Higgs fields:
      \be 
      |\mu|^2 (|H^+_{\rm u}|^2 +|H^-_{\rm d}|^2 + |H^0_{\rm u}|^2 + |H^0_{\rm d}|^2).
      \label{eq:muhiggs}
      \ee     
      But these terms all have the (positive) sign appropriate to a standard `$m^2 \phi^\dagger \phi$' 
      bosonic mass term, {\em not} the negative sign needed for electroweak symmetry breaking via the 
      Higgs mechanism (recall the discussion following equation (\ref{eq:HiggsV})). This means that 
      our SUSY-invariant Lagrangian cannot accommodate electroweak symmetry breaking.

      Of course, SUSY itself - in the MSSM application we are considering - cannot be an exact 
      symmetry, since we've not yet observed the s-partners of the SM fields. We shall discuss  
      SUSY breaking briefly in section 15, but it is clear from the above that some   
      SUSY-breaking  terms will be needed in the Higgs potential, in order to allow 
      electroweak symmetry breaking.
      
      This actually poses something of a puzzle \cite{KN}.
       The parameter $\mu$ should presumably lie roughly in the range 
      100 GeV - 1 TeV, or else we'd need delicate cancellations between the positive 
      $|\mu|^2$ terms and the negative SUSY-breaking terms 
      necessary for electroweak symmetry breaking (see a similar argument in section 
      1.1). We saw in section 1.1 that the general `no fine-tuning' argument suggested that 
      SUSY-breaking masses should not be much greater than 1 TeV. But the $\mu$ term 
      doesn't break SUSY! We are faced with an apparent difficulty: where does this 
      low scale for the SUSY-respecting parameter $\mu$ come from? References to some 
      proposed solutions to this `$\mu$ problem' are given in \cite{martin} section 5.1, 
      where some further discussion is also given of the various interactions present in 
      the MSSM; see also \cite{CEKKLW} section 4.2, and particularly the review of the $\mu$ problem in 
      \cite{polonsky}.

      \section{Gauge Coupling Unification in the MSSM}
      
      As mentioned in section 1.2(b), the idea \cite{georgi} that the three scale-dependent (`running') 
      SU(3)$\times$SU(2)$\times$U(1) gauge couplings of the SM should converge to a common 
      value - or {\em unify} - at some very high energy scale does not, in fact, prove to be 
      the case for the SM itself, but it does work very convincingly in the MSSM \cite{gauge}. The 
      evolution of the gauge couplings is determined by the numbers and types of the gauge and matter 
      multiplets present in the theory, which we have just now given for the MSSM; we can therefore 
      proceed to describe this celebrated result. 
      
      The couplings $\alpha_3$ and $\alpha_2$ are defined by 
      \be
      \alpha_3=g_{\rm s}^2/4 \pi, \ \ \ \ \ \alpha_2=g^2/4 \pi 
      \ee
      where $g_{\rm s}$ is the SU(3)$_{\rm c}$ gauge coupling of QCD and $g$ is that of the 
      electroweak SU(2)$_{\rm L}$. The definition of the third coupling $\alpha_1$ is a little 
      more complicated. It obviously has to be related in some way to $g^{\prime 2}$, where 
      $g'$ is the gauge coupling of the U(1)$_y$ of the SM. The constants $g$ and $g'$ appear in 
      the SU(2)$_{\rm L}$ covariant derivative (see equation (22.21) of \cite{AH32} for example) 
      \be 
      D_\mu = \partial_\mu + {\rm i} g ({\bm \tau}/2) \cdot {\bm W}_\mu + {\rm i} g' (y/2) B_\mu.
      \label{eq:su2covD}
      \ee
      The problem is that, strictly within in the SM framework, the scale of `$g'$' is 
      arbitrary: we could multiply the weak hypercharge generator $y$ by an arbitrary constant 
      $c$, and divide $g'$ by $c$, and nothing would change. In contrast to this, the 
      normalization of whatever couplings multiply the three generators $\tau^1, \tau^2$ and 
      $\tau^3$ in (\ref{eq:su2covD}) is fixed by the normalization of the $\tau$'s:
      \be 
      {\rm Tr} ( \frac{\tau^\alpha}{2} \frac{\tau^\beta}{2})= \frac{1}{2} \delta_{\alpha \beta}.
      \ee
      Since each generator is normalized to the same value, the same constant $g$ must multiply 
      each one - no relative rescalings are possible. 
      Within a `unified' framework, therefore, we hypothesize that some multiple of $y$, say 
      $Y=c(y/2)$, is one of the generators of a larger group (SU(5) for instance), which 
      also includes the generators of SU(3)$_{\rm c}$ and SU(2)$_{\rm L}$, all being subject to 
      a common normalization condition; there is then only one (unified) gauge coupling. The quarks and 
      leptons of one family will  all belong to a single representation of the larger group, 
      though this need not necessarily be the fundamental representation. All that matters is that 
      the generators all have a common normalization. For example, we can demand the condition 
      \be 
      {\rm Tr} (c^2 (y/2)^2) = {\rm Tr} (t_3)^2 \label{eq:norm}
      \ee
      say, where $t_3$ is the third SU(2)$_{\rm L}$ generator (any generator will give the 
      same result), and the Trace is over all states in the representation - here, u, d, $\nu_{\rm e}$ and 
      ${\rm e}^-$.  The Traces are simply the sums of the squares of the eigenvalues. On the RHS of 
      (\ref{eq:norm}) we obtain 
      \be 
      3(\frac{1}{4} + \frac{1}{4}) + \frac{1}{4} + \frac{1}{4} = 2
      \ee
      where the `3' comes from colour, while on the LHS we find from Table 1 
      \be 
      c^2 (\frac{3}{36} + \frac{3}{36} + \frac{3.4}{9} + \frac{3.1}{9} + \frac{1}{4} + 1 + \frac{1}{4}) = 
      c^2 \frac{20}{6}.
      \ee
      It follows that 
      \be 
      c=\sqrt{\frac{3}{5}},
      \ee
      so that the correctly normalized generator is 
      \be 
      Y=\sqrt{\frac{3}{5}}\, y/2.
      \ee 
      The $B_\mu$ term in (\ref{eq:su2covD}) is then
      \be 
      {\rm i} g' \sqrt{\frac{5}{3}} \, Y \, B_\mu,
      \ee
      indicating that the correctly normalized $\alpha_1$ is 
      \be 
      \alpha_1=\frac{5}{3} \frac{g^{\prime 2}}{4 \pi} \equiv \frac{g_1^2}{4 \pi}.\label{eq:normcon}
      \ee
      
      Equation (\ref{eq:normcon}) can also be interpreted as a prediction for the weak angle 
      $\theta_{\rm W}$ at the unification scale: since $g \tan \theta_{\rm W} = g' = \sqrt{3/5} g_1$ 
      and $g=g_1$ at unification, we have $\tan \theta_{\rm W} = \sqrt{3/5}$, or 
      \be 
      \sin^2 \theta_{\rm W}({\mbox{unification scale}})=\frac{3}{8}.
      \ee

      We are now ready to consider the running of the couplings $\alpha_i$. To one loop order, the 
      renormalization group equation (RGE) has the form (for an introduction, see chapter 15 of 
      \cite{AH32} for example)
      \be 
      \frac{{\rm d} \alpha_i}{{\rm d} t} = - \frac{b_i}{2 \pi} \alpha_i^2 \label{eq:arun}
      \ee
      where $t=\ln Q$ and $Q$ is the `running' energy scale, and the coefficients $b_i$ are 
      determined by the gauge group and the matter multiplets to which the gauge bosons couple. For 
      SU(N) gauge theories with matter in the fundamental representation, we have (see \cite{PS} for 
      example) 
      \be 
      b_N=\frac{11}{3} N - \frac{1}{3} n_{\rm f} - \frac{1}{6} n_{\rm s} \label{eq:bN}
      \ee
      where $n_{\rm f}$ is the number of left-handed fermions (counting, as usual, right-handed 
      ones as left-handed antiparticles), and $n_{\rm s}$ is the number of complex scalars, which 
      couple to the gauge bosons. For a U(1)$_Y$ gauge theory in which the fermionic matter particles 
      have charges $Y_{\rm f}$ and the scalars have charges $Y_{\rm s}$, the corresponding formula is 
      \be 
      b_1=-\frac{2}{3} \sum_{\rm f} Y^2_{\rm f} - \frac{1}{3} \sum_{\rm s} Y^2_{\rm s}.
      \ee
      To examine unification, it is convenient to rewrite (\ref{eq:arun}) as 
      \be 
      \frac{\rm d}{{\rm d} t} (\alpha_i^{-1}) = \frac{b_i}{2 \pi} 
      \ee
      which can be immediately integrated to give 
      \be 
      \alpha^{-1}_i(Q) = \alpha^{-1}_i(Q_0) + \frac{b_i}{2 \pi} \ln (Q/Q_0),\label{eq:runlin}
      \ee
      where $Q_0$ is the scale at which running commences. We see that  the inverse couplings run linearly 
      with $\ln Q$. $Q_0$ is taken to be $m_{\rm Z}$, where the couplings are well measured. `Unification' 
      is then the hypothesis that, at some higher scale $Q_{\rm U}=m_{\rm U}$, the couplings are equal:
      \be 
      \alpha_1(m_{\rm U})=\alpha_2(m_{\rm U})=\alpha_3(m_{\rm U}) \equiv \alpha_{\rm U}.
      \ee
      This implies that the three equations (\ref{eq:runlin}), for $i=1,2, 3$, become 
      \bea 
      \alpha_{\rm U}^{-1} &=& \alpha^{-1}_3(m_{\rm Z})  + 
      \frac{b_3}{2 \pi} \ln (m_{\rm U}/m_{\rm Z}) \label{eq:run3} \\      
      \alpha_{\rm U}^{-1} &=& \alpha^{-1}_2(m_{\rm Z})  +
       \frac{b_2}{2 \pi} \ln (m_{\rm U}/m_{\rm Z}) \label{eq:run2}\\
      \alpha_{\rm U}^{-1} &=& \alpha^{-1}_1(m_{\rm Z})  +
       \frac{b_1}{2 \pi} \ln (m_{\rm U}/m_{\rm Z}). \label{eq:run1}
      \eea
      Eliminating $\alpha_{\rm U}$ and $\ln (m_{\rm U}/m_{\rm Z})$ from these equations gives one 
      condition relating the measured constants $\alpha^{-1}_i(m_{\rm Z})$ and the calculated numbers 
      $b_i$, which is 
      \be 
      \frac{\alpha_3^{-1}(m_{\rm Z}) - \alpha_2^{-1}(m_{\rm Z})} {\alpha_2^{-1}(m_{\rm Z}) 
      - \alpha_1^{-1}(m_{\rm Z})} = \frac{b_2-b_3}{b_1-b_2}. \label{eq:uncond}
      \ee
      Checking the truth of (\ref{eq:uncond}) is one simple way of testing unification quantitatively 
      (at least, at this one-loop level). 
      
      Let's call the LHS of (\ref{eq:uncond}) $B_{\rm exp}$, and the RHS $B_{\rm th}$. For $B_{\rm exp}$, 
      we use the data 
      \be 
      \sin ^2 \theta_{\rm W}(m_{\rm Z}) = 0.231 \label{eq:sthexp}
      \ee
      \be 
      \alpha_3(m_{\rm Z})=0.119, \ \ \ \ \ {\mbox{or}} \ \ \ \ \ \alpha^{-1}_3(m_{\rm Z})= 8.40\label{eq:a3exp}
      \ee
      \be 
      \alpha^{-1}_{\rm em}(m_{\rm Z})= 128.\label{eq:aemexp}
      \ee
      We are not going to bother with errors here, but the uncertainty in $\alpha_3(m_{\rm Z})$ is 
      about 2\%, and that in $\sin^2 \theta_{\rm W}(m_{\rm Z})$ and $\alpha_{\rm em}(m_{\rm Z})$ is much less. 
      Here $\alpha_{\rm em}$ is defined by $\alpha_{\rm em}=e^2/4 \pi$ where $e=g \sin \theta_{\rm W}$. 
      Hence 
      \be 
       \alpha_2^{-1}(m_{\rm Z}) = \alpha^{-1}_{\rm em}(m_{\rm Z}) \sin^2 \theta_{\rm W}(m_{\rm Z}) = 29.6.
       \ee
       Finally, 
       \be 
       g^{\prime 2}=g^2 \tan^2 \theta_{\rm W}
       \ee
       and hence 
       \be 
       \alpha^{-1}_1(m_{\rm Z})=\frac{3}{5} \alpha^{\prime -1}(m_{\rm Z})= \frac{3}{5} \alpha^{-1}_2(m_{\rm Z}) 
       \cot^2 \theta_{\rm W}(m_{\rm Z}) = 59.12.
       \ee
       From these values we obtain 
       \be 
       B_{\rm exp}=0.72. \label{eq:Bexp}
       \ee
       
       Now let's look at $B_{\rm th}$. First, consider the SM. For SU(3)$_{\rm c}$ we have 
       \be 
       b_3^{\rm SM}= 11 - \frac{1}{3} 12 = 7.
       \ee
       For SU(2)$_{\rm L}$ we have 
       \be 
       b_2^{\rm SM}= \frac{22}{3}-4 - \frac{1}{6}= \frac{19}{6},
       \ee
       and for U(1)$_Y$ we have 
       \bea 
       b_1^{\rm SM}&=& - \frac{2}{3} \, \frac{3}{5} \sum_{\rm f} (y_{\rm f}/2)^2 - \frac{1}{3}\, \frac{3}{5} 
       \sum_{\rm s}(y_{\rm s}/2)^2 \label{eq:b1ferm} \\
       &=& - \frac{2}{5} \,3 \,\frac{20}{6} - \frac{1}{5}\, \frac{1}{2} = - \frac{41}{10}.
       \eea 
       Hence, in the SM, the RHS of (\ref{eq:uncond}) gives 
       \be 
       B_{\rm th} = \frac{115}{218} = 0.528,
       \ee
       which is in very poor accord with (\ref{eq:Bexp}).
       
       What about the MSSM? Formula (\ref{eq:bN}) must be modified to take account of the fact that, in 
       each SU(N), the gauge bosons are accompanied by gauginos in the regular representation of the 
       group. Their contribution to $b_N$ is $-2N/3$. In addition, we have to include the scalar 
       partners of the quarks and of the leptons, in the fundamental representations of 
       SU(3) and SU(2); and we must not forget that we have two Higgs doublets, both 
       accompanied by Higgsinos, all in the fundamental representation of SU(2). These changes give 
       \be 
       b_3^{\rm MSSM}= 7-2-\frac{1}{6} 12 = 3,\label{eq:b3mssm}
       \ee
       and 
       \be 
       b_2^{\rm MSSM}= \frac{19}{6} - \frac{4}{3} - \frac{1}{6} 12 - \frac{1}{3} 2 - \frac{1}{6} = -1.
       \label{eq:b2mssm}
       \ee   
      It is interesting that the sign of $b_2$ has been reversed. For $b_1^{\rm MSSM}$, there is no 
      contribution from the gauge bosons or their fermionic partners. The left-handed fermions 
      contribute as in (\ref{eq:b1ferm}), and are each accompanied by corresponding scalars, so that
      \be 
      b_1^{\rm MSSM}({\mbox{fermions and sfermions}})= - \frac{3}{5} 10 = -6.
      \ee
      The Higgs and Higginos contribute 
      \be 
      b_1^{\rm MSSM}({\mbox{Higgs and Higgsinos}})= - \frac{3}{5} \, 4\,  \frac{1}{4}= - \frac{3}{5}.
      \ee
      In total, therefore, 
      \be 
      b_1^{\rm MSSM}= - \frac{33}{5}.\label{eq:b1mssm}
      \ee
      From (\ref{eq:b3mssm}), (\ref{eq:b2mssm}) and (\ref{eq:b1mssm}) we obtain 
      \be 
      B_{\rm th}^{\rm MSSM}= \frac{5}{7}= 0.74 \label{eq:Bpredmssm}
      \ee
      which is in excellent agreement with (\ref{eq:Bexp}). 
      
      This has been by no means a `professional' calculation. One should consider two-loop 
      corrections. Also, SUSY must be broken, presumably at a scale of order 1 TeV 
      or less, and the resulting mass differences between the particles and their 
      s-partners will lead to `threshold' corrections. Similarly, the details of the 
      theory at the high scale (in particular, its breaking) may be expected to lead to 
      (high energy) threshold corrections. A recent analysis by Pokorski \cite{pok} 
      concludes that the present data are in good agreement with the predictions of 
      supersymmetric unification, for reasonable estimates of such corrections. 
      
      Returning to (\ref{eq:run2}) and (\ref{eq:run1}), and inserting the values of 
      $\alpha^{-1}_2(m_{\rm Z})$ and $\alpha^{-1}_1(m_{\rm Z})$, we can 
      obtain an estimate of the unification scale $m_{\rm U}$. We find 
      \be 
      \ln (m_{\rm U}/m_{\rm Z})= \frac{10 \pi}{28} [ \alpha^{-1}_1(m_{\rm Z}) - \alpha^{-1}_2(m_{\rm 
      Z})]  \simeq 33.1,
      \ee
      which implies 
      \be 
      m_{\rm U} \simeq 2.2 \times 10^{16} {\rm GeV},
      \ee
      \
      a value relatively close to the Planck scale $m_{\rm P} \simeq 1.2 \times 10^{19}$ {\rm GeV}. 
      
      Of course, one can  make up any number of  models yielding the experimental value $B_{\rm exp}$. But there 
      is no denying that the correct prediction (\ref{eq:Bpredmssm}) is an unforced consequence 
      simply of the matter content of the MSSM, and agreement was clearly not inevitable. It does seem to 
      provide support both for the inclusion of supersymmetric particles in the RGE, and for gauge unification. 
      Grand unified theories are reviewed by Raby in \cite{RPP}.

      \section{$R$-parity}

      As stated in the  section 12, the `minimal' supersymmetric extension of 
      the SM is specified by the choice (\ref{eq:WMSSM}) for the superpotential. There are, 
      however, other gauge-invariant and renormalizable  terms which could also 
      be included in the superpotential, namely (\cite{martin} section 5.2) 
      \be 
      W_{\Delta L =1} = \lambda_{e}^{ijk}L_i \cdot L_j {\bar{e}}_k + 
      \lambda_{L}^{ijk} L_i \cdot Q_j {\bar{d}}_k 
      + \mu_{L}^i L_i \cdot H_{\rm u}\label{eq:Lviol}
      \ee
      and \be 
      W_{\Delta B=1} = \lambda_{B}^{ijk} {\bar{u}}_i {\bar{d}}_j {\bar{d}}_k.\label{eq:Bviol}
      \ee  
       The superfields $Q$ carry baryon number $B=1/3$ and  ${\bar{u}}$,  ${\bar{d}}$ carry $B=-1/3$, 
       while $L$ carries lepton  number $L=1$ and ${\bar{e}}$ carries $L=-1$. Hence the terms in 
       (\ref{eq:Lviol}) violate lepton number conservation by one unit of $L$, and those in 
       (\ref{eq:Bviol}) violate baryon number conservation by one unit of $B$. Now, $B$- and 
       $L$-violating processes have never been seen experimentally. If both the couplings 
       $\lambda_{L}$ and $\lambda_{B}$ were present,  the proton could decay via channels 
       such as ${\rm e}^+ {\pi}^0, \ \mu^+ \pi^0, \ldots$ etc. The non-observance of such 
       decays places strong limits on the strengths of such couplings, which would have to 
       be extraordinarily small (being renormalizable, the couplings are dimensionless, 
       and there is no natural suppression by a high scale such as would occur in a non-renormalizable 
       term). It is noteworthy that in the SM, there are no possible renormalizable terms in the 
       Lagrangian which violate $B$ or $L$ - this is indeed a nice bonus provided by the SM. 
       We could of course just impose $B$ and $L$ conservation as a principle, thus 
       forbidding (\ref{eq:Lviol}) and (\ref{eq:Bviol}). But in fact both are known to be 
       violated by non-perturbative electroweak effects, which are negligible at ordinary 
       energies but which might be relevant in the early universe. Neither $B$ nor $L$ 
       can therefore be regarded as a fundamental symmetry. Instead, people have come up 
       with an alternative symmetry, which forbids (\ref{eq:Lviol}) and 
       (\ref{eq:Bviol}), while allowing all the interactions of the MSSM.  
          
          This symmetry is called $R$-parity, which is multiplicatively conserved, 
          and is defined by 
          \be 
          R=(-)^{3B+L+2s}\label{eq:R}
          \ee
          where $s$ is the spin of the particle. One quickly finds that $R$ is +1 for all 
          conventional matter particles, and (because of the $(-)^{2s}$ factor) -1 for all their 
          s-partners. Since the product of $(-)^{2s}$ is +1 for the particles involved in any 
          interaction vertex, by angular momentum conservation, it's clear that both (\ref{eq:Lviol}) 
          and (\ref{eq:Bviol}) do not conserve $R$-parity, while the terms in (\ref{eq:WMSSM}) do. 
          In fact, every interaction vertex in (\ref{eq:WMSSM}) contains an even number of $R=-1$ 
          sparticles, which has important phenomenological consequences:
          \begin{itemize}
          \item The lightest sparticle (`LSP') is absolutely stable, and if electrically 
          uncharged it could be an attractive candidate for non-baryonic dark matter. 
          \item The decay products of all other sparticles must contain an odd number of 
          LSP's.
          \item In accelerator experiments, sparticles can only be produced in pairs. 
          \end{itemize}
          
          In the context of the MSSM, the LSP must lack electromagnetic and strong interactions;  
          otherwise, LSP's surviving from the Big Bang era would have bound to nuclei forming 
          objects with highly unusual charge to mass ratios, but  searches for such exotics have excluded 
          all models with stable charged or strongly interacting particles unless their mass exceeds 
          several TeV, which is unacceptably high for the LSP. An important implication is that in collider 
          experiments LSP's will carry away energy and momentum while escaping detection. Since all 
          sparticles will decay into at least one LSP (plus SM particles), and since in the MSSM 
          sparticles are pair-produced, it follows that at least $2 m_{{\tilde{\chi}}^0_1}$ missing 
          energy will be associated with each SUSY event, where $m_{{\tilde{\chi}}^0_1}$ is the mass 
          of the LSP (often taken to be a neutralino - 
          see section 17). In ${\rm e}^- {\rm e}^+$ machines, the total 
          visible energy and momentum can be well measured, and the beams have very small spread, so that 
          the missing energy and momentum can be well correlated with the energy and momentum of the 
          LSP's. In hadron colliders, the distribution of energy and longitudinal momentum of the 
          partons (i.e. quarks and gluons) is very broad, so in practice only the missing transverse momentum 
          (or missing transverse energy ${\not{\!\!E}}_{\rm T}$) is useful. 
          
          Further aspects of $R$-parity are discussed in \cite{martin}.

         \section{SUSY breaking} 
          
         Since SUSY is manifestly not an exact symmetry of the known particle spectrum, 
         the issue of SUSY-breaking must be addressed before the MSSM can be applied 
         phenomenologically. We know only two ways in which a symmetry can be broken: 
         either (a) by explicit symmetry-breaking terms in the Lagrangian, or (b) by spontaneous 
         symmetry breaking, such as occurs in the case of the chiral symmetry of QCD, and 
         is hypothesized to occur for the electroweak symmetry of the SM via the Higgs mechanism. 
         In the electroweak case,  the introduction of 
         explicit mass terms for the fermions and massive gauge bosons would spoil renormalizability, 
         which is why in this case spontaneous symmetry breaking (which preserves 
         renormalizability) is preferred theoretically - and indeed is strongly indicated 
         by experiment, via the precision measurement of finite radiative corrections.         
          We shall give a brief introduction to spontaneous SUSY-breaking, 
         since it presents some novel features as compared, say, to the more `standard' 
         QCD and electroweak cases. But in fact there is no consensus on how `best' to break 
         SUSY spontaneously, and  in practice one is reduced to introducing explicit 
         SUSY-breaking terms as in approach (a) after all, which parametrize the low-energy 
         effects of the unknown breaking mechanism presumed  (usually) 
         to operate at some high mass scale. We shall see in section 15.2 that these SUSY-breaking 
         terms are quite constrained by the requirement that they do not re-introduce divergences which would  
         spoil the SUSY solution to the hierarchy problem; nevertheless, over 100 parameters are 
          needed to characterize them.  
         
         \subsection{Breaking SUSY Spontaneously}  
         
         The fundamental requirement for a symmetry in field theory 
         to be spontaneously broken (see for example \cite{AH32} Part 7) 
         is that a field, which is not invariant under the 
         symmetry, should have a non-vanishing vacuum expectation value. That is, if the 
         field in question is denoted by $\phi'$, then we require $\langle 0 | \phi'(x) | 0 \rangle 
         \neq 0$. Since $\phi'$ is not invariant, it must belong to a symmetry multiplet of 
         some kind, along with other fields, and it must be possible to express $\phi'$ as 
         \be 
         \phi'(x) = {\rm i}[Q, \phi(x)] \label{eq:phipQ}
         \ee
         where $Q$ is a generator of the symmetry group, and $\phi$ is a suitable field in the 
         multiplet to which $\phi'$ belongs. So then we have 
         \be
        \langle 0 | \phi' | 0 \rangle = \langle 0 |{\rm i}  [Q, \phi] | 0 \rangle 
        = \langle 0 | {\rm i}Q \phi - {\rm i}\phi Q | 0 \rangle \neq 0.\label{eq:SSB}
        \ee
         Now the vacuum state $|0 \rangle$ is usually assumed to be such that $Q |0 \rangle =0$, 
         since this implies that $|0 \rangle$ is invariant under the transformation generated 
         by $Q$. But if we take $Q |0 \rangle =0$, we violate (\ref{eq:SSB}). Hence for 
         spontaneous symmetry breaking we have to assume $Q |0 \rangle \neq 0$ - that is, 
         the vacuum is not invariant under the symmetry. 
         
         In the case of SUSY, this means that we require 
         \be 
         Q_a | 0 \rangle \neq 0, \ \ \ Q^\dagger_b |0 \rangle \neq 0 \label{eq:SSBSUSY}
         \ee
         for the SUSY generators of section 5. The condition (\ref{eq:SSBSUSY}) has 
         a remarkable consequence in SUSY, which is strikingly different from all other 
         examples of spontaneous symmetry breaking. The SUSY algebra (\ref{eq:QQstar1}) is 
         \be 
         \{ Q_a, Q^\dagger_b \} = (\sigma^\mu)_{ab} P_\mu.
         \ee
         So we have 
         \bea Q_1 Q^\dagger_1 + Q^\dagger_1 Q_1 = (\sigma^\mu)_{11} P_\mu &=& P_0 + P_3 
         \nonumber \\
         Q_2 Q^\dagger_2 + Q^\dagger_2 Q_2 = (\sigma^\mu)_{22}P_\mu  &=& P_0 -P_3.
         \eea
         It follows that 
         \be 
         P_0 = \frac{1}{2}(Q_1Q^\dagger_1 + Q^\dagger_1 Q_1 + Q_2 Q^\dagger_2 + Q^\dagger_2 Q_2) = H,
         \ee
         where $H$ is the Hamiltonian of the theory considered. Hence we find 
         \bea 
         \langle 0 | H | 0 \rangle &=& \frac{1}{2}(\langle 0 | Q_1 Q^\dagger_1 | 0 \rangle 
         + \langle 0 | Q^\dagger_1 Q_1 | 0 \rangle + \ldots )\nonumber \\
         &=& \frac{1}{2}(\  |(Q^\dagger_1 |0 \rangle) |^2 + | (Q_1 |0 \rangle )|^2 + \ldots )\nonumber \\
         &>& 0,
         \eea
         the strict inequality in the last step following from the basic symmetry-breaking 
         assumption (\ref{eq:SSBSUSY}). We deduce the remarkable result: {\em when SUSY is 
         spontaneously broken,  the vacuum energy is necessarily positive}.\footnote{This is 
         true for global SUSY - i.e. the case in which the parameters $\xi, \xi^\dagger$ in 
         SUSY transformations don't depend on the space-time coordinate $x$. In the local 
         case, which is essentially supergravity, it turns out that the vacuum has exactly 
         zero energy in the spontaneously broken case.} On the other hand, when SUSY 
         is exact, so that $Q_a |0 \rangle = Q^\dagger_b |0 \rangle =0$, we obtain 
         $\langle 0 | H | 0 \rangle = 0$ - the vacuum energy of a (globally) SUSY-invariant theory is 
         zero.
         
         For SUSY to be spontaneously broken, therefore, the scalar potential ${\cal{V}}$ must have 
         no SUSY-respecting minimum (assuming the kinetic bits of the Hamiltonian don't 
         contribute in the vacuum). For,  if it did, such a SUSY-respecting configuration 
         would necessarily have zero energy, and since by hypothesis this is  
          the minimum value of ${\cal{V}}$, SUSY-breaking (which requires ${\cal{V}} > 0$) 
          will simply not happen, on energy grounds.   
        
         What kinds of field $\phi'$ could have a non-zero value in the SUSY case? Returning 
         to (\ref{eq:phipQ}), with $Q$ now a SUSY generator, we consider all such possible 
         commutation relations, beginning with those for the chiral supermultiplet. The 
         commutation relations of the $Q$'s with the fields are determined by the 
         SUSY transformations which are   
         \bea 
         \delta_\xi \phi &=& {\rm i} [\xi \cdot Q, \phi] = \xi \cdot \chi \nonumber \\
         \delta_\xi \chi &=& {\rm i} [\xi \cdot Q, \chi] = 
         -{\rm i} \sigma^\mu {\rm i} \sigma_2 \xi^* \partial_\mu \phi + \xi F \nonumber \\
         \delta_\xi F  &=& {\rm i} [\xi \cdot Q, F] = -{\rm i} \xi^\dagger {\bar{\sigma}}^\mu 
         \partial_\mu \chi. \label{eq:Qcoms}
         \eea
         Considering the terms on the RHS of each of the three relations in (\ref{eq:Qcoms}), 
         we cannot have $\langle 0 | \chi | 0 \rangle \neq 0$ since $\chi$ isn't a scalar 
         field, and such a vev for a spinor would break Lorentz invariance; we can't have 
         $\langle 0 | \partial_\mu \phi | 0 \rangle \neq 0$ either, because $\phi$ is assumed 
         constant in the vacuum; so this leaves 
         \be 
         \langle 0 | F | 0 \rangle \neq 0
         \ee
         as the only possibility! This is called `F-type SUSY breaking', since it is the auxiliary 
         field $F$ which acquires a vev. 
         
         Recall now that in W-Z models, with superpotentials of the form (\ref{eq:W}) 
         such as are used in the MSSM, we had 
         \be 
         F_i = -\left( \frac{\partial W}{\partial \phi_i} \right)^\dagger= (M_{ij} \phi_j 
         + \frac{1}{2}y_{ijk}\phi_j \phi_k)^\dagger,
        \ee
        and ${\cal{V}}(\phi)=|F_i|^2$, which has an obvious minimum when all the $\phi$'s are zero. 
        Hence with this form of $W$, SUSY can't be spontaneously broken. To get spontaneous 
        SUSY breaking, we must add a constant to $F_i$, that is a linear 
        term in $W$ (see footnote 10). Even this is tricky, and it needs ingenuity to 
        produce a simple working model . One (due to O'Raifeartaigh \cite{OR}) employs three 
        chiral supermultiplets, and takes $W$ to be  
        \be 
        W=m \phi_1 \phi_3 + g \phi_2(\phi_3^2 -M^2).
        \ee
        This produces 
        \be 
        -F_1^\dagger = m \phi_3, \ -F_2^\dagger = g(\phi_3^2 - M^2), \ 
        -F_3^\dagger= m \phi_1 + 2 \phi_2 \phi_3.
        \ee 
        Hence 
        \bea 
        {\cal{V}}&=& |F_1|^2 + |F_2|^2 + |F_3|^2 \nonumber \\
        &=& m^2 |\phi_3|^2 + g^2 |\phi_3^2 - M^2|^2 + | m \phi_1 + 2 \phi_2 \phi_3|^2.\label{eq:VOR}
        \eea 
         The first two terms in (\ref{eq:VOR}) cannot both vanish at once, and so there is no 
         possible field configuration giving ${\cal{V}}=0$, which is the SUSY-preserving solution. 
         Instead, there is a SUSY-breaking  minimum at 
         \be 
         \phi_1=\phi_3=0,
         \ee
         which are interpreted as the corresponding vev's, 
         with $\phi_2$ undetermined (a so-called `flat' direction in field space). 
         This solution gives 
         \be
         \langle 0 | F^\dagger_1 | 0 \rangle = \langle 0 | F^\dagger_3 | 0 \rangle =0, 
         \ee
         but 
         \be 
         \langle 0 | F^\dagger_2 | 0 \rangle = gM^2. 
         \ee

         The minimum value of ${\cal{V}}$ is 
         $g^2 M^4$, which is strictly positive, as expected. 
         Note that the parameter $M$ does indeed have the dimensions of a mass: it can be understood 
         as signifying the scale of spontaneous SUSY breaking, via $\langle 0 | F_2^\dagger | 
         0 \rangle \neq 0$, much as the Higgs vev sets the scale of electroweak symmetry breaking. 
        
         In the SM, or MSSM, all the terms in $W$ must be gauge-invariant - but there is no 
         field in the SM which is  itself gauge-invariant (i.e. all its gauge quantum 
         numbers are zero). Hence in the SM or MSSM we cannot have a linear term in $W$, and must 
         look beyond these models if we want to pursue this form of SUSY breaking. 
         
         In this F-type SUSY breaking, then, we have 
         \be 
         0 \neq \langle 0 | [Q, \chi(x)] | 0 \rangle = \sum_n \langle 0 | Q | n \rangle 
         \langle n | \chi(x) | 0 \rangle - \langle 0 | \chi(x) | n \rangle \langle n | 
         Q | 0 \rangle ,\label{eq:SUSYgold}
         \ee
         where $|n\rangle$ is a complete set of states. It can be shown that (\ref{eq:SUSYgold}) 
         implies that there must exist among the states $|n\rangle$ a {\em massless} state 
         $|{\tilde{g}}\rangle$ which couples to the vacuum via the generator $Q$: $\langle 
         0 | Q | {\tilde{g}} \rangle \neq 0$. This is the SUSY version of Goldstone's 
         theorem - see for example section 17.4 of \cite{AH32}. The theorem states 
         that when a symmetry is spontaneously broken, one or more massless particles 
         must be present, which couple to the vacuum via a symmetry generator. 
         In the non-SUSY case, they are (Goldstone) bosons; in the SUSY case, since the 
         generators are fermionic, they are fermions - `Goldstinos'. You can check that 
         the fermion spectrum in the above model contains a massless field $\chi_2$ - it is in fact 
         in a supermultiplet  along  with $F_2$, the auxiliary field which gained a vev, and the 
         scalar field $\phi_2$, where $\phi_2$ is  the field direction along which the 
         potential is  `flat' -  a situation analogous to that for the standard Goldstone `wine-bottle' 
         potential, where the massless mode is associated with excitations round the flat 
         rim of the bottle.

          {\bf Exercise} Show that the mass spectrum of the 
          O'Raifeartaigh  model consists of (a) 6 real scalar 
          fields with tree-level masses 0,0 (the real and imaginary parts of the 
          complex field $\phi_2$) $m^2, m^2$ (ditto for the complex 
          field $\phi_1$) and $m^2-2g^2M^2, m^2+2g^2M^2$ (the no longer degenerate real and imaginary 
          parts of the complex field $\phi_3$); (b) 3 L-type fermions with masses 0 (the Goldstino  
          $\chi_2$), $|\mu|, |\mu|$ (linear combinations of the fields $\chi_1$ and $\chi_3$). 
          (Hint: for the scalar masses, take $\langle \phi_2 \rangle =0$ for convenience,  
          expand the potential about the point $\phi_1=\phi_2=\phi_3=0$, and examine the quadratic 
          terms. For the fermions, the mass matrix of (\ref{eq:WZ2}) is $W_{13}=W_{31}=m$, all other 
          components vanishing; diagonalize the mass term by introducing the linear 
          combinations $\chi_-=(\chi_1-\chi_3)/\sqrt{2}, \chi_+=(\chi_1+\chi_3)/\sqrt{2}$. 
          See also \cite{bailin} section 2.10).)
          
          In the absence of SUSY breaking, a single massive chiral supermultiplet 
          consists (as in the W-Z model of section 8) of a complex scalar field 
          (or equivalently two real scalar fields) degenerate in mass with an L-type spin-1/2 field. It 
          is interesting that in the O'Raifeartaigh model the masses of the `3' supermultiplet,  
          after SUSY breaking, obey the relation 
          \be 
          (m^2-2g^2M^2)+(m^2+2g^2M^2)=2m^2=2m^2_{{\chi_{3}}},\label{eq:ORmass}
          \ee
           which is evidently a generalization of the relation that would hold in the SUSY-preserving 
           case $g=0$. Indeed, there is a general sum rule for the tree-level (mass)$^2$ values of 
           scalars and chiral fermions in theories with spontaneous SUSY breaking \cite{FGP}:
           \be 
           \sum m^2_{\rm real \,  scalars} = 2 \sum m^2_{\rm chiral \,   fermions},\label{eq:FGPmass}
           \ee
           where it is understood that the sums are over sectors with the same electric 
           charge, colour charge, baryon number and lepton number. 
           Unfortunately, (\ref{eq:FGPmass}) implies that this kind of  SUSY breaking cannot 
           be phenomenologically viable, since it requires the existence of (for example) 
           scalar partners of right-handed d-type quarks, with masses of at most a few GeV - and 
           this is excluded experimentally.

          We also need to consider possible SUSY breaking via terms in a gauge supermultiplet. 
          This time the transformations are 
          \bea
          \delta_\xi W^{\mu \alpha} &=& {\rm i} [\xi \cdot Q, W^{\mu \alpha}] = -\frac{1}{\sqrt{2}}
          (\xi^\dagger {\bar{\sigma}}^\mu \lambda^\alpha + 
          \lambda^{\alpha \dagger} {\bar{\sigma}}^\mu \xi) \nonumber \\
          \delta_\xi \lambda^\alpha &=& {\rm i} [\xi \cdot Q, \lambda^\alpha] = 
          -\frac{{\rm i}}{2\sqrt{2}} \sigma^\mu {\bar{\sigma}}^\nu \xi F^\alpha_{\mu \nu} + 
          \frac{1}{\sqrt{2}}\xi D^\alpha 
          \nonumber \\
          \delta_\xi D^\alpha &=& {\rm i} [\xi \cdot Q, D^\alpha] = 
          \frac{\rm i}{\sqrt{2}}(\xi^\dagger {\bar{\sigma}}^\mu (D_\mu \lambda)^\alpha - 
          (D_\mu \lambda)^{\alpha \dagger}{\bar{\sigma}}^\mu \xi).\label{eq:Qcomg}
          \eea
          Inspection of (\ref{eq:Qcomg}) shows that, as for the chiral supermultiplet, only the  
          auxiliary fields can have a non-zero vev:
          \be 
          \langle 0 | D^\alpha | 0 \rangle,
          \ee
          which is called D-type symmetry breaking.  
          
          At first sight, however, such a mechanism can't operate in the MSSM, for which  the scalar 
          potential is as given in (\ref{eq:VDF}). `F-type' SUSY breaking comes from the first term 
          $|W_i|^2$, D-type from the second, and the latter clearly has a SUSY-preserving minimum 
          at ${\cal{V}}=0$ when all the fields vanish. But there is another possibility, rather 
          like the `linear term in $W$' trick used for F-type breaking, which was discovered 
          by Fayet and Iliopoulos \cite{FI} for the U(1) gauge case. The auxiliary field $D$ of a 
          U(1) gauge supermultiplet is gauge-invariant, and a term in the Lagrangian proportional to $D$ 
          is SUSY-invariant too, since (see (\ref{eq:delDU(1)})) it transforms by a total derivative.   
          Suppose, then, that  we add  a term $M^2D$, the {\em Fayet-Iliopoulos term}, to the Lagrangian 
          (\ref{eq:Lgch}). The part involving $D$ is now   
          \be 
          {\cal{L}}_{D} = M^2  D + \frac{1}{2}D^2 -g_1D \sum_i e_i \phi_i^\dagger \phi_i,
          \ee
           where $e_i$ are the U(1) charges of the scalar fields $\phi_i$ in 
           units of $g_1$, the U(1) coupling constant.  Then the equation of 
           motion for $D$ is 
           \be 
           D= -M^2 +  g_1 \sum_i e_i \phi_i^\dagger \phi_i.
           \ee 
           The corresponding potential   is now 
           \be 
           {\cal{V}}_D= \frac{1}{2}(-M^2 +  g_1 \sum_i  e_i \phi_i^\dagger \phi_i)^2.\label{eq:VD}
           \ee
           Consider for simplicity the case of just one scalar field $\phi$, with charge $eg_1$. 
           If $eg_1>0$ there will be a SUSY-preserving solution, i.e. with ${\cal{V}}_D=0$, 
           at $|\langle 0 | \phi | 0 \rangle | = (M^2/eg_1)^{1/2}$. This is actually a Higgs-type 
           breaking of the U(1) symmetry, and it will also generate a mass for the U(1) gauge field. 
           On the other hand, if $eg_1<0$, we find ${\cal{V}}_D=\frac{1}{2}M^4$ when 
           $\langle 0 | \phi | 0 \rangle =0$, which is a U(1)-preserving and SUSY-breaking 
           solution. In fact, we then have 
           \be 
           {\cal{L}}_D = -\frac{1}{2}M^4 - |eg_1| M^2 \phi^\dagger \phi + \ldots 
           \ee
           showing that the $\phi$ field has a mass $M(|eg_1|)^{1/2}$. The gaugino field 
           $\lambda$ and the gauge field $A^\mu$ remain massless, and $\lambda$ can be 
           interpreted as a Goldstino. 
           
           This mechanism can't be used in the non-Abelian case, because no term of the 
           form $M^2 D^\alpha$ can be gauge-invariant. Could we have D-term breaking 
           in the U(1)$_y$ sector of the MSSM? Unfortunately not. What we want is a situation 
           in which the scalar fields in (\ref{eq:VD}) do not acquire  vev's (for example, 
           because they have large mass terms in the superpotential), so that the minimum of 
           (\ref{eq:VD}) forces $D$ to have a non-zero (vacuum) value, thus breaking SUSY.  
           In the MSSM, however,  the squark and slepton fields 
           have no superpotential mass terms, and so wouldn't be prevented from acquiring 
           vev's {\em en route} to minimizing (\ref{eq:VD}).  
             But this would imply the breaking of any symmetry associated with 
           quantum numbers carried by these fields, for example colour, which is not acceptable.   
                    
            One common viewpoint seems to be that SUSY breaking could occur in a sector that is 
            weakly coupled to the chiral supermultiplets of the MSSM. For example, it could 
            be (a) via gravitational interactions (presumably at the Planck scale, so that 
            SUSY-breaking mass terms would enter as (the vev of an F- or D-type parameter 
            having dimension $M^2$)/$M_{\rm P}$, which gives $\sqrt({\rm vev}) \sim 10^{10} $ GeV, say), 
            or (b) via electroweak gauge interactions. These possibilities are  discussed 
            in \cite{martin} section 6. A more recent review, with additional SUSY-breaking 
            mechanisms, is contained in \cite{CEKKLW} section 3.

            \subsection{Soft SUSY-breaking Terms} 
            
            In any case, however the necessary breaking of SUSY is effected, we can always 
            look for a parametrization of the SUSY-breaking terms which should be present 
            at `low' energies, and do phenomenology with them. It is a vital point that such 
            phenomenological SUSY-breaking terms in the (now effective) Lagrangian should 
            be `soft', as the jargon has it - that is, they should have positive mass 
            dimension, for example `$M^2 \phi^2$', `$M \phi^3$', `$M \chi \cdot \chi$', etc. 
             The reason 
            is that such terms (which are super-renormalizable) will not introduce new divergences 
            into, for example, the relations between the dimensionless coupling constants which 
            follow from SUSY, and which guarantee the stability of the mass hierarchy, 
            which was one of the prime motivations for SUSY in the first place. As we saw     
            section 1.1, a typical one-loop radiative correction to a scalar mass$^2$ is 
            \be 
            \delta m^2 \sim (\lambda_{\rm scalar} - g_{\rm fermion}^2) \Lambda^2 \label{eq:delmsq}
            \ee
            where $\Lambda$ is the u-v cutoff. In SUSY we essentially have $\lambda_{\rm scalar} 
            = g_{\rm fermion}^2$, and the dependence on $\Lambda$ becomes safely logarithmic. 
            Suppose, on the other hand, that  the dimensionless couplings $\lambda_{\rm scalar}$ or 
            $g_{\rm fermion}$ themselves received  divergent one-loop corrections,
            arising from renormalizable (rather than super-renormalizable) SUSY-breaking 
            interactions.\footnote{One example of such a renormalizable SUSY-breaking 
            interaction would be the Standard Model Yukawa interaction that generates 
            mass for `up' fermions and which involves the charge-conjugate of the Higgs 
            doublet that generates mass for the `down' fermions (see footnote 11). The argument 
            being given here implies that we do {\em not} want to generate `up' masses 
            this way, but rather via a second, independent, Higgs field.}  
             Then $\lambda_{\rm scalar}$ and $g_{\rm fermion}$ would 
            differ by terms of order $\ln \Lambda$, 
            with the result that the mass shift (\ref{eq:delmsq}) becomes very large indeed, once more. 
            In general, soft SUSY-breaking terms maintain the cancellations of quadratically 
            divergent radiative corrections to scalar mass$^2$, to all orders in perturbation theory
             \cite{gris}. 
            This means that corrections $\delta m^2$ go 
            like $m_{\rm soft}^2 \ln (\Lambda/m_{\rm soft})$, where $m_{\rm soft}$ is the typical 
            mass scale of the soft SUSY-breaking terms. This is a stable shift in the sense 
            of the hierarchy problem, provided of course that (as remarked in section 1.1) the 
            new mass scale $m_{\rm soft}$ is not much greater than 1 TeV, say. The origin of 
            this mass scale remains unexplained. 
            
            The forms of possible soft SUSY breaking terms are quite limited. They are as follows.
            
            (a) Gaugino masses for each gauge group:
            \be 
            - \frac{1}{2}(M_3 {\tilde{g}}^\alpha \cdot {\tilde{g}}^\alpha + M_2 {\tilde{W}}^\alpha 
            \cdot {\tilde{W}}^\alpha + M_1 {\tilde{B}}\cdot {\tilde{B}} + {\rm h.c.}) \label{eq:ginomass}
            \ee
            where  in the first (gluino) term $\alpha$ runs from 1 to 8 and in the second (wino) 
            term it runs from 1 to 3, the dot here signifying the Lorentz invariant spinor product.
            
            (b) Squark (mass)$^2$ terms:
           \be 
           -m^2_{{\tilde{\rm Q}}ij} {\tilde{Q}}^\dagger_i \cdot {\tilde{Q}}_j - 
           m^2_{{\tilde{\bar{\rm u}}}ij} {\tilde{\bar{u}}}^\dagger_i {\tilde{\bar{u}}}_j - 
           m^2_{{\tilde{\bar{\rm d}}}ij} {\tilde{\bar{d}}}^\dagger_i {\tilde{\bar{d}}}_j, \label{eq:sqmass}
           \ee
            where $i$ and $j$ are family labels, and the first term is an SU(2)$_{\rm L}$-invariant 
            dot product of scalar doublets partnering L-type quark doublets (see footnote 13 and 
            the `In parenthesis' paragraph following that footnote). 
            
            (c) Slepton (mass)$^2$ terms: 
            \be 
            -m^2_{{\tilde{\rm L}}ij} {\tilde{L}}^\dagger_i \cdot {\tilde{L}}_j - 
            m^2_{{\tilde{\bar{\rm e}}}ij} {\tilde{\bar{e}}}^\dagger_i {\tilde{\bar{e}}}_j. \label{eq:slepmass}
            \ee

            (d) Higgs (mass)$^2$ terms:
            \be 
            -m^2_{{\rm H}_{\rm u}} H_{\rm u}^\dagger \cdot H_{\rm u} - m^2_{{\rm H}_{\rm d}} 
            H^\dagger_{\rm d} \cdot H_{\rm d} - (b H_{\rm u} \cdot H_{\rm d} + {\rm h.c.}) \label{eq:higgsmass} 
            \ee
            where the SU(2)$_{\rm L}$ invariant dot products are 
            \be 
            H^\dagger_{\rm u} \cdot H_{\rm u} = |H^{+}_{\rm u}|^2 + | H^0_{\rm u}|^2 
            \ee
            and similarly for $H^\dagger_{\rm d} \cdot H_{\rm d}$, while 
            \be 
            H_{\rm u} \cdot H_{\rm d} = H^+_{\rm u} H^-_{\rm d} - H^0_{\rm u} H^0_{\rm d}.
            \ee
            
            (e) Triple scalar couplings 
            \be 
            -a_{\rm u}^{ij} {\tilde{\bar{u}}}_i {\tilde{Q}}_j \cdot H_{\rm u} 
            + a_{\rm d}^{ij} {\tilde{\bar{d}}}_i {\tilde{Q}}_j \cdot H_{\rm d} + 
            a_{\rm e}^{ij} {\tilde{\bar{e}}}_i {\tilde{L}}_j \cdot H_{\rm d} + {\rm h.c.} \label{eq:triplesc} 
            \ee

            The five (mass)$^2$ matrices are in general complex, but must be Hermitian so that the 
            Lagrangian is real. All the terms (\ref{eq:ginomass})-(\ref{eq:triplesc}) manifestly 
            break SUSY: the mass terms only involve part of the relevant supermultiplets, and the 
            triple scalar couplings involve three `$\phi$' or `$\phi^\dagger$' fields, rather than 
            (c.f. (\ref{eq:cubic})) the combinations `$\phi^2 \phi^\dagger$' or `$\phi^{2\dagger} \phi$'. 
            
            On the other hand, it is important to emphasize that the terms (\ref{eq:ginomass}) - 
            (\ref{eq:triplesc}) do respect the SM gauge symmetries. 
            The $b$ term in (\ref{eq:higgsmass}) and the triple 
            scalar couplings in (\ref{eq:triplesc}) have the same form as the `$\mu$' and `Yukawa' 
            couplings in the (gauge-invariant) superpotential (\ref{eq:WMSSM}), but here involving 
            just the scalar fields, of course. It is particularly noteworthy that gauge-invariant 
            mass terms are possible for all these superpartners, in marked contrast to the situation for 
            the known SM particles. Consider (\ref{eq:ginomass}) for instance.   
            The gluinos are in the regular representation of a 
         gauge group, like their gauge boson superpartners: for example, in SU(2) the winos  are 
         in the $t=1$ (`vector') representation. In this representation, the transformation 
         matrices can be chosen to be real (the generators are pure imaginary, $(T^{(1)}_i)_{jk}
         =-{\rm i} \epsilon_{ijk}$), which means that they are orthogonal rather than unitary, 
         just like rotation matrices in ordinary 3-D space. Thus quantities of the form 
         `${\widetilde{{\bm W}}} \cdot {\widetilde{{\bm W}}}$' are invariant under 
         SU(2) transformations, 
         including local ones since no derivatives are involved; similarly for the gluinos 
         and the bino. Coming to  (\ref{eq:sqmass}) and (\ref{eq:slepmass}), squark and slepton mass 
         terms of this form are allowed if $i$ and 
         $j$ are family indices, and the $m^2_{ij}$'s are Hermitian matrices in family space, 
         since under a gauge transformation $\phi_i \to U \phi_i, \ 
         \phi_j \to U \phi_j$, where $U^\dagger U =1$, and the $\phi$'s stand for a squark or 
         slepton flavour multiplet. Higgs mass terms like $-m^2_{{\rm H}_{\rm u}} H^\dagger_{\rm u} 
         H_{\rm u}$ are of course present in the SM already, and  (as we saw in section 12 - 
         see the remarks following equation (\ref{eq:muhiggs}))   
         from the perspective of the MSSM we need to 
         include such SUSY-violating terms in order to have any chance of breaking electroweak 
         symmetry spontaneously (the parameter `$m^2_{{\rm H}_{\rm u}}$' can of course have either sign). 
          The $b$ term in (\ref{eq:higgsmass}) is  like the SUSY-invariant $\mu$ term of (\ref{eq:muterm}), 
         but it only involves the Higgs, not the Higgsinos, and is hence SUSY-breaking.  
         
         The upshot of these considerations is that  mass terms which break SUSY, but 
         preserve electroweak symmetry,  can be written down for all 
         the so-far unobserved  particles of the MSSM. 
         By contrast, of course,  similar  mass terms for the known particles of the SM 
          would all break electroweak symmetry explicitly, which is unacceptable 
          (non-renormalizability/unitarity violations):  the masses of the known 
          SM particles must all arise via spontaneous breaking of electroweak symmetry.  Thus it 
          could be argued that,  
          from the viewpoint of the MSSM, it is natural that the known particles have 
          been found, since they are `light', with a scale associated with electroweak 
          symmetry breaking. The masses  of the undiscovered particles, on the 
          other hand,  are  associated with 
          SUSY breaking, which can be significantly higher.\footnote{The   
          Higgs is an interesting special case (taking it to be unobserved as yet). In the SM 
          its mass is arbitrary (though see footnote 1), but in the MSSM the lightest Higgs 
          particle is predicted to be no heavier than about 140 GeV (see 
          the following section).} As against this, it must be repeated that 
          electroweak symmetry breaking is not possible while preserving SUSY: the Yukawa-like 
          terms in (\ref{eq:WMSSM}) do respect SUSY, but will not generate fermion masses unless some Higgs 
          fields have a non-zero vev, and this won't happen with a potential of the form 
          (\ref{eq:VDF}) (see also (\ref{eq:muhiggs})); similarly, the gauge-invariant couplings (\ref{eq:covDphi}) 
          are part of a SUSY-invariant theory,  but the electroweak gauge boson masses 
          require a Higgs vev in (\ref{eq:covDphi}).  
          So some, at least, of the SUSY-breaking parameters must 
          have values not too far from the scale of electroweak symmetry breaking, if we 
          don't want fine tuning. From this point of view, then,   there seems no very clear 
           distinction between the scales of electroweak and of SUSY breaking. 
          
          Unfortunately, although the terms (\ref{eq:ginomass}) - (\ref{eq:triplesc}) are restricted in 
          form, there are nevertheless quite a lot of possible terms in total, when all 
          the fields in the MSSM are considered, and this implies very many new parameters. 
          In fact, Dimopoulos and Sutter \cite{dim} counted a total of 105 
          parameters describing masses, mixing angles 
          and phases, after allowing for all allowed redefinitions of bases. It is worth 
          emphasizing that this massive increase in parameters is entirely to do with SUSY breaking,   
          the SUSY-invariant (but unphysical) MSSM Lagrangian having only one new parameter ($\mu$) 
          with respect to the SM.

          One may well be dismayed by such an apparently huge arbitrariness in the theory, but this impression 
          is in a sense misleading since extensive regions of parameter space are in fact excluded 
          phenemenologically. This is because generic values of most of the new parameters allow 
          flavour changing neutral current (FCNC) processes, or new sources of CP violation, at levels 
          which are excluded by experiment. For example, if the matrix ${\bf m}^2_{{\bar{\rm e}}}$ in 
          (\ref{eq:slepmass}) has a non-suppressed off-diagonal term such as\footnote{Perhaps more 
          physically, ${\tilde{\bar{\rm e}}}_{\rm L}$ is the slepton partner of the ${\rm e}_{\rm R}$, 
          and ${\tilde{\bar{\mu}}}_{\rm L}$ that of $\mu_{\rm R}$.} 
          \be 
          (m^2_{{\bar{\rm e}}})_{{\rm e} \mu} {\tilde{\bar{e}}}^\dagger_{\rm L} {\tilde{\bar{\mu}}}_{\rm L}
          \label{eq:emu}
          \ee
          (in the basis in which the lepton masses are diagonal), then unacceptably large lepton 
          flavour changing ($\mu \to {\rm e}$) will be generated. We can, for instance, 
          envisage a loop diagram contributing to $\mu \to {\rm e} + \gamma$, in which the $\mu$ 
          first decays virtually to ${\tilde{\bar{\mu}}}_{\rm L}$ + bino through one of the 
          couplings in (\ref{eq:Lgch}), the ${\tilde{\bar{\mu}}}_{\rm L}$ then changing to 
          ${\tilde{\bar{\rm e}}}_{\rm L}$ via (\ref{eq:emu}), followed by  ${\tilde{\bar{\rm e}}}_{\rm L}$ 
          re-combining with the bino to make an electron, after emitting a photon. The upper limit 
          on the branching ratio for $\mu \to {\rm e} + \gamma$ is $1.2 \times 10^{-11}$, and our loop 
          amplitude will be many orders of magnitude larger than this, even for sleptons as heavy 
          as 1 TeV. Similarly, the squark (mass)$^2$ matrices are tightly constrained both as to 
          flavour mixing and as to CP-violating complex phases by data on ${\rm K}^0 - {\bar{\rm K}}^0$ 
          mixing, ${\rm D}^0 - {\bar{\rm D}}^0$ and ${\rm B}^0 - {\bar{\rm B}}^0$ 
          mixing, and the decay b $\to {\rm s} \gamma$. For a recent survey, with further references, 
          see \cite{CEKKLW} section 5.  
          
          The existence of these strong constraints on the SUSY-breaking parameters at the SM scale 
          suggests that whatever the actual SUSY-breaking mechanism might be, it should be such as to 
          lead naturally to the suppression of such dangerous off-diagonal terms. One framework 
          which guarantees this is the `minimal supergravity (mSUGRA)' theory 
          \cite{nilles2} \cite{CAN} \cite{BFS}, in which the parameters 
          (\ref{eq:ginomass}) - ({\ref{eq:triplesc}) take a particularly simple form at the Planck 
          scale: 
          \be 
          M_3=M_2=M_1=m_{1/2};\label{eq:ginoun}
          \ee
          \be 
          {\bf m}^2_{{\tilde{\rm Q}}}={\bf m}^2_{{\tilde{\bar{\rm u}}}}={\bf m}^2_{{\tilde{\bar{\rm d}}}} = 
          {\bf m}^2_{\tilde{\rm L}}= {\bf m}^2_{{\tilde{\bar{\rm e}}}}= m^2_0 \, {\bf 1}, \label{eq:sqslun}
          \ee
          where `${\bf 1}$' stands for the unit matrix in family space;
          \be
          m^2_{{\rm H}_{\rm u}} = m^2_{{\rm H}_{\rm d}}= m^2_0;\label{eq:higgsun}
          \ee
          and 
          \be 
          {\bf a}_{\rm u} = A_0 {\bf y}_{\rm u}, \ {\bf a}_{\rm d}=A_0 {\bf y}_{\rm d}, \ {\bf a}_{\rm e} 
          = A_0 {\bf y}_{\rm e} \label{eq:yukun}
          \ee
          where the ${\bf y}$ matrices are those appearing in (\ref{eq:WMSSM}). Relations (\ref{eq:sqslun}) 
          imply that at $m_{\rm P}$ all squark and sleptons are degenerate in mass (independent of 
          both flavour and family, in fact) and so, in particular, squarks and sleptons with the same 
          electroweak quantum numbers can be freely transformed into each other by  unitary 
          transformations. All mixings can then be eliminated, apart from that originating via the 
          triple scalar terms. But conditions (\ref{eq:yukun}) ensure that only the squarks 
          and sleptons of the (more massive) third family can have large triple scalar couplings. If 
          $m_{1/2}, A_0$ and $b$ of (\ref{eq:higgsmass}) all have the same complex phase, the only CP-violating 
          phase in the theory will be the usual CKM one (leaving CP-violation in the 
          neutrino sector aside). Somewhat weaker conditions than (\ref{eq:ginoun}) - (\ref{eq:yukun}) would 
          also suffice to accommodate the phenomenological constraints. (For completeness, we mention 
          other SUSY-breaking mechanisms that have been proposed: gauge-mediated \cite{GMSB}, 
          gaugino-mediated \cite{gaMSB} and anomaly-mediated \cite{AMSB} symmetry breaking.)

          We must now remember, of course, that if we use this kind of effective Lagrangian to 
          calculate quantities at the electroweak scale, in perturbation theory, the results 
          will involve logarithms of the form\footnote{Expression (\ref{eq:highlow}) may be thought of 
          in the context either of running the quantities `down' in scale - i.e. from a supposedly 
          `fundamental' high scale $Q_0 \sim m_{\rm P}$ to a low scale $\sim m_{\rm Z}$; or - 
          as in (\ref{eq:run3}) - (\ref{eq:run1}) - of running `up' from a low scale $Q_0 \sim 
          m_{\rm Z}$ to a high scale $\sim m_{\rm P}$ (in order, perhaps, to try and infer high-scale 
          physics from weak-scale input). Either way, a crucial hypothesis is, of course, that 
          no new physics intervenes between $\sim m_{\rm Z}$ and $\sim m_{\rm P}$.}   
          \be 
          \ln[({\mbox{high scale}}, {\mbox{for example }} m_{\rm P})  
          / {\mbox{low scale }} m_{\rm Z}], \label{eq:highlow}
          \ee
          coming from loop diagrams, which can be large enough to invalidate 
          perturbation theory. As usual, such `large logarithms' must be re-summed by the 
          renormalization group technique (see chapter 15 of \cite{AH32} for example). This amounts to treating 
          all couplings and masses as running parameters, which evolve as the energy scale changes according to 
          RG equations, whose coefficients can be calculated perturbatively. Conditions such as 
          (\ref{eq:ginoun}) - (\ref{eq:yukun}) are then interpreted as boundary conditions on the 
          parameters at the high scale. 
          
          This implies that after evolution to the SM scale the relations (\ref{eq:ginoun}) - 
          (\ref{eq:yukun}) will no longer hold, in general. However, RG corrections due to 
          gauge interactions will not introduce flavour-mixing or CP-violating phases, while 
          RG corrections due to Yukawa interactions are quite small except for the third family. It 
          seems to be generally the case that if FCNC and CP-violating terms are suppressed at a 
          high $Q_0$, then supersymmetric contributions to FCNC and CP-violating observables are 
          not in conflict with present bounds, though this may change  as the bounds are 
          improved.

          \subsection{RGE Evolution of the Parameters in the (Softly Broken) MSSM} 
          
          It is fair to say that the apparently successful gauge unification 
          in the MSSM (section 13) encourages us to apply a similar RG analysis to the other 
          MSSM couplings and  to the soft parameters (\ref{eq:ginoun}) - (\ref{eq:yukun}). One-loop 
          RGEs for the   MSSM  are given in \cite{CEKKLW} Appendix C.6; see also \cite{martin} section 7.1.  
          
          A simple example is provided by the gaugino mass parameters $M_i$ ($i=1,2,3$) whose evolution 
          (at 1-loop order) is determined by an equation very similar to (\ref{eq:arun}) for the running 
          of the $\alpha_i$, namely 
          \be 
          \frac{{\rm d} M_i}{{\rm d} t}=- \frac{b_i}{2 \pi} \alpha_i M_i. \label{eq:Mrun}
          \ee
          From (\ref{eq:arun}) and (\ref{eq:Mrun})  we obtain 
          \be 
          \frac{1}{\alpha_i} \frac{{\rm d}M_i}{{\rm d} t} - M_i \frac{1}{\alpha_i^2} \frac{{\rm d}\alpha_i} 
          {{\rm d} t} =0,
          \ee
          and hence 
          \be 
          \frac{{\rm d}}{{\rm d} t} (M_i/\alpha_i)=0.
          \ee
          It follows that the three ratios ($M_i/\alpha_i$) are RG-scale independent at 1-loop order. 
          In mSUGRA-type models, then, we can write 
          \be 
          \frac{M_i(Q)}{\alpha_i(Q)}=\frac{m_{1/2}}{\alpha_i(m_{\rm P})},\label{eq:gauguni1}
          \ee
          and since all the $\alpha_i$'s are already unified below $m_{\rm P}$ we obtain 
          \be 
          \frac{M_1(Q)}{\alpha_1(Q)}=\frac{M_2(Q)}{\alpha_2(Q)}=\frac{M_3(Q)}{\alpha_3(Q)} \label{eq:Mrats}
          \ee
          at any scale $Q$, up to small 2-loop corrections and possible threshold effects at high scales. 
          
          Applying (\ref{eq:Mrats}) at $Q=m_{\rm Z}$ we find  
          \be 
          M_1(m_{\rm Z})= \frac{\alpha_1(m_{\rm Z})}{\alpha_2(m_{\rm Z})} M_2(m_{\rm Z}) = 
          \frac{5}{3} \tan^2 \theta_{\rm W}(m_{\rm Z}) \simeq 0.5 M_2(m_{\rm Z}) \label{eq:M1M2}
          \ee
          and 
          \be 
          M_3(m_{\rm Z})= \frac{\alpha_3(m_{\rm Z})}{\alpha_2(m_{\rm z})} M_2(m_{\rm Z}) = 
          \frac{\sin^2 \theta_{\rm W}(m_{\rm Z})}{\alpha_{\rm em}(m_{\rm Z})} \alpha_3(m_{\rm Z}) 
          M_2(m_{\rm Z}) \simeq 3.5 M_2(m_{\rm Z}) \label{eq:M2M3}
          \ee
          where we have used (\ref{eq:sthexp}) - (\ref{eq:aemexp}). Equations (\ref{eq:M1M2}) and 
          (\ref{eq:M2M3})  may be summarized as 
          \be 
          M_3(m_{\rm Z}) : M_2(m_{\rm Z}) : M_1(m_{\rm Z}) \simeq 7:2:1.
          \ee   
          This simple prediction is common to most supersymmetric phenomenology. It implies that the 
          gluino is expected to be heavier than the states associated with the electroweak sector. 
          (The latter are `neutralinos', which are mixtures of the neutral Higgsinos 
          (${{\tilde{H}}}^0_{\rm u},{{\tilde{H}}}^0_{\rm d}$) and neutral gauginos 
          (${\tilde{B}}, {\tilde{W}}^0$), and `charginos', which are mixtures of the 
          charged Higgsinos (${\tilde{H}}^+_{\rm u}, {\tilde{H}}^-_{\rm d}$) and winos 
          (${\tilde{W}}^+, {\tilde{W}}^-$).)

          A second significant example concerns the running of the scalar masses. Here the gauginos 
          contribute to the RHS of `${\rm d} m^2/{\rm d} t$'  
           with a negative coefficient, which tends to increase the mass as the scale $Q$ is lowered. 
          On the other hand, the contributions from fermion loops have the opposite sign, tending to 
          decrease the mass at low scales. The dominant such contribution is provided by 
          top quark loops since $y_{\rm t}$ is so much larger than the other Yukawa couplings. 
          If we retain only the top quark Yukawa coupling, the 1-loop evolution equations 
          for $m^2_{{\rm H}_{\rm u}}$, $m^2_{{\tilde{\rm Q}}_3}$ and $m^2_{{\tilde{\bar{\rm u}}}_3}$ are 
          \be 
          \frac{{\rm d} m^2_{{\rm H}_{\rm u}}}{{\rm d} t} = [\frac{3 X_{\rm t}}{4 \pi} - 6 \alpha_2 M_2^2 
          - \frac{6}{5}\alpha_1 M_1^2]/4 \pi \label{eq:mhurun}
          \ee
          \be 
          \frac{{\rm d} m^2_{{\tilde{\rm Q}}_3}}{{\rm d} t} = [\frac{X_{\rm t}}{4 \pi} - \frac{32}{3} \alpha_3 
          M_3^2 -6 \alpha_2 M_2^2 - \frac{2}{15}\alpha_1 M_1^2]/4 \pi \label{eq:mQrun}
          \ee
          \be
          \frac{{\rm d} m^2_{{\tilde{\bar{\rm u}}}_3}}{{\rm d} t} = [ \frac{2 X_{\rm t}}{4 \pi} - 
          \frac{32}{3} \alpha_3 M_3^2 - \frac{32}{15} \alpha_1 M_1^2]/4 \pi,
          \label{eq:mubarrun}
          \ee
          where 
          \be 
          X_{\rm t} = 2 |y_{\rm t}|^2(m^2_{{\rm H}_{\rm u}}+m^2_{{\tilde{\rm Q}}_3} + 
          m^2_{{\tilde{\bar{\rm u}}}_3} + A^2_0]
          \ee
          and we have used (\ref{eq:yukun}). In contrast, the corresponding equation for 
          $m^2_{{\rm H}_{\rm d}}$, to which the top quark loop does not contribute, is 
          \be 
          \frac{{\rm d} m^2_{{\rm H}_{\rm d}}}{{\rm d}t}= [ -6\alpha_2 M^2_2 - \frac{6}{5} \alpha_1 
          M_1^2]/4 \pi. \label{eq:mhdrun}
          \ee
          Since the quantity $X_{\rm t}$ is  positive, its effect is always to decrease the 
          appropriate (mass)$^2$ parameter at low scales. From (\ref{eq:mhurun}) - (\ref{eq:mubarrun}) we can 
          see that, of the three masses, $m^2_{{\rm H}_{\rm u}}$ is (a) decreased the most because 
          of the factor 3, and (b) increased the least because the gluino contribution (which is 
          larger than those of the other gauginos) is absent. On the other hand, $m^2_{{\rm H}_{\rm d}}$ 
          will always tend to increase at low scales. The possibility then arises that $m^2_{{\rm H}_{\rm u}}$ 
          could run from a positive value at $Q_0 \sim 10^{16}$ GeV to a negative value at the 
          electroweak scale, while all the other (mass)$^2$ parameters of the scalar particles remain 
          positive\footnote{Negative values for the squark (mass)$^2$ parameters would have the 
          undesirable consequence of spontaneously breaking the colour SU(3).}. 
          This can indeed happen, thanks to the large value of the top quark mass (or equivalently 
          the large value of $y_{\rm t}$): see \cite{inoue2} \cite{ib} \cite{ib2} \cite{ellis3} \cite{AG} \cite{kane2}.  
           Such a negative (mass)$^2$ value would tend to destabilize the point $H^0_{\rm u} =0$, 
          providing a strong (though not  conclusive - see the next section) indication that this 
          is the trigger for electroweak symmetry breaking. 
          
          The parameter $y_{\rm t}$ in (\ref{eq:mhurun})-(\ref{eq:mubarrun}), and the other 
          Yukawa couplings in (\ref{eq:WMSSM}), all run too; consideration of the RGEs for these 
          couplings provides some further interesting results. If (for simplicity) we make the 
          approximations that only third-family couplings are significant, and ignore 
          contributions from $\alpha_1$ and $\alpha_2$, the 1-loop RGEs for the parameters 
          $y_{\rm t}$, $y_{\rm b}$ and $y_\tau$ are 
          \bea
          \frac{{\rm d}y_{\rm t}}{{\rm d} t} &=&
          \frac{y_{\rm t}}{4 \pi} [(6y^2_{\rm t} + y^2_{\rm b})/4\pi - \frac{16}{3} \alpha_{\rm s}]
          \label{eq:ytrun} \\
          \frac{{\rm d}y_{\rm b}}{{\rm d} t} &=& 
          \frac{y_{\rm b}}{4 \pi} [(6y^2_{\rm b}+y^2_{\rm t}+y^2_{\tau})/4 \pi - 
          \frac{16}{3} \alpha_{\rm s}] \label{eq:ybrun}\\
          \frac{{\rm d} y_{\tau}}{{\rm d}t} &=& 
          \frac{y_{\tau}}{16 \pi^2} [4y^2_{\tau}+3y^2_{\rm b}].\label{eq:ytaurun}
          \eea 
          As in equations (\ref{eq:mhurun}) - (\ref{eq:mubarrun}) the Yukawa couplings and the 
          gauge coupling $\alpha_{\rm s}$ enter the RHS of (\ref{eq:ytrun}) - (\ref{eq:ytaurun}) with 
          opposite signs; the former tend to increase the $y$'s at high scales, while $\alpha_{\rm s}$ 
          tends to reduce $y_{\rm t}$ and $y_{\rm b}$. It is then conceivable that, starting at low scales 
          with $y_{\rm t} > y_{\rm b} > y_{\tau}$, the three $y$'s might unify at or around $m_{\rm U}$. 
          Indeed, there is some evidence that the condition $y_{\rm b}(m_{\rm U})=y_{\tau}(m_{\rm U})$, 
          which arises naturally in many GUT models, leads to good low-energy phenomenology  
          \cite{arasony} \cite{bargery} \cite{carenay} \cite{langackery}.
          
          Further unification with $y_{\rm t}(m_{\rm U})$ must be such as to be consistent with 
          the known top quark mass at low scales. To get a rough idea of how this works, we return 
          to the relation (\ref{eq:muvy}), and similar ones for $m_{{\rm d}ij}$ and $m_{{\rm e}ij}$, 
          which in the mass-diagonal basis give 
          \be 
          y_{\rm t} = \frac{m_{\rm t}}{v_{\rm u}}, \ \ \ y_{\rm b}= \frac{m_{\rm b}}{v_{\rm d}}, 
          \ \ \ y_{\tau} = \frac{m_\tau}{v_{{\rm d}}},
          \ee
          where $v_{\rm d}$ is the vev of the field $H^0_{\rm d}$. It is clear that the viability of 
          $y_{\rm t} \approx y_{\rm b}$ will depend on the value of the additional parameter 
          $v_{\rm u}/v_{\rm d}$ (denoted by $\tan \beta$ - see the following section). It seems that 
          `Yukawa unification' at $m_{\rm U}$ may work in the parameter regime where $\tan \beta 
          \approx m_{\rm t}/m_{\rm b}$ \cite{olechowski} \cite{ananth} \cite{dimopoulos} \cite{carena} 
          \cite{hall} \cite{hempfling} \cite{rattazi}.

         In the following  
          section we shall discuss the Higgs sector of the MSSM where - even without 
          assumptions such as (\ref{eq:ginoun}) - (\ref{eq:yukun}) - only a few parameters 
          enter, and one important prediction can be made: namely, an upper bound 
          on the mass of the lightest Higgs boson, which is well within reach of the LHC.

          \section{The Higgs Sector and Electroweak Symmetry Breaking in the MSSM}
          
          \subsection{The scalar potential and the conditions for electroweak symmetry breaking}  
          
          We largely follow the treatment in Martin \cite{martin} section 7.2. The first task is 
          to find the potential for the scalar Higgs fields in the MSSM. As 
          frequently emphasized, there are two complex Higgs SU(2)$_{\rm L}$ doublets which we 
          are denoting by $H_{\rm u}=(H^+_{\rm u}, H^0_{\rm u})$ which has weak hypercharge 
          $y=1$, and $H_{\rm d}=(H^0_{\rm d}, H^-_{\rm d})$ which has $y=-1$. The classical 
          (tree-level) potential for these scalar fields is made up of several terms. First, 
          quadratic terms arise from the SUSY-invariant (`F-term') contribution 
           (\ref{eq:muhiggs}) which involves  
          the $\mu$ parameter from (\ref{eq:WMSSM}), and also from SUSY-breaking terms of the type 
          (\ref{eq:higgsmass}). The latter two contributions are
          \be 
          m^2_{{\rm H_{\rm u}}}(|H^+_{\rm u}|^2 +|H^0_{\rm u}|^2)+ 
          m^2_{{\rm H_{\rm d}}}(|H^0_{\rm d}|^2 +|H^-_{\rm d}|^2),
          \ee   
         where despite appearances it must be remembered that the arbitrary parameters 
         `$m^2_{{\rm H_{\rm u}}}$' and `$m^2_{{\rm H_{\rm d}}}$' may have either sign, and 
         \be 
         b(H^+_{\rm u}H^-_{\rm d}-H^0_{\rm u}H^0_{\rm d}) + {\rm h.c.}
         \ee
         To these must be added the quartic SUSY-invariant `D-terms' of (\ref{eq:VDF}), 
         of the form (Higgs)$^2$ (Higgs)$^2$, which we need 
         to evaluate for the electroweak sector  of the MSSM.

         There are two groups G, SU(2)$_{\rm L}$ with coupling $g$ and U(1)$_y$ with coupling $g'/2$ 
         (in the convention of \cite{AH32} - see equation (22.21) of that reference). 
         For the first, the matrices $T^\alpha$ are just $\tau^\alpha/2$, and we must evaluate 
         \bea
         &&\sum_\alpha (H^\dagger_{\rm u} (\tau^\alpha/2) H_{\rm u} + H^\dagger_{\rm d} 
         (\tau^\alpha/2) H_{\rm d})(H^\dagger_{\rm u} (\tau^\alpha/2) H_{\rm u} + 
         H^\dagger_{\rm d} (\tau^\alpha/2) H_{\rm d}) \nonumber \\
         &&=(H^\dagger_{\rm u} ({\bm \tau}/2) H_{\rm u}) \cdot (H^\dagger_{\rm u} 
         ({\bm \tau}/2)  H_{\rm u}) + (H^\dagger_{\rm d} ({\bm \tau}/2) H_{\rm d}) \cdot 
         (H^\dagger_{\rm d} ({\bm \tau}/2) H_{\rm d}) \nonumber \\
         &&+ 2 
         (H^\dagger_{\rm u} ({\bm \tau}/2) H_{\rm u}) \cdot (H^\dagger_{\rm d} ({\bm \tau}/2)  
         H_{\rm d}).\label{eq:Dud}
         \eea 
         If we write 
         \be 
         H_{\rm u} = \left( \begin{array}{c} a \\ b \end{array} \right), H_{\rm d}= \left( 
         \begin{array}{c} c \\ d \end{array} \right), 
         \ee
         then brute force evaluation of the matrix and dot products in (\ref{eq:Dud}) yields the result 
         \be 
         \frac{1}{4}\{[(|a|^2 +|b|^2)-(|c|^2 + |d|^2)]^2 +4(ac^*+bd^*)(a^*c+b^*d)\},\label{eq:Dud2}
         \ee
         so that the SU(2) contribution is (\ref{eq:Dud2}) multiplied by $g^2/2$. The U(1) 
         contribution is  
         \be 
         \frac{1}{2}(g^{\prime }/2)^2 [  H^\dagger_{\rm u} H_{\rm u} - 
         H^\dagger_{\rm d} H_{\rm d}]^2 = \frac{g^{\prime 2}}{8}[(|a|^2 + |b|^2)-(|c|^2+|d|^2)]^2.
         \label{eq:Dud1}
         \ee
         Re-writing (\ref{eq:Dud2}) and (\ref{eq:Dud1}) in the notation of the fields, and including the 
         quadratic pieces, the complete potential for the scalar fields in the MSSM is
         \bea 
         &&{\cal V}= (|\mu|^2 +m^2_{{\rm H}_{\rm u}})(|H^+_{\rm u}|^2+|H^0_{\rm u}|^2) +    
          (|\mu|^2 +m^2_{{\rm H}_{\rm d}})(|H^0_{\rm d}|^2+|H^-_{\rm d}|^2) + \nonumber \\
          && [b(H^+_{\rm u}H^-_{\rm d}-H^0_{\rm u}H^0_{\rm d}) + {\rm h.c.}] + 
          \frac{(g^2+g^{\prime 2})}{8}(|H^+_{\rm u}|^2 +|H^0_{\rm u}|^2-|H^0_{\rm d}|^2 
          -|H^-_{\rm d}|^2)^2 \nonumber \\
          &&+ \frac{g^2}{2}|H^+_{\rm u}H^{0\dagger}_{\rm d}+H^0_{\rm u}H^{-\dagger}_{\rm d}|^2.
          \label{eq:VscalH}
          \eea
          We prefer not to re-write $(|\mu|^2 +m^2_{{\rm H}_{\rm u}})$ and 
          $(|\mu|^2 +m^2_{{\rm H}_{\rm d}})$ as $m_1^2$ and $m^2_2$, say, so as to retain a 
          memory of the fact that $|\mu|^2$ arises from a SUSY-invariant term, and 
          is necessarily positive, 
          while $m^2_{{\rm H}_{\rm u}}$ and $m^2_{{\rm H}_{\rm d}}$ are SUSY-breaking and of 
          either sign {\em a priori}. 
          
          We must now investigate whether - and if so under what conditions - this potential can 
          have a minimum which (like that of the simple Higgs potential (\ref{eq:HiggsV}) of the SM) breaks 
          the SU(2)$_{\rm L}\times$U(1)$_y$ electroweak symmetry down to U(1)$_{\rm em}$.
          
          We can use the gauge symmetry to simplify the algebra somewhat. As in the SM 
          (see for example sections 17.6 and 19.6 of \cite{AH32}) we can reduce a possible vev 
          of one component of either $H_{\rm u}$ or $H_{\rm d}$ to zero by an SU(2)$_{\rm L}$ 
          transformation. We choose $H^+_{\rm u} =0$ at the minimum of ${\cal V}$. The conditions 
          $H^+_{\rm u}=0$ and $\partial {\cal V}/ \partial H^+_{\rm u}=0$ then imply that, at 
          the minimum of the potential,  either  
          \be
          H^-_{\rm d}=0 \label{eq:H-cond}
          \ee
          or 
          \be 
          b+\frac{g^2}{2}H^{0\dagger}_{\rm d} H^{0 \dagger}_{\rm u} = 0.\label{eq:bcond}
          \ee
          The second condition (\ref{eq:bcond}) implies that the $b$ term in (\ref{eq:VscalH}) becomes 
          \be
          g^2|H^0_{\rm u}|^2 |H^0_{\rm d}|^2
          \ee
          which is definitely positive, and unfavourable to symmetry-breaking. As we shall see, 
          condition (\ref{eq:H-cond}) leads to a negative $b$-contribution. Accepting  
          alternative  (\ref{eq:H-cond}) then, it follows that neither $H^+_{\rm u}$ nor 
          $H^-_{\rm d}$ acquire a vev, which means (satisfactorily) that electromagnetism is 
          not spontaneously broken. We can now ignore the charged components, and concentrate 
          on the potential for the neutral fields which is 
          \bea 
          &&{\cal V}_{\rm n}=(|\mu|^2+m^2_{{\rm H}_{\rm u}})|H^0_{\rm u}|^2 +   
          (|\mu|^2+m^2_{{\rm H}_{\rm d}})|H^0_{\rm d}|^2  \nonumber \\    
          &&-(bH^0_{\rm u}H^0_{\rm d} + {\rm h.c.}) + \frac{(g^2+g^{\prime 2})}{8}(|H^0_{\rm u}|^2 
          -|H^0_{\rm d}|^2)^2. \label{eq:Vneut}
          \eea
          
          This is perhaps an appropriate point to note that the coefficient of the quartic term 
          is {\em not} a free parameter, but is determined by the known electroweak couplings 
          ($(g^2+g^{\prime 2})/8 \approx 0.065$). This is of course in marked contrast to the 
          case of the SM, where the coefficient $\lambda/4$ in (\ref{eq:HiggsV}) is a free parameter. 
          Recalling from (\ref{eq:MH}) that, in the SM, the mass of the Higgs boson is proportional 
          to $\sqrt{\lambda}$, for given Higgs vev, this suggests that in the MSSM there should be 
          a relatively light Higgs particle. As we shall see, this is indeed the case, 
          though the larger field content of the Higgs sector in the MSSM makes the analysis 
          more involved.   
          
          Consider now the $b$-term in (\ref{eq:Vneut}), which is the only one that depends on 
          the phases of the fields. Without loss of generality, $b$ may be taken to be real and 
          positive, any possible phase of $b$ being absorbed into the relative phase of 
          $H^0_{\rm u}$ and $H^0_{\rm d}$. For a minimum of ${\cal V}_{\rm n}$, the product 
          $H^0_{\rm u}H^0_{\rm d}$ must be real and positive too, which means that (at the 
          minimum) the vev's of $H^0_{\rm u}$ and $H^0_{\rm d}$ must have equal and opposite 
          phases. Since these fields have equal and opposite hypercharges,  we can make a U(1)$_y$ 
          gauge transformation to reduce both their phases to zero. All vev's and couplings can 
          therefore be chosen to be real, which means that {\bf CP} is not spontaneously broken 
          by the 2-Higgs potential of the MSSM, any more than it is in the 1-Higgs potential of 
          the SM.\footnote{While this is true at tree level, {\bf CP} symmetry could be broken 
          significantly by radiative corrections, specifically via loops involving third generation 
          squarks \cite{carenacp}; this would imply that the three neutral Higgs eigenstates 
          would not have well defined {\bf CP} quantum numbers (for the usual, {\bf CP} conserving, 
          case, see comments following equation (\ref{eq:bA0}) below).} 
          
          The scalar potential now takes the more manageable form 
          \be 
          {\cal V}_{\rm n}=(|\mu|^2+m^2_{{\rm H}_{\rm u}})x^2 + (|\mu|^2+m^2_{{\rm H}_{\rm d}})y^2 - 
          2bxy + \frac{(g^2 +g^{\prime 2})}{8}(x^2-y^2)^2, \label{eq:Vxy}
          \ee
          where $x=|H^0_{\rm u}|, y=|H^0_{\rm d}|$; it depends on three parameters, 
          $|\mu|^2+m^2_{{\rm H}_{\rm u}}$, $|\mu|^2+m^2_{{\rm H}_{\rm d}}$ and $b$. 
          We want to identify the conditions required for 
          the stable minimum of ${\cal V}_{\rm n}$ to occur at non-zero values of $x$ and $y$. 
          First note that, along the special (`flat') direction $x=y$, the potential will be 
          unbounded from below (no minimum) unless 
          \be 
          2|\mu|^2 +m^2_{{\rm H}_{\rm u}} + m^2_{{\rm H}_{\rm d}} > 2b >0.\label{eq:flatcond}
          \ee
          Hence $(|\mu|^2 +m^2_{{\rm H}_{\rm u}})$ and $(|\mu|^2 +m^2_{{\rm H}_{\rm d}})$ cannot 
          both be negative. This implies, referring to (\ref{eq:Vxy}), that the point 
          $x=y=0$ cannot be a maximum of ${\cal V}_{\rm n}$. If $(|\mu|^2 +m^2_{{\rm H}_{\rm u}})$       
           and $(|\mu|^2 +m^2_{{\rm H}_{\rm d}})$ are both positive, then the origin is a minimum 
           (which would be an unwanted symmetry-preserving solution) unless 
           \be 
           (|\mu|^2 +m^2_{{\rm H}_{\rm u}})(|\mu|^2 +m^2_{{\rm H}_{\rm d}})  < b^2, \label{eq:spcond}
           \ee
           which is the condition for the origin to be a saddle point. (\ref{eq:spcond}) is 
           automatically satisfied if either $(|\mu|^2 +m^2_{{\rm H}_{\rm u}})$ 
            or $(|\mu|^2 +m^2_{{\rm H}_{\rm d}})$ is negative. 
            
            The $b$-term favours electroweak symmetry breaking, but it is not required to 
            be non-zero. What can be said about $m^2_{{\rm H}_{\rm u}}$ and $m^2_{{\rm H}_{\rm d}}$? 
            A glance at conditions (\ref{eq:flatcond}) and (\ref{eq:spcond}) shows 
            that they cannot both be satisfied if $m^2_{{\rm H}_{\rm u}}=m^2_{{\rm H}_{\rm d}}$, 
            a condition that is typically taken to hold at a high scale $\sim 10^{16}$ GeV. However, 
            the parameter $m^2_{{\rm H}_{\rm u}}$ is, in fact, the one whose renormalization group 
            evolution can drive it to negative values at the electroweak 
            scale, as discussed at the end of the previous section. It is clear that a negative value of 
            $m^2_{{\rm H}_{\rm u}}$ will tend to help condition (\ref{eq:spcond}) to be 
            satisfied, but it is neither necessary nor sufficient ($|\mu|$ may be too large or 
            $b$ too small). A `large' negative value for $m^2_{{\rm H}_{\rm u}}$ is a significant 
            factor, but it falls short of a demonstration that electroweak symmetry breaking {\em will} 
            occur via this mechanism. 
            
            Having established the conditions (\ref{eq:flatcond}) and (\ref{eq:spcond}) required for 
            $|H^0_{\rm u}|$ and $|H^0_{\rm d}|$ to have non-zero vevs, say $v_{\rm u}$ and 
            $v_{\rm d}$ respectively, we can now proceed to write down the equations determining 
            these vevs which follow from imposing the stationary conditions 
            \be 
            \frac{\partial {\cal V}_{\rm n}}{\partial x}= 
             \frac{\partial {\cal V}_{\rm n}}{\partial y}=0.\label{eq:vuvdcond}
             \ee
             Performing the differentiations and setting $x=v_{\rm u}$ and $y=v_{\rm d}$ we obtain 
             \bea 
            (|\mu|^2 + m^2_{{\rm H}_{\rm u}})v_{\rm u}&=& b v_{\rm d} + \frac{1}{4}(g^2+g^{\prime 2}) 
            (v^2_{\rm d}-v^2_{\rm u}) \label{eq:mincon11}\\
            (|\mu|^2 + m^2_{{\rm H}_{\rm d}})v_{\rm d}&=& b v_{\rm u} - \frac{1}{4}(g^2+g^{\prime 2}) 
            (v^2_{\rm d}-v^2_{\rm u}). \label{eq:mincon22} 
            \eea

             One  combination of $v_{\rm u}$ and 
             $v_{\rm d}$ is fixed by experiment,  since it determines the mass of the W 
             and Z bosons, just as in the SM. The relevant terms in the electroweak sector are 
             \be 
             (D_\mu H_{\rm u})^\dagger (D^\mu H_{\rm u}) +  (D_\mu H_{\rm d})^\dagger (D^\mu H_{\rm d})
             \label{eq:higgsKE}        
             \ee
             where (see equation (22.21) of \cite{AH32})
             \be 
             D_\mu=\partial_\mu +{\rm i}g ({\bm \tau}/2) \cdot {\bm W}_\mu +{\rm i}(g'/2)y B_\mu.
             \label{eq:higgscovder}
             \ee The mass terms for the vector particles come (in unitary gauge) from inserting 
             the vevs for $H_{\rm u}$ and $H_{\rm d}$, and defining 
             \be 
             Z^\mu = (-g'B^\mu +g W^\mu_3)/(g^2+g^{\prime 2})^{1/2}.
             \ee
             One finds 
             \bea 
             m_{\rm Z}^2 &=& \frac{1}{2}(g^2+g^{\prime 2})(v^2_{\rm u} + v^2_{\rm d}) \label{eq:MZsq} \\
             m_{\rm W}^2&=& \frac{1}{2}g^2(v^2_{\rm u} + v^2_{\rm d}). \label{eq:MWsq}
             \eea 
             Hence  (see equations (22.29)-(22.32) of \cite{AH32}) 
             \be 
             (v^2_{\rm u}+v^2_{\rm d})^{1/2} = \left( \frac{2 m_{\rm W}^2}{g^2} \right)^{1/2} = 
             174 {\rm GeV}.\label{eq:vuvdcon}
             \ee

            Equations (\ref{eq:mincon11}) and (\ref{eq:mincon22}) may now be written as 
            \bea
            (|\mu|^2 + m^2_{{\rm H}_{\rm u}})&=&b \cot \beta +(m^2_{\rm Z}/2) \cos 2 \beta 
            \label{eq:mincon1}\\
            (|\mu|^2 + m^2_{{\rm H}_{\rm d}})&=&b \tan \beta -(m^2_{\rm Z}/2) \cos 2 \beta,
            \label{eq:mincon2}
            \eea
            where  
             \be 
             \tan \beta \equiv v_{\rm u}/v_{\rm d}.\label{eq:tanbeta}
             \ee
             It is easy to check that (\ref{eq:mincon1}) and (\ref{eq:mincon2}) satisfy the 
             necessary conditions (\ref{eq:flatcond}) 
             and (\ref{eq:spcond}). We may use (\ref{eq:mincon1}) and (\ref{eq:mincon2}) to 
             eliminate the parameters $|\mu|$ and $b$ in favour of $\tan \beta$, but the phase 
             of $\mu$ is not determined. Because both $v_{\rm u}$ and $v_{\rm d}$ are real and 
             positive, the angle $\beta$ lies between 0 and $\pi/2$.

              We are now ready to calculate the mass spectrum.   
            
            \subsection{The tree-level masses of the scalar Higgs states in the MSSM}
            
            In the SM, there are four real scalar degrees of freedom in the Higgs doublet 
            (\ref{eq:higgsdoubSM}); after electroweak symmetry breaking (i.e. given a 
            non-zero Higgs vev), three of them become the longitudinal modes of the massive 
            vector bosons ${\rm W}^{\pm}$ and ${\rm Z}^0$, while the fourth is the neutral Higgs 
            boson of the SM, the mass of which is found by considering quadratic deviations away from 
            the symmetry-breaking minimum (see chapter 19 of \cite{AH32} for example). In the 
            MSSM, there are 8 real scalar degrees of freedom. Three of them are massless, and 
            just as in the SM they get `swallowed' by the ${\rm W}^{\pm}$ and ${\rm Z}^0$. The masses 
            of the other five are again calculated by expanding the potential about the minimum, 
            up to second order in the fields. Though straightforward, the work is complicated 
            by the fact that the quadratic deviations are not diagonal in the fields, so that 
            some diagonalization has to be done before the physical masses can be extracted.
            
            To illustrate the procedure, consider the Lagrangian 
            \be
            {\cal L}_{12}=\partial_\mu \phi_1 \partial^\mu \phi_1 +
            \partial_\mu \phi_2 \partial^\mu \phi_2 -V(\phi_1, \phi_2),\label{eq:L121}
            \ee
            where $V(\phi_1, \phi_2)$ has a minimum at $\phi_1=v_1, \phi_2=v_2$. We expand $V$ about 
            the minimum, retaining only quadratic terms, and discarding an irrelevant constant; this 
            yields 
            \bea
            &&{\cal L}_{12,{\rm quad}}=\partial_\mu \phi_1 \partial^\mu \phi_1 + \partial_\mu \phi_2 
            \partial^\mu \phi_2  -
            \frac{1}{2}\frac{\partial^2 V}{\partial \phi_1^2} (\phi_1-v_1)^2 \nonumber  \\
            && -\frac{1}{2}\frac{\partial^2 V}{\partial \phi_2^2} (\phi_2-v_2)^2 -
            \frac{\partial^2 V}{\partial \phi_1 \partial \phi_2} (\phi_1-v_1)(\phi_2-v_2) \label{eq:L12quad}
            \eea
            where the derivatives are evaluated at the minimum $(v_1, v_2)$. Defining 
            \be 
            {\tilde{\phi_1}}=\sqrt{2}(\phi_1-v_1), \ \ \ {\tilde{\phi_2}}=\sqrt{2}(\phi_2-v_2),
            \ee 
            (\ref{eq:L12quad}) can be written as 
            \be
            {\cal L}_{12, {\rm quad}}= \frac{1}{2}\partial_\mu {\tilde{\phi_1}}\partial^\mu{\tilde{\phi_1}}   
           + \frac{1}{2}\partial_\mu {\tilde{\phi_2}}\partial^\mu{\tilde{\phi_2}} -
           \frac{1}{2} ({\tilde{\phi_1}} \, {\tilde{\phi_2}}) {\bf M}^{\rm sq} \left( \ba{c} 
           {\tilde{\phi_1}}\\{\tilde{\phi_2}} \ea \right), \label{eq:L12qp} 
           \ee
           where the $({\rm mass})^2$ matrix ${\bf M}^{\rm sq}$ is given by
           \be
           {\bf M}^{\rm sq}=\frac{1}{2} \left( \ba{cc} V^{\prime \prime}_{11} & V^{\prime \prime}_{12} 
           \\ V^{\prime \prime}_{12} & V^{\prime \prime}_{22} \ea \right) 
           \ee
           where 
           \be 
           V^{\prime \prime}_{ij}=\frac{\partial^2 V}{\partial \phi_i \partial \phi_j}(v_1, v_2).
           \ee
           The matrix ${\bf M}^{\rm sq}$ is real and symmetric, and can be diagonalized via an orthogonal 
           transformation of the form 
           \be
           \left( \ba{c} \phi_+ \\\phi_- \ea \right) 
           = \left( \ba{cc} \cos \alpha & -\sin \alpha \\ \sin \alpha & \cos \alpha \ea \right) 
           \left( \ba{c} {\tilde{\phi_1}} \\ {\tilde{\phi_2}} \ea \right).
           \ee  
            If the eigenvalues of ${\bf M}^{\rm sq}$ are $m^2_+$ and $m^2_-$, we see that  in 
            the new basis (\ref{eq:L12qp}) becomes 
            \be 
            {\cal L}_{12, {\rm quad}}= \frac{1}{2}\partial_\mu \phi_+ 
            \partial^\mu \phi_+ +\frac{1}{2}\partial_\mu \phi_-  
            \partial^\mu \phi_- -\frac{1}{2}(\phi_+)^2 m^2_+ - \frac{1}{2} (\phi_-)^2 m^2_-,
            \label{eq:L12diag}
            \ee
            from which it follows (via the equations of motion for $\phi_+$ and $\phi_-$) that 
            $m^2_+$ and $m^2_-$ are the squared masses of the modes described by $\phi_+$ and $\phi_-$.
            
            We apply this formalism first to the pair of fields 
            $({\rm Im} H^0_{\rm u}, {\rm Im} H^0_{\rm d})$. The part of our scalar potential involving 
            this pair is 
            \bea
            &&{\cal V}_{\rm A}=(|\mu|^2 +m^2_{{\rm H}_{\rm u}})({\rm Im}H^0_{\rm u})^2 + 
            (|\mu|^2 +m^2_{{\rm H}_{\rm d}})({\rm Im}H^0_{\rm d})^2 + 
            2b({\rm Im}H^0_{\rm u})({\rm Im}H^0_{\rm d})\nonumber \\
            && + \frac{(g^2+g^{\prime 2})}{8} 
            [({\rm Re}H^0_{\rm u})^2 +({\rm Im}H^0_{\rm u})^2 -
            ({\rm Re}H^0_{\rm d})^2 -({\rm Im}H^0_{\rm d})^2]^2.
            \eea 
            Evaluating the second derivatives at the minimum point, we find the elements 
            of the $({\rm mass})^2$ matrix:
            \be 
            M_{11}^{\rm sq}= |\mu|^2 +m^2_{{\rm H}_{\rm u}}+ 
            \frac{(g^2+g^{\prime 2})}{4}(v^2_{\rm u}-v^2_{\rm d}) = b \cot \beta,
            \ee
            where we have used (\ref{eq:mincon11}), and similarly  
            \be 
            M_{12}^{\rm sq}=b, \ \ \ M_{22}^{\rm sq}=b \tan \beta.
            \ee
            The eigenvalues of ${\bf M}^{\rm sq}$ are easily found to be 
            \be 
            m^2_+=0, \ \ \ m^2_-=2b/\sin 2 \beta. 
            \ee
            The eigenmode corresponding to the massless state is 
            \be 
            \sqrt{2}[\sin \beta ({\rm Im}H^0_{\rm u}) - \cos \beta ({\rm Im}H^0_{\rm d})],\label{eq:physHZ}
            \ee
            and this will become the longitudinal state of the ${\rm Z}^0$. The orthogonal 
            combination 
            \be 
            \sqrt{2}[\cos \beta ({\rm Im} H^0_{\rm u}) + \sin \beta ({\rm Im} H^0_{\rm d})]\label{eq:physA0} 
            \ee
            is the field of a scalar particle `${\rm A}^0$', with mass
            \be  
            m_{{\rm A}^0}=(2b/\sin 2 \beta)^{1/2}.\label{eq:massA0}
            \ee
            In discussing the parameter space of the Higgs sector of the MSSM, the pair of parameters  
            ($b, \tan \beta$)  is usually replaced by the pair ($m_{{\rm A}^0}, \tan \beta$).  
            
            Next, consider the charged pair $(H^+_{\rm u}, H^{-\dagger}_{\rm d})$. In this case the 
            relevant part of the Lagrangian is 
            \bea 
            &&{\cal L}_{\rm ch,\, quad}=(\partial_\mu H^+_{\rm u})^\dagger (\partial^\mu H^+_{\rm u}) + 
             (\partial_\mu H^-_{\rm d})^\dagger (\partial^\mu H^-_{\rm d}) - 
             \frac{\partial^2 {\cal V}}{\partial H^{+\dagger}_{\rm u} \partial H^+_{\rm u}}
             H^{+\dagger}_{\rm u}H^+_{\rm u}  \nonumber \\
             &&- \frac{\partial^2 {\cal V}}{\partial H^{-\dagger}_{\rm d} \partial H^-_{\rm d}}
             H^{-\dagger}_{\rm d}H^-_{\rm d} - 
             \frac{\partial^2 {\cal V}}{\partial H^{+}_{\rm u} \partial H^-_{\rm d}}H^+_{\rm u}H^-_{\rm d} - 
             \frac{\partial^2 {\cal V}}{\partial H^{+\dagger}_{\rm u} \partial H^{-\dagger}_{\rm d}}
             H^{+\dagger}_{\rm u}H^{-\dagger}_{\rm d},\label{eq:Lchquad}
             \eea
             where we use (\ref{eq:VscalH}) for ${\cal V}$, and the derivatives are evaluated at 
             $H^0_{\rm u}=v_{\rm u}, \, H^0_{\rm d}=v_{\rm d}, \, H^+_{\rm u}= H^-_{\rm d}=0.$
             We write the potential terms as 
             \be 
             (H^{+\dagger}_{\rm u} \, H^-_{\rm d}) {\bf M}^{\rm sq}_{\rm ch} \left( \ba{c} 
             H^+_{\rm u} \\ H^{-\dagger}_{\rm d} \ea \right) 
             \ee
             where 
             \be 
             {\bf M}^{\rm sq}_{\rm ch}=\left( \ba{cc} M^{\rm sq}_{++} & M^{\rm sq}_{+-} 
             \\ M^{\rm sq}_{-+} & M^{\rm sq}_{--} \ea \right) 
             \ee
             with $M^{\rm sq}_{++}=\partial^2 {\cal V}/\partial H^{+\dagger}_{\rm u} \partial H^+_{\rm u}$ 
             etc. Performing the differentiations and evaluating the results at the minimum, we obtain 
             \be 
             {\bf M}^{\rm sq}_{\rm ch}= \left( \ba{cc} b \cot \beta + \frac{g^2}{2} v^2_{\rm d} & 
             b + \frac{g^2}{2}v_{\rm u} v_{\rm d} \\ b+ \frac{g^2}{2}v_{\rm u} v_{\rm d} & 
             b \tan \beta + \frac{g^2}{2} v^2_{\rm u} \ea \right).
             \ee
             This matrix has eigenvalues 0 and $m^2_{\rm W}+m^2_{{\rm A}^0}$. The massless state 
             corresponds to the superposition 
             \be 
             G^+=\sin \beta H^+_{\rm u} - \cos \beta H^{-\dagger}_{\rm d},
             \ee
             and it provides the longitudinal mode of the ${\rm W}^+$ boson. There is a similar state 
             $G^-\equiv(G^+)^\dagger$, which goes into the ${\rm W}^-$. The massive (orthogonal) state is 
             \be 
             H^+=\cos \beta H^+_{\rm u} + \sin \beta H^{-\dagger}_{\rm d},
             \ee
             which has mass $m_{{\rm H}^+}=(m^2_{{\rm W}} +m^2_{{\rm A}^0})^{1/2}$, 
             and there is a similar state $H^-\equiv (H^+)^\dagger$. Note that  after 
             diagonalization (\ref{eq:Lchquad}) becomes 
             \be
             (\partial_\mu G^+)^\dagger (\partial^\mu G^+) + (\partial H^+)^\dagger(\partial^\mu H^+) 
             - m^2_{{\rm H}^+} H^{+\dagger} H^+
             \ee
             and the equation of motion for $H^+$ shows that $m^2_{{\rm H}^+}$ is correctly identified 
             with the physical squared mass, without the various factors of 2 that appeared in 
             our example (\ref{eq:L12quad})-(\ref{eq:L12diag}) of two neutral fields.
             
             Finally, we consider the coupled pair $({\rm Re}H^0_{\rm u}-v_{\rm u}, {\rm Re}H^0_{\rm d} 
             -v_{\rm d})$, which is of the same type as our example, and as the pair 
             $({\rm Im} H^0_{\rm u}, {\rm Im} H^0_{\rm d})$. The $({\rm mass})^2$ matrix is 
             \be
             {\bf M}^{\rm sq}_{{\rm h, H}}= \left( \ba{cc} 
             b \cot \beta + m^2_{{\rm Z}} \sin^2 \beta & -b - \frac{1}{2} m^2_{\rm Z} \sin 2 \beta \\
             -b -\frac{1}{2} m^2_{\rm Z} \sin 2 \beta & b \tan \beta + m^2_{\rm Z} \cos^2 \beta \ea 
             \right) \label{eq:msqhH}
             \ee
             which has eigenvalues 
             \be 
             m^2_{{\rm h}^0}=\frac{1}{2}\{m^2_{{\rm A}^0} +m^2_{\rm Z} -[(m^2_{{\rm A}^0}+m^2_{\rm Z})^2 
             -4 m^2_{{\rm A}^0} m^2_{\rm Z} \cos^2 2 \beta]^{1/2} \} \label{eq:mh0}
             \ee
             and 
             \be 
             m^2_{{\rm H}^0}=\frac{1}{2}\{m^2_{{\rm A}^0} +m^2_{\rm Z} +[(m^2_{{\rm A}^0}+m^2_{\rm Z})^2 
             -4 m^2_{{\rm A}^0} m^2_{\rm Z} \cos^2 2 \beta]^{1/2} \}. \label{eq:mH0}
             \ee
             Equation (\ref{eq:mh0}) and (\ref{eq:mH0}) display the dependence of $m_{{\rm h}^0}$ and 
             $m_{{\rm H}^0}$ on the parameters $m_{{\rm A}^0}$ and  $\beta$. 
             The corresponding eigenmodes will be given in the following subsection. 
             
             The crucial point now is that, whereas the masses $m_{{\rm A}^0}, \, m_{{\rm H}^0}$ and 
             $m_{{\rm H}^{\pm}}$ are unconstrained (since they all grow as $b / \sin \beta$ which can in 
             principle be arbitrarily large), the mass $m_{{\rm h}^0}$ is bounded from above. Let us write 
             $x=m^2_{{\rm A}^0}, \, a=m^2_{\rm Z}$; then 
             \be 
             m^2_{{\rm h}^0}= \frac{1}{2}\{x+a-[(x+a)^2-4ax\cos^2 2 \beta]^{1/2}\}.
             \ee
             It is easy to verify that this function has no stationary point for finite values of $x$. 
             Further, for small $x$ we find 
             \be 
             m^2_{{\rm h}^0} \approx x \cos^2 2 \beta, 
             \ee 
             while for large $x$ 
             \be 
             m^2_{{\rm h}^0} \to a \cos^2 2 \beta -(a^2/4x) \sin^2 4 \beta.
             \ee 
             Hence the maximum value of $m^2_{{\rm h}^0}$, reached as $m^2_{{\rm A}^0} \to \infty$, is 
             $a \cos^2 2 \beta$, that is 
             \be 
             m_{{\rm h}^0} \leq m_{\rm Z} |\cos 2 \beta|. \label{eq:higgsbd}
             \ee
             Note that the RHS actually vanishes for $\beta = \pi/4$ i.e. for $\tan \beta = 1$.  
              
              This is the promised upper bound on the mass of one of the neutral Higgs bosons in 
              the MSSM, and it is surely a remarkable result \cite{inoue}, \cite{flores}. 
              The bound (\ref{eq:higgsbd}) 
              has, of course\footnote{Well, maybe not! Drees \cite{drees2} has recently suggested that 
              the 2.3 $\sigma$ excess of events around 98 GeV and the 1.7 $\sigma$ excess around 
              115 GeV reported by the four LEP experiments 
              \cite{LEP} might actually be the ${\rm h}^0$ and ${\rm H}^0$ respectively.},  
               already been exceeded by the current experimental lower bound \cite{LEP} 
              \be 
              m_{\rm H} \geq 114.4 {\mbox{ GeV (95\% c.l.)}}. \label{eq:higgsexp}
              \ee
              Fortunately for the MSSM, the tree-level mass formulae derived above receive significant 
              1-loop corrections, particularly in the case of the ${\rm h}^0$, whose mass is shifted 
              upwards by a substantial amount \cite{okada} \cite{barbieri} \cite{haber} \cite{ellis2}. 
              However, $m_{{\rm h}^0}$ is still minimized for $\tan \beta \approx 1$. The 
              quantitative mass shift depends on additional MSSM parameters entering in the loops, 
              but if these are tuned so as to maximize $m_{{\rm h}^0}$ for 
              each value of $m_{{\rm A}^0}$ and $\tan \beta$ \cite{CHWW}, 
              the experimental lower bound (\ref{eq:higgsexp}) on $m_{\rm H}$ (assuming it to be so) can in 
              principle be used to obtain exclusion limits on $\tan \beta$. This depends rather 
              sensitively on the top quark mass. A recent summary \cite{deg} 
              which includes leading two-loop 
              effects and takes the average top squark squared mass to be $(2 {\rm Tev})^2$, concludes 
              that in the `$m^{\rm max}_{{\rm h}^0}$' scenario \cite{CHWW}, with $m_{\rm t}=179.4$ GeV, there is no
              constraint on $\tan \beta$, and $m_{{\rm h}^0} \leq 140$ GeV (with an accuracy of 
              a few GeV).  
               This is still an extremely interesting result. In the words of Drees \cite{drees}:
              ``If experiments....fail to find at least one Higgs boson [in this energy region], the 
              MSSM can be completely excluded, independent of the value of its 100 or so free parameters.''

            \subsection{Tree-level Couplings of the ${\rm h}^0$, ${\rm H}^0$ and ${\rm A}^0$ Bosons.}

            The phenomenolgy of the Higgs-sector particles depends, of course, not only on their 
            masses but also on their couplings, which enter into production and decay processes. 
            In this section we shall derive some of the more important couplings, for illustrative 
            purposes. 
            
            First, note that after transforming to the mass-diagonal basis, the relation 
            (\ref{eq:muvy}) and similar ones for $m_{{\rm d}ij}$ and $m_{{\rm e}ij}$ become 
            \bea
            m_{{\rm u}, {\rm c}, {\rm t}} &=& v_{\rm u} y_{{\rm u}, {\rm c}, {\rm t}} \label{eq:muct}\\
            m_{{\rm d}, {\rm s}, {\rm b}} &=& v_{\rm d} y_{{\rm d}, {\rm s}, {\rm b}} \label{eq:mdsb}\\ 
             m_{{\rm e}, {\mu}, {\tau}} &=& v_{\rm d} y_{{\rm e}, {\mu}, {\tau}}.\label{eq:memutau}
             \eea
             In this basis, the Yukawa couplings in the superpotential are therefore (making use of 
             (\ref{eq:MWsq})) 
             \bea
             y_{{\rm u},{\rm c}, {\rm t}} &=& \frac{m_{{\rm u},{\rm c},{\rm t}}}{v_{\rm u}} = 
             \frac{g m_{{\rm u}, {\rm c}, {\rm t}}}{{\sqrt{2}} m_{\rm W} \sin \beta } \label{eq:yuct}\\
               y_{{\rm d},{\rm s}, {\rm b}} &=& \frac{m_{{\rm d},{\rm s},{\rm b}}}{v_{\rm d}} = 
             \frac{g m_{{\rm d}, {\rm s}, {\rm b}}}{{\sqrt{2}} m_{\rm W} \cos \beta } \label{eq:ydsb} \\
               y_{{\rm e},{\mu}, {\tau}} &=& \frac{m_{{\rm e},{\mu},{\tau}}}{v_{\rm d}} = 
             \frac{g m_{{\rm e}, {\mu}, {\tau}}}{{\sqrt{2}} m_{\rm W} \cos \beta }.\label{eq:yemutau}
             \eea
             Relations (\ref{eq:yuct}) and (\ref{eq:ydsb}) suggest that very rough upper and lower 
             bounds may be placed on $\tan \beta$ by requiring that neither $y_{\rm t}$ nor $y_{\rm b}$ 
             is non-perturbatively large. For example, if $\tan \beta \geq 1$ then $y_{\rm t} \leq 1.4$, 
             and if $\tan \beta \leq 50$ then $y_{\rm b} \leq 1.25$. Some GUT models can unify the running 
             values of $y_{\rm t}, y_{\rm b}$ and $y_{\tau}$ at the unification scale; this requires 
             $ \tan \beta \approx m_{\rm t}/m_{\rm b} \simeq 40.$                    
            
            To find the couplings of the MSSM Higgs bosons to fermions, we return to the 
            Yukawa couplings (\ref{eq:umass}) (together with the analogous ones for 
            $y^{ij}_{\rm d}$ and $y^{ij}_{\rm e}$), and expand $H^0_{\rm u}$ and $H^0_{\rm d}$ 
            about their vacuum values. In order to get the result in terms of the physical 
            fields $h^0$, $H^0$, however, we need to know how the latter  are related to 
            ${\rm Re} H^0_{\rm u}$ and ${\rm Re} H^0_{\rm d}$ - that is, we require expressions 
            for the eigenmodes of the (mass)$^2$ matrix (\ref{eq:msqhH}) corresponding to the 
            eigenvalues $m^2_{{\rm h}^0}$ and $m^2_{{\rm H}^0}$ of (\ref{eq:mh0}) and (\ref{eq:mH0}). 
            We can write (\ref{eq:msqhH}) as 
            \be
            {\bf M}^{\rm sq}_{{\rm h}, {\rm H}}= \frac{1}{2} \left( 
            \ba{cc} A+Bc & -As \\ -As & A-Bc \ea \right) \label{eq:mhHABC}
            \ee
            where $A=(m^2_{{\rm A}^0} + m^2_{\rm Z}), \ B=(m^2_{{\rm A}^0} - m^2_{\rm Z}), \ 
            c= \cos 2 \beta, \ s=\sin 2 \beta$, and we have used (\ref{eq:massA0}). Expression 
            (\ref{eq:mhHABC}) is calculated in the basis $(\sqrt{2}({\rm Re} H^0_{\rm u} - v_{\rm u}), 
            \sqrt{2}({\rm Re} H^0_{\rm d} - v_{\rm d}))$. Let us denote the normalized eigenvectors 
            of (\ref{eq:mhHABC}) by  $u_{\rm h}$ and $u_{\rm H}$ where 
            \be 
            u_{\rm h} = \left( \ba{c} \cos \alpha \\ - \sin \alpha \ea \right), \ \ \ 
            u_{\rm H} = \left( \ba{c} \sin \alpha \\ \cos \alpha \ea \right),
            \ee
            with eigenvalues $m^2_{{\rm h}^0}$ and $m^2_{{\rm H}^0}$ respectively where 
            \bea 
            m^2_{{\rm h}^0} &=& \frac{1}{2}(A-C) \\
            m^2_{{\rm H}^0}&=& \frac{1}{2} (A+C),
            \eea
            with $C=[A^2-(A^2-B^2)c^2]^{1/2}$. The equation determining $u_{\rm h}$ is then 
            \be
            \left( \ba{cc} A+Bc & -As \\ -As & A-Bc \ea \right) \left( \ba{c} \cos \alpha \\
            - \sin \alpha \ea \right) = (A-C) \left( \ba{c} \cos \alpha \\ - \sin \alpha \ea 
            \right),
            \ee
            which leads to 
            \bea 
            (C+Bc) \cos \alpha &=& -A s \sin \alpha \label{eq:ev1}\\
            (-C+Bc) \sin \alpha &=& A s \cos \alpha. \label{eq:ev2}
            \eea   
             It is conventional to rewrite (\ref{eq:ev1}) and (\ref{eq:ev2}) more 
             conveniently, as follows. Multiplying (\ref{eq:ev1}) by $\sin \alpha$ and 
             (\ref{eq:ev2}) by $\cos \alpha$ and then subtracting the results, we obtain 
             \be 
             \sin 2 \alpha = - \frac{A s}{C} = - \frac{(m^2_{{\rm A}^0} + m^2_{\rm Z})}
             {(m^2_{{\rm H}^0} -m^2_{{\rm h}^0})} \sin 2 \beta. \label{eq:s2a}         
             \ee
             Again, multiplying (\ref{eq:ev1}) by $\cos \alpha$ and (\ref{eq:ev2}) by 
             $\sin \alpha$  and adding the results gives 
             \be 
             \cos 2 \alpha = - \frac{Bc}{C} = - \frac{(m^2_{{\rm A}^0} -m^2_{\rm Z})} 
             {(m^2_{{\rm H}^0} - m^2_{{\rm h}^0})} \cos 2 \beta. \label{eq:c2a}
             \ee
             Equations (\ref{eq:s2a}) and (\ref{eq:c2a}) serve to define the correct 
             quadrant for the mixing angle $\alpha$, namely $- \pi/2 \leq \alpha \leq 0.$ 
             Note that in the limit $m^2_{{\rm A}^0} \gg m^2_{\rm Z}$ we have 
             $\sin 2 \alpha \approx - \sin 2 \beta$ and $\cos 2 \alpha \approx - \cos 2 \beta$, 
             and so 
             \be 
             \alpha \approx \beta - \pi/2 \ \ {\mbox{for $m^2_{{\rm A}^0} \gg m^2_{\rm Z}$}}. 
             \label{eq:Alimit}
             \ee
             The physical states are defined by 
             \be 
             \left( \ba{c} h^0 \\H^0 \ea \right) = \sqrt{2} \left( \ba{cc} \cos \alpha & - \sin \alpha \\
             \sin \alpha & \cos \alpha \ea \right) \left( \ba{c} {\rm Re} H^0_{\rm u} -v_{\rm u} \\
             {\rm Re} H^0_{\rm d} - v_{\rm d} \ea \right), 
             \ee
             which we can write as 
             \bea 
             {\rm Re} H^0_{\rm u} &=& [v_{\rm u} + \frac{1}{\sqrt{2}}(\cos \alpha \, h^0 + 
             \sin \alpha \, H^0)] 
             \label{eq:physHu0} \\
             {\rm Re} h^0_{\rm d} &=& [ v_{\rm d} + \frac{1}{\sqrt{2}}(-\sin \alpha \, h^0 + \cos \alpha\,  H^0)].
             \label{eq:physHd0}
             \eea
             We also have, from (\ref{eq:physHZ}) and (\ref{eq:physA0}), 
             \bea 
             {\rm Im} H^0_{\rm u}= \frac{1}{\sqrt{2}}(\sin \beta \, H_{\rm Z} + \cos \beta \, A^0)
              \label{eq:physImu}\\
             {\rm Im} H^0_{\rm d}= \frac{1}{\sqrt{2}}(- \cos \beta \, H_{\rm Z} + \sin \beta \, A^0) 
             \label{eq:physImd}
             \eea
             where $H_{\rm Z}$ is the massless field `swallowed' by the ${\rm Z}^0$.
             
             We can now derive the couplings to fermions. For 
             example, the Yukawa coupling (\ref{eq:umass}) in the mass eigenstate basis, and for the 
             third generation, is 
             \be 
             -y_{\rm t}[\chi_{{\bar{\rm t}}{\rm L}} \cdot \chi_{{\rm t}{\rm L}} ({\rm Re} H^0_{\rm u} 
             + {\rm i}\, {\rm Im}\, H^0_{\rm u}) + \chi^\dagger_{{\rm t}{\rm L}} \cdot 
             \chi^\dagger_{{\bar{\rm t}}{\rm L}} ({\rm Re} H^0_{\rm u}-{\rm i} \,{\rm Im} \,H^0_{\rm u})].
             \label{eq:tyuk}
             \ee
             Substituting (\ref{eq:physHu0}) for ${\rm Re}H^0_{\rm u}$, the `$v_{\rm u}$' part 
             simply produces the Dirac mass $m_{\rm u}$ via (\ref{eq:mudirmass}), 
             while the remaining part gives
             \bea
             &&- \frac{m_{\rm t}}{\sqrt{2}v_{\rm u}} ( \chi_{{\bar{\rm t}}{\rm L}} \cdot 
             \chi_{{\rm t}{\rm L}} + \chi^\dagger_{{\rm t}{\rm L}} \cdot 
             \chi^\dagger_{{\bar{\rm t}}{\rm L}})(\cos \alpha \, h^0 + \sin \alpha \, H^0) \nonumber \\
             &&= - \left( \frac{g m_{\rm t}}{2 m_{\rm W}}\right) \, {\bar{t}}t \, \left( 
             \frac{\cos \alpha}{\sin \beta} \, h^0 + \frac{\sin \alpha}{\sin \beta} \, H^0
             \right) , \label{eq:thig}
             \eea
             where `${\bar{t}}t$' is the four-component Dirac bilinear. The corresponding expression 
             in the SM would be just 
             \be 
             - \left( \frac{g m_{\rm t}}{2 m_{\rm W}} \right) {\bar{t}} t \, H_{\rm SM},
             \ee
             where $H_{\rm SM}$ is the SM Higgs boson. Equation (\ref{eq:thig}) shows how the SM 
             coupling is modified in the MSSM. Simliarly, the coupling to the b quark  is 
             \be 
             - \left( \frac{g m_{\rm b}}{2 m_{\rm W}}\right) \, {\bar{b}}b \, \left( 
            - \frac{\sin \alpha}{\cos \beta} \, h^0 + \frac{\cos \alpha}{\cos \beta} \, H^0
             \right), \label{eq:bhig}
             \ee
             which is to be compared with the SM coupling 
             \be 
               - \left( \frac{g m_{\rm b}}{2 m_{\rm W}} \right) {\bar{b}} b \, H_{\rm SM}.
               \ee
               Finally the t-A$^0$ coupling is found by substituting (\ref{eq:physImu}) into 
               (\ref{eq:tyuk}), with the result 
               \bea
               && - {\rm i} \frac{m_{\rm t}}{v_{\rm u}}(\chi_{{\bar{\rm t}}{\rm L}} \cdot 
             \chi_{{\rm t}{\rm L}} - \chi^\dagger_{{\rm t}{\rm L}} \cdot 
             \chi^\dagger_{{\bar{\rm t}}{\rm L}}) \frac{1}{\sqrt{2}} \cos \beta \, A^0 \nonumber \\
             &&= {\rm i} \left(\frac{g m_{\rm t}}{2 m_{\rm W}}\right) \cot \beta \, 
             {\bar{t}} \gamma_5 t \, A^0 
             \label{eq:tA0}
             \eea 
             where we have used (\ref{eq:gam5cplng}); and similarly the b-A$^0$ coupling is 
             \be 
             {\rm i}  \left(\frac{g m_{\rm b}}{2 m_{\rm W}}\right) \tan \beta \, 
             {\bar{b}} \gamma_5 b \, A^0.\label{eq:bA0}
             \ee
             Incidentally, the $\gamma_5$ coupling in (\ref{eq:tA0}) and (\ref{eq:bA0}) 
             shows that the A$^0$ is a pseudoscalar boson ({\bf CP} = $-1$), while the couplings (\ref{eq:thig}) 
             and (\ref{eq:bhig}) show that h$^0$ and H$^0$ are scalars 
             ({\bf CP}=+1). The limit of large $m_{{\rm A}^0}$ 
             is interesting: in this case, $\alpha$ and $\beta$ are related by (\ref{eq:Alimit}), 
             which implies 
             \bea 
             \sin \alpha &\approx& - \cos \beta \\
             \cos \alpha &\approx& \sin \beta.\label{eq:cosalap}
             \eea
             It then follows from (\ref{eq:thig}) and (\ref{eq:bhig}) that in this limit the couplings 
             of h$^0$ become those of the SM Higgs, while the couplings of H$^0$ are the same as 
             those of the A$^0$. For small $m_{{\rm A}^0}$ and large $\tan \beta$ on the other hand, 
             the couplings can differ substantially from the SM couplings, b-states being relatively 
             enhanced and t-states being relatively suppressed. 
             
             The couplings of the Higgs bosons to the gauge bosons are determined by the SU(2)$_{\rm L}
             \times$U(1)$_y$ gauge invariance, that is by the terms (\ref{eq:higgsKE}) with $D_\mu$ 
             given by (\ref{eq:higgscovder}). The terms involving $W^1_\mu, W^2_\mu, {\rm Re} H^0_{\rm u}$ 
             and ${\rm Re} H^0_{\rm d}$ are easily found to be 
             \be 
             \frac{g^2}{4}(W^1_\mu W^{1 \mu} + W^2_\mu W^{2 \mu}) [({\rm Re} H^0_{\rm u})^2 
             +({\rm Re} H^0_{\rm d})^2]. \label{eq:W1W2}
             \ee
             Substituting (\ref{eq:physHu0}) and (\ref{eq:physHd0}), the $v_{\rm u}^2$ and $v_{\rm d}^2$ parts 
             generate the W-boson (mass)$^2$ term via (\ref{eq:MWsq}), while the W-W-(h$^0$,H$^0$)
              couplings are
             \bea
             &&\frac{g^2}{4}(W^1_\mu W^{1 \mu} + W^2_\mu W^{2 \mu}) \sqrt{2} [v_{\rm u} ( 
             \cos \alpha \, h^0 + \sin \alpha \, H^0) + v_{\rm d}(- \sin \alpha \, h^0 +
              \cos \alpha \, H^0)] \nonumber \\
              &&= \frac{g m_{\rm W}}{2}(W^1_\mu W^{1 \mu} + W^2_\mu W^{2 \mu}) [ \sin (\beta - \alpha) 
              \, h^0 + \cos (\beta - \alpha) \, H^0]. \label{eq:higgsW1}
              \eea
              Similarly, the Z-Z-(h$^0$,H$^0$) couplings are 
              \be
              \frac{g m_{\rm Z}}{2 \cos \theta_{\rm W}} Z_\mu Z^\mu [\sin(\beta - \alpha) \, h^0 + 
              \cos (\beta-\alpha) \, H^0]. \label{eq:higgsZ}
              \ee
              Again, these are the same as the couplings of the SM Higgs to W and Z, but modified 
              by a factor $\sin (\beta - \alpha)$ for the h$^0$, and a factor $\cos (\beta-\alpha)$ for 
              the H$^0$.\footnote{This is essential for the viability of 
              Drees's suggestion \cite{drees2}: the excess of events near 98 GeV amounts to about 10\% 
              of the signal for a SM Higgs with that mass, and hence interpreting it as the h$^0$ 
              requires that $\sin^2(\beta - \alpha) \approx 0.1$. It then follows that ZH$^0$ 
              production at LEP would occur with nearly SM strength, if allowed kinematically. 
              Hence the identification of the excess at around 115 GeV with the H$^0$.} Once again, 
              there is a simple large $m_{{\rm A}^0}^2$ limit: 
              \be 
              \sin(\beta - \alpha) \approx 1, \ \ \ \ \cos (\beta - \alpha) \approx 0,
              \ee
              showing that in this limit h$^0$ has SM couplings to gauge bosons, while the 
              H$^0$ decouples from them entirely. At tree level, the A$^0$ has no coupling to pairs of 
              gauge bosons.
              
              The total widths of the MSSM Higgs bosons depend sensitively on $\tan \beta$. 
              The $h^0$ decays mainly to fermion-antifermion pairs, with a width 
              generally  comparable to that of the SM Higgs, while the 
              H$^0$ and A$^0$ are generally narrower than an SM Higgs of the same mass. The 
              production rate at the LHC also depends on $\tan \beta$. The dominant production 
              mechanism, as in the SM, is expected to be gluon fusion, proceeding via quark 
              (or squark) loops. In the SM case, the top quark loop dominates; in the MSSM, 
              if $\tan \beta$ is large and $m_{{\rm A}^0}$ not too large, the ${\bar{\rm b}}{\rm b}
              {\rm h}$ coupling is relatively enhanced, as noted after equation (\ref{eq:cosalap}), 
              and the bottom quark loop becomes important.  
              Searches for MSSM Higgs bosons are reviewed by Igo-Kemenes in \cite{RPP}.

                 \section{SUSY Particles in the MSSM}
                 
                 In this final section, we shall give a very brief introduction to the physics of 
                 the various SUSY particle states in  the MSSM. As in the scalar Higgs sector, 
                 the discussion is complicated by mixing phenomena. In particular, 
                 after SU(2)$_{\rm L}\times$U(1)$_y$ breaking, mixing will in general occur 
                 between any two (or more) fields which have the same colour, charge and spin.

                \subsection{Neutralinos}
                
                We consider first the sector consisting of the neutral higgsinos ${\tilde{H}}^0_{\rm u}$ 
                and ${\tilde{H}}^0_{\rm d}$, and the neutral gauginos ${\tilde{B}}$ 
                (bino)  and ${\tilde{W}}^0$ (wino)  (see Tables 1 and 2). These are all L-type 
                spinor fields in our presentation (but they can equivalently be represented as 
                Majorana fields, as discussed in section 2.3). In the absence of electroweak 
                symmetry breaking, the ${\tilde{B}}$ and ${\tilde{W}}^0$ fields would have masses 
                given by just the soft SUSY-breaking mass terms of (\ref{eq:ginomass}):
                \be
                - \frac{1}{2} M_1 {\tilde{B}} \cdot {\tilde{B}} - \frac{1}{2} M_2 {\tilde{W}}^0 \cdot 
                {\tilde{W}}^0 + {\rm h.c.}
                \ee
                However, bilinear combinations of one of $({\tilde{B}}, {\tilde{W}}^0)$ with 
                one of $({\tilde{H}}^0_{\rm u}, {\tilde{H}}^0_{\rm d})$ are generated by the term 
                `$- \sqrt{2}g [......]$' in (\ref{eq:Lgch}), when the neutral scalar Higgs fields 
                acquire a vev. Such bilinear terms will, as in the Higgs sector, appear as 
                non-zero off-diagonal entries in the $4 \times 4$ 
                mass matrix for the four  fields ${\tilde{B}}, {\tilde{W}}^0, {\tilde{H}}^0_{\rm u}$, 
                and ${\tilde{H}}^0_{\rm d}$  - that is, 
                they will cause mixing. After the mass matrix is diagonalized, the resulting 
                four neutral mass eigenstates are called neutralinos, usually denoted by 
                ${\tilde{\chi}}^0_i$ ($i=1,2,3,4$), with the convention that the masses are 
                ordered as $m_{{\tilde{\chi}}^0_1} < m_{{\tilde{\chi}}^0_2} < m_{{\tilde{\chi}}^0_3} < 
                m_{{\tilde{\chi}}^0_4}$.

                Consider for example the SU(2) contribution in (\ref{eq:Lgch}) from the $H_{\rm u}$ 
                supermultiplet, with $\alpha=3, T^3 \equiv \tau^3/2, \lambda^3 \equiv {\tilde{W}}^0$, 
                which is 
                \be - \sqrt{2} g (H^{+ \dagger}_{\rm u} \, H^{0 \dagger}_{\rm u}) \frac{\tau^3}{2} 
                \left( \ba{c} {\tilde{H}}^+_{\rm u} \\ {\tilde{H}}^0_{\rm u} \ea \right) 
                \cdot {\tilde{W}}^0 + {\rm h.c.} \label{eq:shigWmix}
                \ee
                When the field $H^{0 \dagger}_{\rm u}$ acquires a vev $v_{\rm u}$ (which we have 
                already chosen to be real), expression (\ref{eq:shigWmix}) contains the piece 
                \be 
                + \frac{g}{\sqrt{2}} v_{\rm u} {\tilde{H}}^0_{\rm u} \cdot {\tilde{W}}^0 + {\rm h.c.},
                \ee
                which we shall re-write as 
                \be 
                - \frac{1}{2} [ - \sin \beta \sin \theta_{\rm W} m_{\rm Z}]({\tilde{H}}^0_{\rm u} \cdot 
                {\tilde{W}}^0 + {\tilde{W}}^0 \cdot {\tilde{H}}^0_{\rm u}) + {\rm h.c.},
                \ee
                using (\ref{eq:tanbeta}) and (\ref{eq:MZsq}), and the result of the first Exercise in
                 Notational Aside (1). In a gauge-eigenstate basis 
                 \be 
                 {\tilde{G}}^0=\left( \ba{c} {\tilde{B}} \\ {\tilde{W}}^0\\ {\tilde{H}}^0_{\rm d} \\
                 {\tilde{H}}^0_{\rm u} \ea \right),
                 \ee
                  this will contribute a mixing between the (2,4) and (4,2) components. Similarly, 
                 the U(1) contribution from the $H_{\rm u}$ supermultiplet, after electroweak 
                 symmetry breaking, leads to the mixing term 
                 \bea 
                 && - \frac{g'}{\sqrt{2}}v_{\rm u} {\tilde{H}}^0_{\rm u} \cdot {\tilde{B}} + {\rm h.c.} \\
                 && = - \frac{1}{2} [ \sin \beta \sin \theta_{\rm W} m_{\rm Z}] 
                 ({\tilde{H}}^0_{\rm u} \cdot {\tilde{B}} + {\tilde{B}} \cdot {\tilde{H}}^0_{\rm u} ) 
                 + {\rm h.c.},
                 \eea
                 which involves the (1,4) and (4,1) components. The SU(2) and U(1) contributions of the 
                 $H_{\rm d}$ supermultiplet to such bilinear terms can be evaluated similarly.  
                 
                 In addition to this mixing caused by electroweak symmetry breaking, mixing between 
                 ${\tilde{H}}^0_{\rm u}$ and ${\tilde{H}}^0_{\rm d}$ is induced by the 
                 SUSY-invariant `$\mu$ term' in (\ref{eq:mutermF}), namely 
                 \be 
                 - \frac{1}{2} (- \mu) ({\tilde{H}}^0_{\rm u} \cdot {\tilde{H}}^0_{\rm d} 
                 + {\tilde{H}}^0_{\rm d} \cdot {\tilde{H}}^0_{\rm u}) + {\rm h.c.}
                 \ee 
                  Putting all this together, mass terms involving the fields in $G^0$ can be written as 
                  \be 
                  - \frac{1}{2} {\tilde{G}}^{0 {\rm T}} {\bf M}_{{\tilde{G}}^0} {\tilde{G}}^0 + {\rm h.c.} 
                  \ee
                  where 
                  \be 
                  {\bf M}_{{\tilde{G}}^0}=\left( \ba{cccc} M_1&0&-c_{\beta} s_{\rm W} 
                  m_{\rm Z} & s_{\beta} s_{\rm W} 
                  m_{\rm Z} \\ 0&M_2&c_{\beta} c_{\rm W} m_{\rm Z} &-s_{\beta} c_{\rm W} m_{\rm Z} \\
                  -c_{\beta} s_{\rm W} m_{\rm Z} & c_{\beta} c_{\rm W} m_{\rm Z} & 0 & -\mu \\
                  s_{\beta} s_{\rm W} m_{\rm Z} & -s_{\beta} c_{\rm W} m_{\rm Z} & - \mu & 0 \ea 
                  \right), \label{eq:neutralinoM}
                  \ee
                  with $c_{\beta}\equiv \cos \beta, s_{\beta} \equiv \sin \beta, c_{\rm W} \equiv \cos 
                  \theta_{\rm W}$, and $s_{\rm W} \equiv \sin \theta_{\rm W}$.

               In general, the parameters $M_1$, $M_2$ and $\mu$ can have arbitrary phases. Most 
               analyses, however, assume the `gaugino unification' condition (\ref{eq:gauguni1}) which 
               implies (\ref{eq:M1M2}) at the electroweak scale, so that one of $M_1$ 
                and $M_2$ is fixed in terms of the other. A redefinition of the phases of 
                ${\tilde{B}}$ and ${\tilde{W}}^0$ then allows us to make both $M_1$ and $M_2$ real 
                and positive. The entries proportional to $m_{\rm Z}$ are real by virtue of the 
                phase choices made for the Higgs fields in section 16.1, which made $v_{\rm u}$ and 
                $v_{\rm d}$ both real. It is usual to take $\mu$ to be real (so as to avoid 
                unacceptably large {\bf CP}-violating effects), but the sign of $\mu$ is unknown - and 
                not fixed by Higgs-sector physics 
                (see the sentence following equation (\ref{eq:tanbeta})). 
                The neutralino sector is then determined by 
                three real parameters, $M_1$, $\tan \beta$ and $\mu$ (as well as by 
                $m_{\rm Z}$ and $ \theta_{\rm W}$, of course).

                Clearly there is not a lot to be gained by pursuing the algebra of this $4 \times 4 $ 
                mixing problem, in general. A simple special case is that in which the 
                $m_{\rm Z}$-dependent terms in (\ref{eq:neutralinoM})     
                are a relatively small perturbation on the other entries, which would imply 
                that the neutralinos ${\tilde{\chi}}^0_1$ and ${\tilde{\chi}}^0_2$ are close to the 
                weak eigenstates bino and wino respectively, with masses 
                approximately equal to $M_1$ and $M_2$,  
                while the higgsinos are mixed by the 
                $\mu$ entries to form (approximately) the combinations 
                \be 
                {\tilde{H}}^0_{\rm S}= \frac{1}{\sqrt{2}}({\tilde{H}}^0_{\rm d}+{\tilde{H}}^0_{\rm u}), 
                \ \ \ {\mbox{and}} \ \ \ \ {\tilde{H}}^0_{\rm A}= \frac{1}{\sqrt{2}}({\tilde{H}}^0_{\rm d}-
                {\tilde{H}}^0_{\rm u}),
                \ee
                each having mass $\sim |\mu|$. 
                
                Assuming it is the LSP, the lightest neutralino, ${\tilde{\chi}}^0_1$,  is an attractive 
                candidate for non-baryonic dark matter \cite{EHDNOS}.\footnote{Other possibilities exist. 
                For example, in gauge-mediated SUSY breaking, the gravitino 
                is naturally the LSP. For this and other dark matter candidates within a softly-broken SUSY framework,  
                see \cite{CEKKLW} section 6.} Taking account of the 
                 restricted range of $\Omega_{\rm CDM} h^2$ consistent with  the WMAP data, calculations 
                 show \cite{EOSS} \cite{Baetal} \cite{Boetal} that ${\tilde{\chi}}^0_1$'s provide the 
                 desired thermal relic density in certain quite well-defined regions in the  
                 space of the mSUGRA parameters ($m_{1/2}, m_0, \tan \beta$ and the sign of $\mu$; 
                 $A_0$ was set to zero). Dark matter is reviewed      
                 by Drees and Gerbier in \cite{RPP}.
                
                \subsection{Charginos} 
                
                The charged analogues of neutralinos are called `charginos': there are two positively 
                charged ones associated (before mixing) with (${\tilde{W}}^+, {\tilde{H}}^+_{\rm u}$), 
                and two negatively charged ones associated with (${\tilde{W}}^-, {\tilde{H}}^-_{\rm d}$). 
                Mixing between ${\tilde{H}}^+_{\rm u}$ and ${\tilde{H}}^-_{\rm d}$ occurs via the 
                $\mu$ term in (\ref{eq:mutermF}). Also, as in the neutralino case, mixing between the 
                charged gauginos and higgsinos will occur via the `$-\sqrt{2}g [....]$' term in 
                (\ref{eq:Lgch}) after electroweak symmetry breaking. Consider for example the $H_{\rm u}$ 
                supermultiplet terms in (\ref{eq:Lgch}) involving ${\tilde{W}}^1$ and ${\tilde{W}}^2$, 
                after the scalar Higgs $H^0_{\rm u}$ has acquired a vev $v_{\rm u}$. These terms are 
                \bea
                && - \frac{g}{\sqrt{2}} \{ (0 \, v_{\rm u}) [ \tau^1 \left( \ba{c} {\tilde{H}}^+_{\rm u} \\
                {\tilde{H}}^0_{\rm u} \ea \right) \cdot {\tilde{W}}^1 + \tau^2 \left( \ba{c} 
                {\tilde{H}}^+_{\rm u} \\ {\tilde{H}}^0_{\rm u} \ea \right) 
                \cdot {\tilde{W}}^2]\} + {\rm h.c.} \\
                &&=- \frac{g}{\sqrt{2}}v_{\rm u} {\tilde{H}}^+_{\rm u}
                 \cdot ({\tilde{W}}^1 - {\rm i} {\tilde{W}}^2)  
                + {\rm h.c.} \\
                &&\equiv - g v_{\rm u} {\tilde{H}}^+_{\rm u} \cdot {\tilde{W}}^- + {\rm h.c.} \\
                &&=- \frac{1}{2} \sqrt{2} s_{\beta} m_{\rm W} ({\tilde{H}}^+_{\rm u} \cdot 
                {\tilde{W}}^- + {\tilde{W}}^- \cdot {\tilde{H}}^+_{\rm u}) + {\rm h.c.} 
                \eea
                The corresponding terms from the $H_{\rm d}$ supermultiplet are 
                \bea 
                && -g v_{\rm d} {\tilde{H}}^-_{\rm d} \cdot {\tilde{W}}^+ + {\rm h.c.} \\
                && = - \frac{1}{2} \sqrt{2} c_\beta m_{\rm W} ({\tilde{H}}^-_{\rm d} \cdot {\tilde{W}}^+ 
                + {\tilde{W}}^+ \cdot {\tilde{H}}^-_{\rm d}) + {\rm h.c.}
                \eea 
                If we define a gauge-eigenstate basis 
                \be 
                {\tilde{g}}^+=\left( \ba{c} {\tilde{W}}^+ \\ {\tilde{H}}^+_{\rm u} \ea \right)
                \ee
                for the positively charged states, and similarly 
                \be 
                {\tilde{g}}^-=\left( \ba{c} {\tilde{W}}^- \\ {\tilde{H}}^-_{\rm d} \ea \right) 
                \ee
                for the negatively charged states, then the chargino mass terms can be written as 
                \be 
                -\frac{1}{2}[{\tilde{g}}^{+{\rm T}} {\bf X}^{\rm T} \cdot {\tilde{g}}^- 
                + {\tilde{g}}^{-{\rm T}} {\bf X} \cdot {\tilde{g}}^+] + {\rm h.c.}, 
                \label{eq:chinomass}
                \ee
                where 
                \be 
                {\bf X} = \left( \ba{cc} M_2 & \sqrt{2} s_\beta m_{\rm W} \\
                \sqrt{2} c_\beta m_{\rm W} & \mu \ea \right).
                \ee

                 Since ${\bf X}^{\rm T} \neq {\bf X}$ (unless $\tan \beta =1$),
                  two distinct $2 \times 2$ matrices are 
                 needed for the diagonalization. Let us define the mass-eigenstate bases by 
                 \bea
                 {\tilde{\chi}}^+&=&{\bf V}{\tilde{g}}^+, \ \ \ {\tilde{\chi}}^+=\left( \ba{c} 
                 {\tilde{\chi}}^+_1 \\{\tilde{\chi}}^+_2 \ea \right) \\
                  {\tilde{\chi}}^-&=&{\bf U}{\tilde{g}}^-, \ \ \ {\tilde{\chi}}^-=\left( \ba{c} 
                 {\tilde{\chi}}^-_1 \\{\tilde{\chi}}^-_2 \ea \right),
                 \eea
                 where {\bf U} and {\bf V} are unitary. Then the second term in (\ref{eq:chinomass}) 
                 becomes 
                 \be 
                 -\frac{1}{2}{\tilde{\chi}}^{- {\rm T}} {\bf U}^* {\bf X} {\bf V}^{-1} \cdot {\tilde{
                 \chi}}^+,
                 \ee
                 and we require 
                 \be 
                 {\bf U}^* {\bf X} {\bf V}^{-1} = \left( \ba{cc} m_{{\tilde{\chi}}^{\pm}_1} & 0 \\
                 0& m_{{\tilde{\chi}}^{\pm}_2} \ea \right).\label{eq:chinomassdiag}
                 \ee
                 What about the first term in (\ref{eq:chinomass})? It becomes 
                 \be 
                 -\frac{1}{2} {\tilde{\chi}}^{+{\rm T}} {\bf V}^* {\bf X}^{\rm T} {\bf U}^\dagger 
                 \cdot {\tilde{\chi}}^-. \label{eq:chinomass1}
                 \ee
                 But since ${\bf V}^* {\bf X}^{\rm T} {\bf U}^\dagger = 
                 ({\bf U}^* {\bf X} {\bf V}^{-1})^{\rm T}$ it follows that the expression (\ref{eq:chinomass1}) 
                 is also diagonal, with the same eigenvalues $m_{{\tilde{\chi}}^{\pm}_1}$ and $m_{{\tilde{\chi}}
                 ^{\pm}_2}$.
                 
                 Now note that the Hermitian conjugate of (\ref{eq:chinomassdiag}) gives 
                 \be 
                 {\bf V} {\bf X}^\dagger {\bf U}^{\rm T} = 
                 \left( \ba{cc} m^*_{{\tilde{\chi}}^{\pm}_1} & 0 \\ 0& m^*_{{\tilde{\chi}}^{\pm}_2} 
                 \ea \right).
                 \ee
                 Hence 
                 \be 
                 {\bf V} {\bf X}^\dagger {\bf X}{\bf V}^{-1} = 
                 {\bf V} {\bf X}^\dagger {\bf U}^{\rm T} {\bf U}^* 
                 {\bf X} {\bf V}^{-1} = \left( \ba{cc} |m_{{\tilde{\chi}}^{\pm}_1}|^2 &0 \\
                 0& |m_{{\tilde{\chi}}^{\pm}_2}|^2 \ea \right),
                 \ee
                 and we see that the positively charged states ${\tilde{\chi}}^+$ diagonalize ${\bf X}^\dagger
                 {\bf X}$. Similarly, 
                 \be 
                 {\bf U}^* {\bf X} {\bf X}^\dagger {\bf U}^{\rm T} = {\bf U}^* {\bf X} {\bf V}^{-1} 
                 {\bf V} {\bf X}^\dagger {\bf U}^{\rm T} = \left( \ba{cc} 
                 |m_{{\tilde{\chi}}^{\pm}_1}|^2 &0 \\
                 0& |m_{{\tilde{\chi}}^{\pm}_2}|^2 \ea \right),
                 \ee
                 and the negatively charged states ${\tilde{\chi}}^-$ diagonalize ${\bf X}{\bf X}^\dagger$. 
                 The eigenvalues of ${\bf X}^\dagger {\bf X}$ (or ${\bf X} {\bf X}^\dagger$) are 
                 easily found to be 
                 \be 
                 \left( \ba{c} |m_{{\tilde{\chi}}^{\pm}_1}|^2 \\|m_{{\tilde{\chi}}^{\pm}_2}|^2 \ea 
                 \right) = \frac{1}{2} [ (M_2^2 +|\mu|^2 +2m^2_{\rm W}) \mp \{(M_2^2 +|\mu|^2 +2m^2_{\rm W})^2 - 
                 4|\mu M_2 - m^2_{\rm W} \sin 2 \beta|^2\}^{1/2} ].\label{eq:chinoev}
                 \ee
                 It may be worth noting that, because {\bf X} is diagonalized by the operation 
                 ${\bf U}^* {\bf X} {\bf V}^{-1}$, rather than by ${\bf V} {\bf X} {\bf V}^{-1}$ or 
                 ${\bf U}^* {\bf X} {\bf U}^{\rm T}$, these eigenvalues are not the squares of 
                 the eigenvalues of {\bf X}.
                 
                 The expression (\ref{eq:chinoev}) is not particularly enlightening, but as in the 
                 neutralino case it simplifies greatly if $m_{\rm W}$ can be regarded as a perturbation. 
                 Taking $M_2$ and $\mu$ to be real, the eigenvalues are then given approximately by 
                 $m_{{\tilde{\chi}}^{\pm}_1} \approx M_2$, and $m_{{\tilde{\chi}}^{\pm}_2} \approx |\mu|$ 
                 (the labelling assumes $M_2 < | \mu |$). In this limit, we have the approximate 
                 degeneracies $m_{{\tilde{\chi}}^{\pm}_1} \approx m_{{\tilde{\chi}}^0_2}$, and 
                 $m_{{\tilde{\chi}}^{\pm}_2} \approx m_{{\tilde{H}}^0_{\rm S}} 
                 \approx m_{{\tilde{H}}^0_{\rm A}}$. In general, the physics is sensitive to the 
                 ratio $M_2/| \mu |.$
                 
                 As an illustration of possible signatures for neutralino and chargino production 
                 (at hadron colliders, for example), we mention the {\em trilepton signal} \cite{AN} \cite{BT} 
                 \cite{BKT} \cite{KLMW} \cite{BCKT} \cite{MKKW}, which arises from the production  
                 \be 
                 {\rm p} {\bar{\rm p}} \ \ ({\rm or}\ \  
                 {\rm p} {\rm p}) \to {\tilde{\chi}}^{\pm}_1 {\tilde{\chi}}^0_2 + X 
                 \ee
                 followed by the decays 
                 \bea 
                 {\tilde{\chi}}^{\pm}_1 &\to& l^{\prime \pm} \nu {\tilde{\chi}}^0_1 \\
                 {\tilde{\chi}}^0_2 & \to & l {\bar{l}} {\tilde{\chi}}^0_1.\label{eq:NLSP}
                 \eea
                 Here the two LSPs in the final state carry away $2m_{{\tilde{\chi}}^0_1}$ of 
                 missing energy, which is observed as missing transverse energy, 
                 ${\not{\!\!E}}_{\rm T}$  (see section 14). 
                 In addition, there are three energetic, isolated leptons, and little jet activity. 
                 The expected SM background is small. Using the data sample collected from the 
                 1992-3 run of the Fermilab Tevatron, D0 \cite{D06} and CDF \cite{CDF6} reported 
                 no candidate trilepton events after applying all selection criteria; the 
                 expected background was roughly 2 $\pm$ 1 events. Upper limits on the 
                 product of the cross section times the branching ratio (single tripleton mode) 
                 were set, for various regions in the space of MSSM parameters. Later 
                 searches using the data sample from the 1994-5 run \cite{D04} \cite{CDF3} 
                 were similarly negative.

                 \subsection{Gluinos}
                 
                 Since the gluino ${\tilde{g}}$ is a colour octet fermion, it cannot mix with 
                 any other MSSM particle, even if $R$-parity is violated. So we get a (unique) break 
                 from mixing phenomena. We have already seen (section 15.3) that most models assume 
                 that the gluino mass is significantly greater than that of the neutralinos and 
                 charginos. 
                 A useful signature for gluino pair  (${\tilde{g}}{\tilde{g}}$) production  
                 is the {\em like-sign dilepton signal} 
                 \cite{BKP} \cite{BGH} \cite{BTW}. This arises if the gluino decays with a 
                 significant branching ratio to hadrons plus a chargino, which then decays to 
                 lepton + $\nu$ + ${\tilde{\chi}}^0_1$. Since the gluino is indifferent to electric 
                 charge, the single lepton from each ${\tilde{g}}$ decay will carry either charge with 
                 equal probability. Hence many events should contain two like-sign leptons (plus jets 
                 plus ${\not{\!\!E}}_{\rm T}$). This has a low SM background, because in the SM isolated 
                 lepton pairs come from ${\rm W}^+ {\rm W}^-$, Drell-Yan or ${\rm t}{\bar{\rm t}}$ 
                 production, all of which give opposite sign dileptons. Like-sign dilepton events can 
                 also arise from ${\tilde{g}}{\tilde{q}}$ and ${\tilde{q}}{\bar{\tilde{q}}}$ production.  
                 
                 CDF \cite{CDF4} reported no candidate events for like-sign dilepton pairs. Other 
                 searches based simply on dileptons (not required to be like-sign) plus two jets 
                 plus ${\not{\!\!{E}}}_{\rm T}$ \cite{CDF5} \cite{D05} reported no sign of any 
                 excess events. Results were expressed in terms of exclusion contours for mSUGRA parameters.     
                 
                 \subsection{Squarks and Sleptons}
                 
                 The scalar partners of the SM fermions form the largest collection of new 
                 particles in the MSSM. Since separate partners are required for each 
                 chirality state of the massive fermions, there are altogether 21 new fields 
                 (the neutrinos are treated as massless here): four squark flavours and 
                 chiralities ${\tilde{u}}_{\rm L}, {\tilde{u}}_{\rm R}, {\tilde{d}}_{\rm L}, 
                 {\tilde{d}}_{\rm R}$ and three slepton flavours and chiralities ${\tilde{\nu}}_{{\rm e}
                 {\rm L}}, {\tilde{e}}_{\rm L}, {\tilde{e}}_{\rm R}$ in the first family, all repeated 
                 for the other two families.\footnote{In the more general family-index notation 
                 of section 15.2 (see equations (\ref{eq:sqmass}), (\ref{eq:slepmass}) 
                 and (\ref{eq:triplesc})), `${\tilde{Q}}_1$' 
                 is the doublet $({\tilde{u}}_{\rm L}, {\tilde{d}}_{\rm L})$, `${\tilde{Q}}_2$' is 
                 $({\tilde{c}}_{\rm L}, {\tilde{s}}_{\rm L})$, `${\tilde{Q}}_3$' is 
                 $({\tilde{t}}_{\rm L}, {\tilde{b}}_{\rm L})$, `${\tilde{\bar{u}}}_1$' is 
                 ${\tilde{u}}_{\rm R}$, `${\tilde{\bar{d}}}_1$' is ${\tilde{d}}_{\rm R}$ (and 
                 similarly for `${\tilde{\bar{u}}}_{2,3}$' and `${\tilde{\bar{d}}}_{2,3}$'), while 
                 `${\tilde{L}}_1$' is $({\tilde{\nu}}_{{\rm e} {\rm L}}, {\tilde{e}}_{\rm L})$, 
                 `${\tilde{\bar{e}}}_1$' is $e_{\rm R}$, etc.} These are all (complex) scalar fields, 
                 and so the `L' and `R' labels do not, of course, here signify chirality, but are 
                 just labels showing which SM fermion they are partnered with (and hence in 
                 particular what their SU(2)$\times$U(1) quantum numbers are - see Table 1). 
                 
                 In principle, any scalars with the same electric charge, $R$-parity and 
                 colour quantum numbers can mix with each other, across families, via the soft SUSY-breaking  
                 parameters in (\ref{eq:sqmass}), (\ref{eq:slepmass}) and (\ref{eq:triplesc}). 
                 This would lead to a $6 \times 6$ mixing problem for the $u$-type squark fields  
                 $({\tilde{u}}_{\rm L}, {\tilde{u}}_{\rm R}, {\tilde{c}}_{\rm L}, {\tilde{c}}_{\rm R}, 
                 {\tilde{t}}_{\rm L}, {\tilde{t}}_{\rm R})$, and for the $d$-type squarks and the charged 
                 sleptons,  and a $3 \times 3$ one for the sneutrinos. However, as we saw in section 
                 15.2, phenomenological constraints imply that interfamily mixing among the SUSY states 
                 must be very small. As before, therefore, we shall adopt the `mSUGRA' form of 
                 the soft parameters as given in equations (\ref{eq:sqslun}) and (\ref{eq:yukun}), 
                 which guarantees the suppression of unwanted interfamily mixing terms 
                 (though one must  remember that other, and more general,  parametrizations 
                 are not excluded). As in the cases considered previously in this section, we shall 
                 also have to include various effects due to electroweak symmetry breaking. 
                 
                 Consider first the soft SUSY-breaking 
                 (mass)$^2$ parameters of the sfermions (squarks and sleptons) of the 
                 first family. In the model of (\ref{eq:sqslun}) they are all degenerate at the 
                 high (Planck?) scale. The RGE evolution down to the electroweak scale is 
                 governed by equations of the same type as (\ref{eq:mQrun}) and (\ref{eq:mubarrun}), but 
                 without the $X_{\rm t}$ terms: the corresponding terms for the first two families 
                 may be neglected because of their much  smaller Yukawa couplings. Thus the soft masses 
                 of the first and second families evolve by purely gauge interactions, which (see 
                 the comment following equation (\ref{eq:mhdrun})) tend to increase the masses at low scales. 
                 Their evolution can be parametrized (following \cite{martin} equations (7.65) - (7.69)) by 
                 \bea
                 m^2_{{\tilde{\rm u}}_{\rm L},{\tilde{\rm d}}_{\rm L}}=
                 m^2_{{\tilde{\rm c}}_{\rm L},{\tilde{\rm s}}_{\rm L}}&=&
                 m^2_0+K_3+K_2+\frac{1}{9}K_1 
                 \label{eq:msf1}\\
                 m^2_{{\tilde{\rm u}}_{\rm R}}=m^2_{{\tilde{\rm c}}_{\rm R}}&=&
                  m^2_0+K_3  \ \ \ \ \ \ \ \ +\frac{16}{9}K_1 \label{eq:msf2} \\
                 m^2_{{\tilde{\rm d}}_{\rm R}}=m^2_{{\tilde{\rm s}}_{\rm R}}&=&
                  m^2_0 +K_3  \ \ \ \ \ \ \ \ + \frac{4}{9}K_1 \label{eq:msf3}\\
                 m^2_{{\tilde{\nu}}_{{\rm e} {\rm L}}, {\tilde{\rm e}}_{\rm L}}=
                 m^2_{{\tilde{\nu}}_{\mu {\rm L}}, {\tilde{\mu}}_{\rm L}}&=& m^2_0  \ \ \ \ \ \ \ \    
                 +K_2 + K_1 \label{eq:msf4}\\
                 m^2_{{\tilde{\rm e}}_{\rm R}}= m^2_{{\tilde{ \mu}}_{\rm R}}&=&
                 m^2_0  \ \ \ \ \ \ \ \ \ \ \ \ \ \ \    + 4K_1.\label{eq:msf5} 
                 \eea     
                 Here $K_3, K_2$ and $K_1$ are the RGE contributions from SU(3), SU(2) and U(1) 
                 gauginos respectively: all the chiral supermultiplets couple to the gauginos 
                 with the same (`universal') gauge couplings. The different numerical coefficients 
                 in front of the $K_1$ terms are the squares of the $y$-values of each field (see Table 
                 1), which enter into the relevant loops. All the $K$'s are positive, and are roughly 
                 of the same order of magnitude as the gaugino (mass)$^2$ parameter $m^2_{1/2}$, but with 
                 $K_3$ significantly greater than $K_2$, which in turn is greater than $K_1$ (this is 
                 because of the relative sizes of the different gauge couplings at the weak scale: 
                 $g_3^2 \sim 1.5, g^2_2 \sim 0.4, g^2_1 \sim 0.2$, see section 13). The large `$K_3$' 
                 contribution is likely to be quite model-independent, and it is therefore reasonable to 
                 expect that squark (mass)$^2$ values will be greater than slepton ones. 
                 
                 Equations (\ref{eq:msf1}) - (\ref{eq:msf5}) give the soft (mass)$^2$ parameters for the  
                 fourteen states involved, in  the first two families (we defer consideration of the third 
                 family for the moment). In addition to these contributions, however, there are further 
                 terms to be included which arise as a result of electroweak symmetry breaking. For the 
                 first two families, the most important such contributions are those coming from SUSY-invariant 
                 $D$-terms (see (\ref{eq:VDF})) of the form (squark)$^2$(higgs)$^2$ and 
                 (slepton)$^2$(higgs)$^2$, after the scalar Higgs fields $H^0_{\rm u}$ and $H^0_{\rm d}$ 
                 have acquired vevs. Returning to equation (\ref{eq:Deqnmot}), the SU(2) contribution to 
                 $D^\alpha$ is 
                 \bea
                 &&D^\alpha=g\{ ({\tilde{u}}^\dagger_{\rm L} {\tilde{d}}^\dagger_{\rm L}) 
                 \frac{\tau^\alpha}{2} \left( \ba{c} {\tilde{u}}_{\rm L}\\{\tilde{d}}_{\rm L} \ea \right) 
                 + ({\tilde{\nu}}^\dagger_{{\rm e}{\rm L}} {\tilde{e}}^\dagger_{\rm L}) 
                 \frac{\tau^\alpha}{2}\left( \ba{c} {\tilde{\nu}}_{{\rm e}{\rm L}} \\ {\tilde{e}}_{\rm L} 
                 \ea \right)  \nonumber \\
                 && +(H^{+ \dagger}_{\rm u} H^{0 \dagger}_{\rm u} ) \frac{\tau^\alpha}{2}\left( 
                 \ba{c} H^+_{\rm u} \\H^0_{\rm u} \ea \right) + (H^{0 \dagger}_{\rm d} H^{-\dagger}_{\rm d}) 
                 \frac{\tau^\alpha}{2} \left( \ba{c} H^0_{\rm d} \\ H^-_{\rm d} \ea \right) \} 
                 \eea 
                 \be  
                  \to g \{ ({\tilde{u}}^\dagger_{\rm L} {\tilde{d}}^\dagger_{\rm L}) 
                 \frac{\tau^\alpha}{2} \left( \ba{c} {\tilde{u}}_{\rm L}\\{\tilde{d}}_{\rm L} \ea \right) 
                 + ({\tilde{\nu}}^\dagger_{{\rm e}{\rm L}} {\tilde{e}}^\dagger_{\rm L}) 
                 \frac{\tau^\alpha}{2}\left( \ba{c} {\tilde{\nu}}_{{\rm e}{\rm L}} \\ {\tilde{e}}_{\rm L} 
                 \ea \right) 
                  - \frac{1}{2}v_{\rm u}^2 \delta_{\alpha 3} + \frac{1}{2}v^2_{\rm d} 
                 \delta_{\alpha 3} \},
                 \ee 
                 after symmetry breaking. When this is inserted into the Lagrangian term $-\frac{1}{2}
                 D^\alpha D^\alpha$, pieces which are quadratic in the scalar fields - and are therefore 
                 (mass)$^2$ terms - will come from cross terms between the `$\tau^\alpha/2$' and 
                 `$\delta_{\alpha 3}$' terms. These cross terms are proportional to $\tau^3/2$, and 
                 therefore split apart the $T^3=+1/2$ weak isospin components from the $T^3=-1/2$ 
                 components, but they are diagonal in the weak eigenstate basis. Their contribution 
                 to the $({\tilde{u}}_{\rm L}, {\tilde{d}}_{\rm L})$ (mass)$^2$ matrix is 
                 therefore
                 \be 
                 +\frac{1}{2} g^2 \, 2 \, \frac{1}{2} (v^2_{\rm d} - v^2_{\rm u}) T^3 \label{eq:msfT3}
                 \ee
                 where $T^3=\tau^3/2$. Similarly, the U(1) contribution to `$D$' is 
                 \be
                 D_y = g'\{\sum_{{\tilde{f}}} \frac{1}{2} {\tilde{f}}^\dagger y_{{\tilde{f}}}
                  {\tilde{f}} - \frac{1}{2} 
                 (v^2_{\rm d}-v^2_{\rm u})\} \label{eq:Dy} 
                 \ee
                 after symmetry breaking, where the sum is over all sfermions (squarks and sleptons).  
                 Expression (\ref{eq:Dy}) leads to the sfermion (mass)$^2$ term 
                 \be 
                 + \frac{1}{2}
                  g^{\prime 2} 2 \, (- \frac{1}{2} y) \frac{1}{2} (v^2_{\rm d}-v^2_{\rm u}).\label{eq:msfD}
                 \ee
                 Since $y/2=Q-T^3$, where $Q$ is the electromagnetic charge, we can combine (\ref{eq:msfT3}) 
                 and (\ref{eq:msfD}) to give a total (mass)$^2$ contribution for each sfermion: 
                 \bea
                 \Delta_{{\tilde{f}}}&=& \frac{1}{2}(v_{\rm d}^2 - v_{\rm u}^2) [ (g^2+g^{\prime 2}) T^3 - 
                 g^{\prime 2} Q]  \nonumber \\
                 &=& m^2_{\rm Z} \cos 2 \beta [T^3-\sin ^2 \theta_{\rm W} Q],\label{eq:Deltamsf}
                 \eea
                 using (\ref{eq:MZsq}). As remarked earlier, $\Delta_{{\tilde{f}}}$ is diagonal in the 
                 weak eigenstate basis, and the appropriate contributions simply have to be added to 
                 the RHS of equations (\ref{eq:msf1}) - (\ref{eq:msf5}). It is interesting to note that the 
                 splitting between the doublet states is predicted to be 
                 \be 
                 -m^2_{{\tilde{\rm u}}_{\rm L}} + m^2_{{\tilde{\rm d}}_{\rm L}}= 
                 -m^2_{{\tilde{\nu}}_{{\rm e}{\rm L}}} 
                 + m^2_{{\tilde{\rm e}}_{\rm L}}= \cos 2 \beta m^2_{\rm W},
                 \ee
                 and similarly for the second family. On the assumption that $\tan \beta$ is most 
                 probably greater than 1 (see the comments following equation (\ref{eq:yemutau})), the 
                 `down' states are heavier. 
                 
                 Sfermion (mass)$^2$ terms are also generated by  SUSY-invariant 
                 $F$-terms, after symmetry breaking - that 
                 is, terms in the Lagrangian of the form 
                 \be 
                 - \left| \frac{\partial W}{\partial \phi_i} \right|^2
                 \ee
                 for every scalar field $\phi_i$ (see equations (\ref{eq:Wi}) and (\ref{eq:WZ2})); for these 
                 purposes we regard $W$ of (\ref{eq:WMSSM}) as being written in terms of the scalar fields, as 
                 in section 8. Remembering that the Yukawa couplings are proportional to the associated 
                 fermion masses (see (\ref{eq:muvy}) and (\ref{eq:muct}) - (\ref{eq:memutau})),
                  we see that on the scale 
                 expected for the masses of the sfermions, only terms involving the Yukawas of the third 
                 family can contribute significantly. Thus to a very good approximation we can write 
                 \bea
                  W &\approx& y_{\rm t} {\tilde{t}}^\dagger_{\rm R}({\tilde{t}}_{\rm L} H^0_{\rm u} 
                 -{\tilde{b}}_{\rm L} H^+_{\rm u}) - y_{\rm b} {\tilde{b}}^\dagger_{\rm R} ({\tilde{t}}_{\rm L} 
                 H^-_{\rm d} -{\tilde{b}}_{\rm L} H^0_{\rm d}) - y_{\tau} {\tilde{\tau}}^\dagger_{\rm R} 
                 ({\tilde{\nu}}_{\tau {\rm L}} H^-_{\rm d} -{\tilde{\tau}}_{\rm L}H^0_{\rm d}) \nonumber \\
                 &+& \mu(H^+_{\rm u}H^-{\rm d}-H^0_{\rm u}H^0_{\rm d})
                 \eea
                 as in (\ref{eq:Wapprox}). Then we have, for example, 
                 \be
                 - \left| \frac{\partial W}{\partial {\tilde{t}}^\dagger_{\rm L}} \right|^2 
                 = - y^2_{\rm t} {\tilde{t}}^\dagger_{\rm L} {\tilde{t}}_{\rm L} |H^0_{\rm u}|^2 
                 \to - y^2_{\rm t} v^2_{\rm u} {\tilde{t}}^\dagger_{\rm L} {\tilde{t}}_{\rm L} = 
                 - m^2_{\rm t} {\tilde{t}}^\dagger_{\rm L} {\tilde{t}}_{\rm L},\label{eq:mLstopsq}
                 \ee
                 after $H^0_{\rm u}$ acquires the vev $v_{\rm u}$. The L-type top squark (`stop') therefore 
                 gets a (mass)$^2$ term equal to the top quark (mass)$^2$. There will be an identical term 
                 for the R-type stop squark, coming from $-|\partial W / \partial {\tilde{t}}_{\rm L}|^2$. 
                 Similarly, there will be (mass)$^2$ terms $m^2_{\rm b}$ for ${\tilde{b}}_{\rm L}$ and 
                 ${\tilde{b}}_{\rm R}$, and $m^2_\tau$ for ${\tilde{\tau}}_{\rm L}$ and 
                 ${\tilde{\tau}}_{\rm R}$, though these are probably negligible in this context. 
                 
                 We also need to consider derivatives of $W$ with respect to the Higgs fields. For example, 
                 we have 
                 \be 
                 - \left| \frac{\partial W}{\partial H^0_{\rm u} } \right|^2 = -|y_{\rm t} 
                 {\tilde{t}}^\dagger_{\rm R} {\tilde{t}}_{\rm L} - \mu H^0_{\rm d}|^2 \to -|y_{\rm t} 
                 {\tilde{t}}^\dagger_{\rm R} {\tilde{t}}_{\rm L} - \mu v_{\rm d} |^2 \label{eq:muytmix} 
                 \ee
                 after symmetry breaking. The expression (\ref{eq:muytmix}) 
                   contains the off-diagonal bilinear term 
                 \be 
                 \mu v_{\rm d} y_{\rm t} ({\tilde{t}}^\dagger_{\rm R}{\tilde{t}}_{\rm L} + 
                 {\tilde{t}}^\dagger_{\rm L} {\tilde{t}}_{\rm R}) = \mu m_{\rm t} \tan \beta 
                 ({\tilde{t}}^\dagger_{\rm R} {\tilde{t}}_{\rm L} + {\tilde{t}}_{\rm L}^\dagger 
                 {\tilde{t}}_{\rm R})
                 \ee
                 which mixes the R and L fields. Similarly, $-|\partial W/\partial H^0_{\rm d} |^2$ 
                 contains the mixing terms 
                 \be 
                 \mu m_{\rm b} \tan \beta ({\tilde{b}}^\dagger_{\rm R} {\tilde{b}}_{\rm L} + 
                 {\tilde{b}}^\dagger_{\rm L} {\tilde{b}}_{\rm R})
                 \ee
                 and 
                 \be 
                 \mu m_{\tau} \tan \beta ({\tilde{\tau}}^\dagger_{\rm R} {\tilde{\tau}}_{\rm L} + 
                 {\tilde{\tau}}^\dagger_{\rm L} {\tilde{\tau}}_{\rm R}). 
                 \ee

                 Finally, bilinear terms can also arise directly from the soft triple scalar couplings 
                 (\ref{eq:triplesc}), after the scalar Higgs fields acquire vevs. Assuming the conditions 
                 (\ref{eq:yukun}), and retaining only the third family contribution as before, the relevant 
                 terms from (\ref{eq:triplesc}) are 
                 \be 
                 -A_0 y_{\rm t} v_{\rm u}({\tilde{t}}^\dagger_{\rm R} {\tilde{t}}_{\rm L} + 
                 {\tilde{t}}^\dagger_{\rm L} {\tilde{t}}_{\rm R}) = -A_0 m_{\rm t}({\tilde{t}}^\dagger_{\rm R} 
                 {\tilde{t}}_{\rm L} + {\tilde{t}}^\dagger_{\rm L} {\tilde{t}}_{\rm R}),
                 \ee
                 together with similar ${\tilde{b}}_{\rm L} - {\tilde{b}}_{\rm L}$ and 
                 ${\tilde{\tau}}_{\rm R} - {\tilde{\tau}}_{\rm L}$ mixing terms. 
                 
                 Putting all this together, then, the (mass)$^2$ values for the squarks and sleptons 
                 of the first two families are given by the expressions (\ref{eq:msf1}) - (\ref{eq:msf5}), 
                 together with the relevant contribution $\Delta_{{\tilde{f}}}$ of (\ref{eq:Deltamsf}). For the 
                 third family, we discuss the ${\tilde{\rm t}}$, ${\tilde{\rm b}}$ and ${\tilde{\tau}}$ sectors 
                 separately. The (mass)$^2$ term for the top squarks is   
                 \be 
                 -({\tilde{ t}}^\dagger_{\rm L} {\tilde{t}}^\dagger_{\rm R}) \, 
                 {\bf M}^2_{{\tilde{\rm t}}} \left( 
                 \ba{c} {\tilde{t}}_{\rm L} \\{\tilde{t}}_{\rm R} \ea \right), 
                 \ee
                 where 
                 \be 
                 {\bf M}^2_{{\tilde{\rm t}}} = \left( \ba{cc} m^2_{{\tilde{\rm t}}_{\rm L}, {\tilde{\rm b}}_{\rm L}}  
                 + m^2_{\rm t} + \Delta_{{\tilde{\rm u}}_{\rm L}} & m_{\rm t}(A_0-\mu \cot \beta) \\
                 m_{\rm t} (A_0-\mu \cot \beta) & m^2_{{\tilde{\rm t}}_{\rm R}} + m^2_{\rm t} + 
                 \Delta_{{\tilde{\rm u}}_{\rm R}} \ea \right),\label{eq:stopmat}
                 \ee
                 with 
                 \be 
                 \Delta_{{\tilde{\rm u}}_{\rm L}}= (\frac{1}{2}-\frac{2}{3} \sin ^2 \theta_{\rm W}) m^2_{\rm Z} 
                 \cos 2 \beta
                 \ee
                 and 
                 \be 
                 \Delta_{{\tilde{\rm u}}_{\rm R}}= -\frac{2}{3} \sin^2 \theta_{\rm W} m^2_{\rm Z} \cos 2 \beta.
                 \ee
                 Here $m^2_{{\tilde{\rm t}}_{\rm L}, {\tilde{\rm b}}_{\rm L}}$ and $m^2_{{\tilde{\rm t}}_{\rm R}}$ 
                 are given approximately by (\ref{eq:mQrun}) and (\ref{eq:mubarrun}) respectively. 
                 In contrast to the 
                 corresponding equations for the first two families, the $X_{\rm t}$ term is now present, and will 
                 tend to reduce the running masses of ${\tilde{\rm t}}_{\rm L}$ and ${\tilde{\rm t}}_{\rm R}$ 
                  at low scales (the second more than the first),  
                   relative to those of the corresponding states in the  
                 first two families;  on the other hand, the $m^2_{\rm t}$ term tends to work in the other 
                 direction.

                  The real symmetric matrix ${\bf M}^2_{{\tilde{\rm t}}}$ can be diagonalized by the 
                 orthogonal transformation 
                 \be 
                 \left( \ba{c} {\tilde{t}}_1 \\{\tilde{t}}_2 \ea \right) = \left( \ba{cc} 
                 \cos \theta_{{\tilde{\rm t}}} & \sin \theta_{{\tilde{\rm t}}} \\
                 - \sin \theta_{{\tilde{\rm t}}} & \cos \theta_{{\tilde{\rm t}}} \ea \right) \left( \ba{c} 
                 {\tilde{t}}_{\rm L} \\ {\tilde{t}}_{\rm R} \ea \right);
                 \ee
                 the eigenvalues are denoted by $m^2_{{\tilde{\rm t}}_1}$ and $m^2_{{\tilde{\rm t}}_2}$, with 
                 $m^2_{{\tilde{\rm t}}_1} < m^2_{{\tilde{\rm t}}_2}$. Because of the large value of $m_{\rm t}$ in the 
                 off-diagonal positions in (\ref{eq:stopmat}), mixing effects in the stop sector are likely to 
                 be substantial, and will probably result in the mass of the lighter stop, 
                 $m_{{\tilde{\rm t}}_1}$, 
                 being significantly smaller than the mass of any other squark. Of course, the mixing effect 
                 must not become too large, or else $m^2_{{\tilde{\rm t}}_1}$ is driven to negative 
                 values, which would imply (as in the electroweak Higgs case) a spontaneous 
                 breaking of colour symmetry. This requirement places a bound on the magnitude of the 
                 unknown parameter $A_0$, which cannot be much greater than $m_{{\tilde{\rm u}}_{\rm L}, 
                 {\tilde{\rm d}}_{\rm L}}$. 
                 
                 At ${\rm e}^+ {\rm e}^-$ colliders the ${\tilde{\rm t}}_1$ production cross section 
                 depends on the mixing angle $\theta_{\tilde{\rm t}}$; for example, the contribution 
                 from Z exchange actually vanishes when $\cos^2 \theta_{{\tilde{\rm t}}} = \frac{4}{3} 
                 \sin^2 \theta_{\rm W}$ \cite{dreeshik}. In contrast, ${\tilde{\rm t}}_1$'s are pair-produced 
                 in hadron colliders with no mixing-angle dependence. Which decay modes of the 
                 ${\tilde{\rm t}}_1$ dominate depends on the masses of charginos and sleptons. For example, 
                 if $m_{{\tilde{\rm t}}_1}$ lies below all chargino and slepton masses, then the dominant 
                 decay is 
                 \be 
                 {\tilde{\rm t}}_1 \to {\rm c} + {\tilde{\chi}}^0_1,
                 \ee
                 which proceeds through loops (a FCNC transition). 
                 If $m_{{\tilde{\rm t}}_1} > m_{{\tilde{\chi}}^\pm}$, 
                 \be 
                 {\tilde{\rm t}}_1 \to {\rm b} + {\tilde{\chi}}^{\pm}
                 \ee
                 is the main mode, with ${\tilde{\chi}}^\pm$ then decaying to 
                 $l \nu {\tilde{\chi}}^0_1$. D0 reported on a search for 
                 such light stops \cite{D01}; their signal was two acollinear jets plus 
                 ${\not{\!\!{E}}}_{\rm T}$ (they did not attempt to identify flavour). Improved 
                 bounds on the mass of the lighter stop 
                 were obtained by CDF \cite{CDF1} using a vertex detector to tag c- and b-quark 
                 jets. More recent searches are reported in \cite{D02} and \cite{CDF2}. The bounds 
                 depend sensitively on the (assumed) mass of the neutralino ${\tilde{\chi}}^0_1$; 
                 data is presented in the form of excluded regions in a $m_{{\tilde{\chi}}^0_1} - 
                 m_{{\tilde{\rm t}}_1}$ plot.

                 Turning now to the ${\tilde{b}}$ sector, the (mass)$^2$ matrix  is 
                 \be         
                   {\bf M}^2_{{\tilde{\rm b}}} = \left( \ba{cc} m^2_{{\tilde{\rm t}}_{\rm L},
                   {\tilde{\rm b}}_{\rm L}} 
                 + m^2_{\rm b} + \Delta_{{\tilde{\rm d}}_{\rm L}} & m_{\rm b}(A_0-\mu \tan \beta) \\
                 m_{\rm b} (A_0-\mu \tan \beta) & m^2_{{\tilde{\rm b}}_{\rm R}} + m^2_{\rm b} + 
                 \Delta_{{\tilde{\rm d}}_{\rm R}} \ea \right),\label{eq:mass2btilde} 
                 \ee
                 with 
                 \be
                 \Delta_{{\tilde{\rm d}}_{\rm L}}=(-\frac{1}{2} + \frac{1}{3} \sin^2 \theta_{\rm W}) m^2_{\rm Z} 
                 \cos 2 \beta
                 \ee
                 and 
                 \be 
                 \Delta_{{\tilde{\rm d}}_{\rm R}}=\frac{1}{3}\sin^2 \theta_{\rm W} m^2_{\rm Z} \cos 2 \beta.
                 \ee
                 Here, since $X_{\rm t}$ enters into the evolution of the mass of ${\tilde{\rm b}}_{\rm L}$ 
                 but not of ${\tilde{\rm b}}_{\rm R}$, we expect that the running 
                 mass of ${\tilde{\rm b}}_{\rm R}$ will 
                 be much the same as those of ${\tilde{\rm d}}_{\rm R}$ and ${\tilde{\rm s}}_{\rm R}$, 
                 but that $m_{{\tilde{\rm b}}_{\rm L}}$ may be less than $m_{{\tilde{\rm d}}_{\rm L}}$ and 
                 $m_{{\tilde{\rm s}}_{\rm L}}$. Similarly, the (mass)$^2$ matrix in the ${\tilde{\tau}}$ sector is 
                  \be         
                   {\bf M}^2_{{\tilde{\tau}}} = \left( \ba{cc} m^2_{{\tilde{\nu}}_{{\tau}{\rm L}},
                   {\tilde{\tau}}_{\rm L}} 
                 + m^2_{\tau} + \Delta_{{\tilde{\rm e}}_{\rm L}} & m_{ \tau}(A_0-\mu \tan \beta) \\
                 m_{ \tau} (A_0-\mu \tan \beta) & m^2_{{\tilde{\tau}}_{\rm R}} + m^2_{\tau } + 
                 \Delta_{{\tilde{\rm e}}_{\rm R}} \ea \right),\label{eq:mass2tautilde} 
                 \ee
                 with 
                 \be
                 \Delta_{{\tilde{\rm e}}_{\rm L}}=(-\frac{1}{2} +  \sin^2 \theta_{\rm W}) m^2_{\rm Z} 
                 \cos 2 \beta
                 \ee
                 and 
                 \be 
                 \Delta_{{\tilde{\rm e}}_{\rm R}}=\frac{1}{3}\sin^2 \theta_{\rm W} m^2_{\rm Z} \cos 2 \beta.
                 \ee

                 Mixing effects in the ${\tilde{\rm b}}$ and ${\tilde{\tau}}$ sectors depend on how large 
                 $\tan \beta$ is (see the off-diagonal terms in (\ref{eq:mass2btilde}) and 
                 (\ref{eq:mass2tautilde})). It seems that for $\tan \beta$ less than about 5(?), mixing 
                 effects will not be large, so that the masses of ${\tilde{\rm b}}_{\rm R}, {\tilde{\tau}}_{\rm R}$ 
                 and ${\tilde{\tau}}_{\rm L}$ will all be approximately degenerate with the corresponding 
                 states in the first two families,  while ${\tilde{\rm b}}_{\rm L}$ will be lighter than 
                 ${\tilde{\rm d}}_{\rm L}$ and ${\tilde{\rm s}}_{\rm L}$. For larger values of 
                 $\tan \beta$, strong mixing may take place, as in the stop sector. In this case, 
                 ${\tilde{\rm b}}_1$ and $ {\tilde{\tau}}_1$ may be   
                  significantly lighter than their analogues in the first two families (also,  
                  ${\tilde{\nu}}_{\tau{\rm L}}$ may be lighter than ${\tilde{\nu}}_{{\rm e}{\rm L}}$ and 
                  ${\tilde{\nu}}_{\mu {\rm L}}$). Neutralinos and charginos will then decay predominantly to 
                  taus and staus, which is more challenging experimentally than (for example) the 
                  dilepton signal from (\ref{eq:NLSP}).

                 The search for a light ${\tilde{\rm b}}_1$ decaying to ${\rm b}+{\tilde{\chi}}^0_1$ is 
                 similar to that for ${\tilde{\rm t}}_1 \to {\rm c} + {\tilde{\chi}}^0_1$. 
                 D0  \cite{D03} tagged b-jets through semi-leptonic decays to muons. 
                 They observed 5 candidate events consistent with the final state ${\rm b} {\bar{\rm b}} + 
                 {\not{\!\!{E}}}_{\rm T}$, as compared to an estimated background of 6.0$\pm$ 1.3 events from 
                 ${\rm t} {\bar{\rm t}}$ and W and Z production; results were presented in the form of an
                 excluded  region in the $(m_{{\tilde{\chi}}^0_1}, m_{{\tilde{\rm b}}_1})$ plane. Improved 
                 bounds were obtained in the CDF experiment \cite{CDF1}. 
                 
                 Searches for SUSY particles are reviewed by Schmitt in \cite{RPP}, including in particular 
                 searches at LEP, which we have not discussed. In rough terms, the present status is that 
                 there is `little room for SUSY particles lighter than $m_{\rm Z}$.' With all LEP 
                 data analysed, and if there is still no signal from the Tevatron collaborations, 
                 it will be left to the LHC to provide definitive tests.
                 
                 We have given  here only a first orientation to the SUSY particle spectrum. Feynman rules 
                 for the interactions of these particles with each other and with the particles of the SM 
                 are given in \cite{HK}, \cite{CEKKLW} and \cite{GH}. Representative calculations of cross sections for 
                 sparticle production at hadron colliders may be found in \cite{DEQ}. Experimental 
                 methods for measuring superparticle masses and cross sections at the LHC are 
                 summarized in \cite{branson}.

                 \subsection{Benchmarks for SUSY Searches}
                 
                 Assuming degeneracy between the first two families of sfermions, 
                 there are 25 distinct masses for the 
                 undiscovered states of the MSSM: 7 squarks and sleptons in the first two families, 
                 7 in the third family, 4 Higgs states, 4 neutralinos, 2 charginos and 1 gluino. Many 
                 details of the phenomenology to be expected (production cross sections, decay branching 
                 ratios) will obviously depend on the precise ordering of these masses. These in turn 
                 depend, in the general MSSM, on a very large number (over 100) of parameters characterizing 
                 the soft SUSY-breaking terms, as noted in section 15.2. Any kind of representative 
                 sampling of such a vast parameter space is clearly out of the question. On the other 
                 hand, in order (for example) to use simulations to assess the prospects for 
                 detecting and measuring these new particles at different accelerators, some 
                 consistent model must be adopted \cite{hinch}. This is because,  very often, 
                  a promising SUSY signal in one channel, which has a small SM background, actually 
                  turns out to have a large background from other SUSY production and decay 
                  processes. Faced with this situation, it seems necessary to reduce drastically 
                  the size of the parameter space, by adopting one of the more restricted models for 
                  SUSY breaking, such as the mSUGRA one. Such models typically have only three or four 
                  parameters; for instance, in mSUGRA they are, 
                  as we have seen, $m_0, m_{1/2}, A_0, \tan \beta$, 
                  and the sign of $\mu$. 
                  
                  But even a sampling of a 3- or 4-dimensional parameter space, in order (say) to 
                  simulate experimental signatures within a detector, is beyond present capabilities. 
                  This is why such studies are performed  only for certain specific points in 
                  parameter space, or in some cases along certain lines. Such parameter sets are 
                  called `benchmark sets'. 
                  
                  Various choices of benchmark have been proposed. To a certain extent, which one is 
                  likely to be useful depends on what is being investigated. For example, the 
                  `$m^{\rm max}_{\rm h}$-scenario' \cite{CHWW} 
                  referred to in section 16.2 is suitable for setting 
                  conservative bounds on $\tan \beta$ and $m_{{\rm A}^0}$, on the basis of the non-observation 
                  of the lightest Higgs state. Another approach is to require that the benchmark points used 
                  for studying collider phenomenology should be compatible with various experimental 
                  constraints - for example \cite{BDEG} the LEP searches for SUSY particles and for the Higgs 
                  boson, the precisely measured value of the anomalous magnetic moment of the muon, the 
                  decay ${\rm b} \to {\rm s} \gamma$, and (on the assumption that ${\tilde{\chi}}^0_1$ is 
                  the LSP) the relic density $\Omega_{{\tilde{\chi}}^0_1}h^2$. The authors of \cite{BDEG} worked 
                  within the mSUGRA model, taking $A_0=0$ and considering 13 benchmark points (subject to 
                  these constraints) in the space of parameters ($m_0, m_{1/2}, \tan \beta, {\rm sign} \ \mu$). 
                  A more recent study \cite{BDEGOP} updates the analysis in the light of the more precise 
                  dark matter bounds provided by the WMAP data.

                  One possible drawback with this approach is that minor modifications to the SUSY-breaking 
                  model might significantly alter the cosmological bounds, or the rate for ${\rm b} \to {\rm s} 
                  \gamma$, while having little effect on the collider phenomenology; thus important regions 
                  of parameter space might be excluded prematurely. In any case, it is clearly desirable to 
                  formulate benchmarks for other possibilities for SUSY-breaking, in particular. The 
                  `Snowmass Points and Slopes' (SPS) \cite{ABBC} are a set of benchmark points and lines 
                  in parameter space, which include seven mSUGRA-type scenarios, 
                  two gauge-mediated symmetry-breaking 
                  scenarios (it should be noted that here the LSP is the gravitino), and one anomaly-mediated 
                  symmetry-breaking scenario. Another study \cite{KLMNWW} concentrates on models which imply 
                  that at least some superpartners are light enough to be detectable at the Tevatron (for 
                  2 fb$^{-1}$ integrated luminosity); such models are apparently common among effective field 
                  theories derived from the weakly coupled heterotic string.  
                  
                  The last two  references conveniently provide diagrams or Tables showing the   
                  SUSY particle spectrum (i.e. the 25 masses) for each of the benchmark points. They are, 
                  in fact, significantly different, and may themselves be regarded as the benchmarks, rather 
                  than the values of the high-scale parameters which led to them. If and when sparticles 
                  are discovered, their masses and other  properties may provide a window into the physics 
                  of SUSY-breaking. However, as emphasized in section 9 of \cite{CEKKLW}, there are 
                  in principle not enough observables at hadron colliders to determine all the 105 
                  parameters of the soft SUSY breaking Lagrangian; for this, data from future 
                  ${\rm e}^+ {\rm e}^-$ colliders will be required. Then again, the MSSM  
                  may not be nature's choice.

      \begin{flushleft}
      {\em{Endnote}}
      
      This is not a review. No serious attempt has been made to compile a representative list 
      of references. The 100 or so which follow have simply come to hand. This is an 
      order of magnitude less than the number of references  included in the review \cite{CEKKLW}, and two orders 
      less than the number of papers on supersymmetry/SUSY/MSSM indicated by SPIRES.
       \end{flushleft}

\end{document}